\documentclass[epj]{svjour}
\usepackage{graphicx,amsmath,amssymb,dcolumn}
\usepackage[hang,bf]{caption}
\usepackage{thumbpdf}
\usepackage[colorlinks=true, linkcolor=blue, citecolor=blue, urlcolor=blue,
pdfauthor={Philip Bechtle, Klaus Desch, Mathias Uhlenbrock, Peter Wienemann},
pdftitle={Constraining SUSY models with Fittino using measurements before, with and beyond the LHC},
pdfkeywords={SUSY LHC Fittino}]{hyperref}

\newcolumntype{d}[1]{D{.}{.}{#1}}

\begin{document}

\title{Constraining SUSY models with Fittino using measurements before, with and beyond the LHC}

\author{Philip Bechtle\inst{1} \and Klaus Desch\inst{2} \and Mathias
  Uhlenbrock\inst{2} \and Peter Wienemann\inst{2}}

\institute{Deutsches Elektronen-Synchrotron, Notkestr. 85, D-22607
  Hamburg, Germany
  \and
  Universit\"at Bonn, Physikalisches Institut, Nussallee~12, D-53115
  Bonn, Germany}

\date{Received: July 15, 2009}

\abstract{ We investigate the constraints on Supersymmetry (SUSY)
  arising from available precision measurements using a global fit
  approach. When interpreted within minimal supergravity (mSUGRA), the
  data provide significant constraints on the masses of supersymmetric
  particles (sparticles), which are predicted to be light enough for
  an early discovery at the Large Hadron Collider (LHC). We provide
  predicted mass spectra including, for the first time, full
  uncertainty bands. The most stringent constraint is from the
  measurement of the anomalous magnetic moment of the muon.  Using the
  results of these fits, we investigate to which precision mSUGRA and
  more general MSSM parameters can be measured by the LHC experiments
  with three different integrated luminosities for a parameter point
  which approximately lies in the region preferred by current data.
  The impact of the already available measurements on these
  precisions, when combined with LHC data, is also studied. We develop
  a method to treat ambiguities arising from different interpretations
  of the data within one model and provide a way to differentiate
  between values of different digital parameters of a model
  (e.~g.~sign$(\mu)$ within mSUGRA).  Finally, we show how
  measurements at a linear collider with up to 1~TeV centre-of-mass
  energy will help to improve precision by an order of magnitude.
  \PACS{ {11.30.Pb}{Supersymmetry} \and {12.60.Jv}{Supersymmetric
      models} \and {14.80.Ly}{Supersymmetric partners of known
      particles} } % end of PACS codes
} %end of abstract

\maketitle

%%%%%%%%%%%%%%%%%%%%%%%%%%%%%%%%%%%%%%%%%%%%%%%%%%%%%%%%%%%%%%%%%%%%%%%%%%%%%%%
%%%                             Introduction                                %%%
%%%%%%%%%%%%%%%%%%%%%%%%%%%%%%%%%%%%%%%%%%%%%%%%%%%%%%%%%%%%%%%%%%%%%%%%%%%%%%%
\section{Introduction}\label{sec:introduction}

The Large Hadron Collider (LHC) will be the first collider to directly
probe physics at the TeV energy scale, the Terascale.  The LHC is
supposed to provide first beam collisions in autumn 2009. Despite its
tremendous success, the Standard Model (SM) of particle physics
exhibits a number of shortcomings which -- according to the belief of
many -- might be remedied by new physics showing up at the Terascale.
One very popular extension of the SM is Supersymmetry
(SUSY)~\cite{Wess:1974tw}.  Among the virtues of SUSY are the
elimination of the hierarchy problem, it can provide natural
candidates to explain dark matter in the Universe and it allows for
the unification of the gauge couplings at the scale of grand
unification.  Since no supersymmetric particles (sparticles) have been
discovered to date, SUSY cannot be an exact symmetry of Nature at
experimentally accessible energies. Unfortunately, the mechanism of
SUSY breaking is unknown.  This ignorance is efficiently parametrised
in the Minimal Supersymmetric Standard Model
(MSSM)~\cite{Nilles:1983ge,Haber:1984rc} by the introduction of all
possible soft SUSY-breaking terms into the Lagrangian with minimal
sparticle content in a phenomenological way.  While the most general
MSSM Lagrangian introduces around 100 new parameters, mild assumptions
on the absence of flavour-non-diagonal and CP-violating terms
(motivated by the absence of strong flavour-changing neutral currents
and electric dipole moments of the electron and neutron) and on the
(effective) universality of the first two generations reduce the
number of parameters to 18 (MSSM18). Still, it is a formidable
experimental challenge to reconstruct 18 parameters simultaneously
from future measurements. An alternative but less rigorous approach is
to confront specific theoretical models of SUSY breaking (which
typically reduce the number of free parameters significantly) directly
with data. Among the most prominent of such models are minimal
Supergravity
(mSUGRA)~\cite{AlvarezGaume:1983gj,Ibanez:1982ee,Ellis:1982wr,Inoue:1982pi,Chamseddine:1982jx}
and Gauge Mediated SUSY Breaking
(GMSB)~\cite{AlvarezGaume:1981wy,Dine:1993yw,Dine:1994vc,Dine:1995ag}.

If new phenomena which are compatible with SUSY are discovered at the
LHC~--~which we assume in this work~--~one of the major challenges
will be to find out the underlying model and to measure its parameters
as precisely as possible. Several studies have already been performed
to investigate the precision with which SUSY model parameters can be
derived from measurements at the LHC and how much is gained by
combining them with data from the International Linear Collider (ILC)
(see
e.~g.~\cite{Weiglein:2004hn,Lafaye:2007vs,Bechtle:2005vt,Lester:2005je}).
So far these studies assume an accuracy for the used observables which
will only be attainable with a fairly large integrated luminosity.
Thus they reflect the situation in which we might be in several years
from now or~--~for studies including ILC measurements~--~even later.

In this paper we test the compatibility of various SUSY models with
presently available data and constrain the corresponding parameters.
Subsequently a projection of the present situation to the LHC era and
beyond is performed to obtain a possible time evolution of the
precision on SUSY parameters for mSUGRA and MSSM18. The
fits are performed using Fittino~\cite{Bechtle:2004pc} version 1.5.0.
The sparticle properties for a given set of Lagrangian parameters
are calculated using SPheno version 3.0beta~\cite{Porod:2003um} which
is interfaced with Fittino via the SUSY Les Houches
Accord~\cite{Skands:2003cj,Allanach:2007zz}.  Previous work into this
direction is found in
\cite{Lafaye:2007vs},\cite{Appelquist:1974tg}-\cite{Flacher:2008zq}.

In this paper, presently measured ``low energy'' (LE) observables are
subjected to a global fit of the mSUGRA and GMSB model based on Markov
Chain Monte Carlo techniques.  To accomplish this, we take advantage
of a recent compilation of up-to-date theoretical calculations of
precision observables within the MSSM~\cite{Buchmueller:2008qe}.
Also, for the first time, we combine future LHC measurements with LE
observables to determine their impact in particular in the early phase
of LHC data taking and within models with a large number of parameters
such as the MSSM18.

This paper is organised as follows: in Section~2 we define and discuss
the present and future measurements which serve as input to the global
fit.  We also describe briefly the computer codes employed to obtain
precise theoretical predictions as a function of the SUSY parameters
the data are confronted with. In Section~3, we outline in some detail
the different methods used to estimate the SUSY parameters from a
global $\chi^2$ variable.  The advantages and disadvantages of the two
main methods, Markov Chains and Toy Fits with Simulated Annealing, are
discussed. We describe an approach to discriminate between different
values for discrete parameters of the models and illustrate this
approach for the parameter sign$(\mu)$ of the mSUGRA model. Also, a
new method to deal with ambiguities arising from different
interpretations of the data within the same model is discussed.  In
Section~4, the results of the different fits are presented. In
Section~4.1, the constraints on mSUGRA and GMSB parameters are derived
from available measurements, including observables from $K$- and
$B$-decays, the anomalous magnetic moment of the muon $(g-2)_\mu$,
precision electro-weak data from colliders and the value of the relic
density of cold dark matter of the Universe, $\Omega_{\mathrm{CDM}}
h^2$.  We also determine the most sensitive observables, $(g-2)_\mu$
and $\Omega_{\mathrm{CDM}} h^2$ and show the effect of their exclusion
from the fit. For the best fit point, we calculate the corresponding
mass spectra of all sparticles. For the first time, the
uncertainties on the parameters are converted into error bands on the
sparticle masses. In Section~4.2, the results from fits to LHC
data with integrated luminosities of 1, 10, 300 fb$^{-1}$ are
displayed for an mSUGRA model point (SPS1a) which leads to a collider
phenomenology similar to that of the best fit point.  In Section~4.3,
we LE data with future LHC data. We also show that by this, a stable
fit of the MSSM18 can be achieved and the masses of most sparticles
can be predicted.  Finally, in Section~4.4 we investigate how
precision measurements of sparticles at a linear electron-positron
collider like the ILC with up to 1~TeV of centre-of-mass energy will
turn SUSY into precision physics.  The paper ends with conclusions in
Section~5.

%%%%%%%%%%%%%%%%%%%%%%%%%%%%%%%%%%%%%%%%%%%%%%%%%%%%%%%%%%%%%%%%%%%%%%%%%%%%%%%
%%%                             Used observables                            %%%
%%%%%%%%%%%%%%%%%%%%%%%%%%%%%%%%%%%%%%%%%%%%%%%%%%%%%%%%%%%%%%%%%%%%%%%%%%%%%%%
\section{Measurements and Predictions}\label{sec:observables}

In this section, we describe the present and future experimental data
which we confront with the SUSY parameter space. We use three different
sets of measurements in an incremental way. These three sets are
\begin{enumerate}
\item "Low energy" observables: existing experimental data which have
the potential to constrain the allowed SUSY parameter space;
\item Simulated LHC measurements: expected SUSY measurements for the parameter
set SPS1a at the LHC experiments ATLAS and CMS for three different integrated luminosities;
\item Simulated ILC measurements: expected SUSY measurements at the ILC running
at $\sqrt{s} = 500 $ GeV and $\sqrt{s} = 1000 $ GeV.
\end{enumerate}

These measurements are briefly discussed in the following sections.
Finally, the codes used for the theoretical
calculations are described in Section~\ref{sec:predictions}.

%%%%%%%%%%%%%%%%%%%%%%%%%%%%%%%%%%%%%%%%%%%%%%%%%%%%%%%%%%%%%%%%%%%%%%%%%%%%%%%
%%%                              LE observables                            %%%
%%%%%%%%%%%%%%%%%%%%%%%%%%%%%%%%%%%%%%%%%%%%%%%%%%%%%%%%%%%%%%%%%%%%%%%%%%%%%%%
\subsection{Low Energy Observables}\label{sec:leobservables}

While no direct evidence for SUSY particles has been found to date, 
these particles contribute to higher order corrections to measured
physical observables in a well-defined and calculable way if SUSY is
realised in Nature.

The measurements which are exploited to obtain constraints on the
allowed SUSY parameter space, can be grouped in four classes:

\begin{enumerate}
\item Rare decays of B- and K-mesons;
\item The anomalous magnetic moment of the muon;
\item Precision measurements and the Higgs boson mass limit from high energy colliders:
LEP, SLC, and Tevatron;
\item The relic density of cold dark matter in the Universe.
\end{enumerate}

For reasons of comparability, the same measured values have been
used for the fit as in~\cite{Buchmueller:2008qe} although some of them,
e.~g.~the mass of the top quark have been updated meanwhile. 
The exploited measurements and their values are summarised in
Table~\ref{tab:leobserables}. In the next sections these measurements are
briefly described and limitations on their interpretation in terms of
SUSY are discussed.

\begin{table*}
  \caption{Available measurements from $B$-factories, kaon
    experiments, LEP, SLC and the Tevatron, as well as the measurement
    of $(g-2)_{\mu}$ and the cold dark matter relic
    density. Correlations amongst the electro-weak precision
    observables as given in~\cite{:2005ema} are studied for the fit to
    the existing measurements.  No effect of the correlations on the
    allowed parameter regions is found.  
}\label{tab:leobserables}
  \begin{center}
    \begin{tabular}{l | c | c c | c}
      \hline\hline
      \multicolumn{1}{c|}{Observable} & \multicolumn{1}{c|}{Experimental} &
      \multicolumn{2}{c|}{Uncertainty} & Exp. Reference \\
      & \multicolumn{1}{c|}{Value} &
      \multicolumn{1}{c}{stat}  & \multicolumn{1}{c|}{syst} & \\
      \hline
      $ {\cal B}(B\to s\gamma)/{\cal B}(B\to s\gamma)_{\mathrm{SM}} $      &      1.117               &  0.076     &   0.096                    &  \cite{Barberio:2006bi} \\
      $ {\cal B}(B_s\to\mu\mu) $                                  & $<$ 4.7$\times10^{-8}$   &            &   &  \cite{Barberio:2006bi} \\
      $ {\cal B}(B_d\to\ell\ell)   $                              & $<$ 2.3$\times10^{-8}$  &            &       &  \cite{Barberio:2006bi} \\
      $ {\cal B}(B\to\tau\nu)/{\cal B}(B\to\tau\nu)_{\mathrm{SM}}  $       &      1.15                &  0.40      &                            &   \cite{Aubert:2004kz}  \\
      $ {\cal B}(B_s\to X_s\ell\ell)/{\cal B}(B_s\to X_s\ell\ell)_{\mathrm{SM}} $ &      0.99         &  0.32      &                            &  \cite{Barberio:2006bi} \\  
      $ \Delta m_{B_s}/\Delta m_{B_s}^{\mathrm{SM}}$                       &      1.11                &  0.01      &   0.32                     &  \cite{Bona:2007vi} \\
      $\frac{\Delta m_{B_s}/\Delta m_{B_s}^{\mathrm{SM}}}{\Delta m_{B_d}/\Delta m_{B_d}^{\mathrm{SM}}}$ & 1.09 &  0.01      &  0.16                      &  \cite{Barberio:2006bi,Bona:2007vi} \\
      $ \Delta \epsilon_K/\Delta \epsilon_K^{\mathrm{SM}} $                &      0.92                &  0.14      &                            &  \cite{Bona:2007vi} \\
      $ {\cal B}(K\to\mu\nu)/{\cal B}(K\to\mu\nu)_{\mathrm{SM}}$           &      1.008               &  0.014     &                            &  \cite{Antonelli:2008jg} \\ 
      $ {\cal B}(K\to\pi\nu\bar\nu)/{\cal B}(K\to\pi\nu\bar\nu)_{\mathrm{SM}} $ & $<$ 4.5            &   &                  &  \cite{Artamonov:2008qb} \\  
      $ a_{\mu}^{\mathrm{exp}}-a_{\mu}^{\mathrm{\mathrm{SM}}} $            &     30.2$\times10^{-10}$ &  8.8$\times10^{-10}$ & 2.0$\times10^{-10}$ &  \cite{Bennett:2004pv,Moroi:1995yh} \\
      $ \sin^2\theta_{\mathrm{eff}}$           &      0.2324              &  0.0012    &                                   &  \cite{:2005ema} \\
      $ \Gamma_Z    $                 &   2.4952 GeV             &  0.0023 GeV &            0.001 GeV             &  \cite{:2005ema} \\
      $ R_l       $                   &     20.767               &  0.025     &                                   &  \cite{:2005ema} \\
      $ R_b       $                   &      0.21629             &  0.00066   &                                   &  \cite{:2005ema} \\
      $ R_c       $                   &      0.1721              &  0.003     &                                   &  \cite{:2005ema} \\
      $ A_{\mathrm{fb}}(b)     $                 &      0.0992              &  0.0016    &                                   &  \cite{:2005ema} \\
      $ A_{\mathrm{fb}}(c)     $                 &      0.0707              &  0.0035    &                                   &  \cite{:2005ema} \\
      $ A_b       $                   &      0.923               &  0.020     &                                   &  \cite{:2005ema} \\
      $ A_c       $                   &      0.670               &  0.027     &                                   &  \cite{:2005ema} \\
      $ A_l       $                   &      0.1513              &  0.0021    &                                   &  \cite{:2005ema} \\
      $ A_{\tau}     $                &      0.1465              &  0.0032    &                                   &  \cite{:2005ema} \\
      $ A_{\mathrm{fb}}(l)     $                 &      0.01714             &  0.00095   &                                   &  \cite{:2005ema} \\
      $ \sigma_{\mathrm{had}}$                 &     41.540 nb            &  0.037 nb  &                                   &  \cite{:2005ema} \\
      $ m_h    $                      & $>$  114.4 GeV           &            &           3.0 GeV                 &  \cite{Barate:2003sz,Schael:2006cr,Degrassi:2002fi} \\
      $ \Omega_{\mathrm{CDM}} h^2  $                 &      0.1099              &  0.0062    &           0.012                   &  \cite{Dunkley:2008ie} \\
      $ 1/\alpha_{em}  $ & 127.925     &  0.016       &                                                    &  \cite{Amsler:2008zzb}  \\
      $ G_F            $ &   1.16637$\times10^{-5} \mathrm{GeV}^{-2} $   &  0.00001$\times10^{-5} \mathrm{GeV}^{-2} $  &                                &  \cite{Amsler:2008zzb}  \\
      $ \alpha_{s}     $ &   0.1176       &  0.0020          &                                                    &  \cite{Amsler:2008zzb}  \\
      $ m_Z            $ &  91.1875  GeV  &  0.0021     GeV  &                                                    &  \cite{:2005ema}  \\
      $ m_W     $      &     80.399 GeV       &  0.025 GeV   &        0.010 GeV                                   &  \cite{Amsler:2008zzb} \\
      $ m_b            $ &   4.20    GeV  &  0.17       GeV  &                                                    &  \cite{Amsler:2008zzb}  \\
      $ m_t    $         & 172.4     GeV  &  1.2        GeV  &                                                    &  \cite{:2008vn}  \\
      $ m_{\tau}        $ &   1.77684 GeV  &  0.00017    GeV  &                                                   &  \cite{Amsler:2008zzb}  \\
      $ m_c  $           &   1.27    GeV  &  0.11        GeV  &                                                    &  \cite{:2005ema}  \\
      
      \hline\hline
    \end{tabular}
  \end{center}
\end{table*}

\subsubsection{Rare Decays of B and K mesons}\label{sec:leobservables:flavour}

A strong constraint on new physics can be derived from flavour physics
experiments, especially at the $B$-factories. The reasons for that are
two-fold: First, the flavour structure of the SM is remarkably exactly
realised in Nature~\cite{Charles:2004jd}. The apparent absence of
CP-violation or flavour changing neutral currents beyond the SM
severely constrains models of new physics with additional flavour
mixing. In this paper, we only study flavour-diagonal SUSY models which
by construction fulfil these constraints.
Second, the exact knowledge of branching fractions of rare decays, which are helicity
suppressed or occur only at loop level with heavy particles in the
loop, strongly constrains also flavour-diagonal models of new physics.

While the observables used here can be precisely measured (within the
statistical limitations of the experiment), their prediction in the SM
or in SUSY is often accompanied with theoretical uncertainties. The assumed
systematical uncertainties on the theoretical predictions are listed in
Table~\ref{tab:leobserables}. They are added in quadrature to the
experimental uncertainties. Amongst the most important constraints are
the recent measurements of $B_s$ oscillations at the Tevatron, the
branching fraction ${\cal B}(B\to\tau\nu)$ and the inclusive branching
fraction of radiative penguin decays, $B\to s\gamma$ of the $B$ meson.

\subsubsection{Anomalous Magnetic Moment of the Muon}\label{sec:leobservables:gmin2mu}

Although the anomalous magnetic moment of the electron $(g-2)_e =2a_e$
is measured approximately a factor of 200 more precisely than the
anomalous magnetic moment of the muon $(g-2)_{\mu}=2a_{\mu}$, the
sensitivity to new physics of the anomalous magnetic moment of the
muon is typically enhanced by a factor of $(m_{\mu}/ m_e)^2 \approx
43\,000$, and represents a much stronger constraint. While its
measurement is undisputed, there is ongoing debate about the exact
value of the SM prediction for $(g-2)_{\mu}$. The reason is the fact
that the non-perturbative contribution from the hadronic vacuum polarisation
has to be extracted from other experiments such as low-energy $e^+e^-$
scattering at BES~\cite{Bai:2001ct} or from $\tau$ lepton
decays~\cite{Lautrup:1971jf,Davier:2003pw,Ghozzi:2003yn}. Due to these
uncertainties, the fit in Section~\ref{sec:leobservables} is performed with
and without using $(g-2)_{\mu}$ as an observable.

\subsubsection{Measurements from High Energy Colliders}\label{sec:leobservables:EWP}

The measurements of the $Z$ boson mass and width and of its couplings to left- and
right-handed fermions in production and decay, the hadronic
cross-section on the $Z$ pole, and the $W$ boson and top quark mass serve to 
constrain the properties of particles contributing at loop level.
Due to their high precision and due the absence of any ambiguity
in the interpretation of the measurement (as e.~g.~in the case of
the relic density of cold dark matter) these measurements represent an
important input to the fit.

As outlined in~\cite{:2005ema}, there are correlations within the LEP
and SLD asymmetry measurements in the heavy flavour sector,
respectively, and within the $Z$ pole observables. The effect of
these correlations on the SUSY fit results have been tested for the
baseline fit to the measurements from Table~\ref{tab:leobserables} as
outlined in Section~\ref{sec:results:LEonly:mSUGRA}.

In addition and for completeness, we also use the measurements of the
bottom and charm quark and tau lepton masses and the measurement of the
strong coupling constant $\alpha_s$ as input to the fit.

\subsubsection{Limit on the SM Higgs Boson Mass}\label{sec:leobservables:higgs}

The exclusion of a Higgs boson with SM-like properties below $m_h =
114.4\,\mathrm{GeV}$ at 95\% C.L. represents an important constraint
on SUSY since at leading order, the lightest CP-even Higgs boson has a
mass below $m_Z$. Only due to radiative corrections, its mass can be
raised up to at most approximately 135-140 GeV~\cite{Heinemeyer:1998np}.

For a fit of the general MSSM to the existing data, the SM limit on
the Higgs boson mass cannot be employed since for a given $m_h$ the
gauge and Yukawa couplings of the lightest CP-even SUSY Higgs boson
can deviate significantly from their SM values. Furthermore,
additional decay channels, e.~g.~$h\to AA$ may occur.

As shown in~\cite{Schael:2006cr}, the experimental limits obtained for
specific parameter choices can be as low as $m_h\geq 90$~GeV within
the general CP-conserving and even lower for the CP-violating
case.

In mSUGRA, however, it has been shown that such deviations cannot be
realised~\cite{Ellis:2001qv}. For GMSB, such a general analysis is not
available, but it has been checked that the model points selected by
the fits in Section~\ref{sec:results:LEonly} do always maintain
$\sin^2(\beta-\alpha)\approx 1$, where $\tan \beta$ is the ratio of
the two Higgs vacuum expectation values and $\alpha$ the mixing angle
of the two CP-even neutral Higgs bosons. This ensures a SM-like
production of the lightest Higgs boson and the absence of additional
decay modes such as $h\to AA$. Therefore, we can safely employ the SM
limit on the Higgs boson mass in this study.

In principle, the full statistical information,
i.~e.~$\mathrm{CL}_{s+b} (m_h)$ on the compatibility of the search
result with a SM Higgs boson of mass $m_h$ could be exploited and
converted into a contribution to the $\chi^2$ function of the global
fit.  However, due to the theoretical uncertainty of 3~GeV on the
prediction of the Higgs boson mass within SUSY, the use of this
information would not have a significant impact on the results.

The recent exclusion of the SM Higgs in a small mass region around
160~GeV by the Tevatron experiments~\cite{:2009pt} is not considered
since the SUSY models under study do not allow $m_h$ above
approximately $135\,\mathrm{GeV}$~\cite{Heinemeyer:1998np}.

\subsubsection{Cold Dark Matter Relic Density}\label{sec:leobservables:CDM}

The results from the WMAP satellite on temperature fluctuations of the
cosmic microwave background together with various other cosmological
constraints have established a cosmological standard model, in which
approximately 23\,\% of the total energy of the Universe is contained
in cold dark matter (CDM). This is expressed in the fit in terms of
$\Omega_{\mathrm{CDM}}h^2$. While the presence of dark matter is
relatively undisputed, its nature is still unknown.  If SUSY is
$R$-parity conserving and the lightest SUSY particle (LSP) is neutral
and sufficiently heavy to contribute to cold dark matter (or if a
metastable neutral sparticle exists with a lifetime comparable to
the lifetime of the Universe), it contributes to dark matter through
its relic density, and it can make up all or part of the
cosmologically observed dark matter. Therefore, in the fits shown
later (see Section~\ref{sec:leobservables}), the observable
$\Omega_{\mathrm{CDM}}h^2$ is used in different ways or not at all as
a constraint.

%%%%%%%%%%%%%%%%%%%%%%%%%%%%%%%%%%%%%%%%%%%%%%%%%%%%%%%%%%%%%%%%%%%%%%%%%%%%%%%
%%%                              LHC observables                            %%%
%%%%%%%%%%%%%%%%%%%%%%%%%%%%%%%%%%%%%%%%%%%%%%%%%%%%%%%%%%%%%%%%%%%%%%%%%%%%%%%
\subsection{LHC Observables}\label{sec:lhcobservables}

As a case study, we assume that SUSY is realised with parameters
as specified in the SPS1a parameter set of~\cite{Allanach:2002nj}.
In Section~~\ref{sec:results:LEonly}), it will be shown that this 
bulk region point leads to a collider phenomenology rather similar
to the best fit point obtained from low energy measurements.

For the SPS1a point, SUSY particles will be copiously produced at the LHC
and a rather rich set of independent observables related to the masses
and branching fractions of SUSY particles can be reconstructed.
Many detailed experimental studies for this point (or phenomenologically 
similar points) exist. 

A direct reconstruction of SUSY particle masses at the LHC is difficult
due to the escaping LSPs. Therefore, where ever possible, 
we use observables which can be directly measured as input to the global fit. 
Such observables are the positions of kinematic edges and endpoint of
invariant mass spectra. Where mass peaks can be reconstructed, those are used as well.
Also, two ratios of branching fractions are employed. 

Measured production rates are not considered in this study for two reasons:
First, the prediction of rates has rather large theoretical uncertainties, in particular
if the production mechanism involves the strong force. Second, the calculation
of the theoretical prediction -- if realistic
experimental cuts are taken into account -- is very time consuming since usually 
Monte Carlo techniques have to be used to obtain these predictions.
Furthermore, the inherent statistical fluctuations of Monte
Carlo predictions easily cause oscillations during the $\chi^2$
minimisation which destabilise the result.

Three different integrated luminosities are considered separately 
to define the sets of accessible observables and their 
statistical and systematic errors: 1~$\text{fb}^{-1}$,
10~$\text{fb}^{-1}$ and 300~$\text{fb}^{-1}$. A centre-of-mass energy
of 14~TeV is assumed throughout. Most of the statistical
uncertainties are taken from~\cite{Weiglein:2004hn}. 
Where ever results for the specified integrated luminosities
are not available, reasonable interpolations/extrapolations are used.
Dominant experimental systematic errors are expected to arise from the
uncertainty of the lepton energy scale (LES) and the jet energy scale (JES).
The LES uncertainty is assumed to be 0.2\% for an
integrated luminosity of 1~$\text{fb}^{-1}$ and 0.1\% for higher integrated luminosity.
For the JES uncertainty, 5\% (1\%) are assumed for  1~$\text{fb}^{-1}$ ($ > 1~\text{fb}^{-1}$).
We assume that the energy scale uncertainties directly translate
into equally large relative uncertainties on the positions of
endpoints in mass spectra in case of fully leptonic or fully hadronic
final states. Following~\cite{Gjelsten:2004ki}, half of the relative
JES uncertainty is assumed as uncertainty on the endpoint for
invariant mass spectra involving both leptons and jets. Uncertainties
on the endpoints related to the JES and the LES are considered 100~\%
correlated between different measurements. Table~\ref{tab:inputs}
summarises all employed LHC observables together with their assumed
uncertainties.

For the SPS1a point, it is possible to
reconstruct sufficiently long decay chains of subsequent two-body
decays, such that mass information can be extracted in a
model independent way. However, it is necessary to assign the observed
decay products to the correct SUSY particles from which they originate.
The full combinatorics for decay chain ambiguities is not yet considered 
in this analysis. We assume that all decay chains are correctly identified. 
As a new approach to check the possible impact of misidentifications,
we study the impact of a wrong assignment of an endpoint to its SUSY particles 
on the global fit as a case study.
A more comprehensive analysis of these effects remains to be done.

Most information on SUSY particle masses within the SPS1a point 
can be obtained from the decay chain
\begin{equation}
\tilde{q}_2 \to q
\tilde{\chi}_2^0 \to q \ell^{\pm} \tilde{\ell}_R^{\mp} \to q \ell^+
\ell^- \tilde{\chi}_1^0,
\label{eq:SquarkDecay}
\end{equation}
where $\ell$ denotes either electrons or
muons. In total there are five different measurable invariant mass
combinations possible for this decay chain: $m_{\ell
  \ell}^{\mathrm{max}}$, $m_{\ell\ell q}^{\mathrm{max}}$, $m_{\ell\ell
  q}^{\mathrm{thr}}$, $m_{\ell q \mathrm{(low)}}^{\mathrm{max}}$ and
$m_{\ell q \mathrm{(high)}}^{\mathrm{max}}$ (for their definition see 
e.~g.~\cite{Gjelsten:2004ki}).

Similar to $m_{\ell \ell}^{\mathrm{max}}$, it is also possible to
measure the endpoint $m_{\tau \tau}$ where the electrons/muons are
replaced by tau leptons. The ratio of the total number of
events in the $m_{\ell \ell}$ and the $m_{\tau \tau}$ distributions
(corrected for efficiency differences) provides a measurement of
\begin{equation}
\frac{\mathcal{B}(\tilde{\chi}_2^0 \rightarrow \tilde{\ell}_R \ell) \times
  \mathcal{B}(\tilde{\ell}_R \rightarrow \tilde{\chi}_1^0 \ell)}
     {\mathcal{B}(\tilde{\chi}_2^0 \rightarrow \tilde{\tau}_1 \tau)
       \times \mathcal{B}(\tilde{\tau}_1 \rightarrow \tilde{\chi}_1^0
       \tau)}.
\label{eq:brratiochi2decays}
\end{equation}

Apart from the $m_{\ell \ell}$ endpoint there is also one additional 
$m_{\ell \ell}$ measurement included. It originates
from $\tilde{\chi}_4^0$ decays instead of $\tilde{\chi}_2^0$. 
Decays of $\tilde{\chi}_3^0$ do not provide a
visible $m_{\ell \ell}$ endpoint since it is mostly
Higgsino for the considered benchmark point and therefore the
couplings are too small. There are several $\tilde{\chi}_4^0$ decay
chains providing two oppositely charged leptons. Of all possibilities,
the chain is chosen which provides the largest $m_{\ell \ell}$
endpoint within the SPS1a scenario. This is the case for $\tilde{\chi}_4^0
\to \ell^{\pm} \tilde{\ell}_L^{\mp} \to \ell^+ \ell^-
\tilde{\chi}_1^0$. The other endpoints are unlikely to be measurable
due to the superimposed spectra from the other di-lepton decay channels.

Similar to the $\ell \to \tau$ replacement it is also possible to
exchange light-flavoured jets $q$ with b-flavoured jets $b$. This
yields a separate $m_{\ell \ell b}^{\text{thres}}$ measurement.

Jets which carry b-flavour also play an important role in obtaining
information about the gluino mass. In SPS1a, the gluino 
decays via $\tilde{g} \to \tilde{q} q$ where $q$ can be any quark flavour.
Due to combinatorial background,
gluinos can be reconstructed best, if one focuses on
\begin{equation}
\tilde{g} \to b \tilde{b}_{1,2} \to b b
\tilde{\chi}_2^0 \to b b \ell^{\pm} \tilde{\ell}_R^{\mp} \to b b \ell^+
\ell^- \tilde{\chi}_1^0
\label{eq:GluinoDecay}
\end{equation}
From this decay chain, the gluino mass can be reconstructed by
calculating the invariant mass of the $\tilde{\chi}_2^0 b b$ system,
provided that the $\tilde{\chi}_2^0$ momentum is known. Due to the
invisible $\tilde{\chi}_1^0$, the $\tilde{\chi}_2^0$ momentum cannot
be measured directly, but in the chosen scenario it can be
approximated reasonably well by~\cite{Weiglein:2004hn}
\begin{equation}
\vec{p}(\tilde{\chi}_2^0) \approx \left( 1 -
\frac{m_{\tilde{\chi}_1^0}}{m_{\ell \ell}} \right) \vec{p}_{\ell \ell}.
\end{equation}
It turns out that the gluino mass estimate from this approach is
highly correlated with the assumed $\tilde{\chi}_1^0$ mass, such that
effectively $m_{\tilde{g}} - m_{\tilde{\chi}_1^0}$ is
measured. Similarly $m_{\tilde{g}} - m_{\tilde{b}_1}$ and
$m_{\tilde{g}} - m_{\tilde{b}_2}$ can be determined by measuring the
difference between the invariant mass of the $\tilde{\chi}_2^0 b b$
and the $\tilde{\chi}_2^0 b$ system. The ratio of the total
number of events in the $\tilde{b}_1$ and $\tilde{b}_2$ mass peaks can
be used to determine
\begin{equation}
\frac{\mathcal{B}(\tilde{g}\rightarrow \tilde{b}_2 b) \times
  \mathcal{B}(\tilde{b}_2 \rightarrow \tilde{\chi}_2^0 b)}
     {\mathcal{B}(\tilde{g}\rightarrow \tilde{b}_1 b) \times
       \mathcal{B}(\tilde{b}_1 \rightarrow \tilde{\chi}_2^0 b)}.
\label{eq:brratiogluinodecays}
\end{equation}

The ``stransverse mass'' $m_{T2}$~\cite{Lester:1999tx,Barr:2003rg} is
used to extract information on the $\tilde{q}_R$ and the
$\tilde{\ell}_L$ mass. The exploited decay chains are $\tilde{q}_R \to
q \tilde{\chi}_1^0$ and $\tilde{\ell}_L \to \ell \tilde{\chi}_1^0$,
respectively. $\tilde{\ell}_L$ is studied in direct electro-weak
di-slepton production via an s-channel Z/$\gamma$ exchange. It turns
out that the endpoint of the $m_{T2}$ spectrum depends on the assumed
$\tilde{\chi}_1^0$ mass in such a way that roughly
$\sqrt{m_{\tilde{q}_R}^2 - 2 m_{\tilde{\chi}_1^0}^2}$ and
$\sqrt{m_{\tilde{\ell}_L}^2 - 2 m_{\tilde{\chi}_1^0}^2}$ are measured,
respectively.

Stop and sbottom sector information is obtained by a measurement of
the endpoint of the invariant mass spectrum of the $tb$ system from
the decay chains
\begin{eqnarray}
\tilde{g} & \to & t \tilde{t}_1 \to t b \tilde{\chi}_1^{\pm} \label{eq:gtt1}\\
\tilde{g} & \to & b \tilde{b}_1 \to t b \tilde{\chi}_1^{\pm} \label{eq:gbb1} .
\end{eqnarray}
The variable $m_{tb}^w$ used in our fits is a branching fraction weighted
average $m_{tb}$ endpoint for decay (\ref{eq:gtt1}) and
(\ref{eq:gbb1}) to account for the possibility that the two endpoints
might be too close to each other to be experimentally distinguishable.

To reconstruct charginos, the decay chain
\begin{equation}
\tilde{q}_L \to q \tilde{\chi}_1^{\pm} \to q W \tilde{\chi}_1^0 \to q
q q \tilde{\chi}_1^0
\end{equation}
is exploited. The chargino mass is obtained from the invariant mass of
the $qq\tilde{\chi}_1^0$ system where the two quarks come from the $W$
decay. The momentum of $\tilde{\chi}_1^0$ is reconstructed (up to a
two-fold ambiguity) using a technique described in detail in
\cite{Nojiri:2003tu}.

The most precise determination of the Higgs boson mass for the
considered mass range is obtained from measurements of the invariant
mass of the four-lepton system in the decay $h \to ZZ \to \ell^+
\ell^- \ell^{+\prime} \ell^{-\prime}$ and the di-photon mass of the
decay $h \to \gamma \gamma$. The top mass is measured from a
combination of several different final states and techniques, the most
precise of which is a kinematic fit for the semi-leptonic final state,
where one $W$ decays hadronically and the other $W$ leptonically.

\begin{table*}
  \caption{LHC observables which serve as input to the fits. The shown
    nominal SPS1a values have been calculated with SPheno. Most of the
    statistical uncertainties are taken
    from~\cite{Weiglein:2004hn}. Where numbers for the specified
    luminosities are not available, some interpolations/extrapolations
    are used. Uncertainties on the endpoints related to the jet energy
    scale (JES) and the lepton energy scale (LES) are considered
    100~\% correlated among different measurements.
}
  \label{tab:inputs}
  \begin{center}
    \begin{tabular}{l | d{3} | d{3} d{3}  d{3} d{2} d{3} d{2} d{3} d{3}}
      \hline\hline \multicolumn{1}{c|}{Observable} & \multicolumn{1}{c|}{Nominal} &
      \multicolumn{7}{c}{Uncertainty} \\ & \multicolumn{1}{c|}{Value} &
      \multicolumn{1}{c}{1 fb$^{-1}$}  & \multicolumn{1}{c}{10 fb$^{-1}$} &
      \multicolumn{1}{c}{300 fb$^{-1}$} & \multicolumn{1}{c}{$\text{LES}_{1}$} &
      \multicolumn{1}{c}{$\text{LES}_{10,300}$} & \multicolumn{1}{c}{$\text{JES}_{1}$} &
      \multicolumn{1}{c}{$\text{JES}_{10,300}$} & \multicolumn{1}{c}{syst.}\\ \hline
      $m_{h}$ & 109.6 & & 1.4 & 0.1 & & 0.1 & & & \\
      $m_{t}$ & 172.4 & 1.1 & 0.05 & 0.01 & & & 1.5 & 1.0 & \\
      $m_{\tilde\chi_1^{\pm}}$ & 180.2 & & & 11.4 & & & & 1.8 & \\
      $\sqrt{m_{\tilde{\ell}_L}^2-2m_{\tilde{\chi}_1^0}^2}$  & 148.8 & & & 1.7 & & 0.1 & & & 6.0\\
      $m_{\tilde{g}}-m_{\tilde{\chi}_1^0}$ & 507.7 & & 13.7 & 2.5 & & & & 5.1 & 10.0 \\
      $\sqrt{m_{\tilde{q}_R}^2-2m_{\tilde{\chi}_1^0}^2}$ & 531.0 & 19.6 & 6.2 & 1.1 & & & 22.7 & 4.5 & 10.0 \\
%      $\langle m_{\tilde{g}} - m_{\tilde{b}_{1,2}}\rangle$ & 522.6 & & 5.4 & & & & & 5.2 & \\
      $m_{\tilde{g}}-m_{\tilde{b}_1}$ & 88.7 & & & 1.5 & & & & 0.9 & \\
      $m_{\tilde{g}}-m_{\tilde{b}_2}$ & 56.8 & & & 2.5 & & & & 0.6 & \\
      $m_{\ell\ell}^{\text{max}}(m_{\tilde{\chi}_1^0},m_{\tilde{\chi}_2^0},m_{\tilde{\ell}_R})$ & 80.4 & 1.7 &0.5 & 0.03 & 0.16 & 0.08 & & & \\
      $m_{\ell\ell}^{\text{max}}(m_{\tilde{\chi}_1^0},m_{\tilde{\chi}_4^0},m_{\tilde{\ell}_L})$ & 280.6 & & 12.6 & 2.3 & & 0.28 & & & \\
      $m_{\tau\tau}^{\text{max}}(m_{\tilde{\chi}_1^0},m_{\tilde{\chi}_2^0},m_{\tilde{\tau}_1})$ & 83.4 & 12.6 & 4.0 & 0.73 & & & 4.2 & 0.8 & 5.7 \\
      $m_{\ell\ell q}^{\text{max}}(m_{\tilde{\chi}_1^0},m_{\tilde{q}_L},m_{\tilde{\chi}_2^0})$ & 452.1 & 13.9 & 4.2 & 1.4 & & & 22.7 & 4.5 & \\
      $m_{\ell q}^{\text{low}}(m_{\tilde{\ell}_R},m_{\tilde{q}_L},m_{\tilde{\chi}_2^0})$ & 318.6 & 7.6 & 3.5 &0.9 & & & 16.2 & 3.2 & \\
      $m_{\ell q}^{\text{high}}(m_{\tilde{\chi}_1^0},m_{\tilde{\chi}_2^0},m_{\tilde{\ell}_R},m_{\tilde{q}_L})$& 396.0 & 5.2 & 4.5 & 1.0 & & & 19.9 & 4.0 & \\
      $m_{\ell \ell q}^{\text{thres}}(m_{\tilde{\chi}_1^0},m_{\tilde{\chi}_2^0},m_{\tilde{\ell}_R},m_{\tilde{q}_L})$ & 215.6 & 26.5 & 4.8 & 1.6 & & & 10.8 & 2.2 & \\
      $m_{\ell \ell b}^{\text{thres}}(m_{\tilde{\chi}_1^0},m_{\tilde{\chi}_2^0},m_{\tilde{\ell}_R},m_{\tilde{b}_1})$ & 195.9 & & 19.7 & 3.6 & & & & 2.0 & \\
      $m_{t b}^{\text{w}}(m_{t}, m_{\tilde t_1},m_{\tilde\chi^{\pm}_1},m_{\tilde g},m_{\tilde b_1})$ & 359.5 & 43.0 & 13.6 & 2.5 & & & 18.0 & 3.6 & \\
      $\frac{\mathcal{B}(\tilde{\chi}_2^0 \rightarrow \tilde{\ell}_R \ell) \times \mathcal{B}(\tilde{\ell}_R \rightarrow \tilde{\chi}_1^0 \ell)}
      {\mathcal{B}(\tilde{\chi}_2^0 \rightarrow \tilde{\tau}_1 \tau) \times \mathcal{B}(\tilde{\tau}_1 \rightarrow \tilde{\chi}_1^0 \tau)}$ & 0.076 & 0.009 & 0.003 & 0.001 & & & & & 0.008 \\
      $\frac{\mathcal{B}(\tilde{g}\rightarrow \tilde{b}_2 b) \times \mathcal{B}(\tilde{b}_2 \rightarrow \tilde{\chi}_2^0 b)}
      {\mathcal{B}(\tilde{g}\rightarrow \tilde{b}_1 b) \times \mathcal{B}(\tilde{b}_1 \rightarrow \tilde{\chi}_2^0 b)}$ & 0.168 & & & 0.078 & & & & & \\
      \hline\hline
    \end{tabular}
  \end{center}
\end{table*}

%%%%%%%%%%%%%%%%%%%%%%%%%%%%%%%%%%%%%%%%%%%%%%%%%%%%%%%%%%%%%%%%%%%%%%%%%%%%%%%
%%%                              ILC observables                            %%%
%%%%%%%%%%%%%%%%%%%%%%%%%%%%%%%%%%%%%%%%%%%%%%%%%%%%%%%%%%%%%%%%%%%%%%%%%%%%%%%
\subsection{ILC Observables}\label{sec:ilcobservables}

At a future linear electron positron collider like the ILC, a huge variety of
precise measurements of SUSY particle properties from their 
electro-weak pair-production processes.

In this paper, in order to illustrate the potential of a linear
collider, a subset of the observables used in~\cite{Bechtle:2005vt} is
used.  All expected mass measurements of~\cite{Weiglein:2004hn} are
used together with the expected measurements of absolute Higgs
branching fractions and a large variety of cross-sections times
branching fraction measurements of all kinematically and statistically
accessible SUSY final states. In contrast to~\cite{Bechtle:2005vt},
only measurements at $\sqrt{s}=500$ and $1000\,\mathrm{GeV}$ and at
polarisations $(P_{e^-}, P_{e^+}) = (\pm80\,\%, \mp 60\,\%)$ are used,
assuming a long running time of the ILC with ${\cal
  L}^{\mathrm{int}}=500\,\mathrm{fb}^{-1}$ on each polarisation at
$\sqrt{s}=500\,\mathrm{GeV}$ and at $\sqrt{s}=1\,\mathrm{TeV}$,
respectively. The criteria used for the selection of expected
cross-sections times branching fraction measurements is outlined
in~\cite{Bechtle:2005vt}.

%%%%%%%%%%%%%%%%%%%%%%%%%%%%%%%%%%%%%%%%%%%%%%%%%%%%%%%%%%%%%%%%%%%%%%%%%%%%%%%
%%%                               PREDICTIONS                               %%%
%%%%%%%%%%%%%%%%%%%%%%%%%%%%%%%%%%%%%%%%%%%%%%%%%%%%%%%%%%%%%%%%%%%%%%%%%%%%%%%
\subsection{Theoretical Predictions}\label{sec:predictions}

Different theoretical codes have been used for the prediction of the
observables. The low energy observables are calculated by a selection
of codes combined in the so-called
Mastercode~\cite{Buchmueller:2008qe}.  The RGE running of the
parameters of the high-scale models down to the SUSY breaking scale
are accomplished with SoftSUSY~\cite{Allanach:2001kg}. Subsequently,
the observables of the Higgs sector and for $(g-2)_{\mu}$ are
accomplished with
FeynHiggs~\cite{Frank:2006yh,Heinemeyer:1998np,Degrassi:2002fi}. The
flavour observables are calculated with SuperIso and other codes based
on~\cite{Isidori:2007jw,Mahmoudi:2008tp}. The electro-weak precision
observables are derived in~\cite{Heinemeyer:2006px,Heinemeyer:2007bw}
and the cold dark matter relic density is calculated by
Micromegas~\cite{Belanger:2006is}.

The SUSY mass spectrum for the LHC measurements, all direct sparticle
decay branching fractions and the cross-sections for ILC are
calculated with SPheno~\cite{Porod:2003um}.

The known systematic uncertainties for the presently available
observables are included and listed in in Table~\ref{tab:leobserables}.

Systematic uncertainties for the LHC predictions are estimated from
the difference of the predictions between different RGE codes and from
scale variations. The differences between RGE codes like SoftSUSY and
SPheno are generally within the statistical and systematical
measurements of the LHC measurements for ${\cal L}^{\mathrm{int}}=1$
and $10\,\mathrm{fb}^{-1}$, effectively making the LHC fits relatively
robust against theoretical uncertainties on the order of
$10-20\,\mathrm{GeV}$~\cite{Allanach:2003jw,AguilarSaavedra:2005pw},
which is in the order of magnitude of the jet energy scale
uncertainties. For ${\cal L}^{\mathrm{int}}=300\,\mathrm{fb}^{-1}$ the
experimental systematic uncertainties are expected to be smaller than
the estimate of the theoretical uncertainties, especially in case of
$m_{h}$, however improvements on the precision of the predictions can
be expected until the LHC has acquired ${\cal
  L}^{\mathrm{int}}=300\,\mathrm{fb}^{-1}$. Theoretical uncertainties
will be included for all luminosities into the fit at a later stage.
Nevertheless we cross-checked the influence of an additional 3~GeV
uncertainty on $m_h$ due to unknown higher-order corrections for some
of our results and found that it does not have a significant effect on
the fit results for high-scale models.

A special case is the prediction of $\Omega_{\mathrm{CDM}}h^2$ in GMSB
models.  Since the gravitino is typically a very light LSP in GMSB
with a mass in the order of several MeV, it represents more hot than
cold dark matter. Therefore $\Omega_{\mathrm{CDM}}h^2$ is not used as
an observable for GMSB.

For the fits with ILC, theoretical uncertainties could play a major
role, because the experimental precision assumed
in~\cite{Bechtle:2005vt} is smaller than the current theoretical
uncertainties even in the gaugino and squark sector, where
uncertainties of around 1~GeV are
expected~\cite{Allanach:2003jw}. However, the possible increase in
theoretical precision until the existence of the ILC is yet unknown,
hence theoretical uncertainties will be introduced into the fits with
ILC at a later time.

Fittino and the calculator programs for the predictions are interfaced
using the SUSY Les Houches Accord~\cite{Skands:2003cj}.

%%%%%%%%%%%%%%%%%%%%%%%%%%%%%%%%%%%%%%%%%%%%%%%%%%%%%%%%%%%%%%%%%%%%%%%%%%%%%%%
%%%                  Applied parameter estimation techniques                %%%
%%%%%%%%%%%%%%%%%%%%%%%%%%%%%%%%%%%%%%%%%%%%%%%%%%%%%%%%%%%%%%%%%%%%%%%%%%%%%%%
\section{Parameter Estimation}\label{sec:techniques}

In order to asses the consistency of a theoretical prediction (defined
by a set of parameters within a specific SUSY model) for a given set
of measurements the following $\chi^2$ is used:
\begin{equation}
\label{eq:chi2}
    \chi^2 = (\vec{M} - \vec{O}(\vec{P}))^T \mbox{cov}_M^{-1} (\vec{M} - \vec{O}(\vec{P})) + \mbox{limits}.
\end{equation}
Here $\vec{M}$ is a vector containing the list of measurements,
$\vec{O}(\vec{P})$ a vector with the theoretical predictions for these
observables for a given point in parameter space
$\vec{P}$. $\mbox{cov}_M$ is the covariance matrix specifying the
uncertainties and correlations of the measurements $\vec{M}$. In
addition to the actual measurements $\vec{M}$, limits on observables
can be specified (e.~g.~the limit on the SM Higgs mass in case of the
fit of a model which ensures the presence of a SM-like Higgs
boson). This is incorporated for $m$ lower (LL) or upper (UL)
limits $L_i^{\mathrm{UL}/\mathrm{LL}}$ in the following way
\begin{equation}
    \mbox{limits} = \sum_{i=1}^{m} \left\{ \begin{array}{c}
        (O_i(\vec{P})-L_i^{\mathrm{UL}})^2/\sigma_i^2\quad\mbox{for}\,\,O_i(\vec{P})>L_i^{\mathrm{UL}}\\
        (L_i^{\mathrm{LL}}-O_i(\vec{P}))^2/\sigma_i^2\quad\mbox{for}\,\,O_i(\vec{P})<L_i^{\mathrm{LL}}\\
        0 \quad\mbox{for}\,\,L_i^{\mathrm{UL}}>O_i(\vec{P})>L_i^{\mathrm{LL}}\\
      \end{array}\right.
\end{equation}
where $\sigma_i$ specifies how steeply the limit is rising once it is
reached.

Being measurements $\vec{M}$ and $\mbox{cov}_M^{-1}$ are independent
of the theoretical model they are confronted with. Contrary to that
$\vec{O}(\vec{P})$ depends on the model and it even depends on the
interpretation of the data within a given model due to ambiguities in
the mapping of an observed final state to its physical origin within
the model (e.~g.~the assignment of kinematic edges in LHC mass spectra
to the decays of the respective SUSY particles). The covariance matrix
$\mbox{cov}_M$ is the sum of the statistical, systematical and
theoretical covariance matrices, where the former is diagonal for
independent measurements and the latter two can contain off-diagonal
elements describing correlations.  Using the $\chi^2$ expression of
Equation~\ref{eq:chi2}, the following tasks can be addressed:
\begin{itemize}
\item find the absolute minimum $\chi^2$, i.~e.~the parameter point
  of a given model which fits the data best;
\item determine the ${\cal P}$-value of the data given a best fit
  parameter point of a model;
\item find secondary minima which could be confused with the
  absolute minimum;
\item derive the probability that a secondary minimum of the $\chi^2$
  surface of the exact observables in a given model turns into the
  absolute minimum of the experimentally observed $\chi^2$ surface due
  to statistical and systematical uncertainties of the experimental
  observables;
\item derive the parameter uncertainties and correlations around the
  absolute minimum, with and without taking ambiguities in the
  interpretation of the data into account;
\item derive the probability that due to the statistical and
  systematical uncertainties of the experimental data the true model of
  new physics is yielding a worse ${\cal P}$-value than an alternative,
  wrong model of new physics;
\item derive predictions for most probable values of observables (and
  their expected variations) which are not used in the fit.
\end{itemize}
In the following, the statistical techniques used for these tasks are
introduced. Their application is described and their advantages and
shortcomings are discussed. We also propose an approach to estimate
the uncertainties on parameters in the presence of different ambiguous
interpretations of the data within the same model.

In this paper, Gaussian uncertainties are assumed both for statistical
and systematic uncertainties. For systematic and theoretical
uncertainties, there are other possible choices. For example,
box-shaped contributions to the $\chi^2$ (instead of a quadratic
function) are investigated elsewhere~\cite{Flacher:2008zq}. Given the
general uncertainty on systematic and theoretical errors, we assume
here that the final result does not depend on such subtle differences.
In fact, a larger effect can arise from unknown correlations among the
systematic and theoretical errors.

The $\chi^2$ hyper-surface for all considered SUSY models is highly
non-trivial.  As already shown in~\cite{Bechtle:2004pc},
gradient-based algorithms for global minimisation like
MINUIT~\cite{James:1975dr} are insufficient for most of the problems
under study.  Rather more elaborate methods, based on Markov Chain
Monte Carlo and Toy Fits algorithms are exploited to efficiently scan
the multi-dimensional parameter space.

\subsection{Minimisation and Scanning Techniques}\label{sec:techniques:minimization}

Two different parameter estimation techniques are used in the
following, a Markov Chain Monte Carlo and Toy Fits. These are briefly
described in the following sections.  Strong emphasis is laid on
ensuring that the global minimum is found, that the errors are
accurate, that the result is stable against different starting values,
and that the sampling of the parameter space is fine-grained
enough. This means that as many $N$-dimensional parameter combinations
(where $N$ is the number of parameters of the problem) as possible are
actually scanned at least within the range of
$\chi^2-\chi^2_{\mathrm{min}}<6$ (approximately corresponding to the
two-dimensional 95\,\% uncertainty interval around the best fit
point). These two techniques are chosen because they are complementary
in the way the uncertainties are defined and in the assumptions made
for the definition of the uncertainties. An agreement in the
uncertainties between the two methods provides a further strong
evidence for the validity of the result.

\subsubsection{Markov Chain Monte Carlo}\label{sec:techniques:minimization:mcmc}

The advantage of the Markov Chain Monte Carlo method is that it allows
to obtain an efficient scan of the $\chi^2$ surface around its minima.
Furthermore, it can be easily arranged that the sampling density in
parameter space directly provides a likelihood distribution for the
SUSY parameters in the Bayesian approach (see
e.~g.~\cite{Allanach:2005kz}, \cite{Allanach:2006cc}).

A Markov chain is a sequence of points $x_i$ $(i=1,\dots,n)$ in
parameter space. Each of these points $x_i$ has an associated
likelihood $\mathcal{L}(x_i)$. For our study we use
\begin{equation}
    \mathcal{L} = \exp\left( - \frac{\chi^2}{2} \right) .
\end{equation}
Using the Metropolis algorithm~\cite{Metropolis:1953am}, a new point
$x_{n+1}$ which is randomly chosen according to a proposal probability
density is added to the chain if $\mathcal{L}(x_{n+1}) >
\mathcal{L}(x_n)$. Otherwise it is accepted with probability
$\mathcal{L}(x_{n+1})/\mathcal{L}(x_n)$. If the new point $x_{n+1}$ is
not accepted, the old point $x_{n}$ is added to the end of the chain
again and the process continues. The result is -- under weak
assumptions -- independent of the specific choice of the proposal
probability density function in the limit of infinite
statistics. However, for finite statistics (even for order of
10~million parameter points for a typical 9-parameter model) the
efficiency of the sampling strongly depends on the proposal
distribution. Fittino implements the choice of box-shaped or Gaussian
proposal distributions, where the width of the box or Gaussian can be
adapted for each parameter. For each model and observable set, a set
of pre-runs with several thousand points per chain is used to adapt
the width parameters of each parameter individually such that the
ratio of accepted and rejected points in the chain lies between $0.8$
and $1.2$, for which the best scanning efficiency is expected.  This
procedure takes the initial uncertainties on each parameter from the
pre-run into account and is repeated manually until the result
converges.  For the results presented in this paper, only Gaussian
proposal distributions are used.

The resulting Markov Chain can be interpreted in two different
ways. In the Bayesian interpretation, it can be shown~\cite{MacKay}
that, if the proposal probability density is properly chosen, the
sampling density of points $x_i$ is proportional to the likelihood
distribution $\mathcal{L}$, which in turn is proportional to the
posterior probability in the case of flat priors (as assumed in this
paper).  Therefore, the best fit is obtained at the parameter point
with the highest sampling density $\propto {\cal L_{\mathrm{max}}}$.
The error on an individual parameter (or a subset of $D$ parameters)
is derived by integrating (``marginalising'') the sampling density
over all parameters apart from the parameter(s) under study.  The
resulting $D$-dimensional distribution can then be interpreted in
terms of $-2\ln{\cal L}+2\ln{\cal L_{\mathrm{max}}}$, where ${\cal
  L_{\mathrm{max}}}$ is the likelihood for the parameter point with
the highest likelihood. The $1\,\sigma$ uncertainty of a one-dimensional
parameter distribution is defined by the region within $-2\ln{\cal
  L}+2\ln{\cal L_{\mathrm{max}}} = 1$. In this interpretation, the
marginalised $\mathcal{L}$ is the probability distribution of the true
parameter value given the measurement.  68\,\% of this distribution is
contained within $1\,\sigma$.

\begin{center}
  \begin{figure*}
    \begin{center}
      \includegraphics[width=0.49\textwidth]{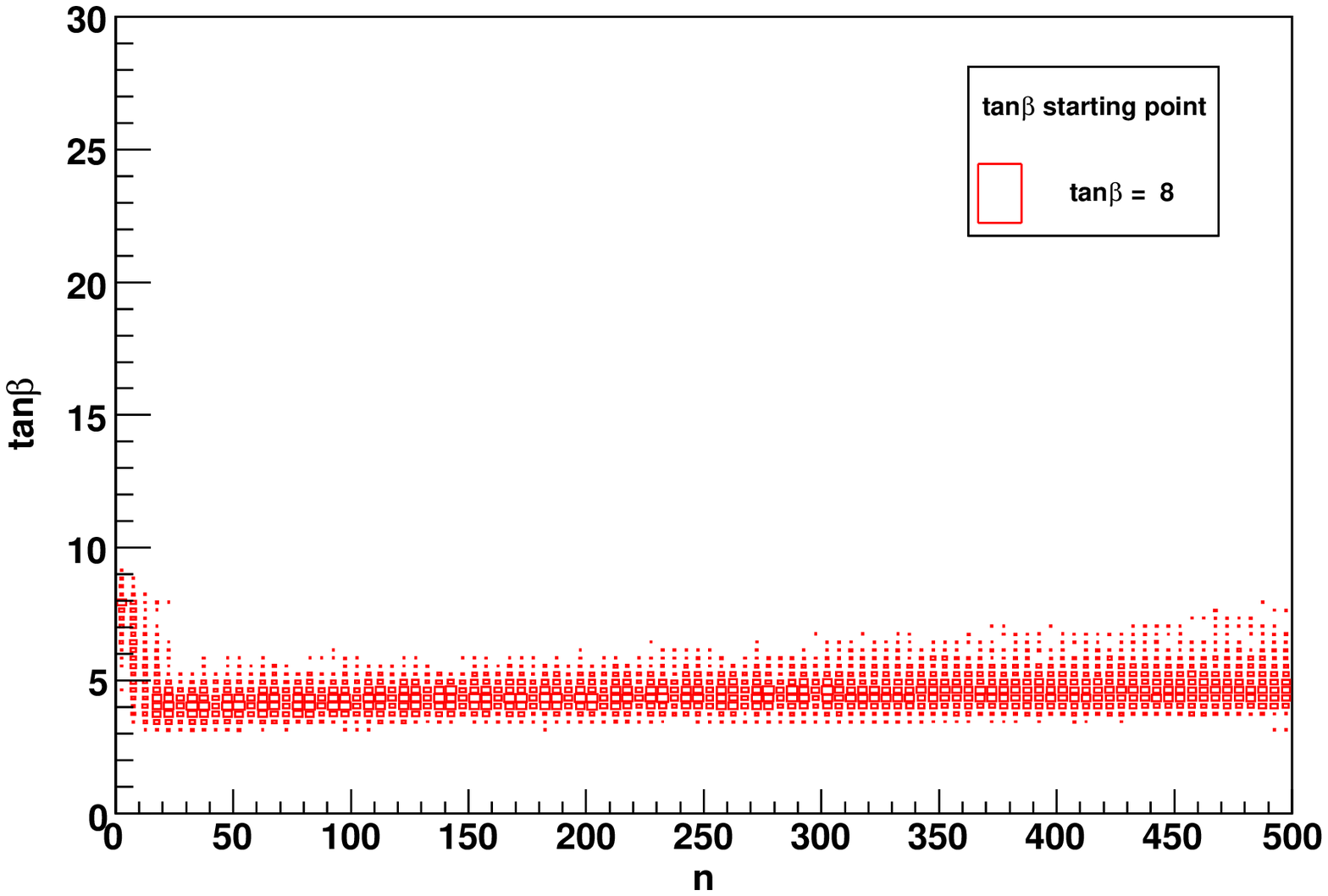}
      \hfill
      \includegraphics[width=0.49\textwidth]{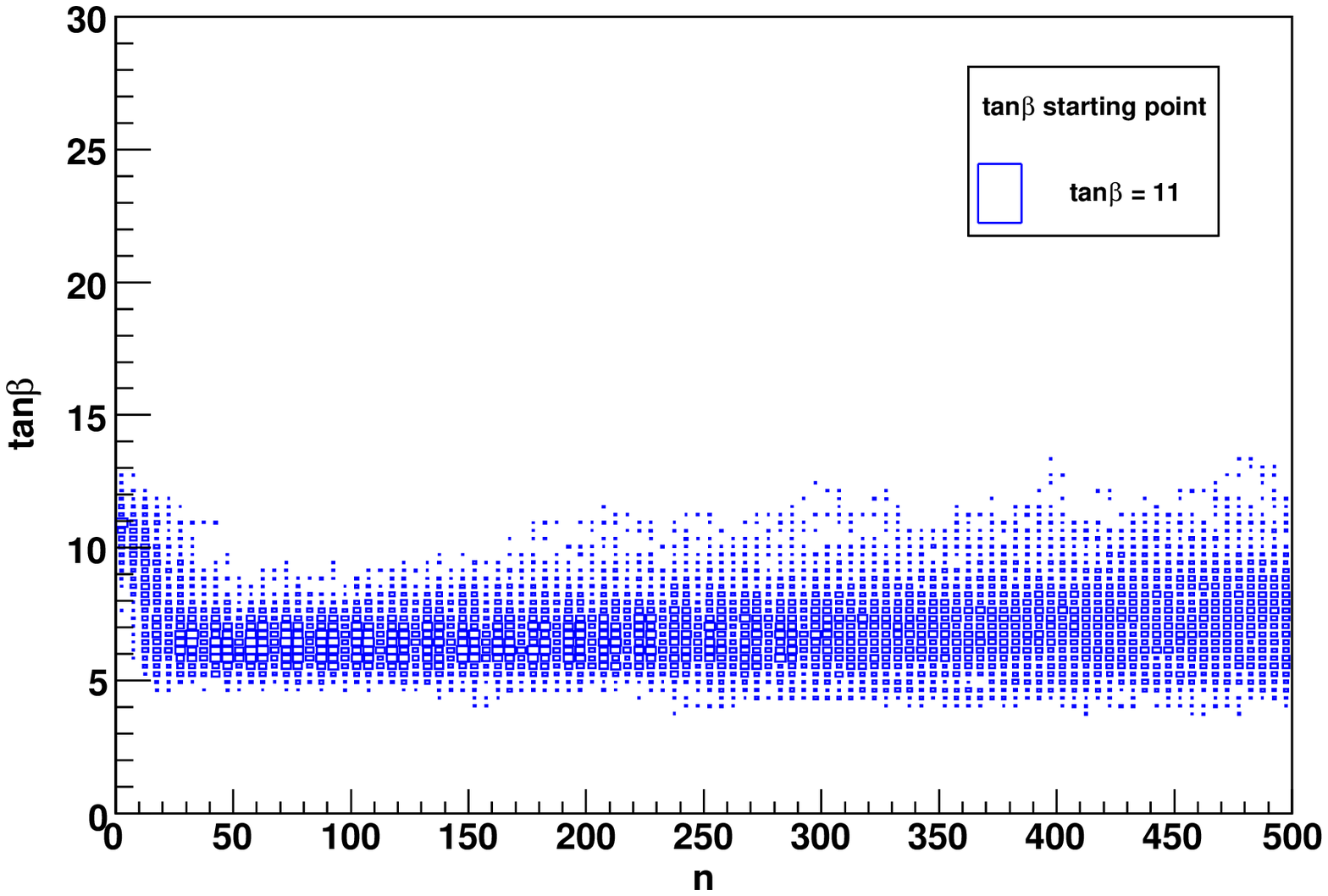}\\
      \includegraphics[width=0.49\textwidth]{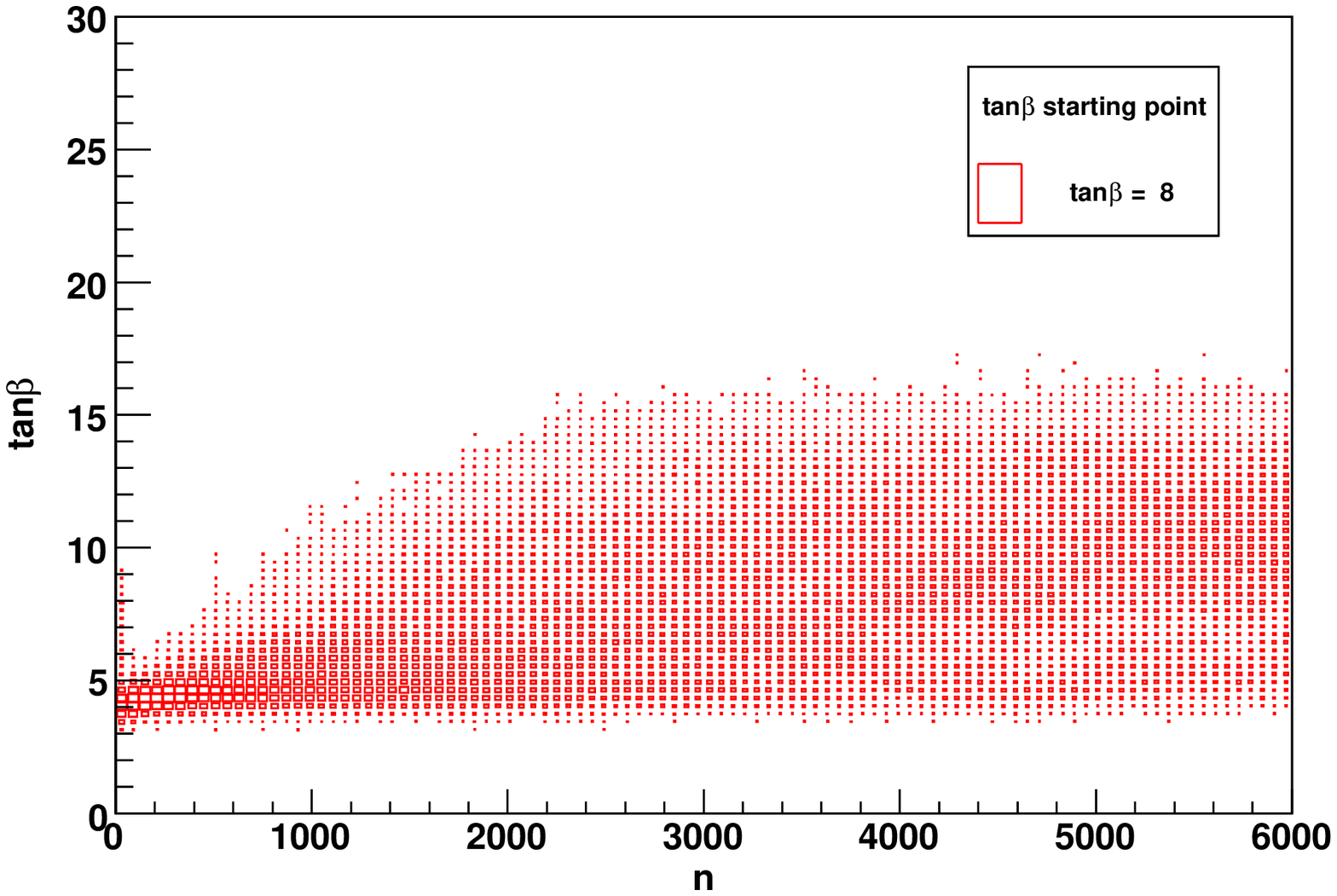}
      \hfill
      \includegraphics[width=0.49\textwidth]{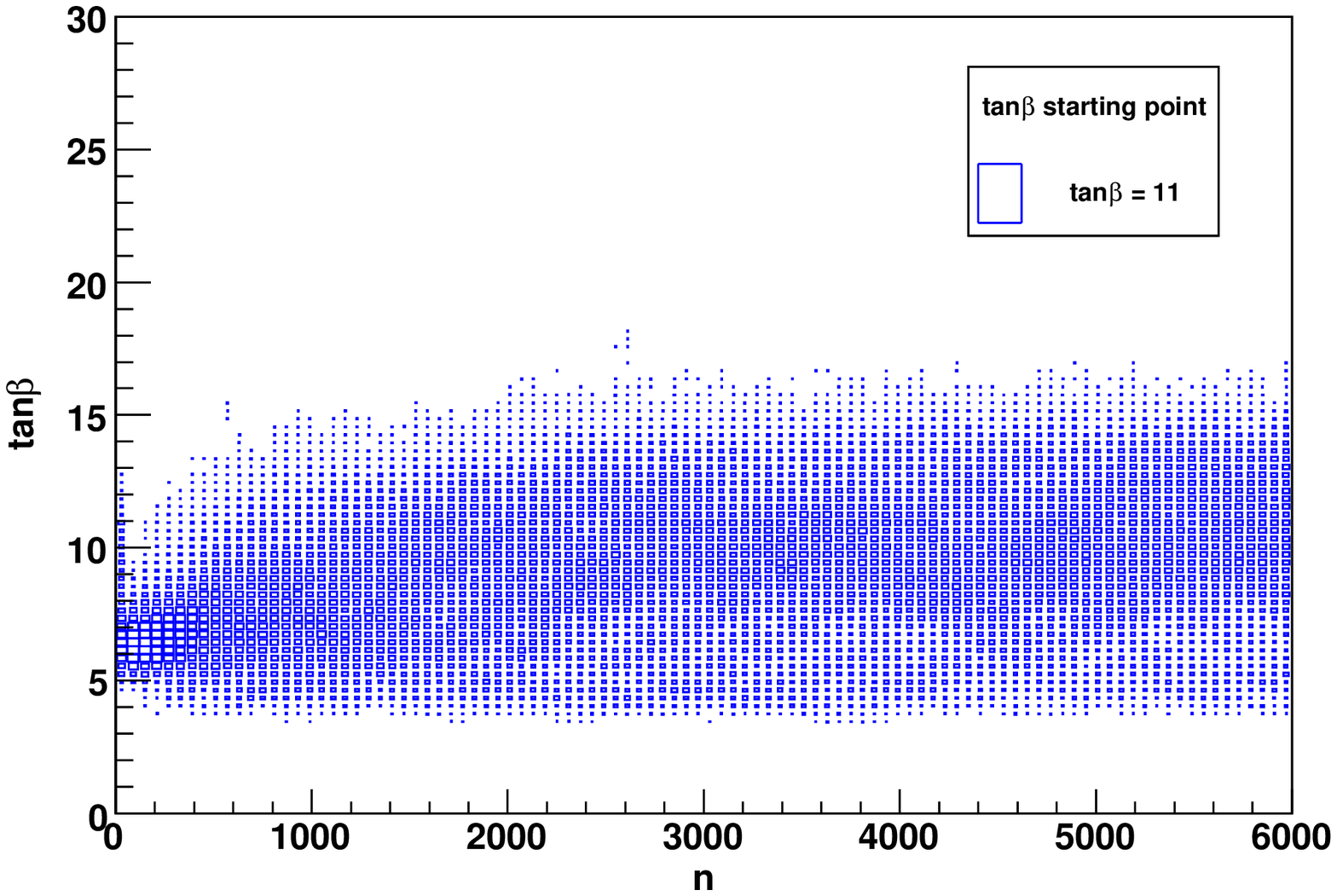}
    \end{center}
    \caption{$\tan \beta$ sampling behaviour of Markov chain using LE
      and LHC measurements for an integrated luminosity of 1~fb$^{-1}$
      within mSUGRA for two different starting points. }
    \label{fig:techniques:burnin}
  \end{figure*}
\end{center}

The Bayesian interpretation has to be handled with care for two
independent reasons. First, the outcome can have a strong dependence
on the chosen prior probability. This is e.~g.~exemplified
in~\cite{AbdusSalam:2009qd}. Second, for very complex parameter spaces
with many parameters (typically $8$ to $18$ in the case of the fits
presented here) one needs to check carefully that the sampling has not
only reached all relevant areas of the parameter space, but that in
addition the sampling is completely in equilibrium, i.~e.~that the
likelihood is really proportional to the sampling density. This
problem is exemplified in Fig.~\ref{fig:techniques:burnin}, which
shows the initial behaviour of two Markov Chains scanning same model
space (mSUGRA) and the same measurements but with different starting
points in the parameter $\tan\beta$. While above $n\approx 3000$ no
dependence on the starting point can be observed in this example,
below $n\approx 3000$ the point density is obviously not proportional
to the likelihood. Therefore including this region into the
calculation of the point density would distort the result unless the
Markov Chain length is so large that this region has negligible
impact. For all results presented in this paper which employ the
Bayesian interpretation of the Markov Chain, two different starting
values have been chosen. Only points for large enough $n$ are included
in the chain analysis such that the projections of each parameter
distribution are consistent within statistics.

While this is possible for cases with well-measured and thus strongly
constraining observables and a small number of parameters (e.~g.~fits
of a high-scale SUSY model like mSUGRA using LE and LHC measurements),
this approach fails for more challenging problems like fits of an
18-dimensional more general MSSM. There are around 20~million points
in a combination of several Markov Chains with different starting
points. These can not be checked efficiently for the effect shown in
Figure~\ref{fig:techniques:burnin}. Neither is it technically possible
to provide sufficiently long chains due to computing limitations.
Therefore, the Bayesian interpretation is used in this paper only in
the case of fits of the mSUGRA model to the LHC data for illustration.

The Frequentist interpretation of the Markov Chain is used as a
default in this paper. It does not make use of the sampling density
directly, but employs only the obtained $\chi^2$ values found in the
chain for each parameter point. The best fit point is directly defined
by the parameter point with the lowest $\chi^2=\chi^2_{\mathrm{min}}$
(or equivalently the point with the largest likelihood $\mathcal{L} =
\mathcal{L}_{\mathrm{max}}$).  To obtain uncertainties for a subset of
parameters this approach scans over all parameters except for those
under study and chooses the scanned parameters such that $\mathcal{L}$
is maximised for each point in the studied parameter subspace. This
procedure yields a profile likelihood. The $1\,\sigma\,(2\,\sigma)$
uncertainties in the one-dimensional case are defined by
$\Delta\chi^2=\chi^2-\chi^2_{\mathrm{min}}=1\,(4)$ and, in the
two-dimensional case, by
$\Delta\chi^2=\chi^2-\chi^2_{\mathrm{min}}=2.3\,(5.99)$~\cite{BarlowBook}. In
the limit of Gaussian parameter distributions, i.~e.~a locally linear
relation between measurements and parameters, the $1\,\sigma$
environment for a given set of measurements covers an area which
contains the true parameter point in 68\,\% of all possible
experimental outcomes.

This approach has several advantages over the Bayesian
interpretation. First, it does not depend on prior distributions,
since the likelihood in the hidden dimensions is not
integrated. Instead the Markov Chain is simply used an efficient
scanning technique for the parameter space: for each bin in the
histogram of the parameter(s) under study the point with the lowest
$\chi^2$ is chosen. Second, it is sufficient to scan different regions
in parameter space with different granularity, as long as the obtained
point density is fine enough to derive a smooth surface.  This
approach requires significantly less points than the Bayesian
approach.

\begin{figure}
  \begin{center}
    \includegraphics[width=0.49\textwidth]{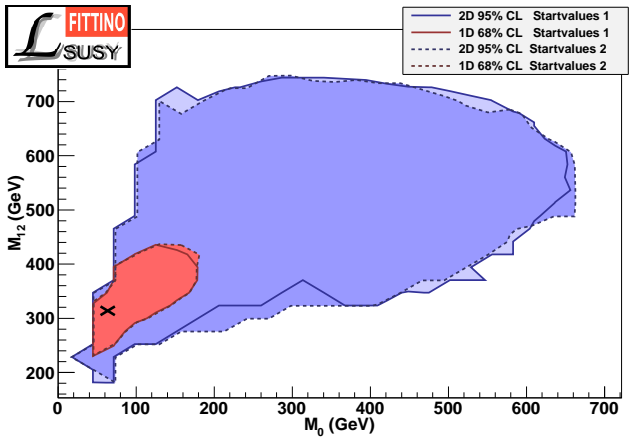}
  \end{center}
  \caption{Frequentist interpretation of Markov chain fitting mSUGRA using only LE
    measurements for two different starting points.}
  \label{fig:techniques:2StartvaluesComp}
\end{figure}

In order to test if the sampling density is sufficient to derive a
smooth $\chi^2$ surface, the following procedure is used. For each
result presented in this paper, at least two different parameter
points in the parameter space are chosen as starting point. These
starting points are chosen individually for each fit after an initial
Markov Chain run such that they lie approximately $2\,\sigma$ away
from the estimated best fit point.  Then, Markov Chains are computed
for each of the two starting points.  An example for the comparison of
the two results can be found in
Figure~\ref{fig:techniques:2StartvaluesComp}.  For a fit to be
accepted, it is required that the differences in the shape and area of
the $1\,\sigma$ and $2\,\sigma$ regions are within the differences of
the binning, which is derived for each chain separately by dividing
the observed $2\,\sigma$ region into $25\times25$ bins.  This
procedure to ensure the robustness of the results turned out to be
very important, in particular in the case of more complex fits,
e.~g.~the MSSM fit to LE and LHC measurements and the mSUGRA fit to LE
measurements. In these cases, several million sampling points turned
out to be required.

The results of the Bayesian and Frequentist interpretations are
compared for the case of mSUGRA fitted to LHC measurements. For more
complex fits there is not enough confidence in the results of Bayesian
interpretation for the reasons explained above (see
also~\cite{AbdusSalam:2009qd}).

As described above, the uncertainties on the model parameters can be
derived from the $\Delta\chi^2$ values of each parameter point. It is
important to note that the identification of $\Delta\chi^2=1$ with the
$68\,\%$ uncertainty region is technically only true for problems with
Gaussian uncertainties of the observables and linear dependencies
between parameters and observables, hence for Gaussian parameter
distributions. As visible already from
Figure~\ref{fig:techniques:2StartvaluesComp}, this assumption is not
fulfilled for all fits. Furthermore, for the interpretation of the fit
result and especially for the derivation of conclusions on RGE running
of parameters or the prediction of other observables which are not yet
used in the fit, it is important to also derive correlations between
the model parameters. The derivation of linear correlation
coefficients can be achieved by interpreting the obtained $\chi^2$
surface of each combination of two parameters of the model as a
numerical version of the Hesse-matrix of the problem. Normally,
e.~g.~in MINUIT, the Hesse matrix is of course an approximation of the
true $\chi^2$ surface. Hence the correlation coefficient can be
calculated from the two-dimensional histogram of $-\Delta\chi^2$ for
each parameter combination. For practical reasons the histogram is
constrained to the region of $\Delta\chi^2<5.99$, corresponding to the
two-dimensional $95\,\%$ uncertainty region around the best fit point.
This is justified because the linear correlation coefficient are an
approximation of the full non-linear correlations and are dominated by
the area around the minimum.

The Frequentist analysis of the parameter space using Markov Chains
could be further refined using MINUIT around the $\chi^2$ minimum
found in the Markov Chains in order to better determine the position
of the absolute minimum, independent of binning effects.  However, in
practice each set of Markov Chains with sufficient sampling of the
$\Delta\chi^2 < 5.99$ region, as required above, ensures a sufficiently
fine sampling around the absolute minimum, such that no significant
improvement of a further refinement of the minimum in a MINUIT
minimisation can be seen.

While Markov Chains provide a powerful tool for the study of complex
parameter fits, there are a few shortcomings which make it desirable
to cross-check the results with an alternative technique and to
provide solutions for tasks which cannot be solved by Markov Chain
Monte Carlos. First, it would be desirable to use a method which is
independent of both the problem of priors (as the case in the Bayesian
interpretation) or assumptions on (local) linearity of measurements
and parameters (as the case in the Frequentist interpretation).

Second, the Markov Chains cannot be used to study how well a certain
model can be distinguished from another model by the data, or how one
interpretation of the observables is distinguished from another
interpretation within the same model. Markov Chains just supply the
${\cal P}$-value of the best fit point of each model or each
interpretation. They do not provide the probability to get a better
${\cal P}$-value for the ``wrong'' model than for the ``correct''
model under the assumption that one of the models is realised.

\subsubsection{Toy Fits and Simulated Annealing}\label{sec:techniques:minimization:toys}

In addition to the Markov Chain analysis, Toy Fits are used to obtain
an independent estimate of the parameter uncertainties and to compare
different models or data interpretations quantitatively.  The Toy Fit
analysis consists of two steps.  First, Markov Chains or minimisation
through Simulated Annealing (for a description of the implemented
algorithm of Simulated Annealing see~\cite{Bechtle:2004pc}) is used to
find the absolute $\chi^2$ minimum of a given problem. Second, Monte
Carlo Toy data are created around the observables corresponding to
best fit point. In the first step, the best fit parameters $\vec{P}_o$
are determined. Then the set of observables $\vec{M}_o$ corresponding
to this set is calculated.  This set is then used to create $N$
different MC Toy sets of pseudo-measurements $\vec{M}_i$ (i.~e.~other
possible experimental outcomes which are consistent with that
parameter set) by smearing around $\vec{M}_o$ according to the
Gaussian uncertainties and correlations as defined in
$\mbox{cov}_M$. For each of the $N$ MC Toy sets $\vec{M}_i$ a fit is
performed using Simulated Annealing followed by a MINUIT minimisation
at the minimum of the Simulated Annealing fit. This procedure yields a
set of $N$ ``best fit'' parameter points $\vec{P}_i$. The
distributions and correlations of $\vec{P}_i$ are then interpreted as
the expected distributions of all possible experimental outcomes given
the best fit parameter set $\vec{P}_o$.

In contrast to the Frequentist interpretation of the Markov Chain
Monte Carlo, the uncertainties and correlations can be directly
calculated from the (co-)variances of $\vec{P}_i$. Therefore, the
results represent an estimate for the expected distribution of all
possible results (including all possible non-Gaussianities).  They
even include possible secondary $\chi^2$ minima, which are turned into
the absolute minimum for a subset of the observable set $\vec{M}_i$,
where statistical and systematic uncertainties of the measurements can
invert the order of the different $\chi^2$ minima.

Toy Fits allow for a robust cross-check of the validity of the
results: For problems which are sufficiently Gaussian the distribution
of the $\chi^2_{\mathrm{min}}$ values has to be consistent with a
$\chi^2$ distribution for $n-m$ degrees of freedom, where $n$ is the
number of observables and $m$ is the number of parameters. This
criterion is checked by fitting the $\chi^2$ distribution to each
$\chi^2_{\mathrm{min}}$ histogram and requiring that the fitted number
of degrees of freedom agrees with the expectation for the problem
within $2\,\sigma$ for a fit to be accepted.

In addition, as for Markov Chains, starting values for the parameters
have been varied and the result is required to be independent of the
starting value up to statistical fluctuations.

However, there is also a disadvantage of the Toy Fits with respect to
the Markov Chains. In case of two different $\chi^2$ minima very far
apart from each other (with respect to the parameter uncertainties,
and with a very high $\chi^2$ barrier in between) which both have
almost identical minimal $\chi^2$, one set of Toy Fits with one given
starting point $\vec{P}_{s1}$ of the fits would not necessarily find
both minima, requiring possibly more than two different starting
points. In the limit of infinite computing power this can be overcome
by starting the individual Toy Fits from randomised positions within
the parameter space in order to scan for additional minima. If
additional minima with acceptable ${\cal P}$-values are found and if
the $\chi^2$ barriers between them are too high to be traversed within
one minimisation, different sets of Toy Fits with starting values
around the different minima can be directly treated as different
models using the prescription proposed in
Section~\ref{sec:techniques:comparisons}.  This also allows to assign
a numerical value to the parameter uncertainties stemming from the
different minima.

\subsection{Model Discrimination and Ambiguities}\label{sec:techniques:comparisons}

All techniques discussed above are directed towards determining the
uncertainties of the parameters of one model in the presence of a
given set of data with one fixed interpretation of the data. In this
paper a new method is proposed to use Toy Fits to solve the two
remaining tasks, namely to measure how often fluctuations of the data
lead to wrongly identifying a different model than the true model as
the model with the best ${\cal P}$-value, and determining how
ambiguities in the mapping between measurements (i.e.\ edges in LHC
mass spectra) and predicted observables, translate into the
uncertainties of the model parameters.

The first of these problems cannot be answered directly by using the
${\cal P}$-value. It represents the probability that the given
experimental data are observed given a certain model with certain best
fit parameters, and hence can be used to determine which models (or
which interpretations of the data in a given model) are acceptable and
which models are rejected by the data due to their low ${\cal
  P}$-value. However, it cannot be used to determine how likely it is
to obtain a certain ${\cal P}$-value for a given model if the data are
actually caused by another model. Such a situation can occur often for
not very well constrained problems, as e.~g.~visible in
Sections~\ref{sec:results:LEonly:mSUGRA},
\ref{sec:results:LEonly:GMSB} and \ref{sec:results:LELHC:mSUGRA}. This
question can be answered by the Toy Fits directly by fitting an
arbitrary number of different models (or different interpretations of
the data in the same model) $j$ with different predictions
$\vec{O}(\vec{P}^j_i)$ to the same set of smeared measurements
$\vec{M}_i$. Then the probability $p_{w}$ to prefer the ``wrong''
models over the ``correct'' model, from which the observable set
$\vec{M}_o$ is derived, can be directly determined by counting how
often one of the ``wrong'' models achieved a better
$\chi^2_{\mathrm{min}}$ than the ``correct'' model. Since in reality
the ``correct'' model is not known, this interpretation can be
repeated for each model which yields a reasonable ${\cal
  P}$-value. Examples using this procedure are presented in
Section~\ref{sec:results:LELHC:mSUGRA}.

The second problem can be solved directly using the simultaneous Toy
Fit method proposed above. The uncertainty of a model parameter is
defined and measured as the square root of the variance of the
parameter distribution.  Each entry $\vec{P}_i$ in the Toy Fit
parameter distributions corresponds to one individual best fit point
for one model, one interpretation of the data and one individual set
of smeared measurements $\vec{M}_i$.

In the presence of several interpretations of the data (or
e.~g.~different possible values of a discrete parameter of the model,
i.~e.~in all cases where different model/data combination can be
fitted with the same parameters and the same measurements) the
distributions of the $\vec{P}_i$ can be exchanged against the
distribution of the $\vec{P}^j_i$, where for each smeared simulated
Toy measurement $i$ the model or interpretation $j$ yielding the best
fit is chosen which yields the best $\chi^2$. This is the natural
extension of the method exploited usually in Toy Fits and described in
Section~\ref{sec:techniques:minimization:toys}, since in the presence
of one given set of measurements $\vec{M}_i$ all remaining ambiguities
of the interpretation of the data would be tested and for the final
result the interpretation with the lowest $\chi^2$ (i.~e.~highest
${\cal P}$-value) would be chosen. The uncertainty stemming from the
ambiguities is then taken into account by not citing the parameter
uncertainties of the best fit interpretation of the data only, but by
citing the uncertainty including the possibility that a different
interpretation would have been chosen as the one with the best ${\cal
  P}$-value, as proposed here. Note that this interpretation naturally
leaves the parameter distribution $\vec{P}_i$ unchanged if only one
model is always yielding the best ${\cal P}$-value, i.~e.~if the
method proposed above determines that there is 0\,\% probability to
prefer a ``wrong'' model over the ``correct'' model. An application of
this method is presented in Section~\ref{sec:results:LELHC:mSUGRA}.

\section{Results}\label{sec:results}

In this chapter, results are presented in terms
of allowed areas in the SUSY parameter space. Also, allowed regions
for SUSY particle masses are calculated from the fitted parameters.

In Section~\ref{sec:results:LEonly}, the high-scale SUSY
models mSUGRA and GMSB are tested against presently available
measurements. Predictions for discovery at the LHC and the expected
range of SUSY masses are presented. 
In Section~\ref{sec:results:LHConly}, the constraint from LHC observables 
are studied alone. In Section~\ref{sec:results:LELHC}, the expected
measurements at the LHC are combined with already available observables.
Here both mSUGRA and a more general 18-parameter MSSM are studied. 
Finally, in Section~\ref{sec:results:LELHCILC}, the impact of SUSY precision
measurements at the ILC together with
${\cal L}^{\mathrm{int}}=300\,\mathrm{fb}^{-1}$ of LHC data
is studied, in particular for the general MSSM.

\subsection{Present Low-Energy Observables}\label{sec:results:LEonly}

Several studies of the mSUGRA parameter space in the light of
different sets of available measurements have been performed
recently~\cite{Buchmueller:2008qe,AbdusSalam:2009qd}. In this section,
parts of these studies are repeated and extended.  Emphasis is laid on
understanding the impact of the most important observables on the
parameter uncertainties, and on the study of the uncertainties of SUSY
particle mass predictions.  The methods outlined in
Section~\ref{sec:techniques} are used to extract parameter
correlations. The impact of the SM parameters is also studied. In
addition, a comparison of the predicted mSUGRA sparticle spectrum
with a GMSB spectrum is presented.

In order to study the effect of the different observables, first a
baseline fit with all observables from Table~\ref{tab:leobserables} is
performed. In order to accommodate the uncertainties of the most
important observables, $(g-2)_{\mu}$ in terms of its SM prediction,
and $\Omega_{\mathrm{CDM}}h^2$ in terms of its origin, several other fits are
then shown. Additionally, these observables are treated in different
ways, removed one by one or removed simultaneously.

%%%%%%%%%%%%%%%%%%%%%%%%%%%%%%%%%%%%%%%%%%%%%%%%%%%%%%%%%%%%%%%%%%%%%%%
%%%%%%%%%%%%%%%%%%%%%%%%%%%%%%%%%%%%%%%%%%%%%%%%%%%%%%%%%%%%%%%%%%%%%%%
%%%%%%%%%%%%%%%%%%%%%%%%%%%%%%%%%%%%%%%%%%%%%%%%%%%%%%%%%%%%%%%%%%%%%%%
\subsubsection{Fits of the mSUGRA Model}\label{sec:results:LEonly:mSUGRA}

\begin{table}
  \caption{Result of the fit of the mSUGRA model with
    $\mathrm{sign}(\mu)=+1$ including four additional SM parameters to
    all measurements listed in Table~\ref{tab:leobserables}. The
    minimum $\chi^2$ value is $20.6$ for 22 degrees of freedom, corresponding to
    a ${\cal P}$-value of 54.4\,\%.}
  \label{tab:results:LEonly:mSUGRA:all}
  \begin{center}
    {\renewcommand{\arraystretch}{1.2}
      \begin{tabular}{lrcl}
        \hline\hline
        Parameter      & Best Fit  &       & Uncertainty \\
        \hline
        sign$(\mu)$      &  +1       &       &                    \\
        $\alpha_s$     & 0.1177  & $\pm$ &   0.0020      \\
        $1/\alpha_{em}$  & 127.924   & $\pm$ &   0.017        \\
        $m_Z$ (GeV)    & 91.1871   & $\pm$ &   0.0020       \\
        $m_t$ (GeV)    & 172.4   & $\pm$ &   1.09          \\ 
        $\tan\,\beta$  & 13.2   & $\pm$ &  7.2               \\ 
        $M_{1/2}$ (GeV) & 331.5   & $\pm$ & 86.6               \\ 
        $M_0$ (GeV)   &  76.2  &       & $^{+79.2}_{-29.1}$ \\ 
        $A_{0}$ (GeV) & 383.8   & $\pm$ &  647               \\
        \hline\hline
      \end{tabular}
    }
  \end{center}
\end{table}

The mSUGRA (or sometimes the CMSSM, which has $\mu$ as additional free
continuous SUSY parameter) scenario is the best studied SUSY scenario
at colliders. 
For the purpose of this paper, it is appealing to study it in detail
using the low energy observables, since experimental studies of
possible LHC measurements are available from ATLAS~\cite{Aad:2009wy}
and CMS~\cite{Ball:2007zza} for several well studied mSUGRA parameter
points. Knowing the allowed parameter space using the already existing
observables allows the selection of a mSUGRA scenario studied at LHC
which is within the currently experimentally allowed parameter region.

The baseline fit is performed using all measurements from
Table~\ref{tab:leobserables}, i.~e.~assuming that the current
predictions for the SM contribution to $(g-2)_{\mu}$ are correct and
that the cold dark matter in the Universe is entirely caused by the
relic density of SUSY LSPs, which in the mSUGRA scenario has to be the
lightest neutralino $\tilde{\chi}^0_1$.
For all fits in this paper, if not stated otherwise, all parameter
points with charged stable LSPs are excluded from the fit. The fit has
been performed using the Markov Chain Monte Carlo technique described
in Section~\ref{sec:techniques:minimization:mcmc} and applying the
Frequentist interpretation. Using two different starting values of the
Markov Chains, approximately 20~million points per chain have been
tested with consistent results for the different starting points. The
sampling density around the minimum is large enough such that an
additional MINUIT minimisation is not necessary.
The $\chi^2$ minimum of $20.6$ for 22 degrees of freedom,
corresponding to a ${\cal P}$-value of 54.4\,\%, can be determined
with good precision from the distribution of the sampled points. The
${\cal P}$-value is significantly larger than the ${\cal P}$-value of
the SM fit~\cite{Collaboration:2008ub} of around $18\,\%$, which does
not include the limits from direct Higgs searches. With the direct
Higgs search limit included, the ${\cal P}$-value is
$17\,\%$~\cite{Flacher:2008zq}.  However the analysis of the variable
pulls explained below shows that this is not directly due to a better
description of the SM~precision variables, which cause the moderate to
low ${\cal P}$-value of the SM fit. It can be seen that mSUGRA
provides an excellent description of the presently available precision
data.

Correlations among the observables of Table~\ref{tab:leobserables}, as
described in Section~\ref{sec:leobservables:EWP}, are studied in a
separate fit. While the minimal $\chi^2$ is lowered by $1.1$ due to
the correlations, the results of the allowed parameter regions for the
fit including correlations are identical to those of the baseline fit
without correlations in the input observables. The reason for that
lies in the fact that $(g-2)_{\mu}$ and $\Omega_{\mathrm{CDM}}h^2$
(which are uncorrelated) constrain the parameter space more strongly
than any other combination of variables (see
Section~\ref{sec:results:LEwithout}).  Therefore there is no
significant impact of the correlations among the electroweak precision
observables.

The result of the baseline fit is given in
Table~\ref{tab:results:LEonly:mSUGRA:all}. Due to the deviation of
$(g-2)_{\mu}$ from the SM prediction towards larger values, a positive
sign of $\mu$ is preferred, which is chosen for this fit. A comparison
with $\mathrm{sign}(\mu)=-1$ is shown below. It can be seen that
moderate values of $\tan\beta$ between 5 and 20 are preferred. The
gaugino mass parameter $M_{1/2}$ is expected in the range of 200~GeV
to 400~GeV, while a low scalar mass parameter $M_0$ between 50~GeV and
150~GeV is preferred. The universal trilinear coupling $A_0$ is not
very well constrained and is expected to be between $-$300~GeV and
1000~GeV. As it is shown below, this region of parameter space is very
favourable for early discoveries at LHC and for a rich phenomenology
at ILC. In Table~\ref{tab:results:LEonly:mSUGRA:all}, as well as in
the following tables, we quote symmetrical uncertainties whenever
upper and lower uncertainties agree within 20\,\%, and asymmetrical
uncertainties are given otherwise. It should be noted that the
probability densities for the fitted parameters are usually not
Gaussian, i.~e.~that the one-dimensional $2\,\sigma$ uncertainty at
$\Delta\chi^2=4$ are not twice the $1\,\sigma$ uncertainty at
$\Delta\chi^2=1$.

Note that the uncertainty of the SM parameters exactly corresponds to
the uncertainties of the measurements to which the given parameter is
100\,\% correlated (see Table~\ref{tab:leobserables}). This means that
the direct measurement of the parameters is so precise that the
additional observables from Table~\ref{tab:leobserables} do not play
any role in their determination, which has consequences for their
correlations with other parameters.

It also has to be noted that $G_F$ is omitted from
Table~\ref{tab:results:LEonly:mSUGRA:all}, since $G_F$ is measured so
precisely that virtually no effect of the inclusion of $G_F$ into the
fit can be observed. This has been checked using a second baseline fit
including $G_F$, which yields identical results for the other SM
parameters and the mSUGRA parameters.  Hence $G_F$ has been omitted
from all subsequent fits to save computing effort.  For completeness
however, it is included in the discussion of the parameter
correlations.

\begin{table*}
  \caption{Correlations of the mSUGRA model and the SM parameters
    (with $\mathrm{sign}(\mu)=+1$) using all measurements listed in
    Table~\ref{tab:leobserables}. Very small correlations between SM
    and mSUGRA parameters is observed while there is significant
    correlation among the mSUGRA parameters.}
  \label{tab:results:LEonly:mSUGRA:corr}
  \begin{center}
    \begin{tabular}{l|ccccccccc}
      \hline\hline
      Parameter        & $\alpha_s$ & $G_F$    & $1/\alpha_{em}$ & $m_Z$    & $m_t$    & $\tan\,\beta$ & $M_{1/2}$ & $M_0$   & $A_{0}$ \\
      \hline                                                                                                       
      $\alpha_s$       & 1.000         & $-$0.005 &  0.006        & $-$0.003 & $-$0.007 &  $-$0.003     & $-$0.005 & $-$0.013 & $-$0.009\\
      $G_F$            &            &  1.000      &  $-$0.003       & $-$0.001& $-$0.003 &  $-$0.002     & $-$0.022  & 0.007 & $-$0.006\\
      $1/\alpha_{em}$  &            &          &    1.000           & $-$0.008 & 0.006  &  $-$0.001     & $-$0.003 & 0.009& 0.007 \\
      $m_Z$            &            &          &                 &  1.000      & $-$0.0348  & 0.053        & 0.035   & 0.046  & 0.029  \\
      $m_t$            &            &          &                 &          &  1.000      & 0.075        & 0.088   & 0.075  & 0.093  \\ 
      $\tan\,\beta$    &            &          &                 &          &          & 1.000            & 0.358    & 0.833   & 0.457   \\ 
      $M_{1/2}$         &            &          &                 &          &          &               & 1.000       & 0.449   & 0.236   \\ 
      $M_0$            &            &          &                 &          &          &               &          &   1.000    & 0.632   \\ 
      $A_{0}$          &            &          &                 &          &          &               &          &         &   1.000    \\
      \hline\hline
    \end{tabular}
  \end{center}
\end{table*}

In addition to the parameter uncertainties, it is important to study
the parameter correlations. Calculating uncertainties of quantities
derived from the parameters does not only depend on the absolute value
of the uncertainties, but also on the correlations. The prescription
proposed for the calculation of correlations from Markov Chain Monte
Carlos in Section~\ref{sec:techniques:minimization:mcmc} is used for
the results in Table~\ref{tab:results:LEonly:mSUGRA:corr}. It is
interesting to note that while there are significant correlations
among the mSUGRA parameters, there is no correlation in excess of
$10\,\%$ between any SM parameter and any mSUGRA parameter. In
addition, there are hardly any correlations among the SM parameters
themselves.  The reason for this is the fact that every SM parameter
can be measured using an observable to which the parameter is
correlated 100\,\%. This ensures that each SM parameter is fixed by
itself to a strong precision (due to the impressive success of the SM
precision measurements) without any impact of any other parameter.
% such that no strong correlation with additional parameters can occur. 
It should be noted however, that the
correlations of Table~\ref{tab:results:LEonly:mSUGRA:corr} are the
linear correlations of the parameters within the two-dimensional
$2\,\sigma$ area. For some choices of observables (e.~g.~without the
use of $(g-2)_{\mu}$, see below) the preferred parameter space
exhibits strong non-linear correlations, restricting the reliability of 
this method to well-constrained fits like the baseline fit described here.

\begin{figure}
  \includegraphics[width=0.49\textwidth]{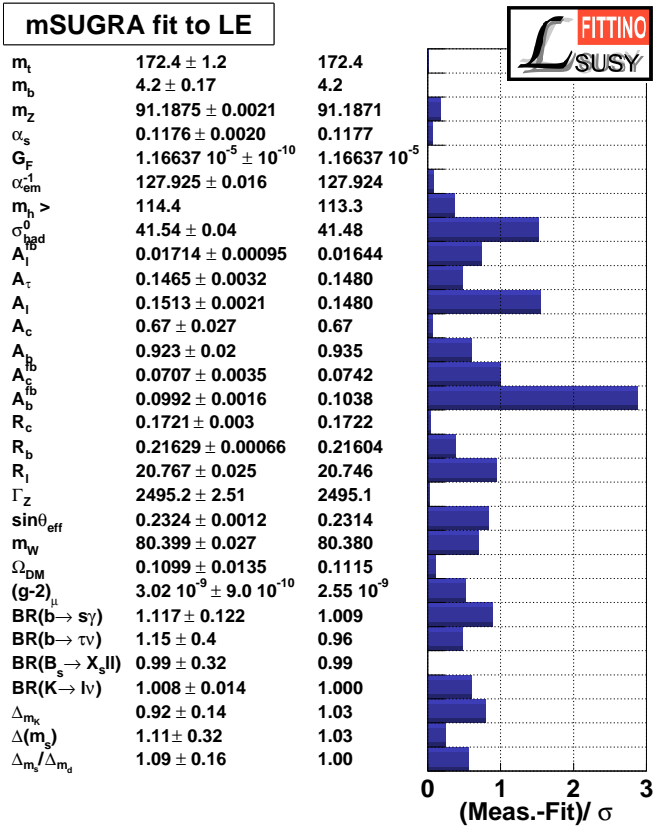}
  \caption{Pull for low energy observables used in the mSUGRA parameter
    fit using the observables from Table~\ref{tab:leobserables} and the
    best fit point from Table~\ref{tab:results:LEonly:mSUGRA:all}.}
  \label{fig:results:LEonly:mSUGRA:pulls}
\end{figure}

The individual pull of each observable is shown in
Figure~\ref{fig:results:LEonly:mSUGRA:pulls}. Again it can be seen that
the SM parameters can all be fitted individually exactly to the values
of their corresponding observables. When comparing the result with the
SM fits in~\cite{Collaboration:2008ub,Flacher:2008zq}, the main
difference is that $m_h$ is not a free parameter anymore, but a
function of the SUSY parameters.  This has the effect that the
electroweak precision observables can be satisfied with a
significantly larger $P$-value than for the SM fit.
The preferred value of $m_h$ is 113.3~GeV and this is below the 95\,\%
CL lower limit of the SM Higgs boson searches, which are applicable to
mSUGRA, as outlined in Section~\ref{sec:leobservables:higgs}. The
preferred value is within 1/3 of $1\,\sigma$ of the theoretical
uncertainty with respect to the 95\,\% CL lower limit at
$m_h>114.4\,\mathrm{GeV}$, and even within approximately $1\,\sigma$
of the lowest mass point where the $CL_s$ of the combined searches
is 0.5 at around $m_h=116.5\,\mathrm{GeV}$. The latter is
corresponding to the mean value of a ``measurement'' of $m_h$, if the
searches are interpreted as such~\cite{Barate:2003sz}.

It can be observed that apart from the change in $m_h$, the fit
results are remarkably similar to the SM fit. This is due to the
decoupling of the sparticles heavier than the electro-weak scale
from the processes contributing to the precision measurements, hence
rendering the result very SM-like. It can be seen that while rendering
$m_h$ naturally heavier, mSUGRA does not mitigate the tension in some
of the SM precision observables.

\begin{center}
  \begin{figure*}
    \begin{center}
      \includegraphics[width=0.49\textwidth]{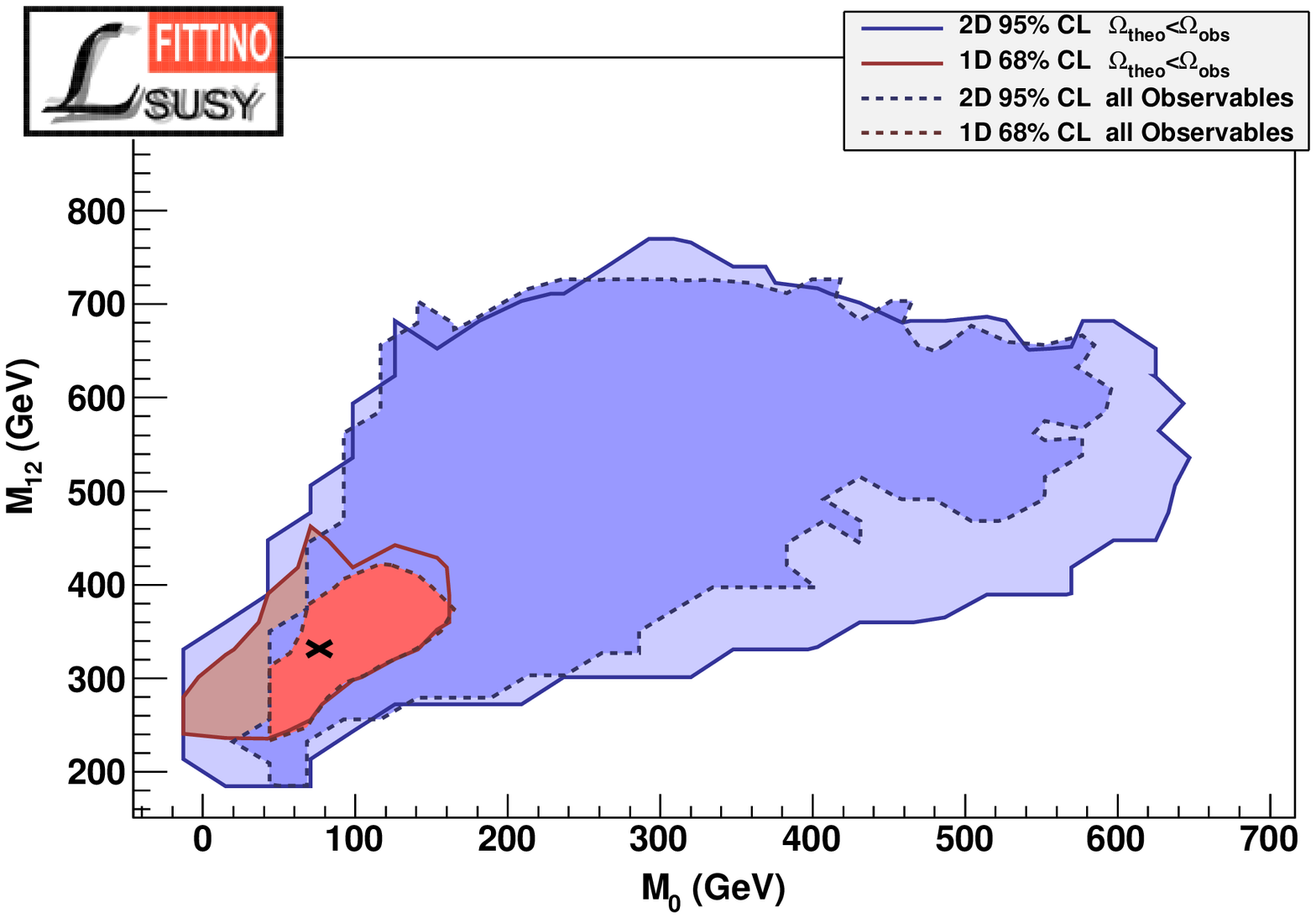}
      \hfill
      \includegraphics[width=0.49\textwidth]{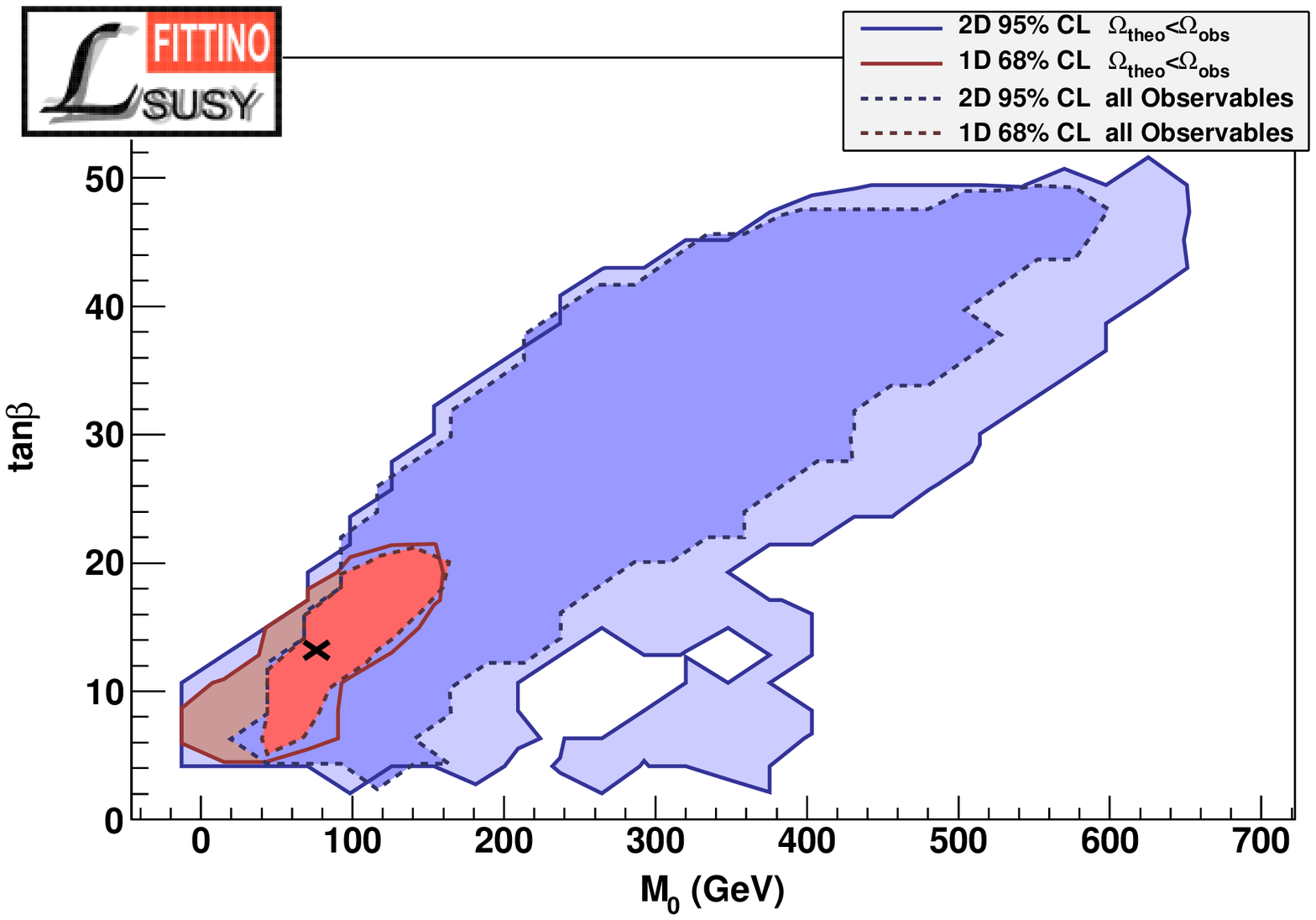}
      \\[1mm]
      \includegraphics[width=0.49\textwidth]{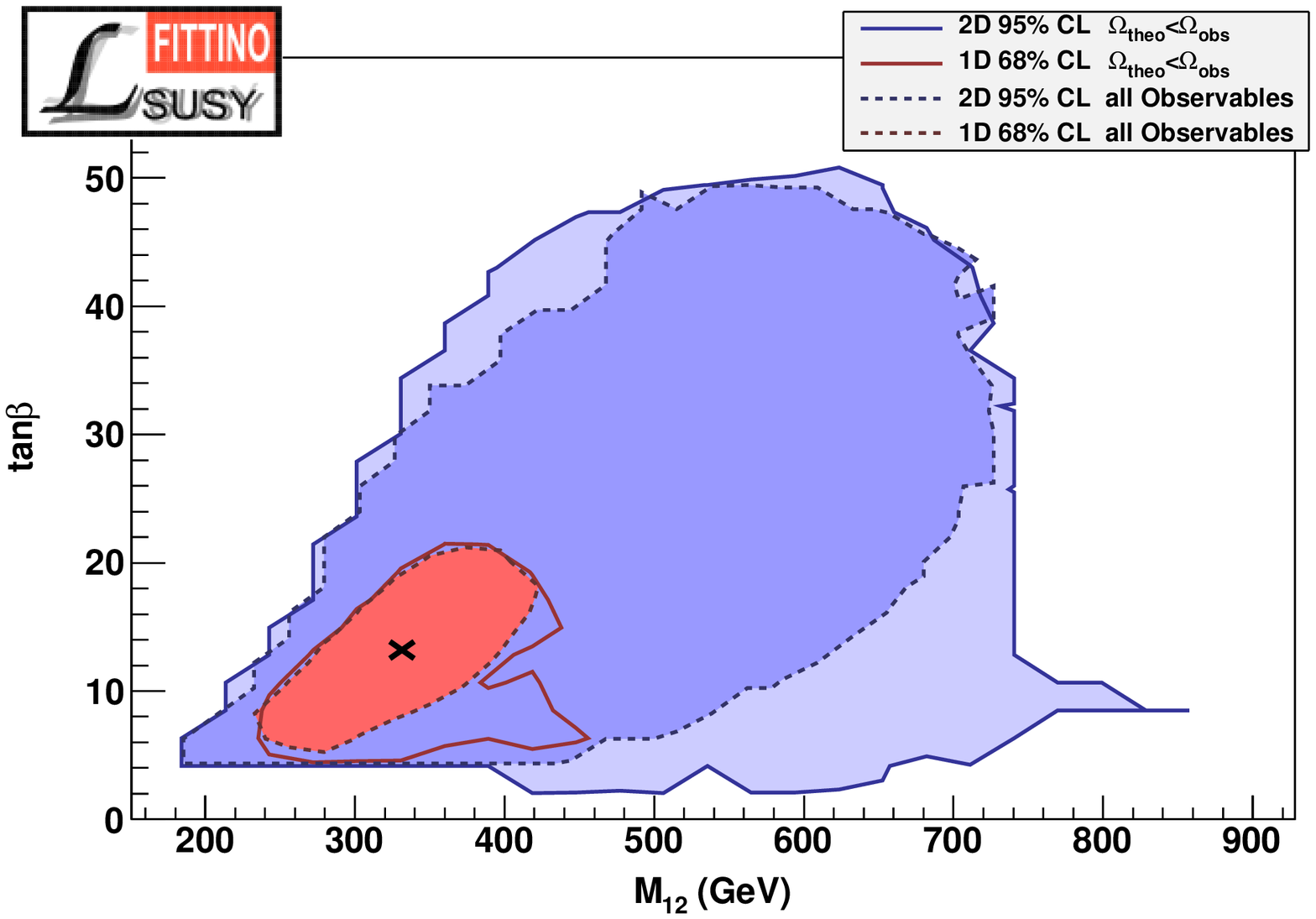}
      \hfill
      \includegraphics[width=0.49\textwidth]{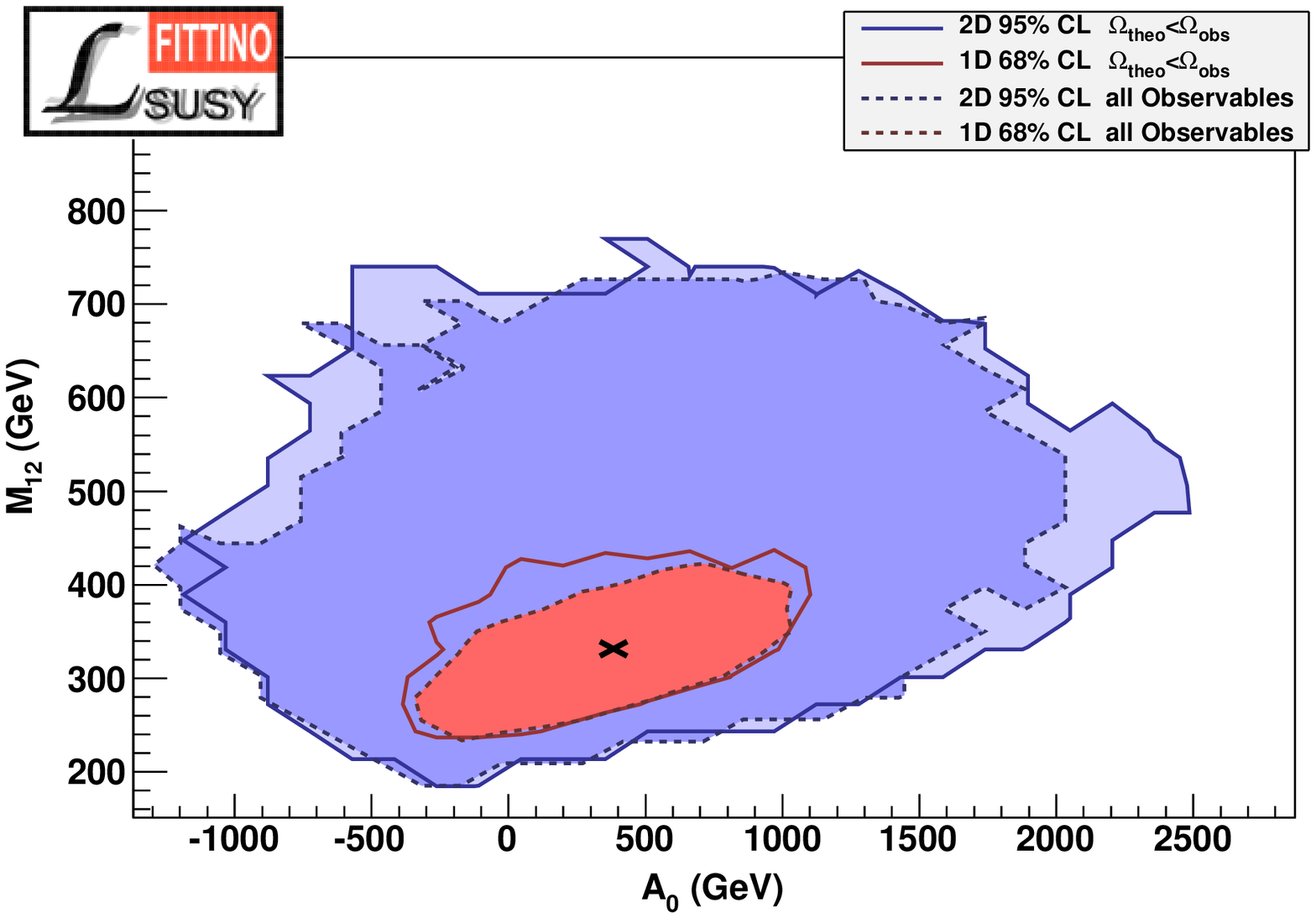}
      \\[1mm]
      \includegraphics[width=0.49\textwidth]{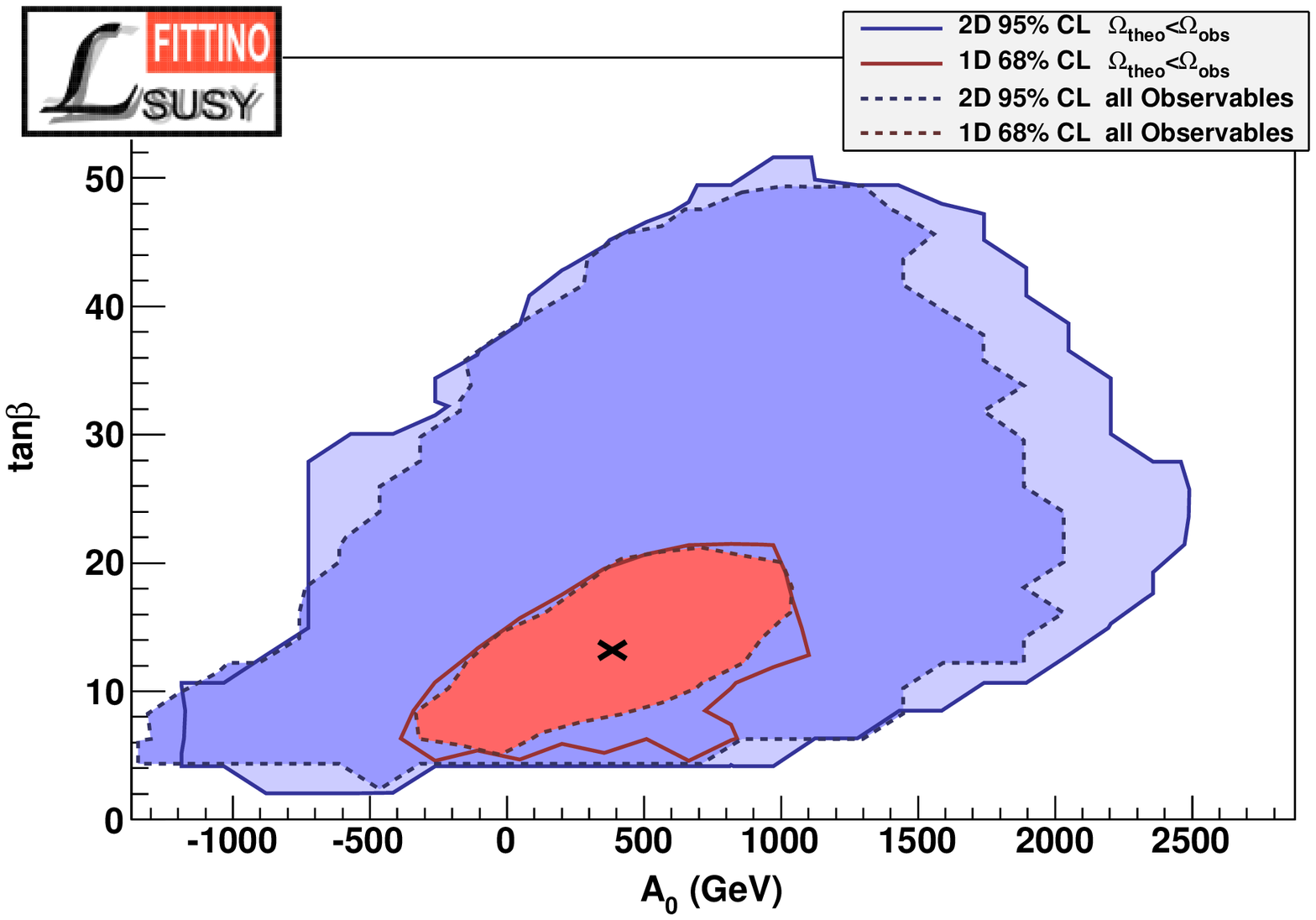}
      \hfill
      \includegraphics[width=0.49\textwidth]{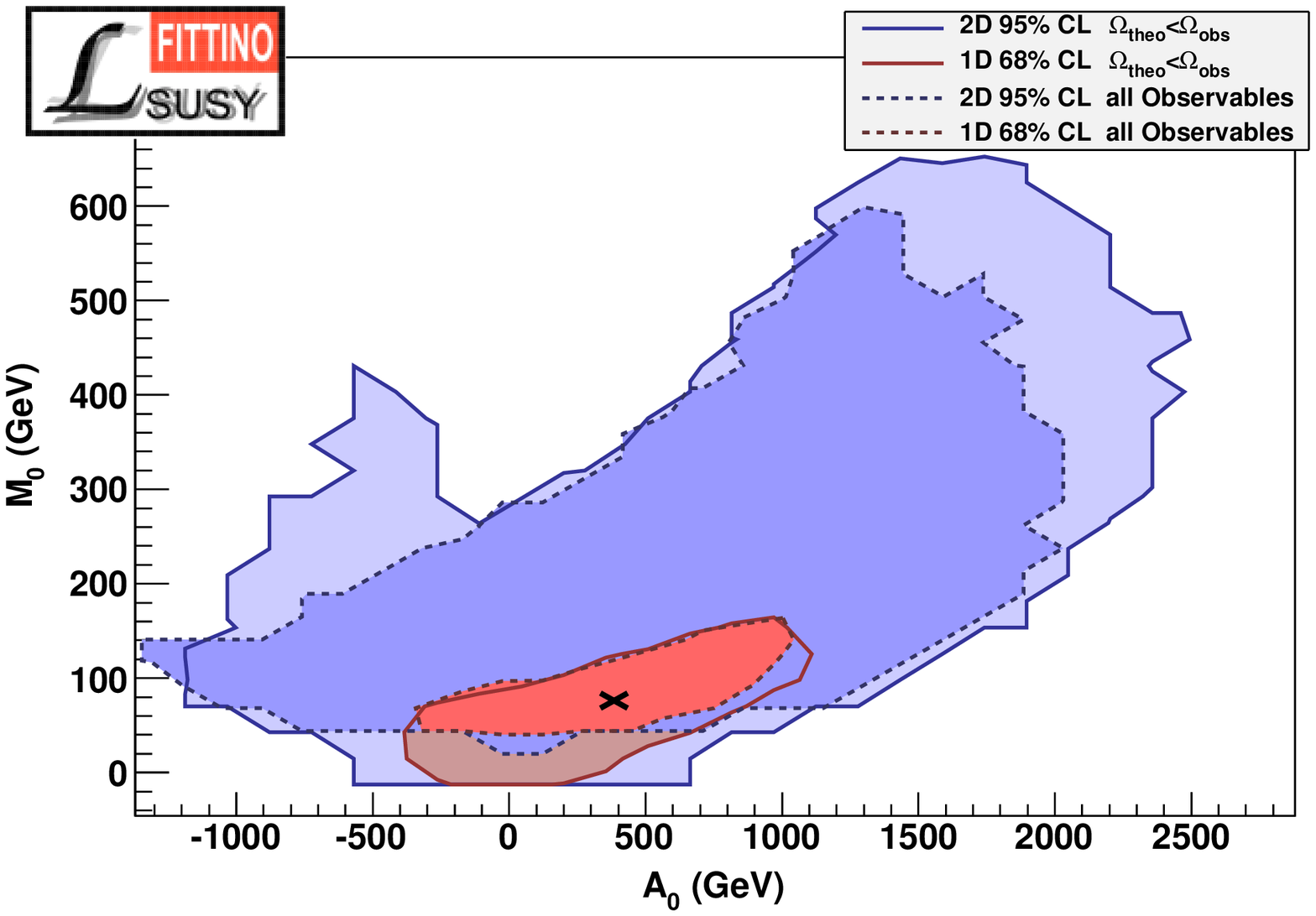}
    \end{center}
    \caption{mSUGRA parameter regions compatible with relic density
      constraint and with all low energy measurements for various
      parameter combinations and sign$(\mu)$ fixed to $+1$. For both
      cases two-dimensional 95~\% confidence level and one-dimensional
      68~\% confidence regions are shown. From the latter $1\,\sigma$
      uncertainties for individual parameters can be derived from a
      projection of the area to the respective axis. The dashed lines
      represent the result when all observables are included. For the
      full lines, the measured relic dark matter density is only
      regarded as an upper limit of the predicted SUSY dark matter
      density. All other observables remain the same.}
    \label{fig:results:LEonly:mSUGRA:2dRes}
  \end{figure*}
\end{center}

Apart from looking at the numerical results of the baseline mSUGRA
fit, it is interesting to study the allowed parameter regions in
detailed two-dimensional projections of each mSUGRA parameter against
another parameter, as shown in the profile likelihood plots obtained
using the Frequentist interpretation of the Markov Chains in
Figure~\ref{fig:results:LEonly:mSUGRA:2dRes}. The SM parameters have
been suppressed due to their negligible correlations with themselves
and with the mSUGRA parameters (see
Table~\ref{tab:results:LEonly:mSUGRA:corr}). The plots show the
two-dimensional 95\,\% CL allowed region of the fits in blue
(corresponding to $\Delta\chi^2=5.99$) and the one-dimensional 68\,\%
region in red (corresponding to $\Delta\chi^2=1$). There are two
reasons why the unusual choice of showing the $\Delta\chi^2=1$ curve
in a two-dimensional plot has been made. First because by this the
one-dimensional uncertainties of the parameters can be directly read
off from the plot, which is not possible for the choice of the
two-dimensional 68\,\% CL area at $\Delta\chi^2=2.3$.  Second, the
two-dimensional 95\,\% CL area gives a good indication of the
experimentally allowed area, while two-dimensional 68\,\% CL area
leaves a large room for parameter points outside the 68\,\% contour.
The most common projection is shown in the upper left plot, which
compares the allowed region in $M_0$ and $M_{1/2}$ between the
baseline fit described above and a fit requiring only
$\Omega_{\mathrm{theo}}h^2\leq\Omega_{\mathrm{obs}}h^2$, as described
below in Section~\ref{sec:results:LEwithout}. It can be seen that on
the upper left side the allowed region is directly adjacent to the
excluded region where $\tilde\tau^{\pm}_1$ is the LSP, which is
excluded from the fit because a stable charged LSP is in conflict with
cosmological measurements. The results also show that on the
$2\,\sigma$ level $M_{1/2}<800\,\mathrm{GeV}$ and
$M_{0}<600\,\mathrm{GeV}$ is expected, yielding relatively light
sparticles and hence good discovery prospects for the LHC. The
upper right plot shows the projection of $\tan\beta$ versus $M_0$,
showing that on the $2\,\sigma$ level $\tan\beta$ values are allowed
between 3 and 50. This means that in a full simultaneous fit of all
parameters $\tan\beta$ is not constrained strongly.  This implies that
a wide variety of different signatures is possible, as outlined in
more detail below. In the middle plot on the right side it can be seen
again that $A_0$ is not constrained very strongly, leaving a lot of
room for the phenomenology of the third generation. However, this has
only a small effect on the discovery prospects at LHC, since both
$\tan\beta$ and $A_0$ have limited effects on the first two
generations and the gauginos, which are most important for the
discovery modes at the LHC.

\begin{figure}
  \includegraphics[width=0.49\textwidth]{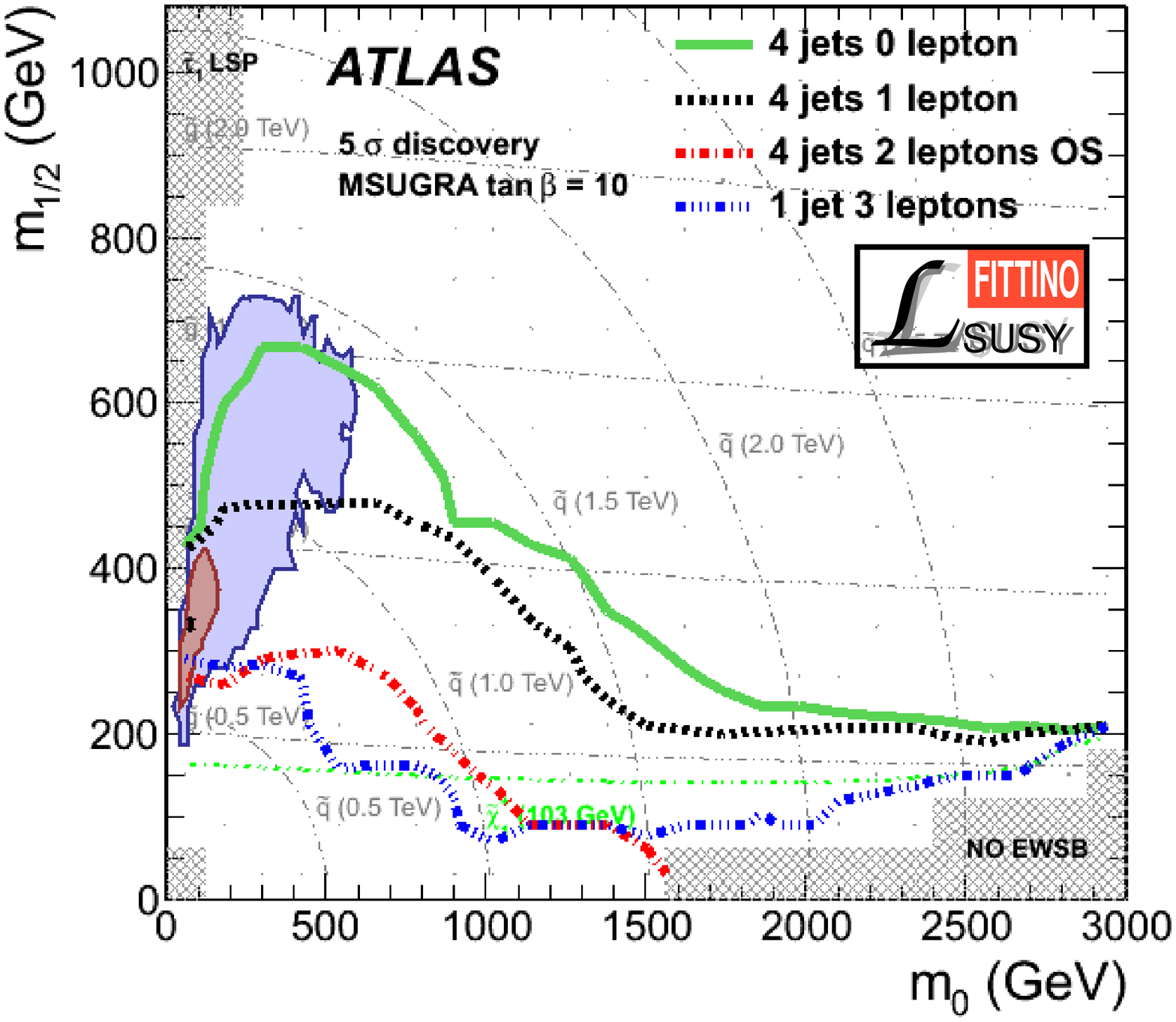}
  \caption{Allowed mSUGRA parameter space from
    Table~\ref{tab:results:LEonly:mSUGRA:all} overlaid upon the
    expected ATLAS discovery reach using $1~\mathrm{fb}^{-1}$ of data
    at $\sqrt{s} = 14$~TeV~\cite{Aad:2009wy}.}
  \label{fig:results:LEonly:mSUGRA:ATLASOverlay}
\end{figure}

The direct comparison of the allowed parameter space of the baseline
fit with the expected discovery reach of the ATLAS experiment is shown
in Figure~\ref{fig:results:LEonly:mSUGRA:ATLASOverlay}. The ATLAS
discovery reach plot is calculated for $\tan\beta=10$ and
$A_0=0\,\mathrm{GeV}$. Although the central values of the baseline fit
of these parameters do not exactly agree with these settings, it is
justified to compare the fit results with the ATLAS discovery reach
plot. This is the case because first the fixed values in the discovery
reach plot are within the uncertainties of the fit and second because
the discovery reach depends mainly on the gluino and first generation
squark masses. These are very insensitive to $A_0$ and not very
sensitive to $\tan\beta$ due to the absence of mixing on tree level.
The lines in Figure~\ref{fig:results:LEonly:mSUGRA:ATLASOverlay}
correspond to the boundaries of the $5\,\sigma$ discovery region. The
most sensitive search is expected to be an inclusive measurement of
the effective mass spectrum of 4 jets and missing
transverse energy. It can be seen that almost the entire mSUGRA
parameter space allowed at 95\,\%~CL is observable already with
1~fb$^{-1}$ of well-understood data at $\sqrt{s}=14\,\mathrm{TeV}$ of
the ATLAS experiment~\cite{Aad:2009wy}. 

\begin{figure}
  \includegraphics[width=0.49\textwidth]{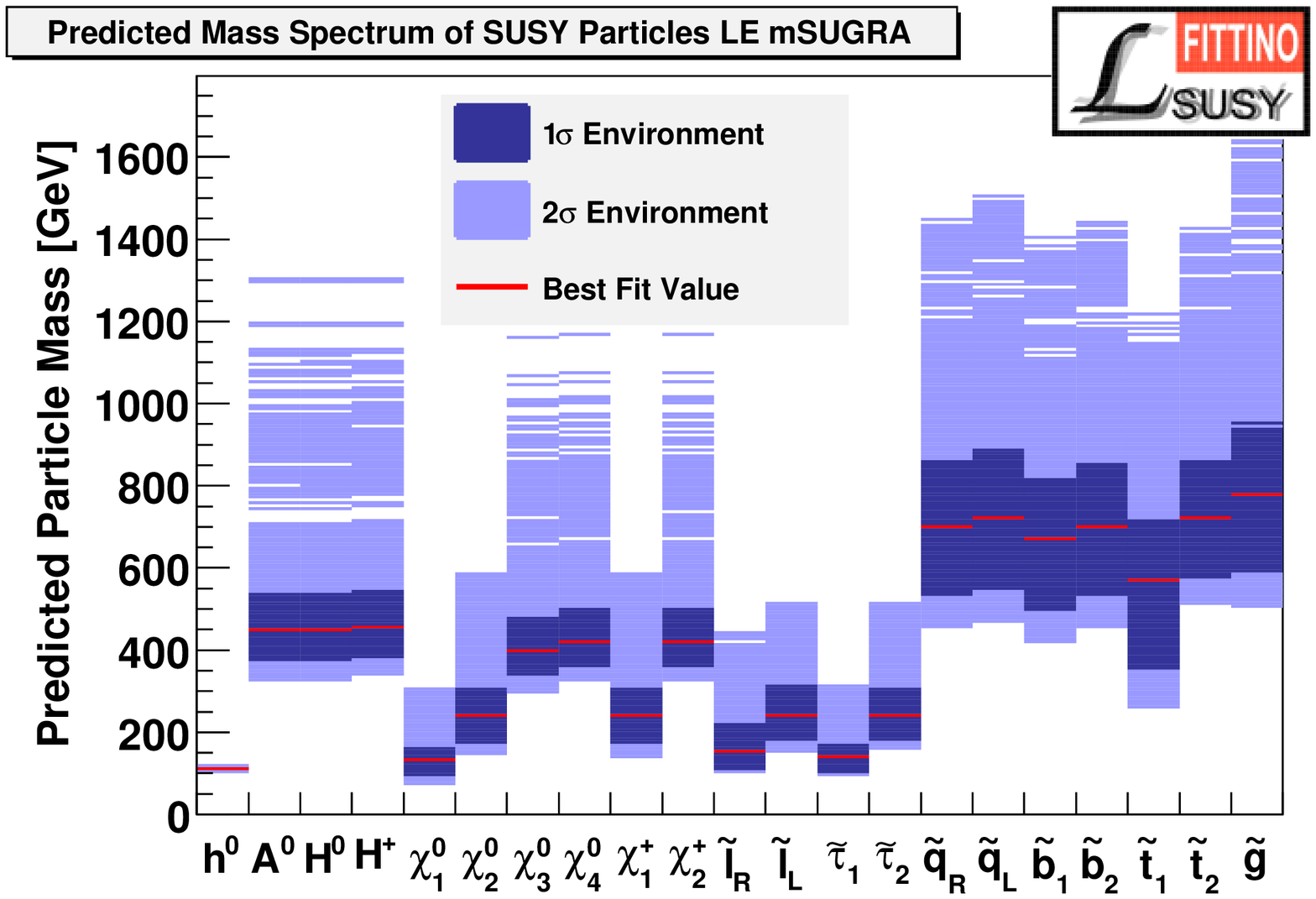}
  \caption{SUSY mass spectrum as predicted by mSUGRA parameter fit to low energy measurements with sign$(\mu)$ fixed to $+1$.}
  \label{fig:results:LEonly:mSUGRA:massDist:allObs}
\end{figure}

From the Markov Chain results, also the probability densities for 
the sparticle masses can be deduced. 
In Figure~\ref{fig:results:LEonly:mSUGRA:massDist:allObs}
the expected masses of the Higgs bosons and sparticles are shown 
for the baseline fit. 
The red lines indicate the masses
corresponding to the best fit parameter point. The dark (light) blue
regions denote the one-dimensional 68\,\% (95\,\%)~CL area,
corresponding to the $\Delta\chi^2<1$ ($\Delta\chi^2<4$) region around
the absolute minimum. The expected spectrum shows several distinct
features: The Higgs boson mass is well constrained to
$m_{h}=113.3\pm2.5\,\mathrm{GeV}$, just above the LEP exclusion bound. The
lightest SUSY particle (LSP) is the stable $\tilde{\chi}^0_1$ at
$m_{\tilde{\chi}^0_1}=130\pm35\,\mathrm{GeV}$ with a relatively small mass
difference to the next-to-lightest SUSY particle (NLSP)
$\tilde{\tau}_1$ at $m_{\tilde{\tau}_1}=140\pm25\,\mathrm{GeV}$, with
the mass difference between NLSP and LSP being rather precisely constrained:
$m_{\tilde{\tau}_1}-m_{\tilde{\chi}^0_1}=9.5\pm2.5\,\mathrm{GeV}$. If mSUGRA
is realised in Nature, this would provide challenging experimental
conditions for precision measurements at the LHC, since the very small
mass difference between NLSP and LSP would provide many final states with
very soft particles, dominated by $\tau$-leptons, as the last
reconstructable particles in the SUSY decay chain. In comparison with
the discovery reach from
Figure~\ref{fig:results:LEonly:mSUGRA:ATLASOverlay} it can be seen that
mSUGRA would allow for relatively easy discovery in inclusive
channels but difficult precision measurements in exclusive
reconstruction of SUSY cascades.
 
Furthermore, all light neutralinos, charginos and sleptons are
expected well below $m\approx600\,\mathrm{GeV}$. The squarks and the
gluino could be found below 1.6\,TeV. The heavy Higgs bosons are
expected below $m_{H/A}<1.2\,\mathrm{TeV}$ and would be challenging
to discover even with more than $10\,\mathrm{fb}^{-1}$ of luminosity
at the LHC, since $\tan\beta$ is not large enough~\cite{Aad:2009wy}. 

\begin{table}
  \caption{Result of the fit of the mSUGRA model with $\mathrm{sign}(\mu)=-1$ including 
    four additional SM parameters to all measurements 
    listed in Table~\ref{tab:leobserables}. The minimum $\chi^2$ value is 
    $31.1$ for 22 degrees of freedom, corresponding to a ${\cal P}$-value of 9.4\,\%.}
  \label{tab:results:LEonly:mSUGRA:negmu:all}
  \begin{center}
    {\renewcommand{\arraystretch}{1.2}
      \begin{tabular}{lrcl}
        \hline\hline
        Parameter     & Best Fit  &       & Uncertainty \\
        \hline
        sign$(\mu)$     &  $-$1       &       &             \\
        $\alpha_s$    &   0.1177   & $\pm$ &  0.0018  \\
        $1/\alpha_{em}$  &   127.924   & $\pm$ &  0.014  \\
        $m_Z$ (GeV)   &     91.188   & $\pm$ &  0.0022  \\
        $m_t$ (GeV)   &    172.5   & $\pm$ &  1.02     \\ 
        $\tan\,\beta$ &   9.6   &       &    $^{+17.8}_{-4.5}$             \\ 
        $M_{1/2}$ (GeV) &  125.1    &       &    $^{+70.0}_{-25.1}$              \\
        $M_0$ (GeV)   &   2313.5    &      &    $^{+622}_{-940}$  \\ 
        $A_{0}$ (GeV) &  $-$29.1    & $\pm$ &   2048                \\
        \hline\hline
      \end{tabular}
    }
  \end{center}
\end{table}

The baseline fit uses a fixed value of the discrete parameter
$\mathrm{sign}(\mu)=+1$, since this choice is preferred by the
positive deviation of the measured anomalous magnetic moment of the
muon $a_{\mu}^{\mathrm{exp}}$ from the SM prediction. However, the
difference in the overall agreement between the choice of
$\mathrm{sign}(\mu)=+1$ and $\mathrm{sign}(\mu)=-1$ has to be assessed
with a global fit to the same observables as used by the baseline fit,
but for $\mathrm{sign}(\mu)=-1$. The result of the this fit (again
using Markov Chains) is shown in
Table~\ref{tab:results:LEonly:mSUGRA:negmu:all}. Due to the tension in
$a_{\mu}^{\mathrm{exp}}-a_{\mu}^{\mathrm{SM}}$ of $3.4\,\sigma$, the
minimal $\chi^2$ of the fit rises to $31.09$, corresponding to a
${\cal P}$-value of 9.4\,\%, compared to 54.4\,\% for the baseline
fit. This result shows that $\mathrm{sign}(\mu)=-1$ is disfavoured but
not excluded.  The preferred parameter regions for both choices of
$\mathrm{sign}(\mu)=-1$ show similar preferred values for $\tan\beta$
and $M_{1/2}$.  For $\mathrm{sign}(\mu)=-1$, much larger values of
$M_0$ are preferred.

\begin{center}
  \begin{figure*}
    \begin{center}
      \includegraphics[width=0.32\textwidth]{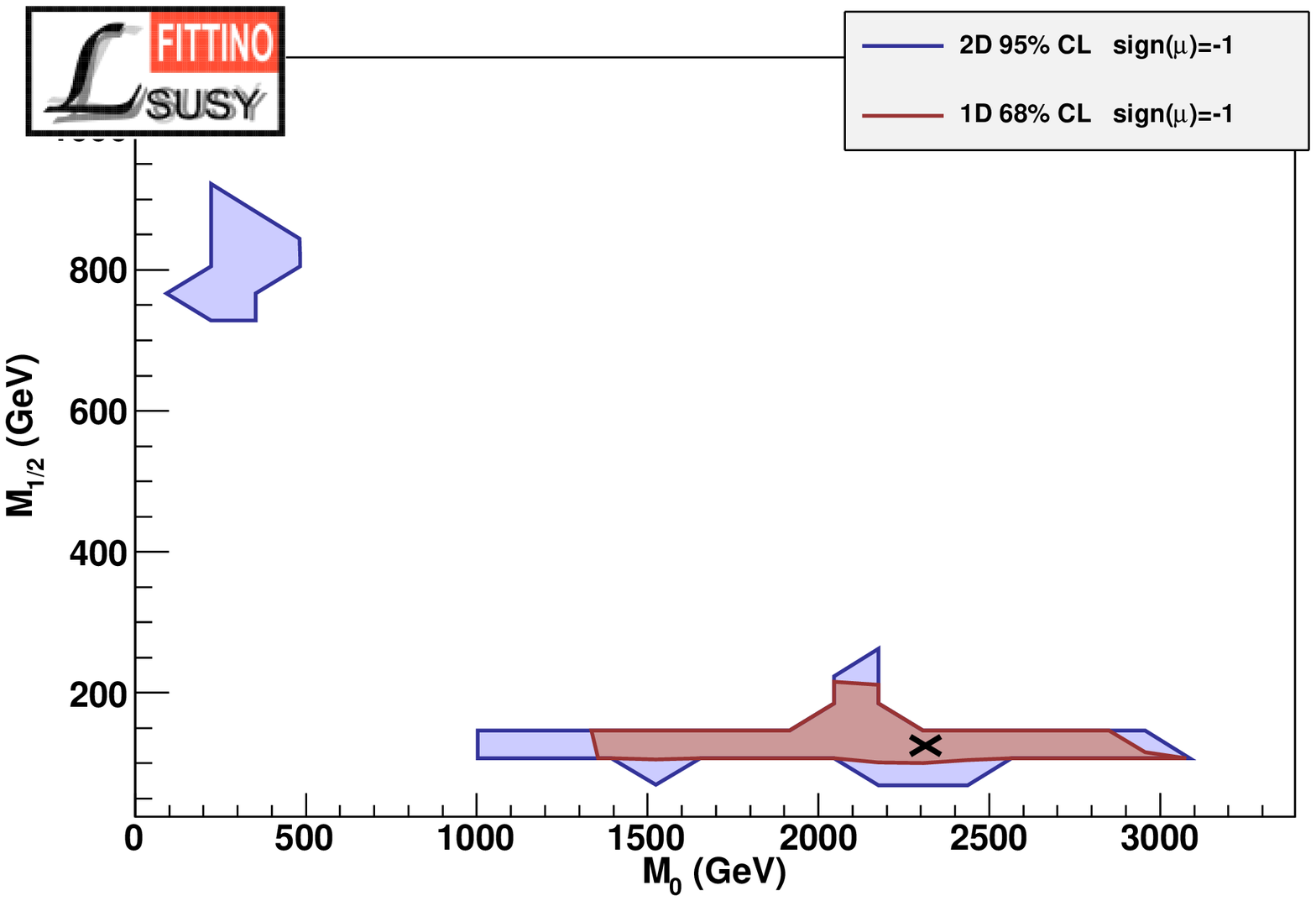}
      \hfill
      \includegraphics[width=0.32\textwidth]{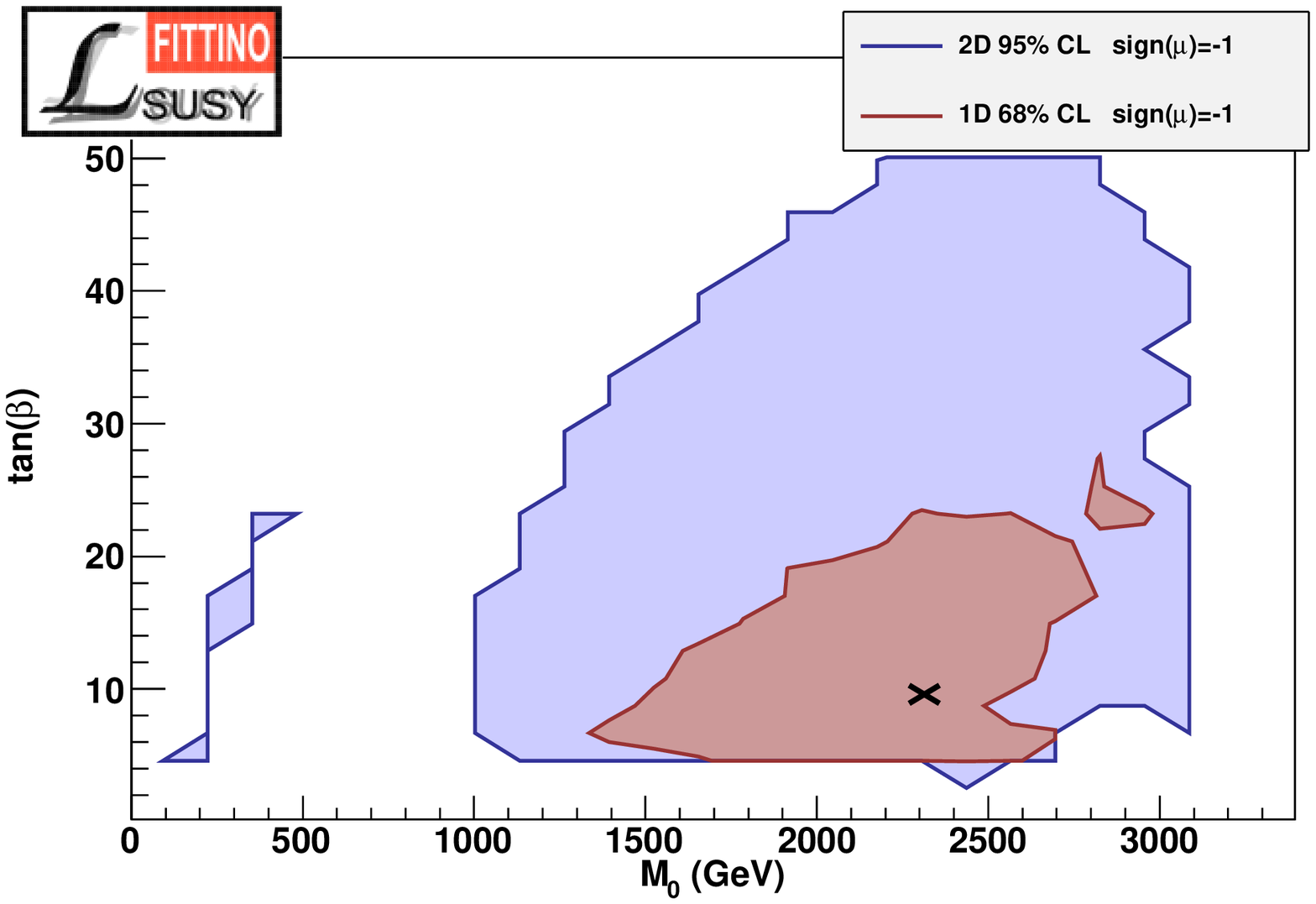}
      \hfill
      \includegraphics[width=0.32\textwidth]{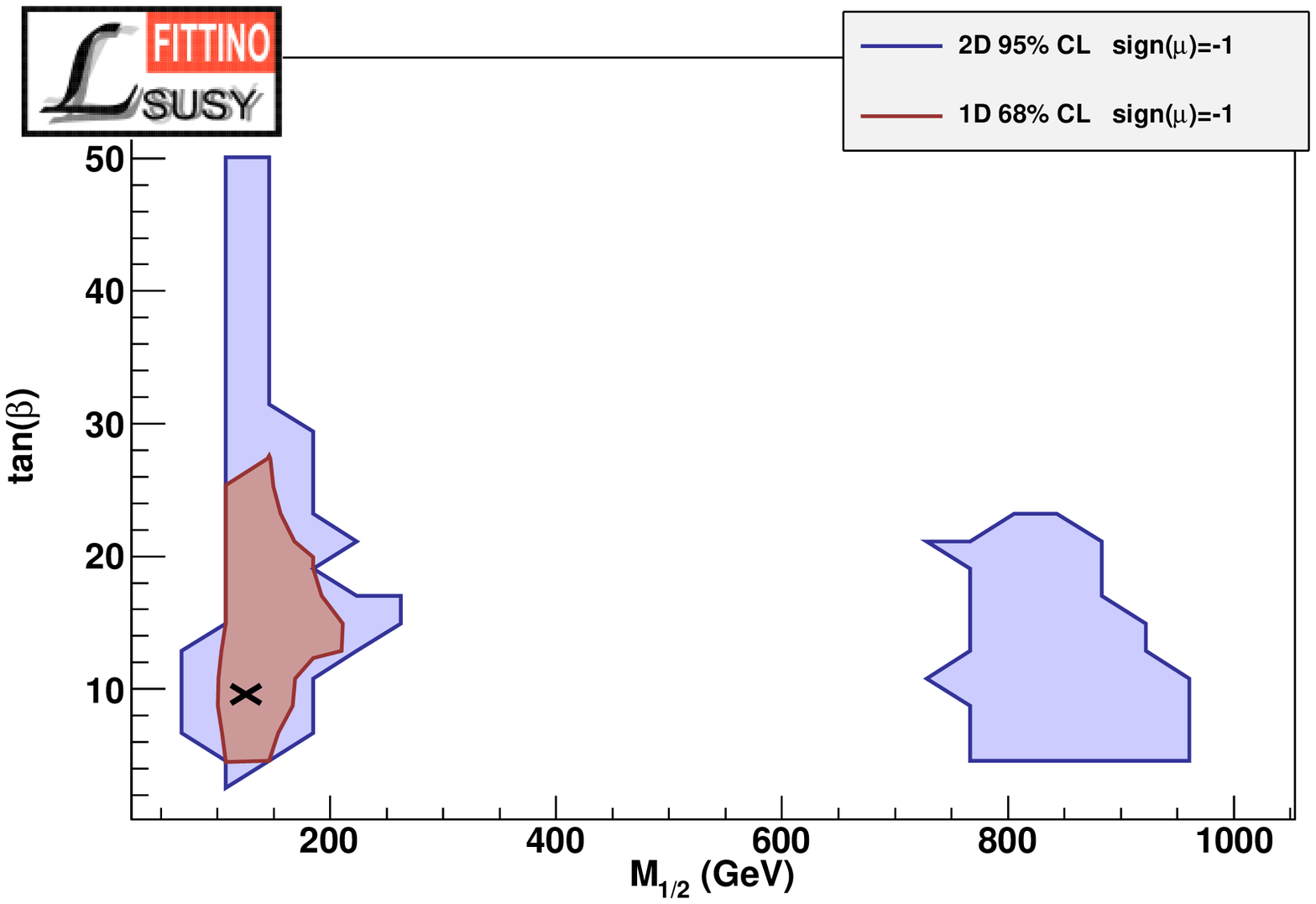}
      \\[1mm]
      \includegraphics[width=0.32\textwidth]{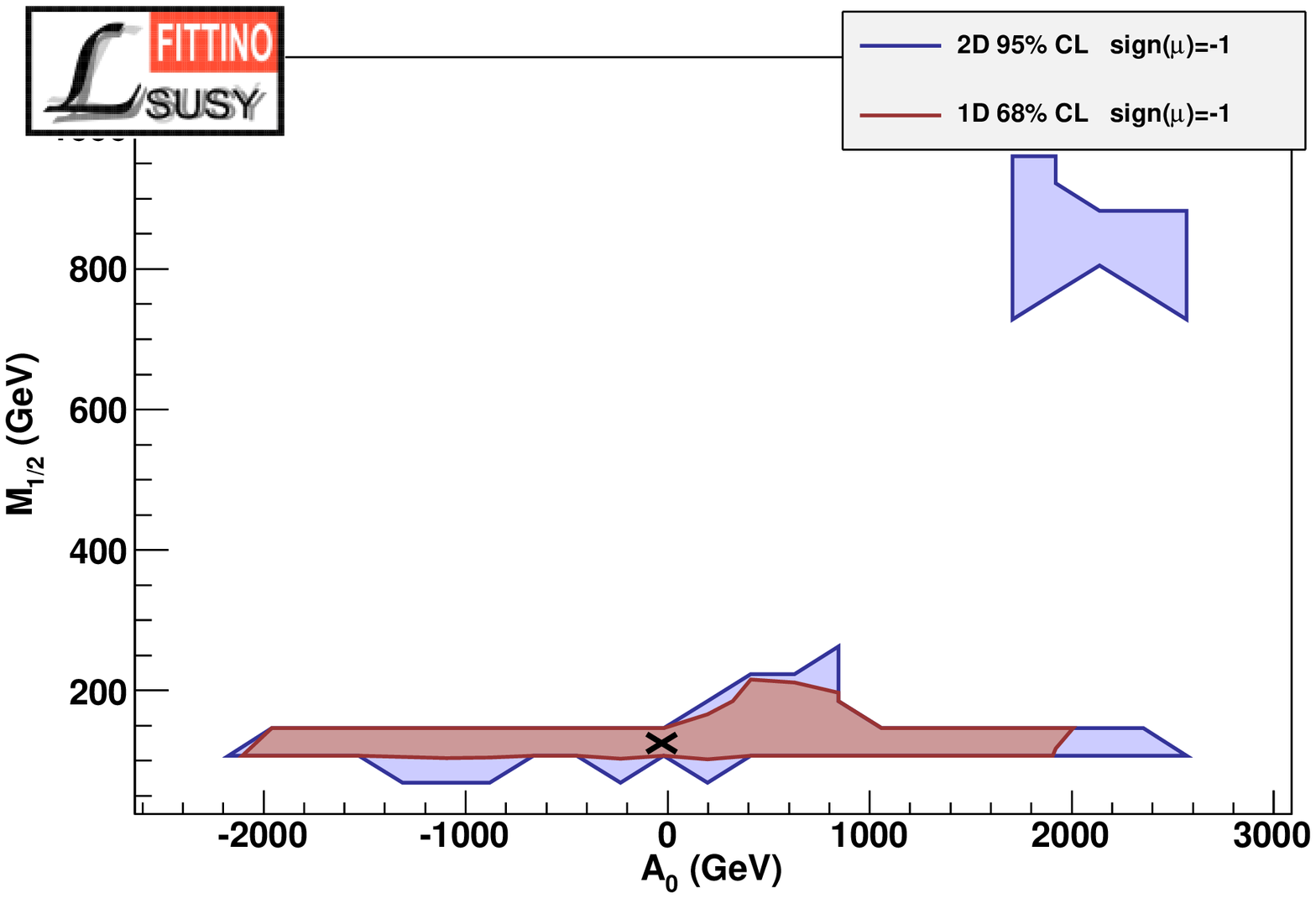}
      \hfill
      \includegraphics[width=0.32\textwidth]{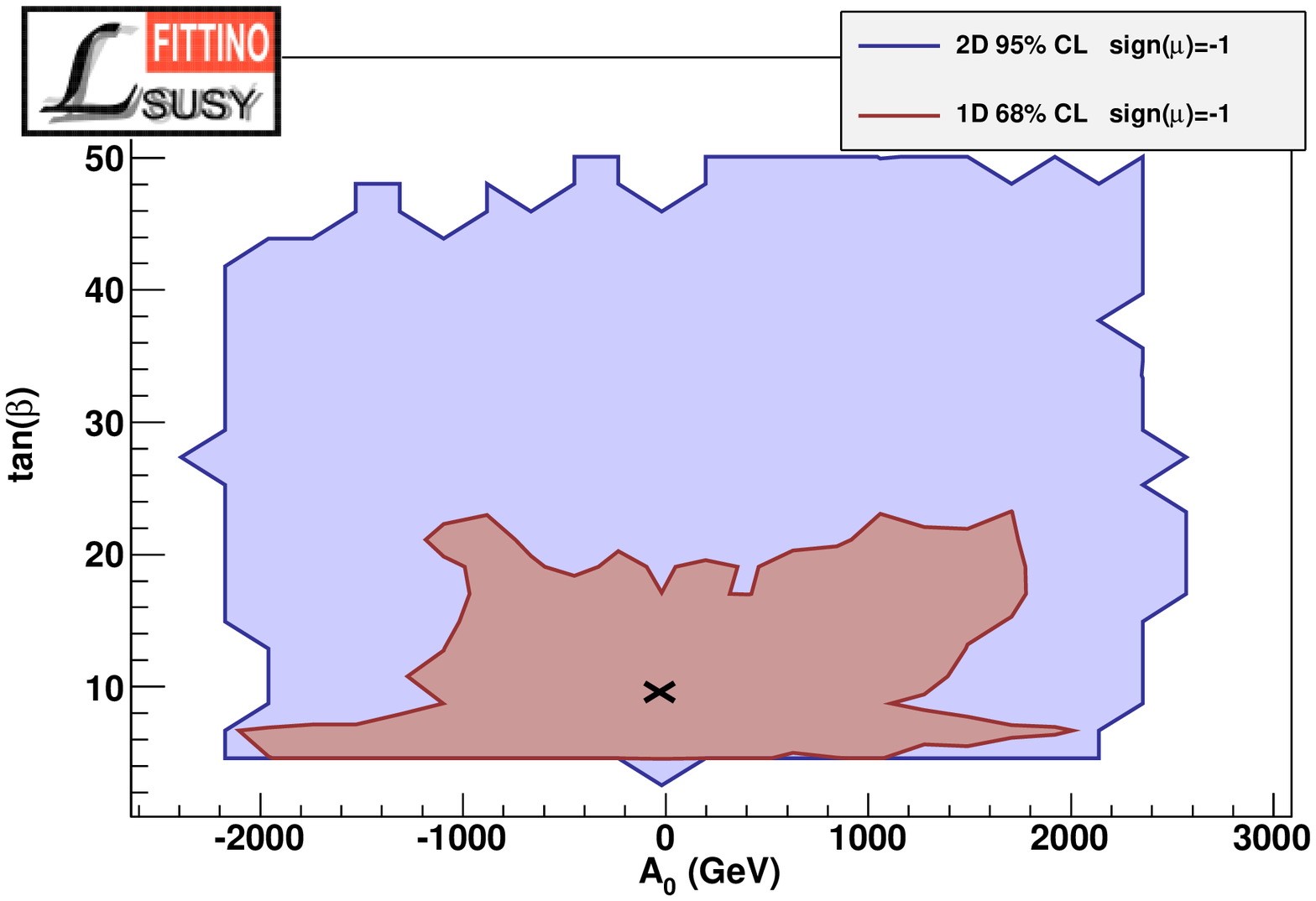}
      \hfill
      \includegraphics[width=0.32\textwidth]{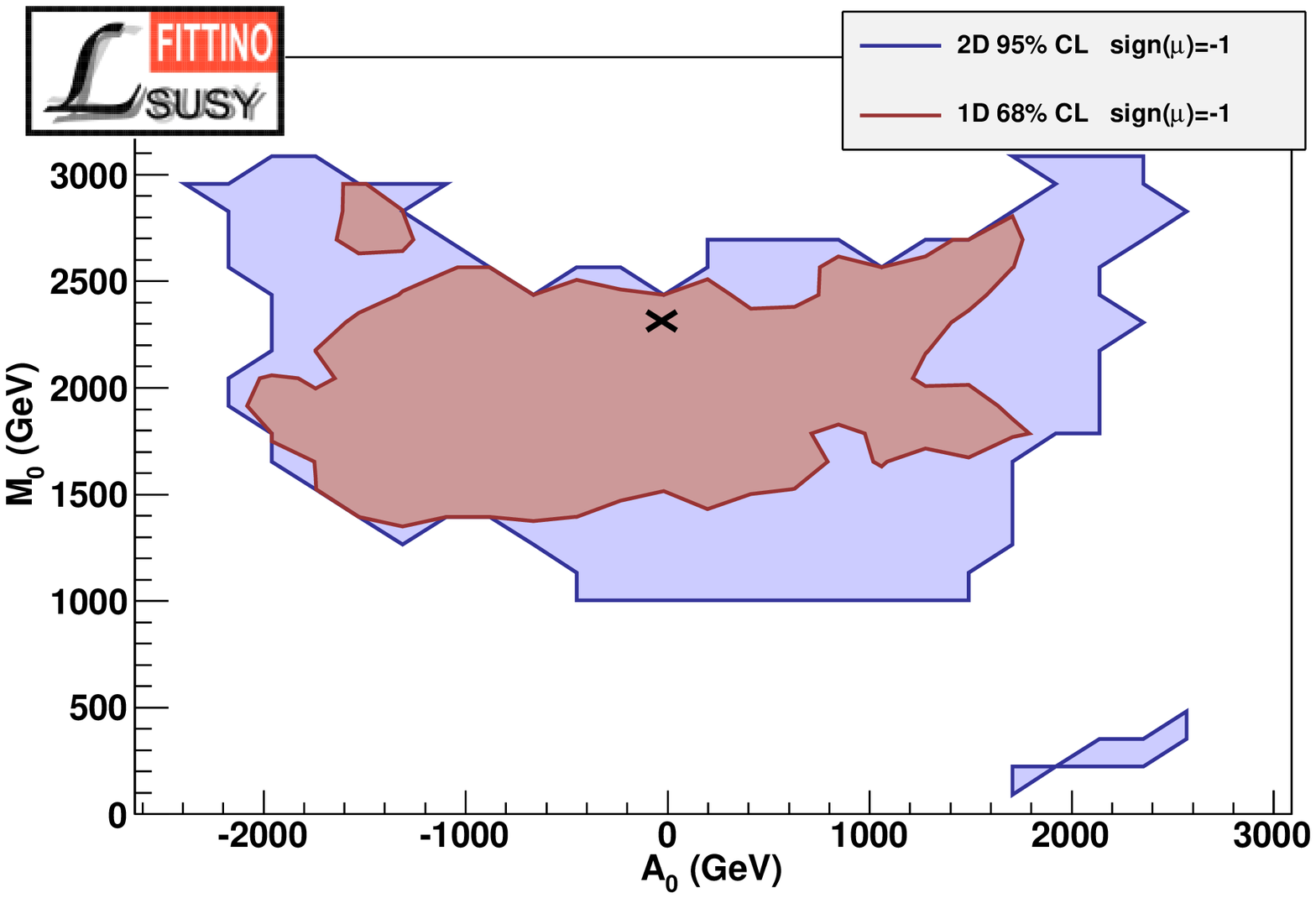}
    \end{center}
    \caption{mSUGRA parameter regions compatible with the relic density
      constraint and with all low energy measurements for various
      parameter combinations and sign$(\mu)$ fixed to $-1$. For both
      cases two-dimensional 95 \% confidence level and one-dimensional 68
      \% confidence regions are shown. From the latter $1\sigma$
      uncertainties for individual parameters can be derived from a
      projection of the area to the respective axis.}
    \label{fig:results:LEonly:mSUGRA:2dRes:negmu}
  \end{figure*}
\end{center}

The two-dimensional projections of the mSUGRA parameters for
$\mathrm{sign}(\mu)=-1$ are given in
Figure~\ref{fig:results:LEonly:mSUGRA:2dRes:negmu}. In contrast to the
baseline fit, where one continuous two-dimensional 95\,\%~CL area is
observed, two distinct areas are found for $\mathrm{sign}(\mu)=-1$. One
is located at large $M_{1/2}$ and small $M_{0}$, while the other one
exhibits the opposite signature. The reason is that
in both regions the negative SUSY contribution to $(g-2)_{\mu}$ is reduced,
while a correct contribution to the cold dark matter
relic density is preserved. The region of both large $M_{1/2}$ and $M_{0}$ is
therefore cut out by the constraint of $\Omega_{\mathrm{CDM}}h^2$. 

\begin{figure}
  \includegraphics[width=0.49\textwidth]{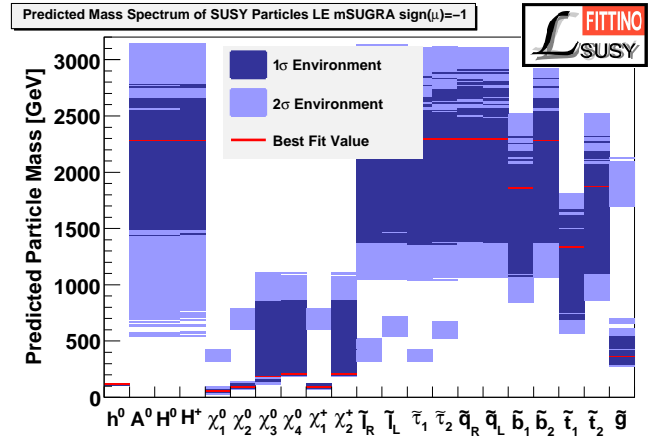}
  \caption{SUSY mass spectrum as predicted by mSUGRA parameter fit to low energy measurements with sign$(\mu)$ fixed to $-1$.}
  \label{fig:results:LEonly:mSUGRAnegmu:massDist:allObs}
\end{figure}

As before, the result can also be expressed in terms of the preferred
sparticle masses.
Fig~\ref{fig:results:LEonly:mSUGRAnegmu:massDist:allObs} shows the
best fit and the 68\,\% and 95\,\%~CL areas of the observable
masses for the mSUGRA fit with $\mathrm{sign}(\mu)=-1$. Since large
$M_0$ and small $M_{1/2}$ is preferred over large $M_{1/2}$ and small
$M_{0}$, as visible in
Figure~\ref{fig:results:LEonly:mSUGRA:2dRes:negmu}, the expected gaugino
masses in this scenario are small. With
$m_{\tilde{\chi}^0_1}=55^{+15}_{-10}\,\mathrm{GeV}$ a very light LSP is
predicted just above the LEP mSUGRA limit of
$m_{\tilde{\chi}^0_1}>47\,\mathrm{GeV}$~\cite{Heister:2003zk}.
In contrast to the baseline fit, heavy sleptons and squarks are
expected due to the large value of $M_{0}$. The gluino is predicted to
be lighter than all squarks and sleptons such that it decays via
three-body decays to two jets and the LSP.

No tension between the predicted lightest Higgs boson mass with the LEP
limits is present for $\mathrm{sign}(\mu)=-1$. The obtained Higgs mass value
is $m_{h}=118.3\pm3\,\mathrm{GeV}$ well above the LEP bound at
$m_{h}>114.4,\mathrm{GeV}$ and also above the preferred LEP Higgs mass
of $m_{h}\approx116.5\,\mathrm{GeV}$.

\begin{table}
  \caption{Expected branching fractions of the SUSY particles in the
    mSUGRA model with $\mathrm{sign}(\mu)=+1$. The results shown for the
    first generation are also valid for the second generation. }
  \label{tab:results:LEonly:mSUGRA:BRexpectations}
  \begin{center}
    {\renewcommand{\arraystretch}{1.3}
      \begin{tabular}{lrl}
        \hline\hline
        Decay Mode                     & Expected Branching Fraction  & Uncertainty \\
        \hline
        $\tilde{\chi}^0_2\to\tilde{\tau}_1 \tau$ & 0.46                         & $^{+0.38}_{-0.44}$    \\
        $\tilde{\chi}^0_2\to\tilde{\nu}_{\tau_1} \nu_{\tau}$ & 0.076            & $^{+0.039}_{-0.067}$  \\
        $\tilde{\chi}^0_2\to\tilde{e}_R e$       & 0.040                        & $^{+0.044}_{-0.038}$  \\
        $\tilde{\chi}^0_2\to\tilde{\chi}^0_1 h$      & 0.036                        & $^{+0.13}_{-0.035}$   \\
        $\tilde{\chi}^0_2\to\tilde{\chi}^0_1 Z$      & 0.018                        & $^{+0.0098}_{-0.018}$ \\
        $\tilde{\chi}^0_2\to\tilde{e}_L e$       & 0.00018                      & $^{+0.14}_{-0.00018}$ \\
        $\tilde{\chi}^0_2\to\tilde{\tau}_2 \tau$ & 0.                           & $^{+0.014}_{-0}$      \\
        $\tilde{\chi}^{\pm}_1\to\tilde{\tau}_1\nu_{\tau}$ & 0.40                & $^{+0.42 }_{-0.39}$  \\
        $\tilde{\chi}^{\pm}_1\to\tilde{\nu}_{e_L} e$      & 0.15                & $^{+0.10 }_{-0.15}$  \\
        $\tilde{\chi}^{\pm}_1\to\tilde{\nu}_{\tau_1} \tau$& 0.15                & $^{+0.10 }_{-0.15}$  \\
        $\tilde{\chi}^{\pm}_1\to\tilde{\chi}^0_1 W^{\pm}$& 0.12                         & $^{+0.079}_{-0.12}$  \\
        $\tilde{\chi}^{\pm}_1\to\tilde{\tau}_2\nu_{\tau}$ & 0                   & $^{+0.14 }_{-0   }$    \\
        $\tilde{g}\to\tilde{u}_L u$               & 0.052               & $^{+0.020}_{-0.023}$           \\
        $\tilde{g}\to\tilde{u}_R u$               & 0.094               & $^{+0.030}_{-0.054}$           \\
        $\tilde{g}\to\tilde{d}_L d$               & 0.051               & $^{+0.018}_{-0.021}$           \\
        $\tilde{g}\to\tilde{d}_R d$               & 0.093               & $^{+0.031}_{-0.055}$           \\
        $\tilde{g}\to\tilde{t}_1 t$               & 0.090               & $^{+0.166}_{-0.066}$           \\
        $\tilde{g}\to\tilde{t}_2 t$               & 0                   & $^{+0.056}_{-0    }$          \\
        $\tilde{g}\to\tilde{b}_1 b$               & 0.179               & $^{+0.022}_{-0.056}$           \\
        $\tilde{g}\to\tilde{b}_2 b$               & 0.11                & $^{+0.009}_{-0.038}$            \\
        $\tilde{e}_R\to\tilde{\chi}^0_1 e$                & 1                   & $^{+0    }_{-0   }$                \\
        $\tilde{e}_L\to\tilde{\chi}^0_1 e$                & 1                   & $^{+0    }_{-0.23}$           \\
        $\tilde{e}_L\to\tilde{\chi}^0_2 e$                & 0                   & $^{+0.085}_{-0   }$           \\
        $\tilde{\tau}_1\to\tilde{\chi}^0_1 \tau$          & 1                   & $^{+0    }_{-0     }$              \\
        $\tilde{\tau}_2\to\tilde{\chi}^0_1 \tau$          & 0.99                & $^{+0.011}_{-0.42  }$    \\
        $\tilde{\tau}_2\to\tilde{\chi}^0_2 \tau$          & 0.0007              & $^{+0.040}_{-0.0007}$    \\
        $\tilde{\tau}_2\to\tilde{\chi}^{\pm}_1 \nu_{\tau}$ & 0.0044          & $^{+0.078}_{-0.0043}$    \\
        $\tilde{u}_L\to\tilde{\chi}^0_1 u$               & 0.008               &  $^{+0.0049}_{-0.0026}$  \\
        $\tilde{u}_L\to\tilde{\chi}^0_2 u$               & 0.31                &  $^{+0.013 }_{-0.0061}$ \\
        $\tilde{u}_L\to\tilde{\chi}^{\pm}_1 d$           &  0.64               &  $^{+0.019 }_{-0.0094}$  \\
        $\tilde{u}_L\to\tilde{\chi}^{\pm}_2 d$           &    0.025            &  $^{+0.0092}_{-0.018 }$   \\
        $\tilde{u}_R\to\tilde{\chi}^0_1 u$               & 0.99                &  $^{+0.012 }_{-0.0080}$  \\
        $\tilde{u}_R\to\tilde{\chi}^0_2 u$               & 0.0078              &  $^{+0.0064}_{-0.0072}$  \\
        $\tilde{t}_1\to\tilde{\chi}^0_1 t$               & 0.18                &  $^{+0.073}_{-0.037}$   \\
        $\tilde{t}_1\to\tilde{\chi}^0_2 t$               & 0.15                &  $^{+0.006}_{-0.069}$      \\
        $\tilde{t}_1\to\tilde{\chi}^{\pm}_1 b$           & 0.51                &  $^{+0.21 }_{-0.28 }$   \\
        $\tilde{t}_1\to\tilde{\chi}^{\pm}_2 b$           & 0.15                &  $^{+0.11 }_{-0.15 }$   \\
        $\tilde{b}_1\to\tilde{\chi}^{\pm}_1 t$           & 0.37                &  $^{+0.13}_{-0.039}$  \\
        $\tilde{b}_1\to\tilde{\chi}^{0}_2 b$             & 0.25                &  $^{+0.12}_{-0.068}$  \\
        $\tilde{b}_1\to\tilde{t}_1 W^{\pm}$            &  0.043              &  $^{+0.29}_{-0.043}$  \\   
        \hline\hline
      \end{tabular}
    }
  \end{center}
\end{table}

In addition to the mass spectrum and the important mass differences,
the expected dominant sparticle decay modes can be studied. A
selection of branching fractions of the mSUGRA parameter points
preferred by the baseline fit is shown in
Table~\ref{tab:results:LEonly:mSUGRA:BRexpectations}.  The central
values correspond to the best fit, the uncertainties are given by
$\Delta\chi^2<1$ with respect to the best fit. It can be generally
observed that the uncertainties of the expected branching fractions
are very large, often close to 100\,\%. Also, there are only few
sparticles with only one relevant decay mode. Therefore a rich
phenomenology with many competing decay modes can be expected at the
LHC.  Branching fractions of the electro-weak gauginos are typically
largest for the third generation, with smaller contributions from the
first and second generation. In connection with the small mass
difference between $m_{\tilde{\tau}_1}$ and $m_{\tilde{\chi}^0_1}$, the larger
branching fractions into $\tau$ leptons compared to other leptons
represent a challenge for the measurement of ratios of branching
fractions and di-tau mass endpoints.

Following the expected decay chains, the gluino decays into squarks
with a slight but not dominant preference for
$\tilde{b}_{1/2}$. Generally, the branching fractions into the
spartners of the right-handed degrees of freedom are larger by
almost a factor of 2 than the branching fractions into the left-handed
counterparts. The right squark $\tilde{q}_R$ decays into $\tilde{\chi}^0_1q$
almost exclusively. As expected, the decay of the $\tilde{q}_L$ is
predicted to be more complex with decays into $\tilde{\chi}^0_2q$ and
$\tilde{\chi}^{\pm}_1q'$. The chargino has a preference for decays into
$\tilde{\tau}_1\nu_{\tau}$, but includes significant contributions of
decays into sneutrinos and into $\tilde{\chi}^0_1W^{\pm}$. The $\tilde{\chi}^0_2$ has
a similarly rich spectrum of decays, with a dominance of
$\tilde{\tau}_1\tau$, but with small contributions of sneutrinos,
other leptons and $\tilde{\chi}^0_1 h^{0}$ and $\tilde{\chi}^0_1 Z^{0}$. The sleptons
exhibit branching fractions of close to 100\,\% into $\tilde{\chi}^0_1$ and
the corresponding lepton.

The results of the baseline mSUGRA fits with $\mathrm{sign}(\mu)=+1$ and
$-1$ provide clear predictions for the expected measurements at
the LHC. However, there are several interesting questions which
remain. First, it remains to be assessed which measured observables
constrain the parameter space and the regions of LHC observables in
which way, i.~e.~which features of the measurements dominate the
prediction of parameters and future observables. Second, as outlined
in Section~\ref{sec:leobservables}, the interpretation of some of the
measurements in terms of mSUGRA is not necessarily unique. Third,
alternative SUSY breaking models may predict other features for
the LHC. These questions are addressed in the next sections.

%%%%%%%%%%%%%%%%%%%%%%%%%%%%%%%%%%%%%%%%%%%%%%%%%%%%%%%%%%%%%%%%%%%%%%%
%%%%%%%%%%%%%%%%%%%%%%%%%%%%%%%%%%%%%%%%%%%%%%%%%%%%%%%%%%%%%%%%%%%%%%%
%%%%%%%%%%%%%%%%%%%%%%%%%%%%%%%%%%%%%%%%%%%%%%%%%%%%%%%%%%%%%%%%%%%%%%%
\subsubsection{Fits of mSUGRA to Low-Energy Observables with reduced Sets of Observables}\label{sec:results:LEwithout}

In this section, the baseline fit with $\mathrm{sign}(\mu)=+1$ 
is modified by fitting the same parameters to
reduced sets of observables. As described in
Section~\ref{sec:leobservables}, both $\Omega_{\mathrm{CDM}}h^2$ and
$(g-2)_{\mu}$ suffer from uncertainties concerning their
interpretation in terms of SUSY. In this section we show that these
observables provide the strongest constraints in the parameter space,
therefore the uncertainty in the interpretation of these observables
has to be evaluated for the predicted parameter space and collider
observables. 

For $\Omega_{\mathrm{CDM}}h^2$ there is very little doubt about the
measurement itself.  However, the cosmological measurement of the cold
dark matter relic density does not imply that the SUSY LSP is solely
responsible for the dark matter.  Therefore, we study three different
possibility: First, a stable neutral SUSY LSP in the context of a
$R$-parity conserving model is the only source of
$\Omega_{\mathrm{CDM}}h^2$, and the process of the LSP production
after the Big Bang and the freeze-out of the LSP is completely
understood, as e.~g.~implemented in~\cite{Belanger:2006is}. This is
assumed for the baseline fit in
Section~\ref{sec:results:LEonly:mSUGRA}. Second, the SUSY LSP
contributes to dark matter, but other unknown additional sources of
dark matter are not excluded. This scenario is tested in this section
by requiring that the LSP is stable and neutral and that the predicted
cold dark matter relic density is smaller or equal than the observed
one $\Omega_{\mathrm{theo}}h^2\leq\Omega_{\mathrm{obs}} h^2$. This
still includes the assumptions that the mechanisms of CDM creation are
understood. Third, additional features like a CDM creation different
from the current understanding, or a meta-stable LSP, could cause
differences between the measured CDM relic density and the predicted
CDM relic density. The maximal effect of such uncertainties are tested
by removing $\Omega_{\mathrm{CDM}}h^2$ from the observables in the
fit.

Similarly, three different scenarios can be distinguished concerning
$(g-2)_{\mu}$, where there is an ongoing debate about uncertainties on
the SM prediction (see
e.~g.~\cite{Stockinger:2008zz,Passera:2009zz,Jegerlehner:2009ry,Davier:2009ag}).
Therefore, in addition to the baseline fit, which uses the current
mean value of measurements and the SM and SUSY prediction
from~\cite{Moroi:1995yh}, deducing the hadronic corrections from
$e^+e^-$ collision data, three other options are tested. First, in
order to show the importance of the $(g-2)_{\mu}$ measurement, it is
removed from the observables used in the fit.  Second, it is assumed
for illustration that the current deviation between SM prediction and
measurement is a statistical deviation or a insufficiency in the SM
prediction, i.~e.~that there is no visible SUSY contribution to
$(g-2)_{\mu}$. Third, the prediction deducing the hadronic corrections
from $\tau$ decay data~\cite{Davier:2009ag} is used. Several
combinations of the above mentioned possibilities are studied in the
following.

\begin{table}
  \caption{Result of the fit of the mSUGRA model with $\mathrm{sign}(\mu)=+1$ including 
    four additional SM parameters to all measurements 
    listed in Table~\ref{tab:leobserables} except $\Omega_{\mathrm{CDM}} h^2$, for which only 
    $\Omega_{\mathrm{\mathrm{theo}}} h^2\leq\Omega_{\mathrm{obs}} h^2$ is required. The minimum $\chi^2$ value is 
    $20.4$ for 21 degrees of freedom, corresponding to a ${\cal P}$-value of 49.3\,\%.}
  \label{tab:results:LEonly:mSUGRA:smallerOmega}
  \begin{center}
    {\renewcommand{\arraystretch}{1.2}
      \begin{tabular}{lrcl}
        \hline\hline
        Parameter     & Best Fit  &       & Uncertainty \\
        \hline
        sign$(\mu)$        &  +1       &       &             \\
        $\alpha_s$       &    0.1176  & $\pm$ &  0.0025       \\
        $1/\alpha_{em}$  & 127.925    & $\pm$ &  0.020        \\
        $m_Z$ (GeV)      &   91.1866   & $\pm$ &  0.0021      \\
        $m_t$ (GeV)      &  172.2    & $\pm$ &  1.1      \\ 
        $\tan\,\beta$    &    9.0     &       & $_{-3.6}^{+11.4}$               \\ 
        $M_{1/2}$ (GeV)   & 303.4       &       & $^{+133}_{-59.0}$               \\
        $M_0$ (GeV)      &   27.6      &      & $^{+122}_{-27.5}$        \\ 
        $A_{0}$ (GeV)    &  143.2       &      &  $^{+850}_{-478}$      \\
        \hline\hline
      \end{tabular}
    }
  \end{center}
\end{table}

Since the composition of the cosmos and its dynamics are a very
dynamic field of study, it is a strong assumption that only the LSP
contributes to cold dark matter. Therefore, a fit of mSUGRA with the preferred
choice of $\mathrm{sign}(\mu)=+1$ is performed requiring
$\Omega_{\mathrm{theo}} h^2\leq\Omega_{\mathrm{obs}} h^2$ and leaving
all other observables unchanged. In this fit, $\Omega_{\mathrm{theo}}
h^2$ does not contribute to the $\chi^2$ for $\Omega_{\mathrm{theo}}
h^2\leq\Omega_{\mathrm{obs}} h^2$ and contributes
$(\Omega_{\mathrm{theo}} h^2-\Omega_{\mathrm{obs}}
h^2)/\sigma_{\Omega_{\mathrm{CDM}}h^2}$ for $\Omega_{\mathrm{theo}}
h^2>\Omega_{\mathrm{obs}} h^2$. The result of this fit is shown in
Table~\ref{tab:results:LEonly:mSUGRA:smallerOmega}. The overall
$\chi^2$ of the fit improves marginally with respect to the baseline
fit. The preferred regions of $\tan\beta$, $M_0$ and $A_0$ move to
slightly smaller values, but the variations are not large compared to
the uncertainties. $M_0$ remains almost unchanged. The uncertainties
increase with respect to the baseline fit, but the order of
magnitude remains the same. The two-dimensional 95\,\%~CL areas of
the parameter projections are increasing, but no qualitatively new
features are observed, as seen in the overlay of parameter regions
with the baseline fit in Figure~\ref{fig:results:LEonly:mSUGRA:2dRes}.
Therefore, for this modification of the baseline fit, the predicted
collider observables do not differ significantly from the baseline
fit.

\begin{figure}
  \includegraphics[width=0.49\textwidth]{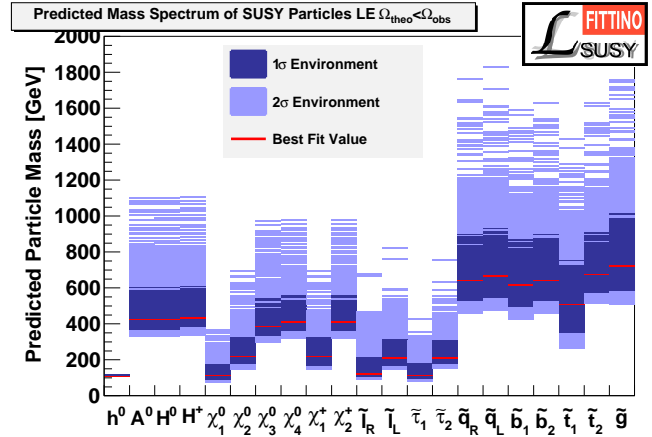}
  \caption{SUSY mass spectrum as predicted by mSUGRA parameter fit to low energy measurements, requiring $\Omega_{\mathrm{theo}} h^2\leq\Omega_{\mathrm{obs}} h^2$ with sign$(\mu)$ fixed to $+1$.}
  \label{fig:results:LEonly:mSUGRA:massDist:smallerOmega}
\end{figure}

In Figure~\ref{fig:results:LEonly:mSUGRA:massDist:smallerOmega}, the
predicted Higgs boson and sparticle mass ranges are shown. The
observed mass ranges are very similar to those observed in
Figure~\ref{fig:results:LEonly:mSUGRA:massDist:allObs}. While the
overall range of masses remains similar, there are subtle changes in
the mass differences, which are explained below.

\begin{table}
  \caption{Result of the fit of the mSUGRA model with $\mathrm{sign}(\mu)=+1$ including 
    four additional SM parameters to all measurements 
    listed in Table~\ref{tab:leobserables} except $\Omega_{\mathrm{CDM}} h^2$. The minimum $\chi^2$ value is 
    $20.4$ for 21 degrees of freedom, corresponding to a ${\cal P}$-value of 49.3\,\%.}
  \label{tab:results:LEonly:mSUGRA:noOmega}
  \begin{center}
    {\renewcommand{\arraystretch}{1.2}
      \begin{tabular}{lrcl}
        \hline\hline
        Parameter     & Best Fit  &       & Uncertainty \\
        \hline
        sign$(\mu)$     &  +1       &       &        \\
        $\alpha_s$    &     0.1177       & $\pm$ &  0.0019      \\
        $1/\alpha_{em}$ &   127.924         & $\pm$ &  0.014       \\
        $m_Z$ (GeV)   &    91.1870         & $\pm$ &  0.0022      \\
        $m_t$ (GeV)   &   172.2         & $\pm$ &  0.83        \\ 
        $\tan\,\beta$ &     10.9       &       &  $_{-5.1}^{+9.9}$      \\ 
        $M_{1/2}$ (GeV) &    316.2        &       &  $_{-69.7}^{+122.9}$      \\
        $M_0$ (GeV)   &     45.1       &       &  $_{-43.8}^{+119.3}$      \\ 
        $A_{0}$ (GeV) &    209.1        &       &  $_{-494.1}^{+973.5}$      \\
        \hline\hline
      \end{tabular}
    }
  \end{center}
\end{table}

\begin{figure}
  \includegraphics[width=0.49\textwidth]{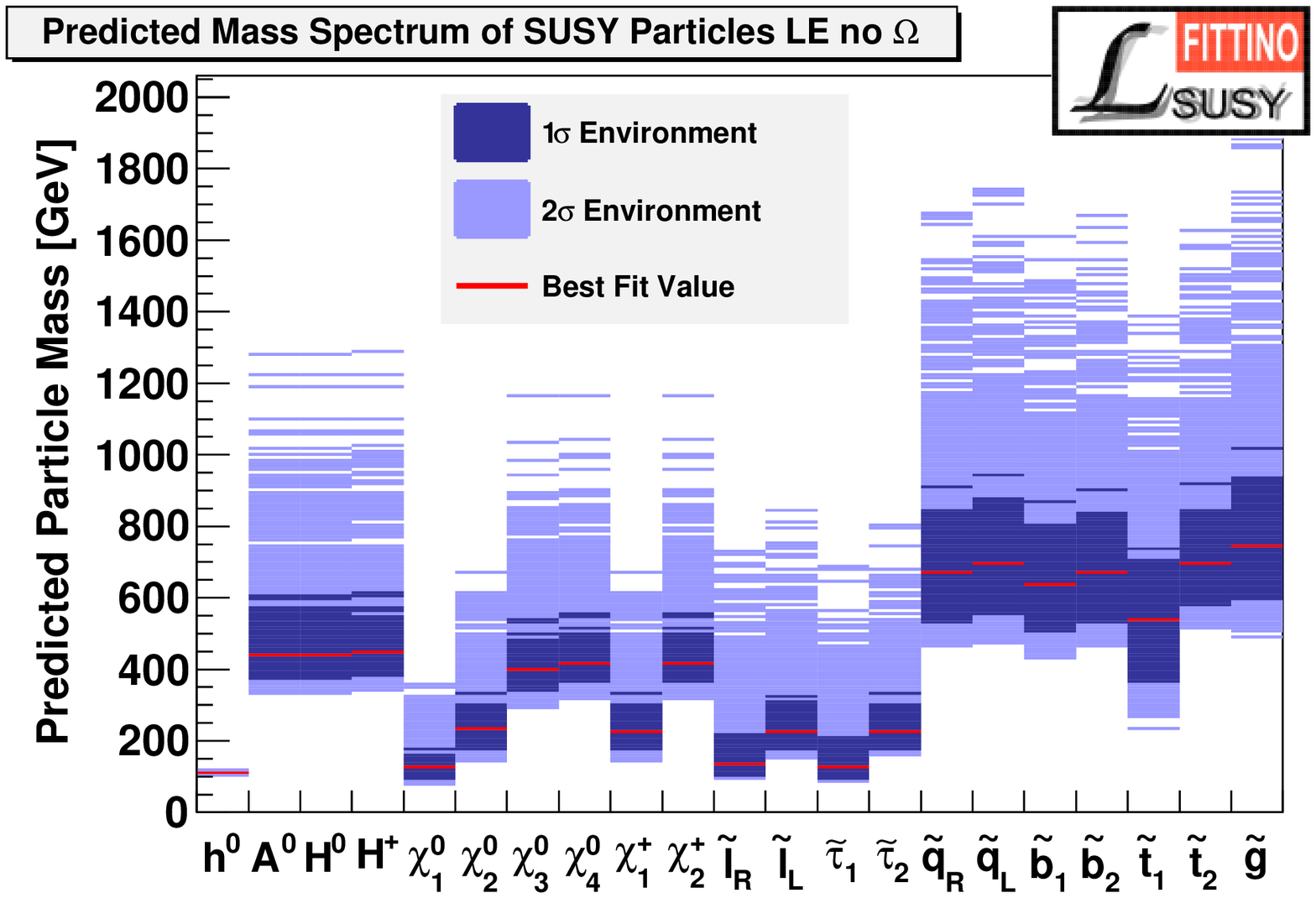}
  \caption{SUSY mass spectrum as predicted by mSUGRA parameter fit to low energy measurements except $\Omega_{\mathrm{CDM}} h^2$ with sign$(\mu)$ fixed to $+1$.}
  \label{fig:results:LEonly:mSUGRA:massDist:noOmega}
\end{figure}

\begin{figure*}
  \includegraphics[width=0.49\textwidth]{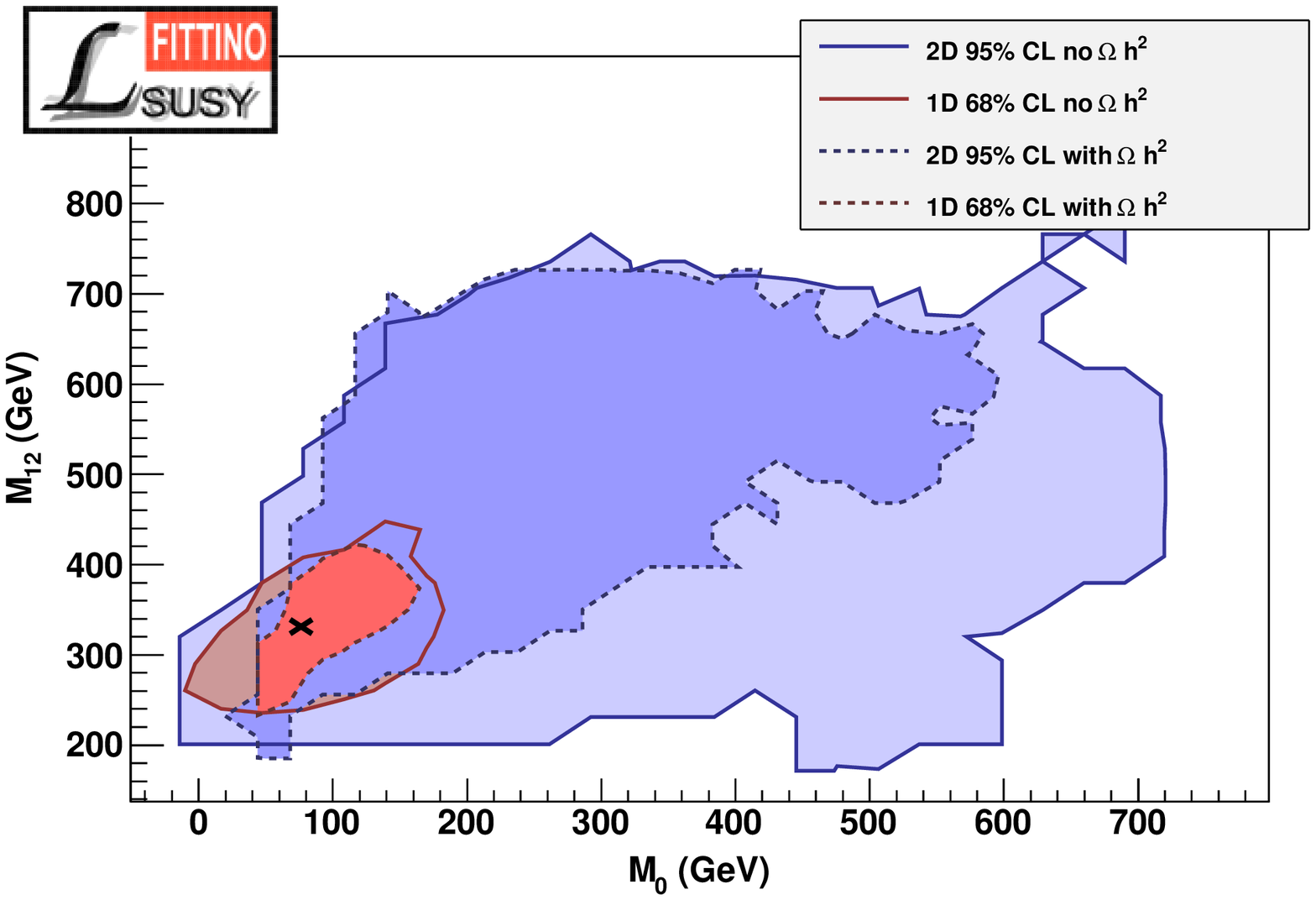}
  \includegraphics[width=0.49\textwidth]{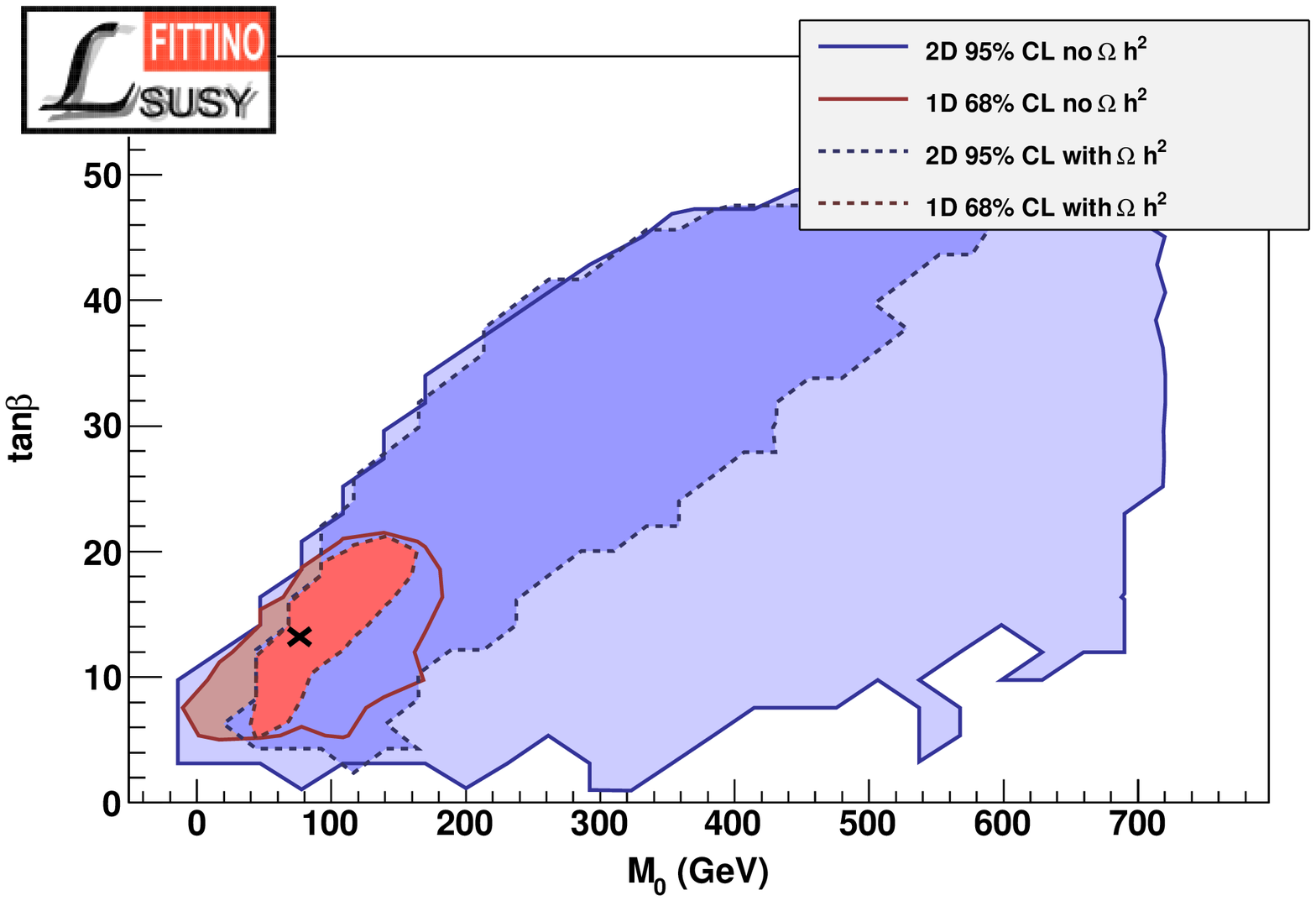}
  \caption{Region in the space of the mSUGRA scenario allowed by fits
    with and without $\Omega_{\mathrm{CDM}}h^2$. It can be seen that
    $\Omega_{\mathrm{CDM}}h^2$ increases the parameter correlations, but leaves
    the one-dimensional projections in a similar range.}
  \label{fig:results:LEonly:mSUGRA:Par:noOmegawithOmega}
\end{figure*}

Removing $\Omega_{\mathrm{CDM}} h^2$ completely from the list of observables does not
change the fit result significantly. The best fit point is identical
within the $1\,\sigma$ uncertainty with the best fit point for
$\Omega_{\mathrm{theo}} h^2\leq\Omega_{\mathrm{obs}} h^2$. The fit
result is presented in
Table~\ref{tab:results:LEonly:mSUGRA:noOmega}. Again the uncertainties
increase with respect to the baseline fit and the fit with
$\Omega_{\mathrm{theo}} h^2\leq\Omega_{\mathrm{obs}} h^2$, but the
order of magnitude of the results and single parameter uncertainties
remain unchanged. This is exemplified in
Figure~\ref{fig:results:LEonly:mSUGRA:Par:noOmegawithOmega}, which shows
the direct comparison of the fit results with and without
$\Omega_{\mathrm{CDM}}h^2$. The increase in correlation through the addition of
$\Omega_{\mathrm{CDM}}h^2$ is clearly visible. 
The predicted mass ranges in
Figure~\ref{fig:results:LEonly:mSUGRA:massDist:noOmega} are similar to
Figure~\ref{fig:results:LEonly:mSUGRA:massDist:smallerOmega}.

\begin{center}
  \begin{figure}
    \begin{center}
      \includegraphics[width=0.49\textwidth]{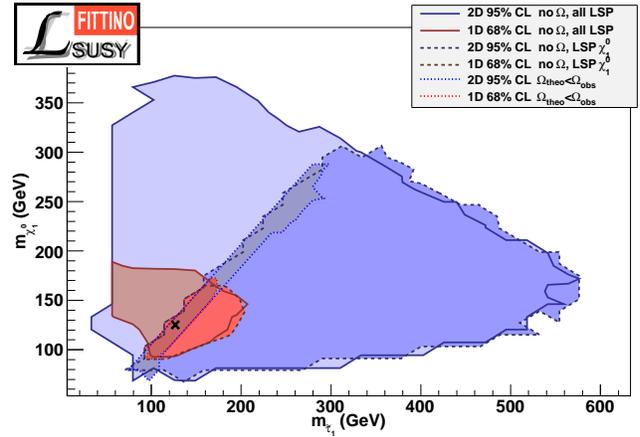}
    \end{center}
    \caption{Region in the space of the LSP mass $m_{\tilde{\chi}^0_1}$ and
      $m_{\tilde{\tau}_1}$, which for most parameter points is the
      NSLP, as predicted by a mSUGRA model compatible with existing
      data. Using all constraints, only a area with small
      $m_{\mathrm{NLSP}}-m_{\mathrm{LSP}}$ remains.}
    \label{fig:results:LEonly:mSUGRA:2dRes:Pheno}
  \end{figure}
\end{center}

While the overall range of the accessible parameter space is not
affected very strongly by including or excluding the constraint on
$\Omega_{\mathrm{CDM}}h^2$ in the presence of the other strong observables from
Table~\ref{tab:leobserables}, there is a strong influence of
$\Omega_{\mathrm{CDM}}h^2$ on the mass differences between the sparticles. This
is shown in Figure~\ref{fig:results:LEonly:mSUGRA:2dRes:Pheno} using the
prediction for $m_{\tilde{\chi}^0_1}$ and $m_{\tilde{\tau}_1}$. If no
requirements on the LSP are made at all, i.~e.~if also charged LSPs
are allowed, a large area of chargino and stau masses are allowed. The
requirement of a neutral LSP excludes the upper left area in
Figure~\ref{fig:results:LEonly:mSUGRA:2dRes:Pheno}, where
$m_{\tilde{\tau}_1}<m_{\tilde{\chi}^0_1}$. The remaining area is allowed
without applying further constraints on $\Omega_{\mathrm{CDM}}h^2$. It exhibits
a wide range of mass differences between LSP and NLSP, between 0~GeV and
450~GeV. The additional constraint of $\Omega_{\mathrm{theo}}
h^2\leq\Omega_{\mathrm{obs}} h^2$ severely constrains this mass
difference. Only a small band close to $\Omega_{\mathrm{theo}}
h^2=\Omega_{\mathrm{obs}} h^2$ remains, with
$m_{\tilde{\tau}_1}-m_{\tilde{\chi}^0_1}<22\,\mathrm{GeV}$ at 95\,\%~CL. As
already outlined in the discussion of the baseline fit, this result
has significant impact on the expected exclusive collider
measurements. This shows that the correlations among the parameters
changes in a significant way if $\Omega_{\mathrm{CDM}}h^2$ is included into the
fit, while the overall range of parameter uncertainties remains
similar.

\begin{table}
  \caption{Result of the Fit of the mSUGRA model with
    $\mathrm{sign}(\mu)=+1$ including four additional SM parameters to
    all measurements listed in Table~\ref{tab:leobserables}, but with
    $a_{\mu}^{\mathrm{exp}}=a_{\mu}^{\mathrm{SM}}$. The minimum
    $\chi^2$ value is $19.7$ for 22 degrees of freedom, corresponding
    to a ${\cal P}$-value of 60.2\,\%.}
  \label{tab:results:LEonly:mSUGRA:moveGmin2:posmu}
  \begin{center}
    {\renewcommand{\arraystretch}{1.2}
      \begin{tabular}{lrcl}
        \hline\hline
        Parameter       & Best Fit             &       & Uncertainty \\
        \hline
        sign$(\mu)$       &  +1                  &       &             \\
        $\alpha_s$      &  0.1174          & $\pm$ &  0.0021  \\
        $1/\alpha_{em}$ &  127.927         & $\pm$ &  0.017  \\
        $m_Z$ (GeV)     &  91.1874         & $\pm$ &  0.0020  \\
        $m_t$ (GeV)     &  172.50          & $\pm$ &  1.19  \\ 
        $\tan\,\beta$   &  7.4            &       &  $_{-4.6}^{+9.1}$            \\ 
        $M_{1/2}$ (GeV)  &  903.6           &       &  $_{-778.4}^{+67.5}$             \\
        $M_0$ (GeV)     &  180.1           &       &  $_{-110.0}^{+1566.0}$      \\ 
        $A_{0}$ (GeV)   &  956.9           &       &  $_{-3048.3 }^{+1515.6}$     \\
        \hline\hline
      \end{tabular}
    }
  \end{center}
\end{table}

The result of a mSUGRA fit to the observables from
Table~\ref{tab:leobserables}, but with
$a_{\mu}^{\mathrm{exp}}=a_{\mu}^{\mathrm{SM}}$ is shown in
Table~\ref{tab:results:LEonly:mSUGRA:moveGmin2:posmu}. This shows the
allowed mSUGRA parameter space for $\mathrm{sign}(\mu)=+1$ in case the
difference between measurement and theoretical SM prediction of
$a_{\mu}=(g-2)_{\mu}/2$ should vanish in the future, either due to
statistical fluctuations or due to systematic shifts in the prediction
because of a different treatment of the hadronic corrections. The
preferred parameter space in this case changes dramatically with
respect to the baseline fit, in contrast to the fits with different
treatments of $\Omega_{\mathrm{CDM}}h^2$. The best fit value of
$M_{1/2}$ moves to 903~GeV, but with large uncertainties towards
smaller values. In contrast, the preferred value of $M_0$ remains
small, but with very large uncertainties towards larger values. The
parameter $\tan\beta$ does not change dramatically, and $A_0$ is
basically unconstrained.

\begin{figure}
  \includegraphics[width=0.49\textwidth]{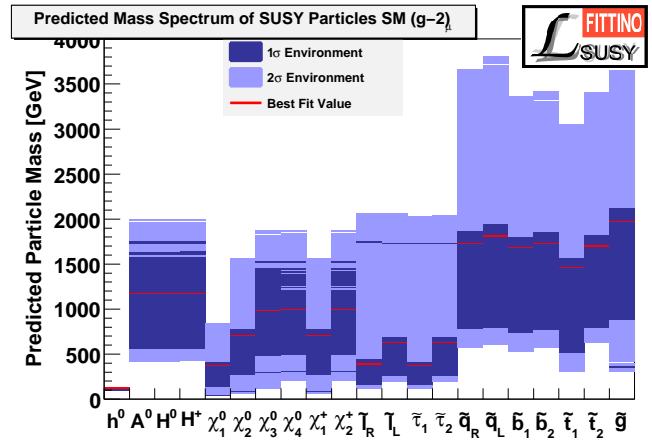}
  \caption{SUSY mass spectrum as predicted by mSUGRA parameter fit 
    to low energy measurements with $a_{\mu}^{\mathrm{exp}}=a_{\mu}^{\mathrm{SM}}$ and with sign$(\mu)$ fixed to $+1$.}
  \label{fig:results:LEonly:mSUGRA:massDist:moveGmin2:posmu}
\end{figure}

This has significant consequences for the expected sparticle
masses, as shown in
Figure~\ref{fig:results:LEonly:mSUGRA:massDist:moveGmin2:posmu}. In
comparison to the prediction for the baseline fit, the situation
changes dramatically. The gluino and squarks tend towards masses
between 500~GeV and 3500~GeV, severely reducing the expected
production cross-section of colour-charged sparticles. In
contrast to the heavy squarks, the slepton and electro-weak gaugino
masses remain in the range below 500~GeV.

\begin{center}
  \begin{figure*}
    \begin{center}
      \includegraphics[width=0.32\textwidth]{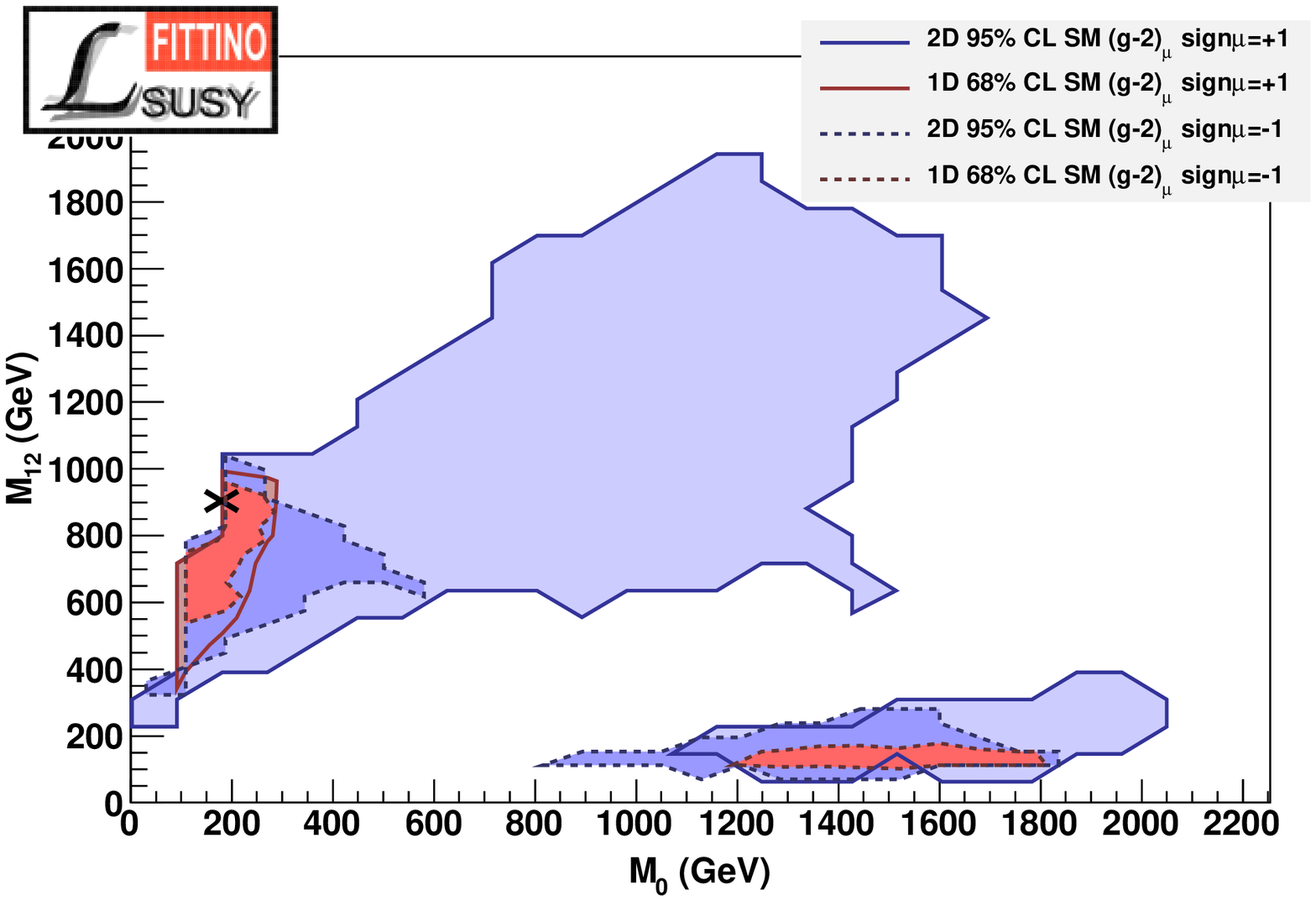}
      \hfill
      \includegraphics[width=0.32\textwidth]{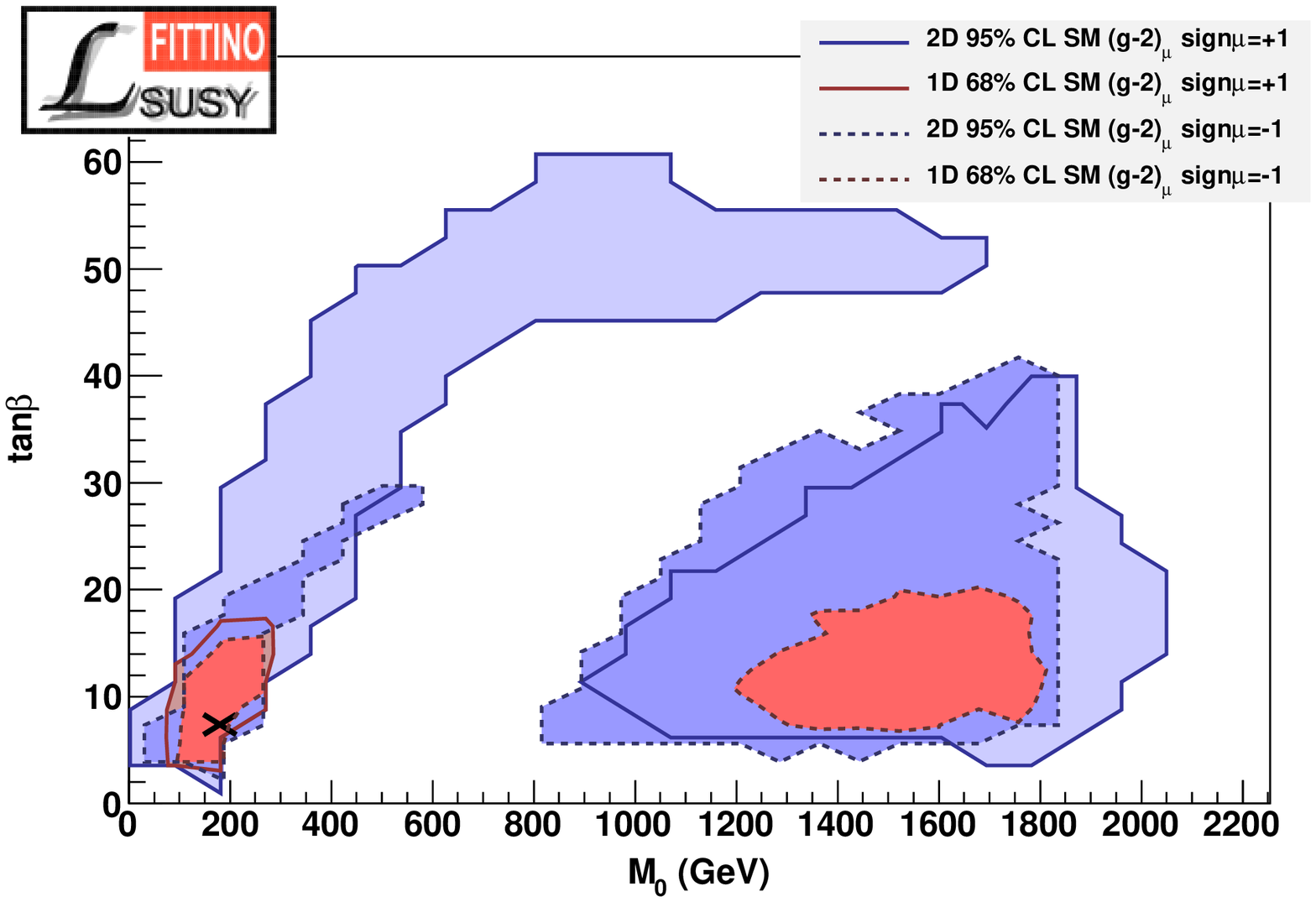}
      \hfill
      \includegraphics[width=0.32\textwidth]{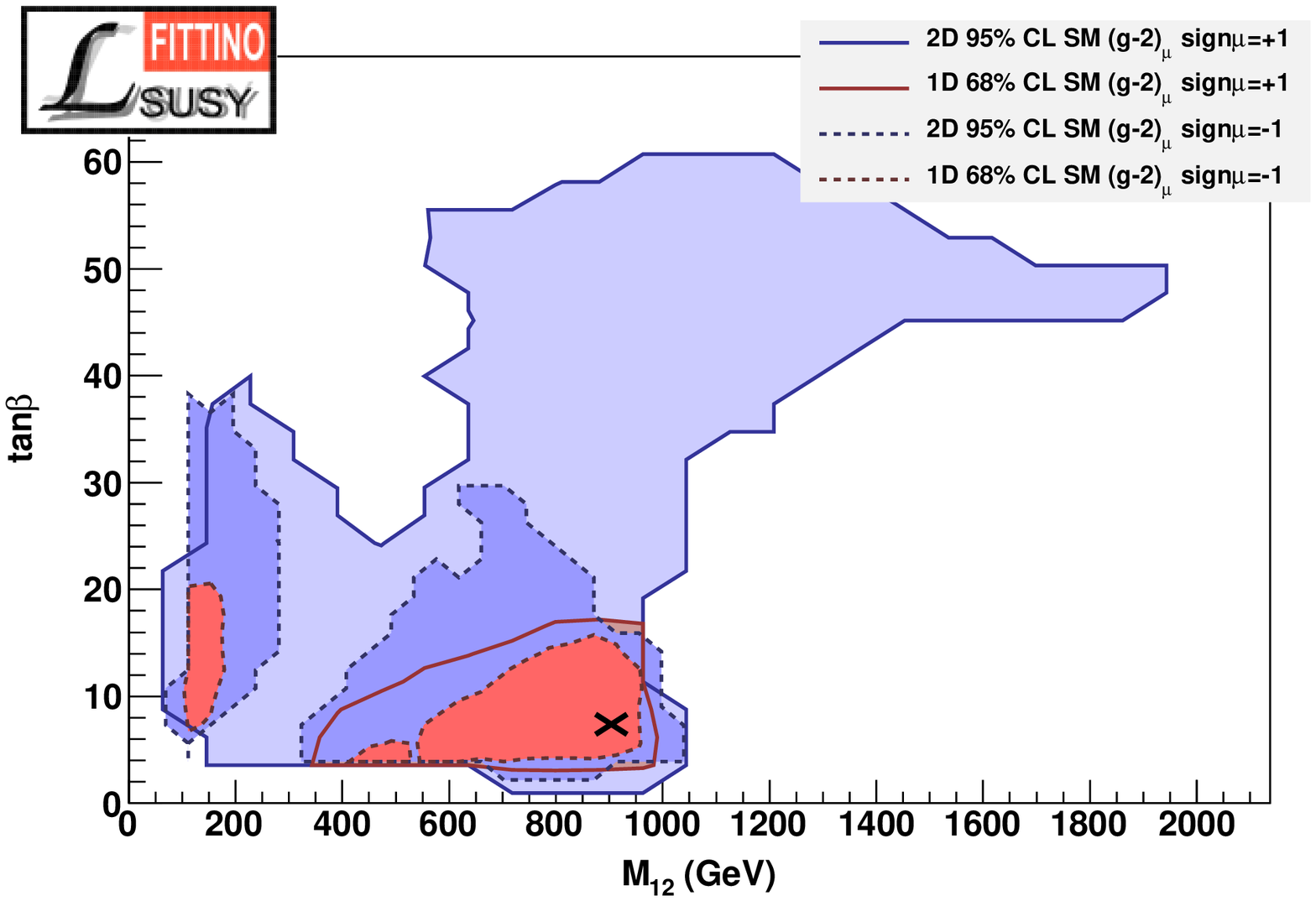}
      \\[1mm]
      \includegraphics[width=0.32\textwidth]{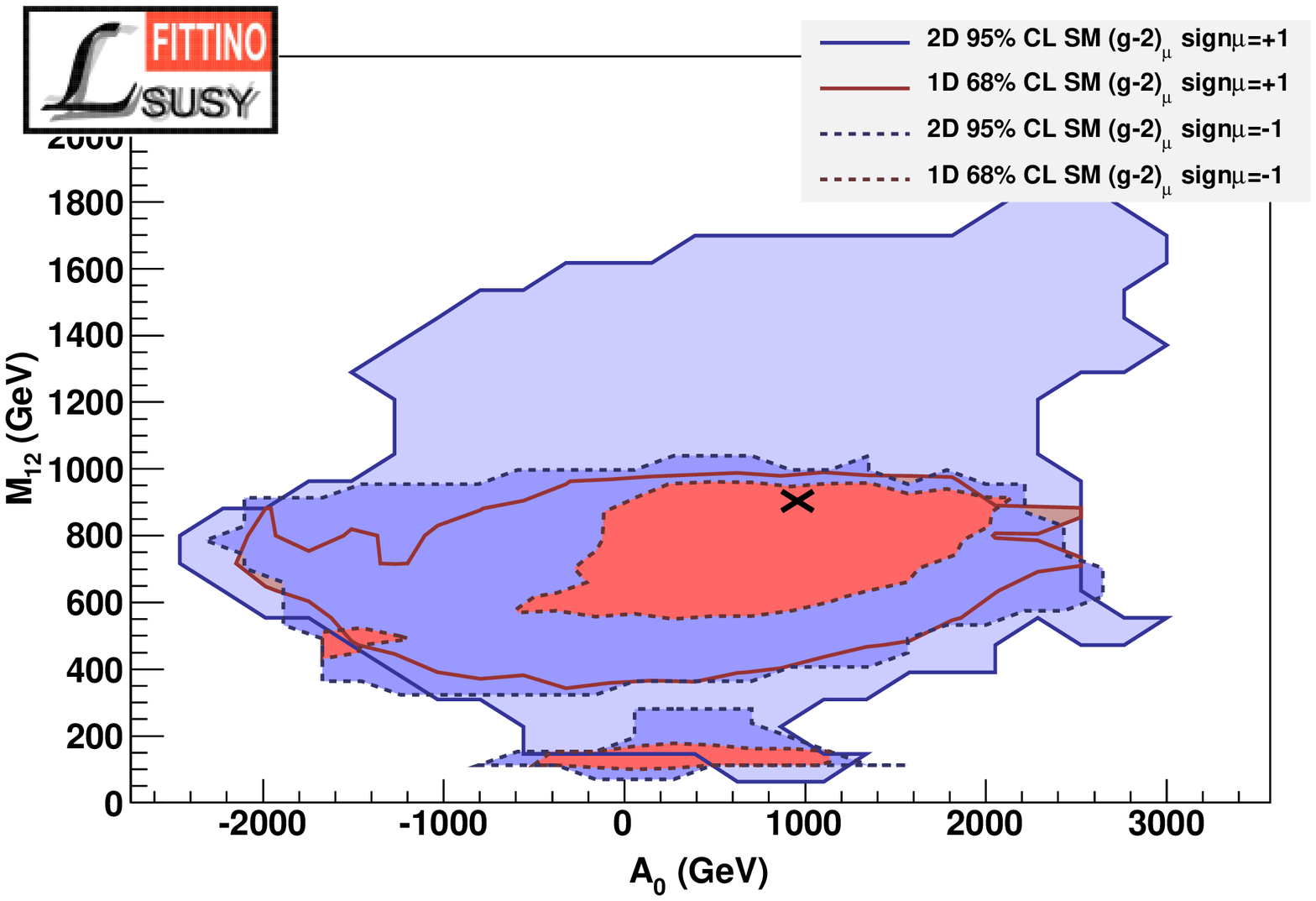}
      \hfill
      \includegraphics[width=0.32\textwidth]{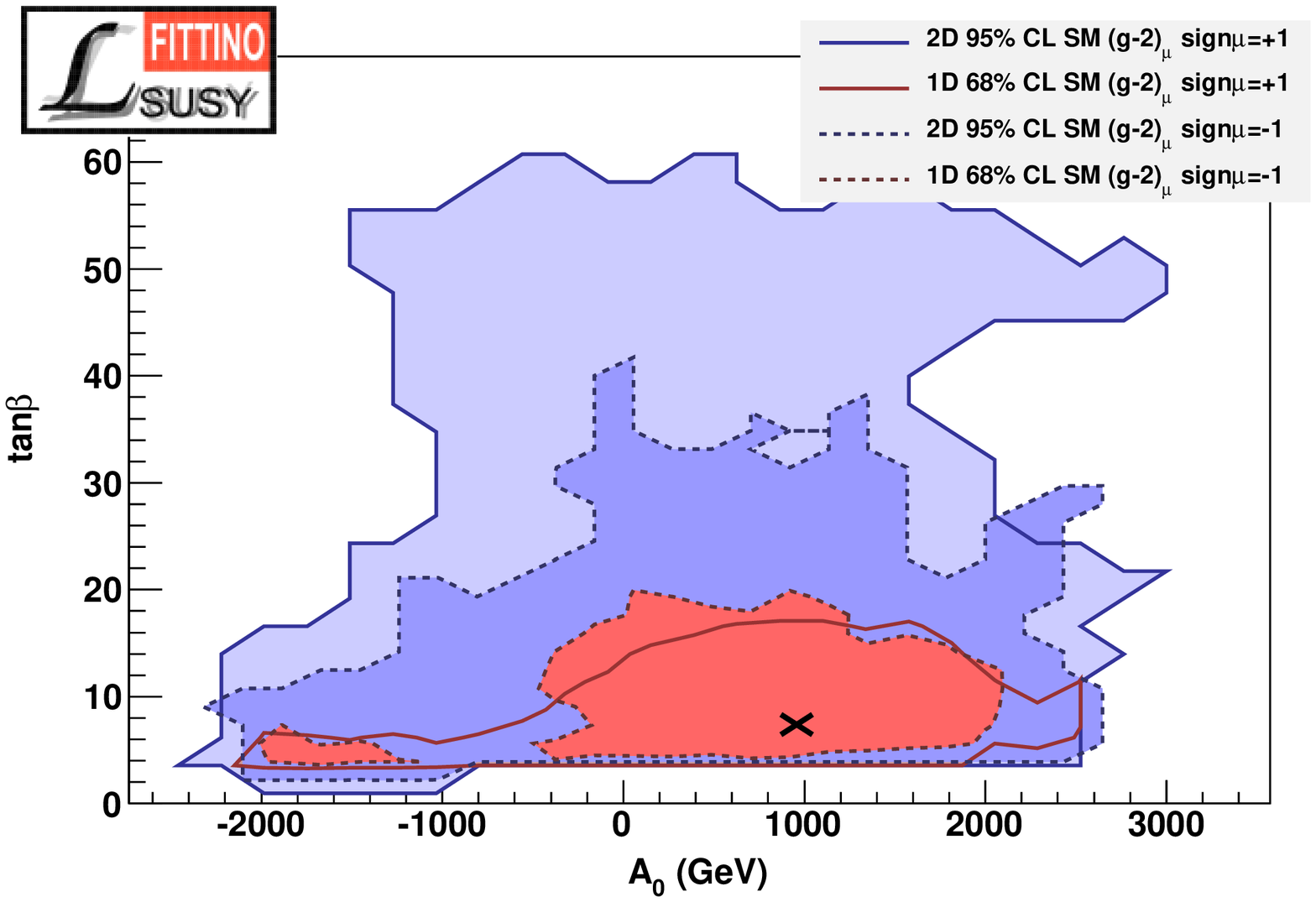}
      \hfill
      \includegraphics[width=0.32\textwidth]{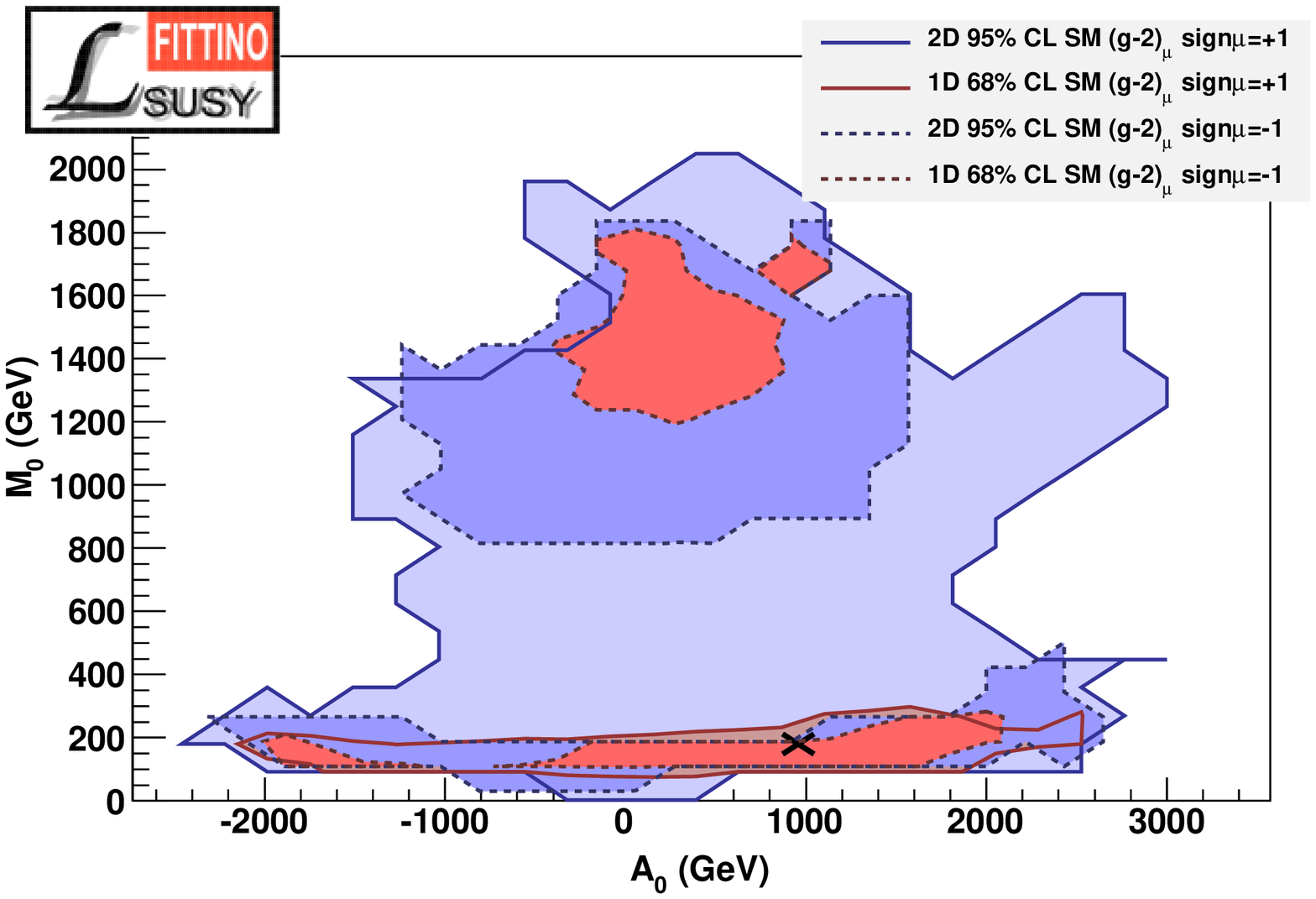}
    \end{center}
    \caption{Expected mSUGRA parameter regions for the fit to low
      energy measurements with $a_{\mu}^{\mathrm{exp}}=a_{\mu}^{\mathrm{SM}}$. Results
      for $\mathrm{sign}(\mu)=+1$ and $\mathrm{sign}(\mu)=-1$ are
      overlaid. The best fit point is shown for the result with
      $\mathrm{sign}(\mu)=+1$}
    \label{fig:results:LEonly:mSUGRA:par:moveGmin2:posmu}
  \end{figure*}
\end{center}

The change with respect to the baseline fit is also visible in the
allowed parameter space in the detailed two-dimensional projections,
as shown in
Figure~\ref{fig:results:LEonly:mSUGRA:par:moveGmin2:posmu}. Two
distinct parameter regions both for the $1\,\sigma$ and the
$2\,\sigma$ environment can be seen in several parameter projections,
together with a tendency towards larger values of the mass parameters
$M_{0}$ and $M_{1/2}$ and hence higher values of the expected
sparticle masses, especially for the case of
$\mathrm{sign}(\mu)=-1$.

\begin{table}
  \caption{Result of the fit of the mSUGRA model with
    $\mathrm{sign}(\mu)=-1$ including four additional SM parameters to
    all measurements listed in Table~\ref{tab:leobserables}, but with
    $a_{\mu}^{\mathrm{exp}}=a_{\mu}^{\mathrm{SM}}$. The minimum
    $\chi^2$ value is $19.1$ for 22 degrees of freedom, corresponding
    to a ${\cal P}$-value of 64.0\,\%.}
  \label{tab:results:LEonly:mSUGRA:moveGmin2:negmu}
  \begin{center}
    {\renewcommand{\arraystretch}{1.2}
      \begin{tabular}{lrcl}
        \hline\hline
        Parameter       & Best Fit             &       & Uncertainty \\
        \hline
        sign$(\mu)$       &  $-$1                  &       &             \\
        $\alpha_s$      &  0.1175         & $\pm$ &   0.0018                 \\
        $1/\alpha_{em}$ &  127.929        & $\pm$ &   0.018                 \\
        $m_Z$ (GeV)     &  91.1872        & $\pm$ &   0.0022                    \\
        $m_t$ (GeV)     &  172.25         & $\pm$ &   1.14                   \\ 
        $\tan\,\beta$   &  11.65          &       &   $_{-7.92}^{+8.03}$           \\ 
        $M_{1/2}$ (GeV)  &  129.9          &       &   $_{-28.8}^{+819.1}$            \\
        $M_0$ (GeV)     &  1760.1         &       &   $_{-1654.7}^{+24.4}$     \\ 
        $A_{0}$ (GeV)   &  62.1           &       &   $_{-2016.3}^{+2000.9}$      \\
        \hline\hline
      \end{tabular}
    }
  \end{center}
\end{table}

If the positive deviation of the measurement of $a_{\mu}$ from the SM
prediction is removed, there is no remaining clear preference among
the measured observables for either $\mathrm{sign}(\mu)=+1$ or
$-1$. The fit of the mSUGRA scenario with
$a_{\mu}^{\mathrm{exp}}=a_{\mu}^{\mathrm{SM}}$ and
$\mathrm{sign}(\mu)=+1$ yields a minimal $\chi^2$ of $19.7$, while the
fit with $\mathrm{sign}(\mu)=-1$ yields $\chi^2_{\mathrm{min}}=19.1$
for 22 degrees of freedom. The results of the latter fit are shown in
Table~\ref{tab:results:LEonly:mSUGRA:moveGmin2:negmu}. The best fit
values of $M_{1/2}$ and $M_{0}$ are strongly different between the two
results, but the uncertainties span similar areas. No significant
differences are obtained for $\tan\beta$ and $A_0$.

\begin{figure}
  \includegraphics[width=0.49\textwidth]{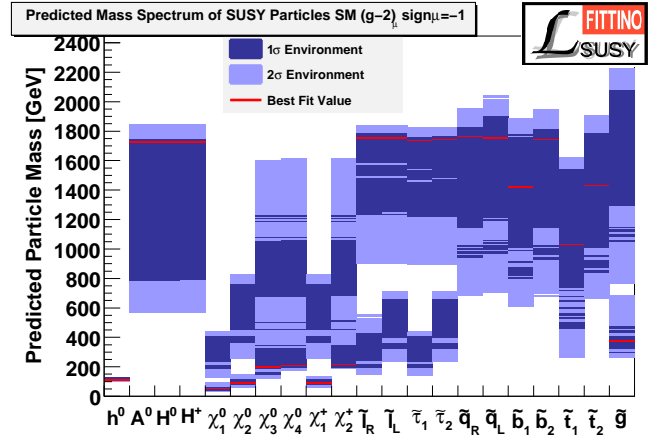}
  \caption{SUSY mass spectrum as predicted by mSUGRA parameter fit 
    to low energy measurements with $a_{\mu}^{\mathrm{exp}}=a_{\mu}^{\mathrm{SM}}$ and with sign$(\mu)$ fixed to $-1$.}
  \label{fig:results:LEonly:mSUGRA:massDist:moveGmin2:negmu}
\end{figure}

The expected sparticle masses and uncertainties for the mSUGRA fit
with $a_{\mu}^{\mathrm{exp}}=a_{\mu}^{\mathrm{SM}}$ and $\mathrm{sign}(\mu)=-1$ are shown
in Figure~\ref{fig:results:LEonly:mSUGRA:massDist:moveGmin2:negmu}. In
contrast to the result for $\mathrm{sign}(\mu)=+1$, very light gauginos
are expected due to the low value of $M_{0}$. The best fit result for
the squark masses is at around 1800~GeV at a similar level, but the
uncertainty does not extend to much higher values. Especially for the
sleptons there are two distinct regions predicted, which is a
consequence of the prominent distinct allowed regions in the $M_0$ and
$M_{1/2}$ distributions shown in
Figure~\ref{fig:results:LEonly:mSUGRA:par:moveGmin2:posmu}. Again, the
gluino is expected to be lighter than the squarks, leading to decay
signatures strongly different from the chain of several two-body
decays expected for the baseline fit.

\begin{table}
  \caption{Result of the fit of the mSUGRA model with $\mathrm{sign}(\mu)=+1$ including 
    four additional SM parameters to all measurements 
    listed in Table~\ref{tab:leobserables} except $(g-2)_{\mu}$. The minimum $\chi^2$ value is 
    $19.47$ for 21 degrees of freedom, corresponding to a ${\cal P}$-value of 55.5\,\%.}
  \label{tab:results:LEonly:mSUGRA:nogmin2}
  \begin{center}
    {\renewcommand{\arraystretch}{1.2}
      \begin{tabular}{lrcl}
        \hline\hline
        Parameter     & Best Fit  &       & Uncertainty \\
        \hline
        sign$(\mu)$     &  +1       &       &             \\
        $\alpha_s$    &    0.1177         & $\pm$ &  0.0019     \\
        $1/\alpha_{em}$  & 127.925           & $\pm$ &  0.015      \\
        $m_Z$ (GeV)   &   91.1875          & $\pm$ &  0.0020      \\
        $m_t$ (GeV)   &  172.5            & $\pm$ &  1.1         \\ 
        $\tan\,\beta$ &    7.5        &       &   $_{-4.6}^{+8.9}$            \\ 
        $M_{1/2}$ (GeV) & 389.2            &       &  $_{-117.5}^{+568.9}$             \\
        $M_0$ (GeV)   &   72.2          &      &   $_{-22.1}^{+145.0}$     \\ 
        $A_{0}$ (GeV) &  270.0           &      &   $_{-1985.9}^{+1492.5}$  \\
        \hline\hline
      \end{tabular}
    }
  \end{center}
\end{table}

In order to explore the significance of $(g-2)_{\mu}$ for the fit
result, Table~\ref{tab:results:LEonly:mSUGRA:nogmin2} shows the result
of the mSUGRA fit with $\mathrm{sign}(\mu)=+1$ to the observables from
Table~\ref{tab:leobserables} excluding $(g-2)_{\mu}$. Due to the lower
minimal $\chi^2$ of $19.5$ than the fit with
$a_{\mu}^{\mathrm{exp}}=a_{\mu}^{\mathrm{SM}}$, no secondary minima are observed,
therefore smaller allowed regions in $M_{0}$ are obtained for the fit
without $a_{\mu}^{\mathrm{exp}}$ than with $a_{\mu}^{\mathrm{exp}}=a_{\mu}^{\mathrm{SM}}$. The
comparison of Table~\ref{tab:results:LEonly:mSUGRA:nogmin2} with the
results of the baseline fit in
Table~\ref{tab:results:LEonly:mSUGRA:all} shows that $(g-2)_{\mu}$
represents an important constraint, but that the remaining observables
still favour SUSY in the mass range below 1~TeV.

\begin{table}
  \caption{Result of the fit of the mSUGRA model with $\mathrm{sign}(\mu)=+1$ including 
    four additional SM parameters to all measurements 
    listed in Table~\ref{tab:leobserables}, using the predicted SM-value 
    of $(g-2)_{\mu}$ from~\cite{Davier:2009ag}. The minimum $\chi^2$ value is 
    $19.6$ for 22 degrees of freedom, corresponding to a ${\cal P}$-value of 60.6\,\%.
    The fit results of the SM parameters are consistent with the other fits 
    in this section.}
  \label{tab:results:LEonly:mSUGRA:taugmin2}
  \begin{center}
    {\renewcommand{\arraystretch}{1.2}
      \begin{tabular}{lrcl}
        \hline\hline
        Parameter       & Best Fit  &       & Uncertainty \\
        \hline
        sign$(\mu)$       &  +1       &       &             \\
        $\tan\,\beta$   &    7.9    &       &   $_{-3.9}^{+10.6}$            \\ 
        $M_{1/2}$ (GeV) & 350.3     &       &   $_{-79.0}^{+331.06}$             \\
        $M_0$ (GeV)     &   65.9    &       &   $_{-15.8}^{+123.8}$     \\ 
        $A_{0}$ (GeV)   &  100.4    &       &   $_{-852.6}^{+1435.1}$  \\
        \hline\hline
      \end{tabular}
    }
  \end{center}
\end{table}

\begin{figure}
  \includegraphics[width=0.49\textwidth]{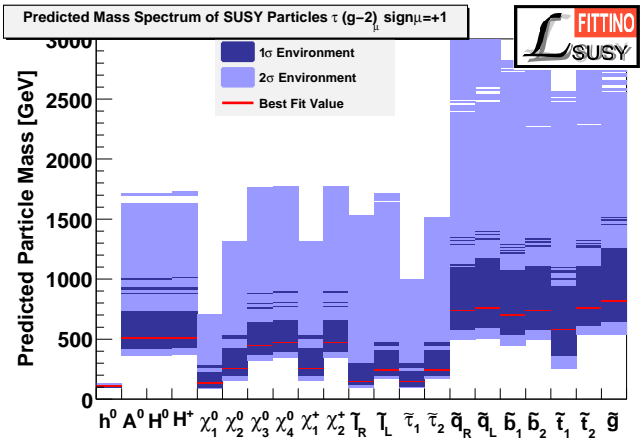}
  \caption{SUSY mass spectrum as predicted by mSUGRA parameter fit to
    low energy measurements using the predicted SM-value of
    $(g-2)_{\mu}$ from~\cite{Davier:2009ag} with sign$(\mu)$ fixed to
    $+1$.}
  \label{fig:results:LEonly:mSUGRA:massDist:taugmin2}
\end{figure}

For illustration of the difference between the SM predictions of
$(g-2)_{\mu}$ from $e^+e^-$ and $\tau$ data, the fit of mSUGRA using
the predicted SM value and uncertainties from~\cite{Davier:2009ag} is
shown in Table~\ref{tab:results:LEonly:mSUGRA:taugmin2} and the
corresponding mass spectrum in
Figure~\ref{fig:results:LEonly:mSUGRA:massDist:taugmin2}. The mean value
of the sparticle masses and mSUGRA parameters is compatible with the
result of the baseline fit within their uncertainties, but the 95\,\%
CL area of the predicted sparticle masses ranges up to around 3~TeV,
in contrast to the baseline fit wit hits upper limit at 1.6~TeV at 95\,\%
CL.

\begin{center}
  \begin{figure*}
    \begin{center}
      \includegraphics[width=0.49\textwidth]{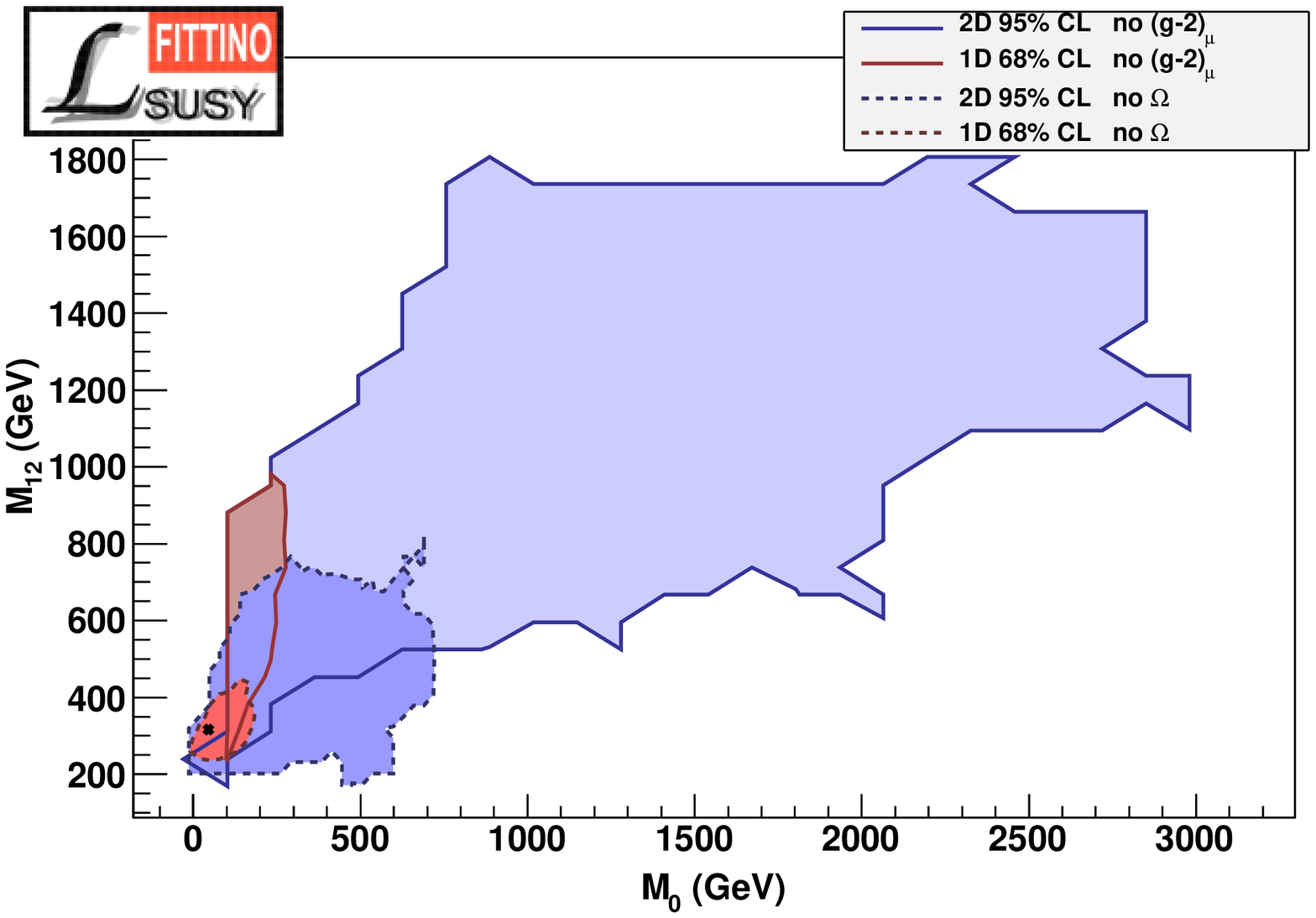}
      \hfill
      \includegraphics[width=0.49\textwidth]{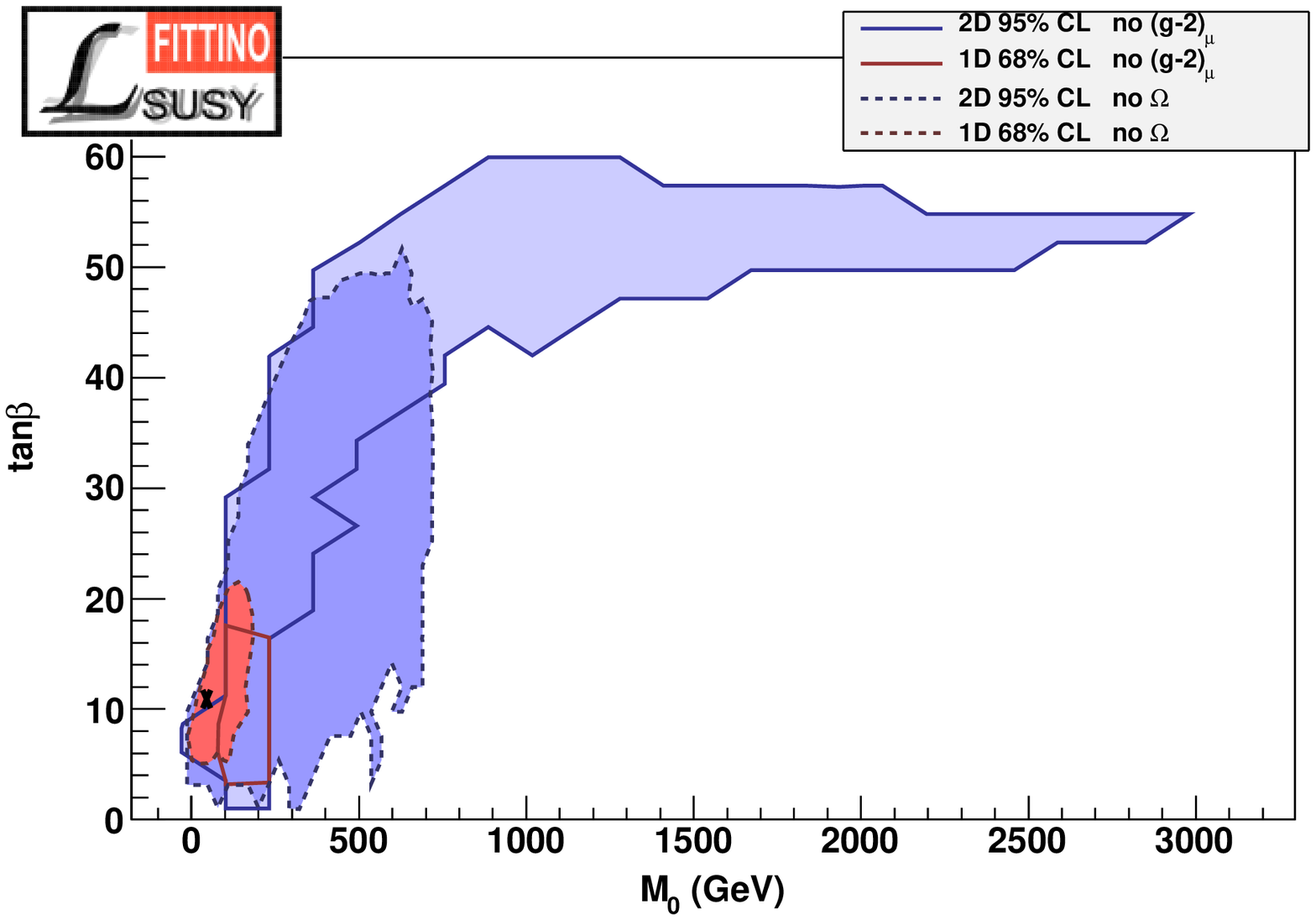}
      \\[1mm]
      \includegraphics[width=0.49\textwidth]{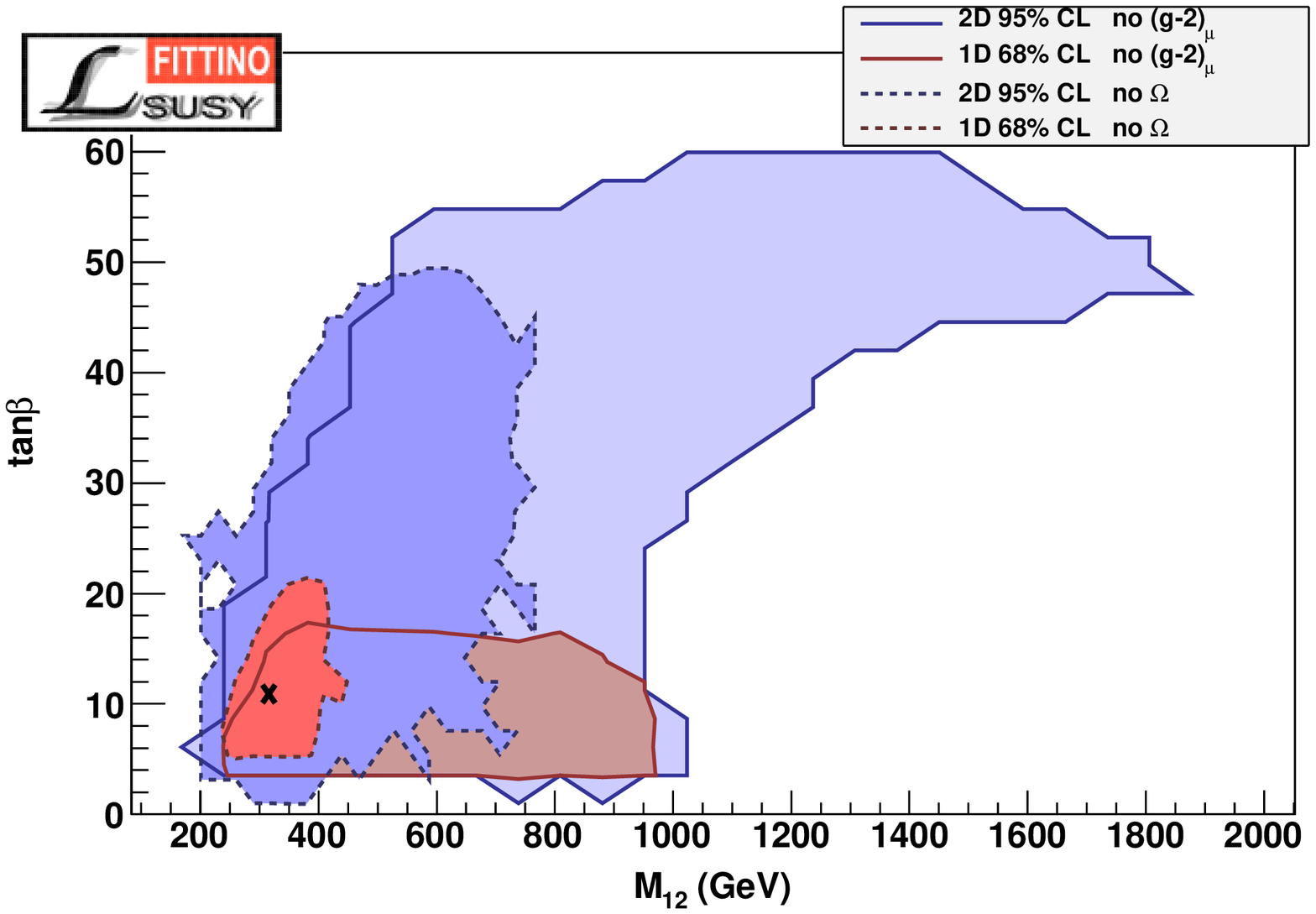}
      \hfill
      \includegraphics[width=0.49\textwidth]{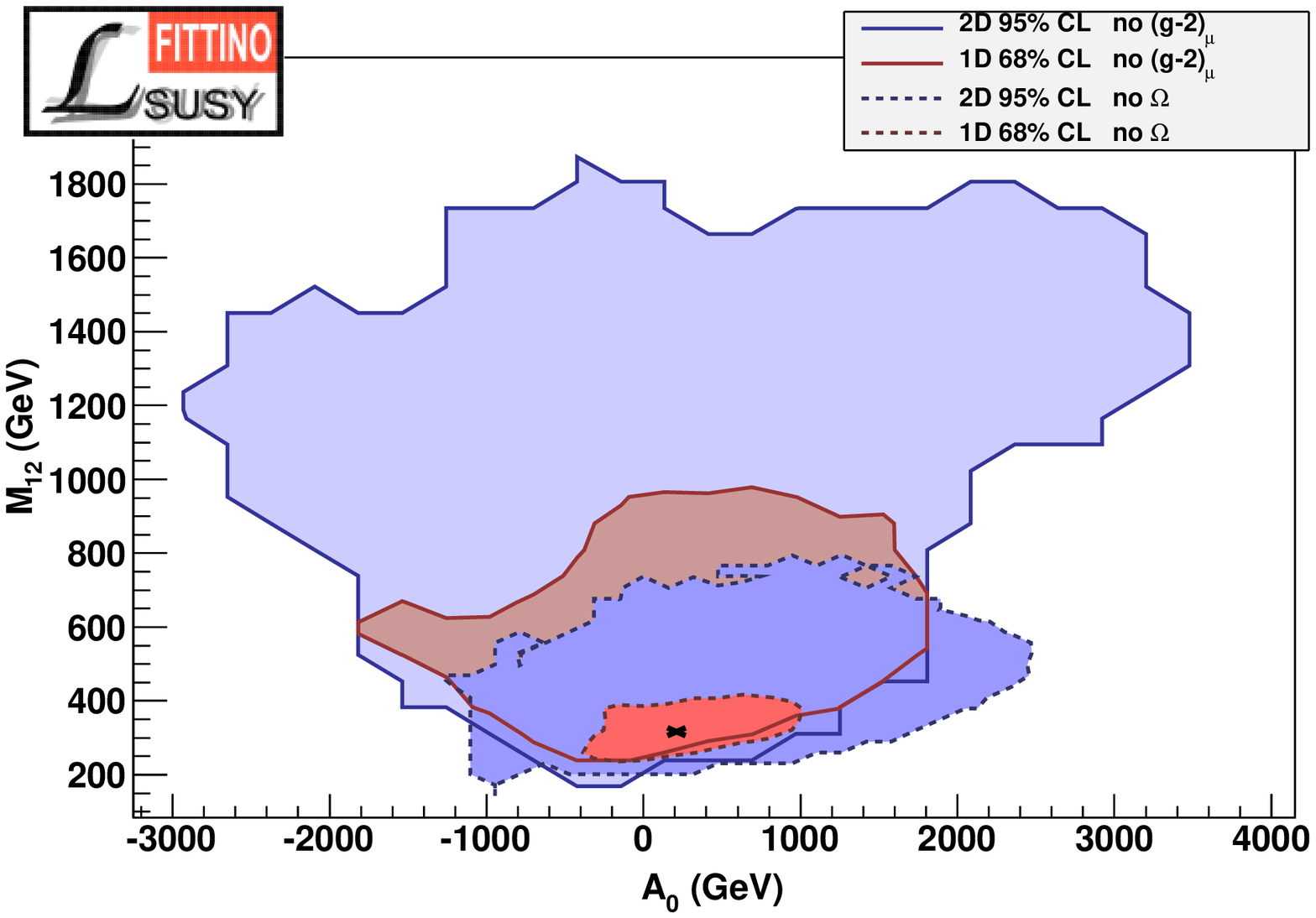}
    \end{center}
    \caption{mSUGRA parameter regions compatible with all low energy measurements
      except (a) $(g-2)_{\mu}$ and (b) $\Omega_{\mathrm{CDM}} h^2$ for various
      parameter combinations and sign$(\mu)$ fixed to $+1$. For both
      cases two-dimensional 95 \% confidence level and one-dimensional 68
      \% confidence regions are shown. From the latter $1\sigma$
      uncertainties for individual parameters can be derived from a
      projection of the area to the respective axis.}
    \label{fig:results:LEonly:mSUGRA:2dRes:NoCosmo}
  \end{figure*}
\end{center}

The impact of the two most stringent observables
$\Omega_{\mathrm{CDM}}h^2$ and $(g-2)_{\mu}$ is compared in
Figure~\ref{fig:results:LEonly:mSUGRA:2dRes:NoCosmo}. The dark blue
areas indicate the two-dimensional 95\,\%~CL allowed parameter region
for the fit without $\Omega_{\mathrm{CDM}}h^2$, showing that the
remaining observables constrain the allowed mSUGRA parameter space to
regions below 700~GeV both in $M_{0}$ and $M_{1/2}$, accesible at LHC.
The light blue region indicates the allowed parameter space for the
fit without $(g-2)_{\mu}$. It can be seen that the two constraints are
complementary. Without $(g-2)_{\mu}$ significantly larger values of
$M_{0}$ and $M_{1/2}$ are allowed, but there is a narrower allowed
band at low $M_{0}$ and $M_{1/2}$, caused by the constraint from
$\Omega_{\mathrm{CDM}}h^2$ favouring the co-annihilation region.

\begin{figure}
  \includegraphics[width=0.49\textwidth]{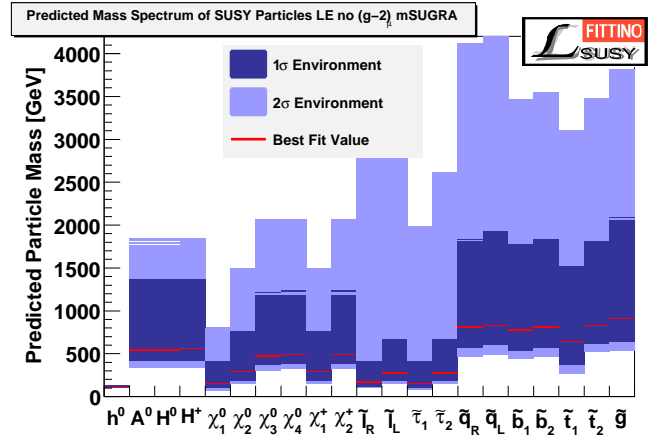}
  \caption{SUSY mass spectrum as predicted by mSUGRA parameter fit to low energy measurements except $(g-2)_{\mu}$ with sign$(\mu)$ fixed to $+1$.}
  \label{fig:results:LEonly:mSUGRA:massDist:nogmin2}
\end{figure}

Since the uncertainties of the mSUGRA parameters for the fit without
$(g-2)_{\mu}$ in Figure~\ref{fig:results:LEonly:mSUGRA:2dRes:NoCosmo}
extend to much larger values of the mass parameters, also the maximal
values of the sparticle masses allowed at 95\,\%~CL are increasing
significantly with respect to the baseline fit or the fit without
$\Omega_{\mathrm{CDM}}h^2$. This is visible in
Figure~\ref{fig:results:LEonly:mSUGRA:massDist:nogmin2}. This result
shows again, similar to the result for $a_{\mu}^{\mathrm{exp}}=a_{\mu}^{\mathrm{SM}}$,
that $(g-2)_{\mu}$ plays a major role in constraining the allowed
mSUGRA sparticle masses to values below 1.6~TeV in the baseline fit. 

\begin{table}
  \caption{Result of the fit of the mSUGRA model with $\mathrm{sign}(\mu)=+1$ including 
    four additional SM parameters to all measurements 
    listed in Table~\ref{tab:leobserables} except $\Omega_{\mathrm{CDM}} h^2$ and $(g-2)_{\mu}$. The minimum $\chi^2$ value is 
    $19.5$ for 20 degrees of freedom, corresponding to a ${\cal P}$-value of 49.1\,\%. 
    Due to the very weak remaining constraint on the SUSY parameter space, in this 
    fit a complete coverage of the possible two-dimensional $2\,\sigma$ parameter space 
    can not be ensured within the available statistics. Hence the uncertainties are 
    to be treated as lower boundaries on the uncertainties.}
  \label{tab:results:LEonly:mSUGRA:noOmeganogmin2}
  \begin{center}
    {\renewcommand{\arraystretch}{1.2}
      \begin{tabular}{lrcl}
        \hline\hline
        Parameter     & Best Fit  &       & Uncertainty \\
        \hline
        sign$(\mu)$     &  +1       &       &             \\
        $\alpha_s$    &    0.1179          & $\pm$ &   0.0021    \\
        $1/\alpha_{em}$ &  127.925             & $\pm$ &   0.015      \\
        $m_Z$ (GeV)   &   91.1876            & $\pm$ &   0.0022     \\
        $m_t$ (GeV)   &  172.4             & $\pm$ &   1.2     \\ 
        $\tan\,\beta$ &     7.4          &       &   $_{-4.7}^{+47.3}$            \\ 
        $M_{1/2}$ (GeV) &   432.6            &       &   $_{-307.1}^{+4405.6}$             \\
        $M_0$ (GeV)   &    64.3           &       &   $_{-62.8 }^{+6816.7}$      \\ 
        $A_{0}$ (GeV) &   387.8            &       &   $_{-10364.9}^{+6341.4}$     \\
        \hline\hline
      \end{tabular}
    }
  \end{center}
\end{table}

The previous results show that $\Omega_{\mathrm{CDM}}h^2$ has a strong role
in constraining the size of the uncertainties, but does not affect the
best fit result or the shape of the predicted sparticle mass
spectrum. On the other hand,
$(g-2)_{\mu}$ has a decisive impact on the shape of the allowed
parameter regions, the best fit values of the parameters, their
uncertainties and hence on the predicted particle spectrum. The
prediction of a rich sparticle spectrum in the kinematic range
accessible by the LHC however remains stable for all explored fits
including or excluding either $\Omega_{\mathrm{CDM}}h^2$ or $(g-2)_{\mu}$,
which is a very encouraging result for the discovery potential of
LHC. The question remains whether there are other observables among
those shown in Table~\ref{tab:leobserables} which constrain the mSUGRA parameter space.
This is addressed by removing both $\Omega_{\mathrm{CDM}}h^2$ and $(g-2)_{\mu}$ from the fit and
looking for remaining constraints. The result of this fit with
$\mathrm{sign}(\mu)=+1$ is shown in
Table~\ref{tab:results:LEonly:mSUGRA:noOmeganogmin2}. The results show
that the uncertainties increase strongly with respect to the baseline
fit as well as with respect to the fits without either
$\Omega_{\mathrm{CDM}}h^2$ or $(g-2)_{\mu}$. This shows both the
complementarity of the $\Omega_{\mathrm{CDM}}h^2$ and $(g-2)_{\mu}$
constraints, respectively, and the lack of other strongly constraining
observables in Table~\ref{tab:leobserables}. Since no significant
constraints on low-energy SUSY can be placed anymore without these
observables, the corresponding fit with $\mathrm{sign}(\mu)=-1$ is
omitted. 

\begin{figure}  
  \includegraphics[width=0.49\textwidth]{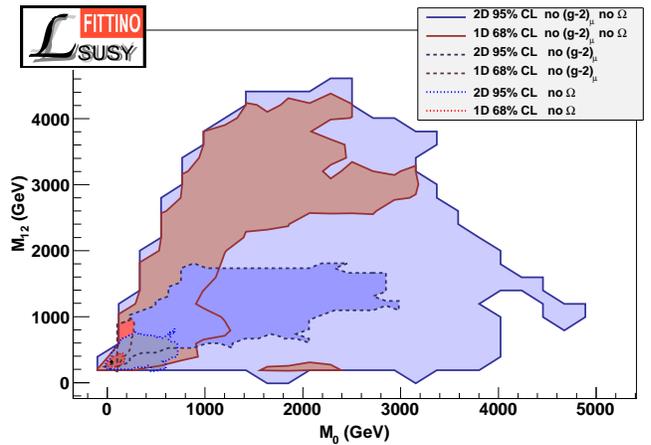}
  \caption{mSUGRA parameter regions compatible with the collider
    precision observables from Table~\ref{tab:leobserables} and with
    the requirement of a neutral LSP, but without any requirement on
    $\Omega_{\mathrm{CDM}} h^2$ and $(g-2)_{\mu}$, overlaid with the allowed
    parameter regions using only no $\Omega_{\mathrm{CDM}} h^2$ or no $(g-2)_{\mu}$.}
  \label{fig:results:LEonly:GMSB:2dRes:noAll}
\end{figure}

A comparison of the size of the allowed parameter space in the $M_{0}$
and $M_{1/2}$ projections, for the fits without
$\Omega_{\mathrm{CDM}}h^2$, $(g-2)_{\mu}$ and without both is shown in
Figure~\ref{fig:results:LEonly:GMSB:2dRes:noAll}. For the fit without
both observables the Markov Chains does not completely explore the
two-dimensional $2\,\sigma$ uncertainty space, hence the uncertainty
region is to be interpreted as a lower bound on the
uncertainties. This could be remedied by using significantly larger
statistics in the Markov Chains. For the result shown here this is not
done, since the results show clearly that SUSY within the reach of the
LHC or the ILC is not ensured without using $\Omega_{\mathrm{CDM}}h^2$
and $(g-2)_{\mu}$.

\begin{figure}
  \includegraphics[width=0.49\textwidth]{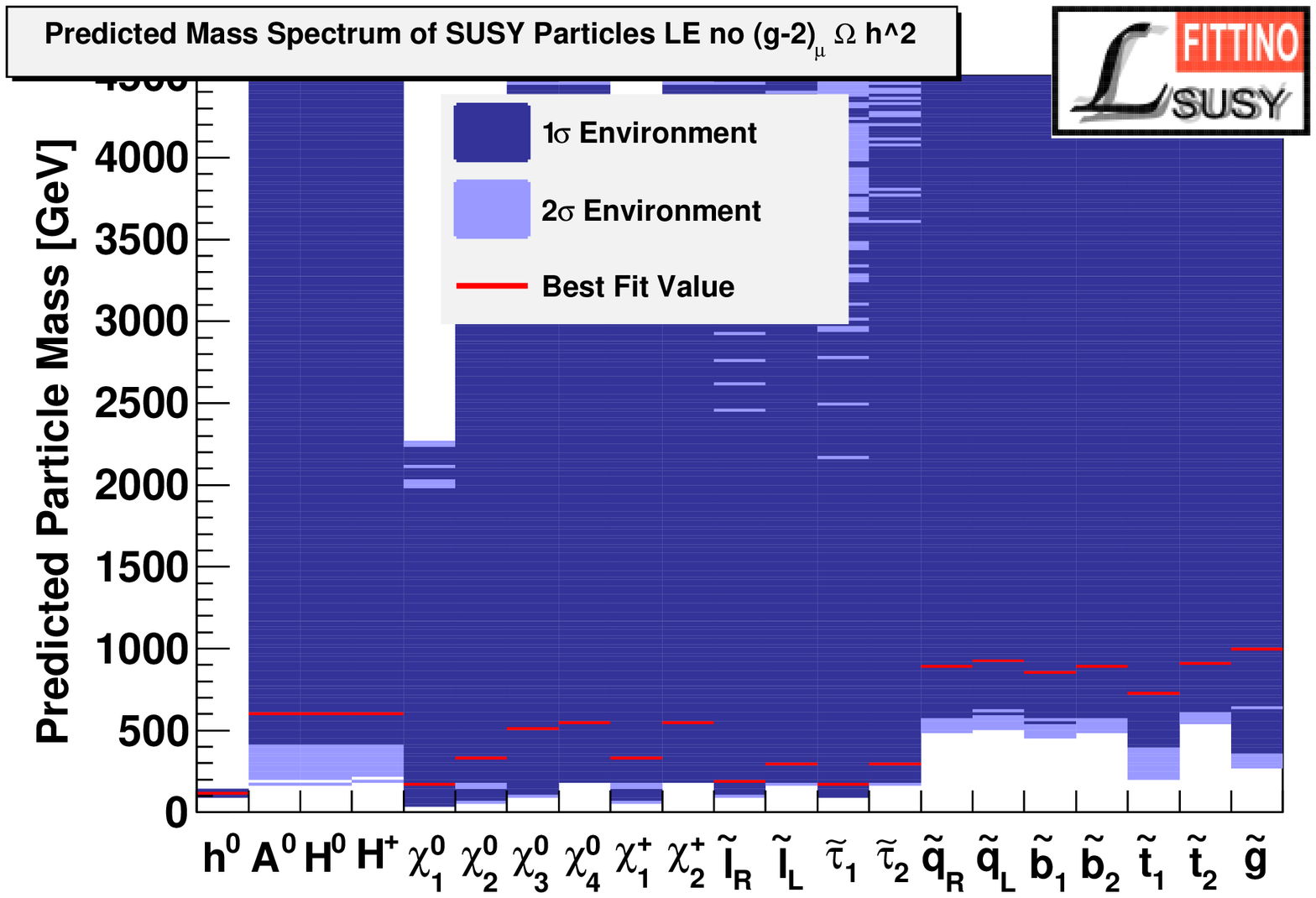}
  \caption{SUSY mass spectrum as predicted by mSUGRA parameter fit 
    to low energy measurements except $(g-2)_{\mu}$ and $\Omega_{\mathrm{CDM}} h^2$ with sign$(\mu)$ fixed to $+1$.}
  \label{fig:results:LEonly:mSUGRA:massDist:nogmin2noOmega}
\end{figure}

A part of the allowed sparticle mass spectrum is shown in
Figure~\ref{fig:results:LEonly:mSUGRA:massDist:nogmin2noOmega}. The
$1\,\sigma$ allowed ranges of all particles apart from the lightest
Higgs boson, neutralino and chargino extend well beyond a mass of
$4\,\mathrm{TeV}$. Hence no discovery at LHC or ILC can be predicted
apart from the SM-like lightest SUSY Higgs boson, which is difficult
to be distinguished from the SM Higgs boson at the
LHC~\cite{AguilarSaavedra:2001rg}.  The allowed mass range of the
lightest Higgs boson is $m_{h}=114.0^{+14}_{-3}\,\mathrm{GeV}$,
showing that the precision data still has a tendency to push the Higgs
mass towards the lowest values allowed by the direct searches. At the
same time, the push is not strong enough to exclude that the light
Higgs boson mass is close to the theoretically allowed upper limit at
around $m_{h}^{\mathrm{theo}}<135\,\mathrm{GeV}$ in this scenario.

No detailed effect of other observables is tested due to the failure
of the remaining observables to constrain the mSUGRA parameter space to
the region accessible with the next generation of collider
experiments. However, this does not mean that the precision
observables or the flavour physics data will not have a decisive role
in helping to understand SUSY once sparticles are discovered. The
reasons for the strong impact of the precision data are
demonstrated in Section~\ref{sec:results:LELHC}. If a Higgs
boson is discovered, the precision of its mass measurement will be
much better than the precision of
$\sigma_{m_{h}}\approx\pm3\,\mathrm{GeV}$ obtained from the
low-energy, cosmological and precision data. Therefore a further
improvement on the uncertainties of the precision observables would be
beneficial for an ultimate cross-check of the collider data below and
above the electro-weak scale. 

In addition,
Figure~\ref{fig:results:LEonly:mSUGRA:massDist:nogmin2noOmega} shows a
very clear impact of the precision and flavour physics observables and
especially of the Higgs searches at LEP in form of an
implicit lower bound on the sparticle masses above or around the
current direct search limits
(e.~g.~\cite{Heister:2003zk,Abazov:2009zi,Aaltonen:2009fm,Heister:2001nk}). The
lightest neutralino mass is expected above 45~GeV, the lightest
chargino mass above 100~GeV, and the $\tilde{\tau}_1$ mass above
110~GeV. A small region with a chargino NLSP is allowed at the
$2\,\sigma$ level, with most points featuring a $\tilde{\tau}_1$ NLSP.

%%%%%%%%%%%%%%%%%%%%%%%%%%%%%%%%%%%%%%%%%%%%%%%%%%%%%%%%%%%%%%%%%%%%%%%
%%%%%%%%%%%%%%%%%%%%%%%%%%%%%%%%%%%%%%%%%%%%%%%%%%%%%%%%%%%%%%%%%%%%%%%
%%%%%%%%%%%%%%%%%%%%%%%%%%%%%%%%%%%%%%%%%%%%%%%%%%%%%%%%%%%%%%%%%%%%%%%
\subsubsection{Fits of the GMSB Model}\label{sec:results:LEonly:GMSB}

The mSUGRA scenario studied so far has the strong advantage that it
solves a very large amount of experimental and theoretical challenges
of the SM. In addition, it is very well studied at colliders, making
it an ideal testing ground. Therefore it is the main SUSY scenario
studied in this paper. However, it is not the only way how SUSY can be
broken at the expected Grand Unified Theory scale of around
$\Lambda_{\mathrm{GUT}}\approx 10^{16}\,\mathrm{GeV}$. As an example of a
different SUSY breaking mechanism, Gauge Mediated Supersymmetry
Breaking (GMSB)~\cite{Dine:1994vc} is explored. It has the
disadvantage that it includes a very light gravitino in the range of
$m_{\tilde{G}}\approx{\cal O}(1-10\,\mathrm{MeV})$ as LSP, which
leads to hot dark matter. This is difficult to be reconciled with
structure formation in the
Universe~\cite{Pierpaoli:1997im,Staub:2009ww}. Therefore,
$\Omega_{\mathrm{CDM}}h^2$ is not included in the analysis.

\begin{table*}
  \caption{Result of the fit of the GMSB model with different values
    of sign$(\mu)$ and $N_5$ including four additional SM parameters
    to all measurements listed in Table~\ref{tab:leobserables} except
    $\Omega_{\mathrm{CDM}} h^2$. The minimum $\chi^2$ value is $19.30$
    for 21 degrees of freedom, corresponding to a ${\cal P}$-value of
    56.5\,\%. The uncertainties correspond to the entry for
    $\mathrm{sign}(\mu)=+1$ and $N_5=1$.}
  \label{tab:results:LEonly:GMSB:noOmega}
  \begin{center}
    {\renewcommand{\arraystretch}{1.35}
      \begin{tabular}{lr|r|r|r|rcl}
        \hline\hline
        Parameter            & Best Fit   & Best Fit   & Best Fit   & Best Fit              & Best Fit   &       & Uncertainty for \\
                             &            &            &            &                       &            &       & $N_5=1,\mathrm{sign}(\mu)=+1$ \\
        \hline                                                                                         
        sign$(\mu)$            &  +1        &  -1        &  +1        &  +1                   &  +1        &       &             \\
        $N_5$                &   1        &   1        &   2        &   3                   &   4        &       &             \\
        $\tan\,\beta$        &  19.2      &   19.4     &  17.9      &       18.3            & 18.5       &       &  $_{-6.7}^{+15.3           }$    \\ 
        $\Lambda$ (GeV)            &  87050   &    307629  &  53284   &      40080          & 32643    &       &  $_{-17151}^{+31970        }$       \\
        $M_{\mathrm{mess}}$ (GeV)  &  431752    &    334662  &  688567    & $1.038\times10^{6}$  &  539328    &       &  $_{-352952 }^{+1.74\times10^{6}}$      \\ 
        $C_{\mathrm{grav}}$  &  411.4     &   446.1    &  885.5     &      460.1            & 3368.1     &       &  $_{-411.3}^{+10042.5    }$ \\
        \hline                                                                                
        $\chi^2_{\mathrm{min}}$ & 19.3    &  31.0      &  19.4      &  19.5                 & 19.5       &       &         \\
        \hline\hline
      \end{tabular}
    }
  \end{center}
\end{table*}

In GMSB, there are four continuous variables: $\tan\beta=v_2/v_1$ is
the ratio of the Higgs vacuum expectation values, $\Lambda$ is
universal mass scale of SUSY particles at the GUT scale,
$M_{\mathrm{mess}}$ denotes the mass scale of the messenger gauge
particles between the SUSY breaking sector and the visible sector, and
$C_{\mathrm{grav}}$ is the scale of the gravitino coupling. In
addition, there are two discrete parameters, namely
$\mathrm{sign}(\mu)$ and the number of messenger fields $N_5$. As
before, separate fits are performed for different values of the
discrete parameters. A selection of the results is shown in
Table~\ref{tab:results:LEonly:GMSB:noOmega}. It can be seen that there
is a similar sensitivity to $\mathrm{sign}(\mu)$ through the positive
value of $a_{\mu}^{\mathrm{exp}}-a_{\mu}^{\mathrm{SM}}$. There is,
however, no sensitivity to $N_5$, since all performed fits with
$N_5=1,2,3,4$ achieve the same value of
$\chi^{2}_{\mathrm{min}}=19.5$, corresponding to a ${\cal P}$-value of
56.5\,\%. It can be seen that no sensitive limit can be placed on
$M_{\mathrm{mess}}$ and $C_{\mathrm{grav}}$, while $\tan\beta$ and
$\Lambda$ can be constrained. Since the prediction for $\tan\beta$ and
$\Lambda$ is stable for different areas in $M_{\mathrm{mess}}$ and
$C_{\mathrm{grav}}$, the remaining large uncertainty does not affect
the constrained regions for $\tan\beta$ and $\Lambda$ given the
existing measurements. This shows an interesting complementarity to
the expected LHC measurements, where constraints cannot be set on
$\tan\beta$ and $C_{\mathrm{grav}}$ or $M_{\mathrm{mess}}$ alone, but
on $C_{\mathrm{grav}}\times
M_{\mathrm{mess}}$~\cite{Culbertson:2000am} through the measurement of
sparticle masses and gaugino lifetimes.

\begin{figure}
\includegraphics[width=0.49\textwidth]{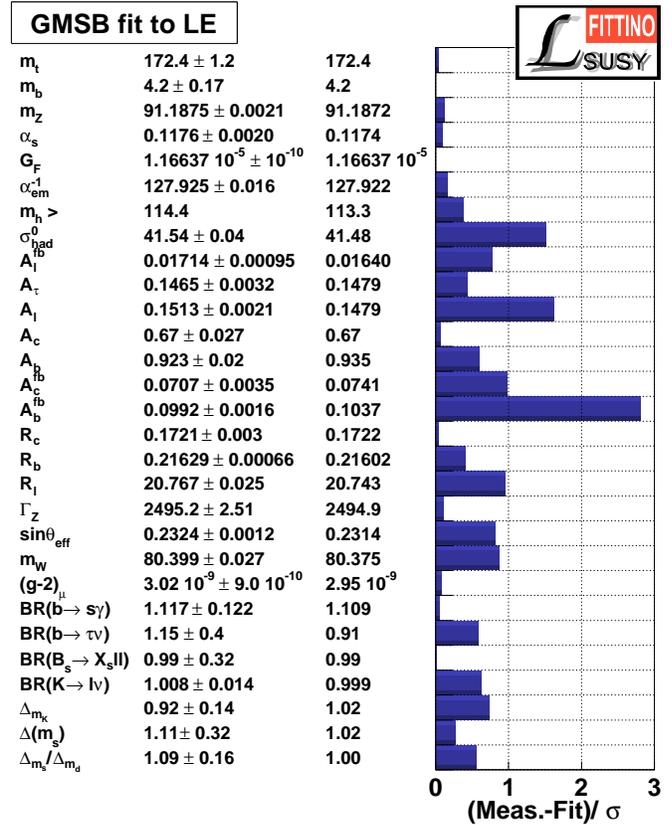}
\caption{Pull for the low energy observables used in the GMSB
  parameter fit with $N_5=1$ and sign$(\mu) = +1$, using the
  observables from Table~\ref{tab:leobserables} and the best fit point
  from Table~\ref{tab:results:LEonly:GMSB:noOmega}.}
\label{fig:results:LEonly:GMSB:pulls}
\end{figure}

The pull of the individual variables with respect to the best fit
result of the fit with $\mathrm{sign}(\mu)=+1$ and $N_5=1$ is shown in
Figure~\ref{fig:results:LEonly:GMSB:pulls}. The obtained pattern is
very similar to the pattern for the mSUGRA scenario, which again
confirms that the tension of the SM with the electro-weak precision
observables cannot be remedied by Supersymmetry, apart from moving the
Higgs boson mass close to or above the experimental limit. The
prediction of GMSB is $m_{h}=(113.5\pm2)\,\mathrm{GeV}$, very similar
to mSUGRA.

\begin{figure}
  \includegraphics[width=0.49\textwidth]{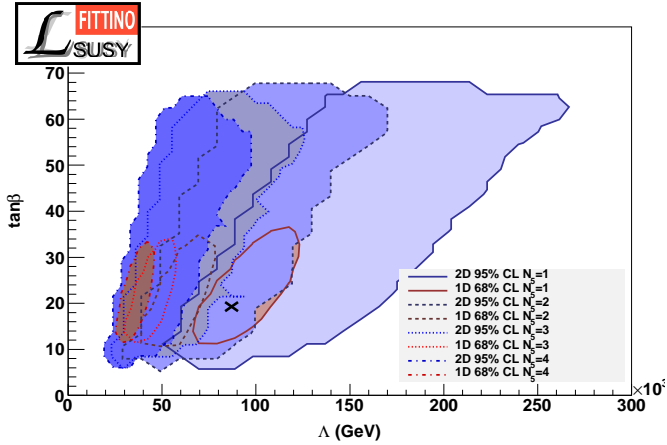}
  \caption{GMSB parameter regions with sign$(\mu) = +1$ and
    different values of $N_5$ compatible with the existing data from
    Table~\ref{tab:leobserables} except $\Omega_{\mathrm{CDM}} h^2$.}
  \label{fig:results:LEonly:GMSB:2dRes}
\end{figure}

The allowed parameter range in $\tan\beta$ and $\Lambda$ is shown in
Figure~\ref{fig:results:LEonly:GMSB:2dRes} for different values of
$N_5$. It is interesting to observe that, as already visible in
Table~\ref{tab:results:LEonly:GMSB:noOmega}, the predictions and hence
$\chi^2_{\mathrm{min}}$ remains unchanged for different values of
$N_5$. There are different preferred parameter regions in $\Lambda$,
leaving $\Lambda\times N_5$ approximately unchanged. Intermediate
values of $\tan\beta$ are preferred, but neither large nor high values
can be excluded at the $2\,\sigma$ level.

\begin{figure}
  \includegraphics[width=0.49\textwidth]{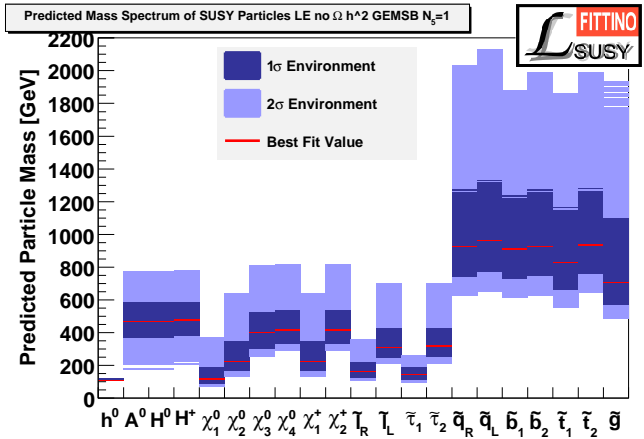}
  \caption{SUSY mass spectrum as predicted by GMSB parameter fit to
    low energy measurements except $\Omega_{\mathrm{CDM}} h^2$ with
    $N_5$ fixed to 1 and sign$(\mu) = +1$.}
  \label{fig:results:LEonly:GMSB:massDist1:noOmega}
\end{figure}

\begin{figure}
  \includegraphics[width=0.49\textwidth]{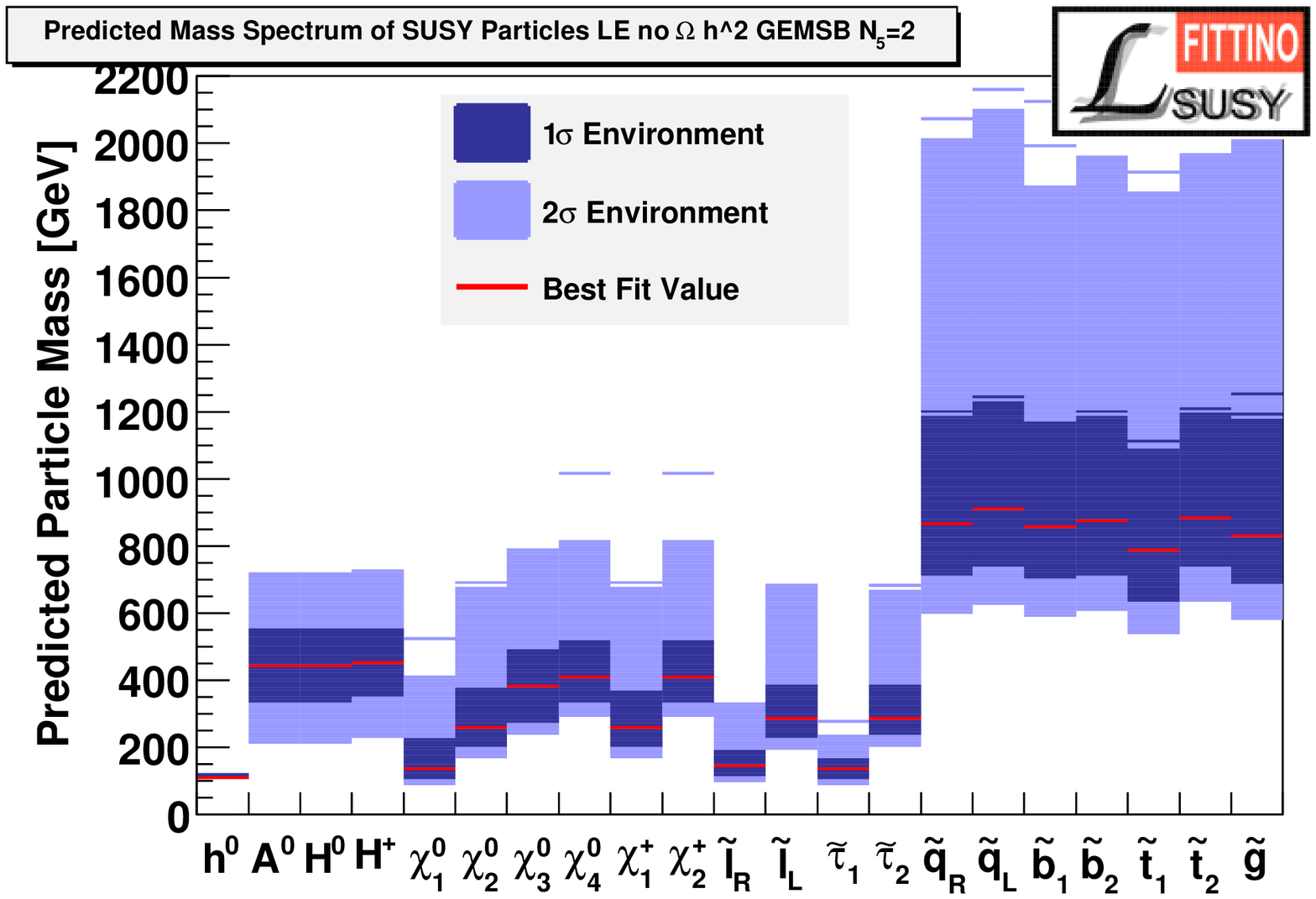}
  \caption{SUSY mass spectrum as predicted by GMSB parameter fit to
    low energy measurements except $\Omega_{\mathrm{CDM}} h^2$ with
    $N_5$ fixed to 2 and sign$(\mu) = +1$.}
  \label{fig:results:LEonly:GMSB:massDist2:noOmega}
\end{figure}

\begin{figure}
  \includegraphics[width=0.49\textwidth]{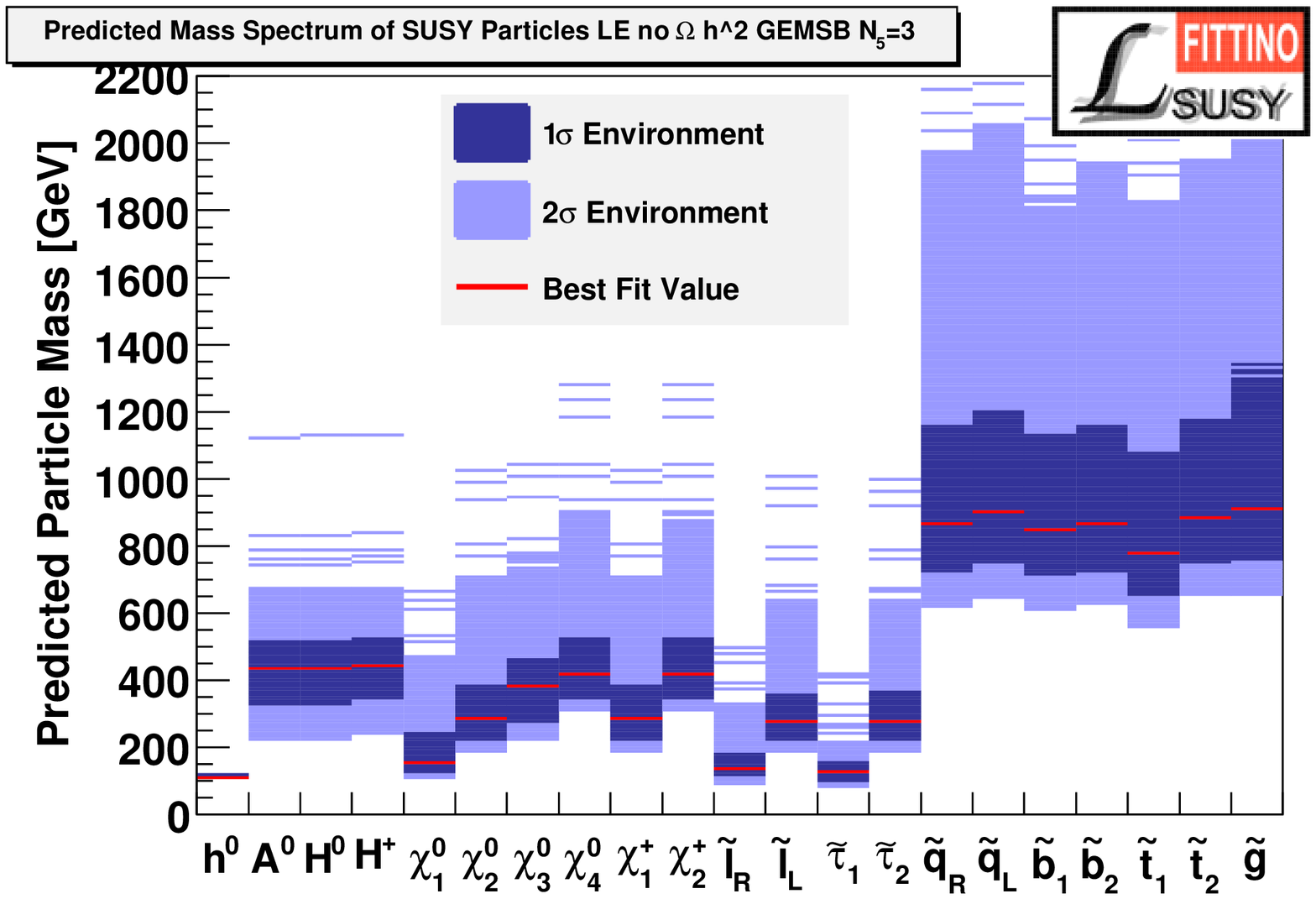}
  \caption{SUSY mass spectrum as predicted by GMSB parameter fit to
    low energy measurements except $\Omega_{\mathrm{CDM}} h^2$ with
    $N_5$ fixed to 3 and sign$(\mu) = +1$.}
  \label{fig:results:LEonly:GMSB:massDist3:noOmega}
\end{figure}

\begin{figure}
  \includegraphics[width=0.49\textwidth]{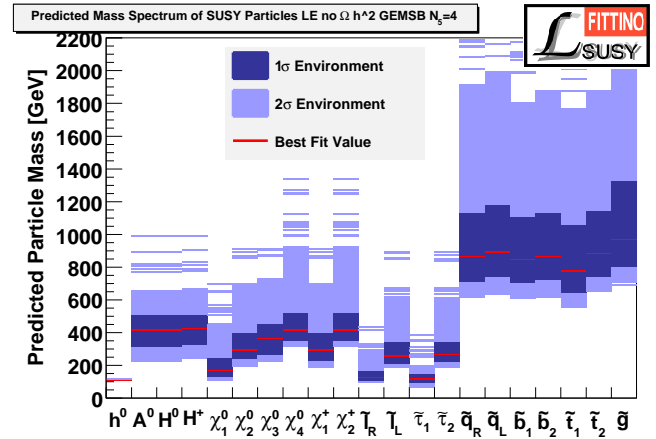}
  \caption{SUSY mass spectrum as predicted by GMSB parameter fit to
    low energy measurements except $\Omega_{\mathrm{CDM}} h^2$ with
    $N_5$ fixed to 4 and sign$(\mu) = +1$.}
  \label{fig:results:LEonly:GMSB:massDist4:noOmega}
\end{figure}

The predicted sparticle spectrum is also insensitive to $N_5$
itself, as shown in
Figures~\ref{fig:results:LEonly:GMSB:massDist1:noOmega}
to~\ref{fig:results:LEonly:GMSB:massDist4:noOmega}, where an almost
perfect agreement between the spectra for different values of $N_5$ is
shown.  Generally the expected spectrum looks similar to the expected
mSUGRA spectrum. This is a strong hint that given the existing
measurements, Supersymmetry generally provides for the prediction of a
rich observable particle spectrum at the LHC, independent of the SUSY
breaking mechanism. In GMSB, there is a tendency towards higher values
of the squark and gluino masses with respect to mSUGRA, but the
difference in the predictions is not decisive enough to base a
distinction between the scenarios on the mass hierarchies for
scenarios with similar visible parts of the decay chains in GMSB and
mSUGRA.

%%%%%%%%%%%%%%%%%%%%%%%%%%%%%%%%%%%%%%%%%%%%%%%%%%%%%%%%%%%%%%%%%%%%%%%
%%%%%%%%%%%%%%%%%%%%%%%%%%%%%%%%%%%%%%%%%%%%%%%%%%%%%%%%%%%%%%%%%%%%%%%
%%%%%%%%%%%%%%%%%%%%%%%%%%%%%%%%%%%%%%%%%%%%%%%%%%%%%%%%%%%%%%%%%%%%%%%
\subsubsection{Conclusions for SUSY Scenarios at Colliders}\label{sec:results:LEonly:Conclusions}

The results show that Supersymmetry, broken at the GUT scale, offers
several possibilities to fit the existing precision data.
In the mSUGRA breaking scenario, all parameters can be constrained and
the existing data from cosmological, low-energy, flavour physics and
precision collider sources clearly prefer parameter ranges which are
accessible at the next generation of collider experiments. It is shown
that as expected $\Omega_{\mathrm{CDM}}h^2$ and $(g-2)_{\mu}$ provide
for the most sensitive constraints among the available
measurements. Even if individual, very sensitive, variables are
removed, or if deviations between data and the SM prediction are
assigned to unknown systematic uncertainties, a clear constraint in
the accessible mass regions remains. Only voluntarily removing both
$\Omega_{\mathrm{CDM}}h^2$ and $(g-2)_{\mu}$ simultaneously from the
list of observables used in the fit, removes the experimental
constraint to the parameter region accessible at the LHC.

The best fit parameter spectrum and the uncertainties from present
data clearly prefers a SUSY scenario with a rich phenomenology both at
the LHC and the ILC. At the $1\,\sigma$ level, all Higgs bosons,
gauginos (apart from the gluino) and all sleptons are expected below
$m\leq 600\,\mathrm{GeV}$. The squarks and gluinos are expected below
$m\leq 900\,\mathrm{GeV}$. While this provides for relatively early
discovery at the LHC, the rich expected spectrum with many concurrent
production and decay modes will contribute to a challenging
reconstruction of the LHC observables sensitive to SUSY masses and
branching fractions. In particular, a very small mass difference
between the neutralino LSP and the NLSP, which is the $\tilde{\tau}_1$
of $m_{\mathrm{NLSP}}-m_{\mathrm{LSP}}<22\,\mathrm{GeV}$ at the
95\,\%~CL level leads to dominating decays of the gauginos into final
states with $\tau$ leptons. However, the exact branching fractions
cannot be predicted with strong precision.

\subsection{Expected LHC Measurements}\label{sec:results:LHConly}

Based on the results of the previous section, we now investigate the
prospects for the determination of SUSY parameters from future LHC
measurements. Within the mSUGRA model, the preferred parameters from
existing LE measurements and constraints clearly point towards rather
light sparticle masses. In order to be consistent with the measured
dark matter relic density, co-annihilation of the LSP and the NLSP has
to contribute to the dark matter annihilation process.  For this
process to be efficient, the mass difference between NLSP and LSP has
to be rather small. For the best fit point, within mSUGRA, the
difference $m(\tilde{\tau}_1) - m(\tilde{\chi}^0_1)$ is only 8~GeV,
and the difference $m(\tilde{e}_R) - m(\tilde{\chi}^0_1)$ is
22~GeV. No detailed experimental studies for LHC prospects are
available for this specific parameter point. However, detailed studies
exist for the SPS1a parameter point with parameters $\tan\beta = 10$,
$A_0 = -100 $~GeV, $M_{1/2} = 250$~GeV, $M_0 = 100$~GeV, sign$(\mu) =
+1$~\cite{Allanach:2002nj}. These parameters taken at face value lead
to a significantly larger dark matter relic density due to mass
differences $m(\tilde{\tau}_1) - m(\tilde{\chi}^0_1) = 37$~GeV and
$m(\tilde{e}_R) - m(\tilde{\chi}^0_1) = 47$~GeV. Apart from this
difference, the collider phenomenology of SPS1a is very similar to
that of the best fit point. The smaller mass differences lead to
softer spectra for the final state leptons, which is a small caveat to
be kept in mind in the following analysis and should serve to trigger
more optimisation of soft lepton identification within the LHC
experiments.  We assume in the following as a plausible scenario that
SUSY is realised with electro-weak scale parameters derived from the
high-scale SPS1a parameters and will be discovered by the LHC
experiments.  As input measurements we use the observables specified
in Section~\ref{sec:lhcobservables} for three different integrated
luminosities. If taken at face value, the lightest Higgs boson mass in
SPS1a calculated with SPheno is 109~GeV, slightly below the LEP
exclusion. Given the theoretical uncertainty as well as the strong
dependence of $m_h$ on the top quark mass we do not consider this as
an inevitable constraint. Technically, we set the LEP Higgs mass limit
slightly below 109~GeV for the luminosity scenarios where no Higgs
boson has been found yet at the LHC and we assume that a 109~GeV Higgs
boson could be discovered at the LHC with similar sensitivity as for
115~GeV.

\subsubsection{mSUGRA Fit with fixed sign$(\mu)$}\label{sec:results:LHConly:fixedmu}

\begin{center}
  \begin{figure*}
    \begin{center}
      \begin{minipage}{0.8\textwidth}
        \includegraphics[width=0.49\textwidth]{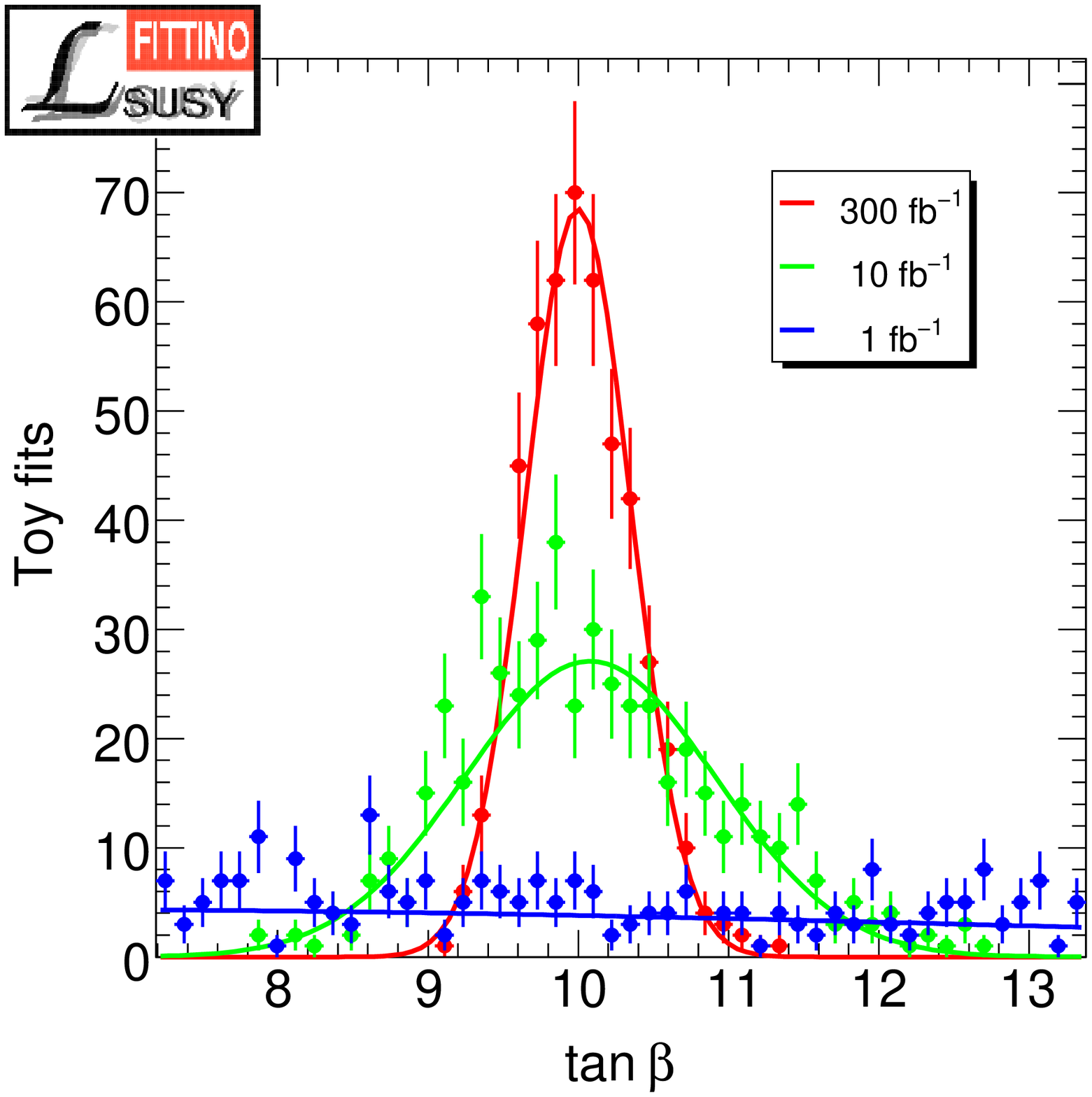}
        \hfill
        \includegraphics[width=0.49\textwidth]{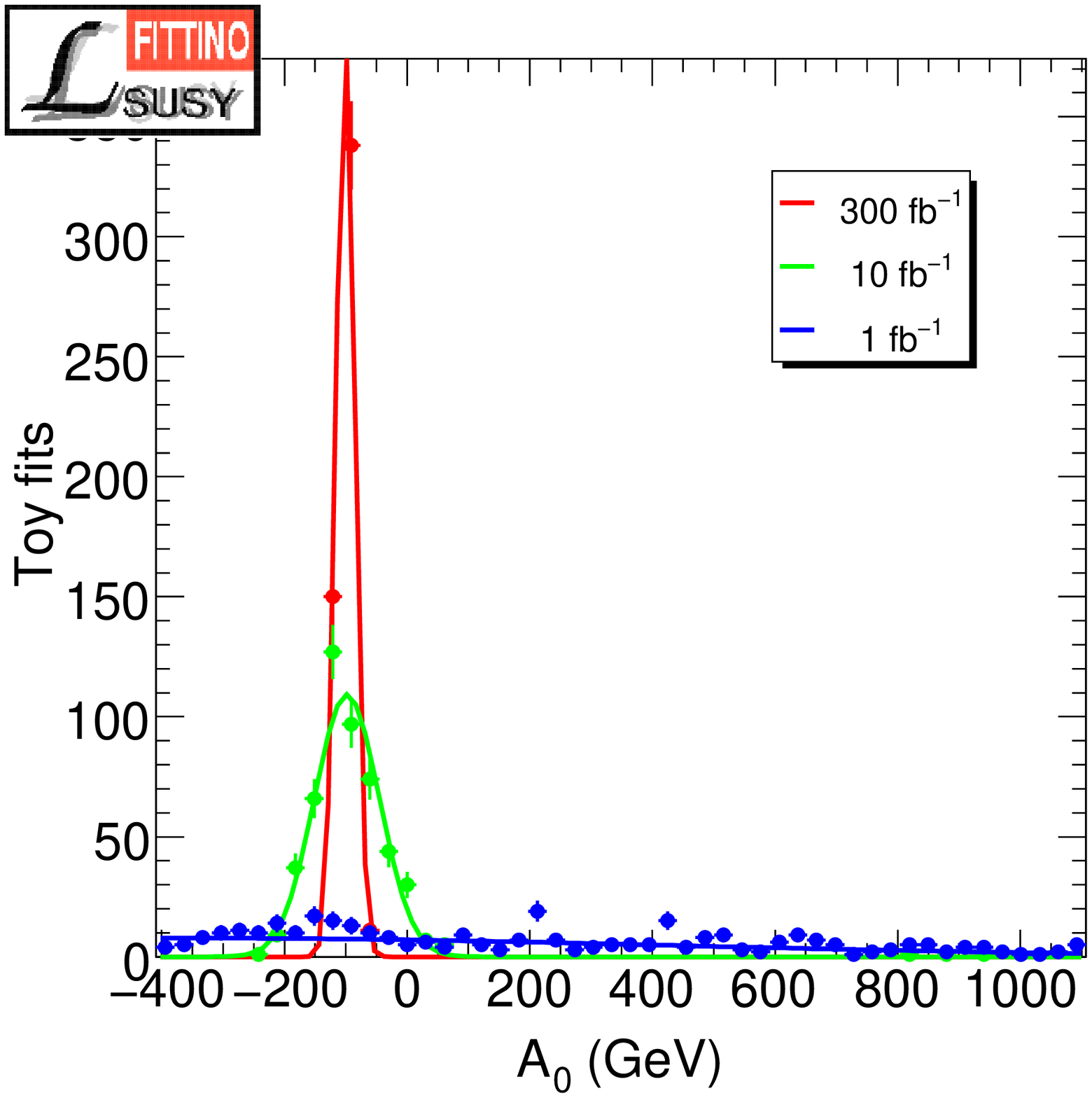}
        \\[1mm]
        \includegraphics[width=0.49\textwidth]{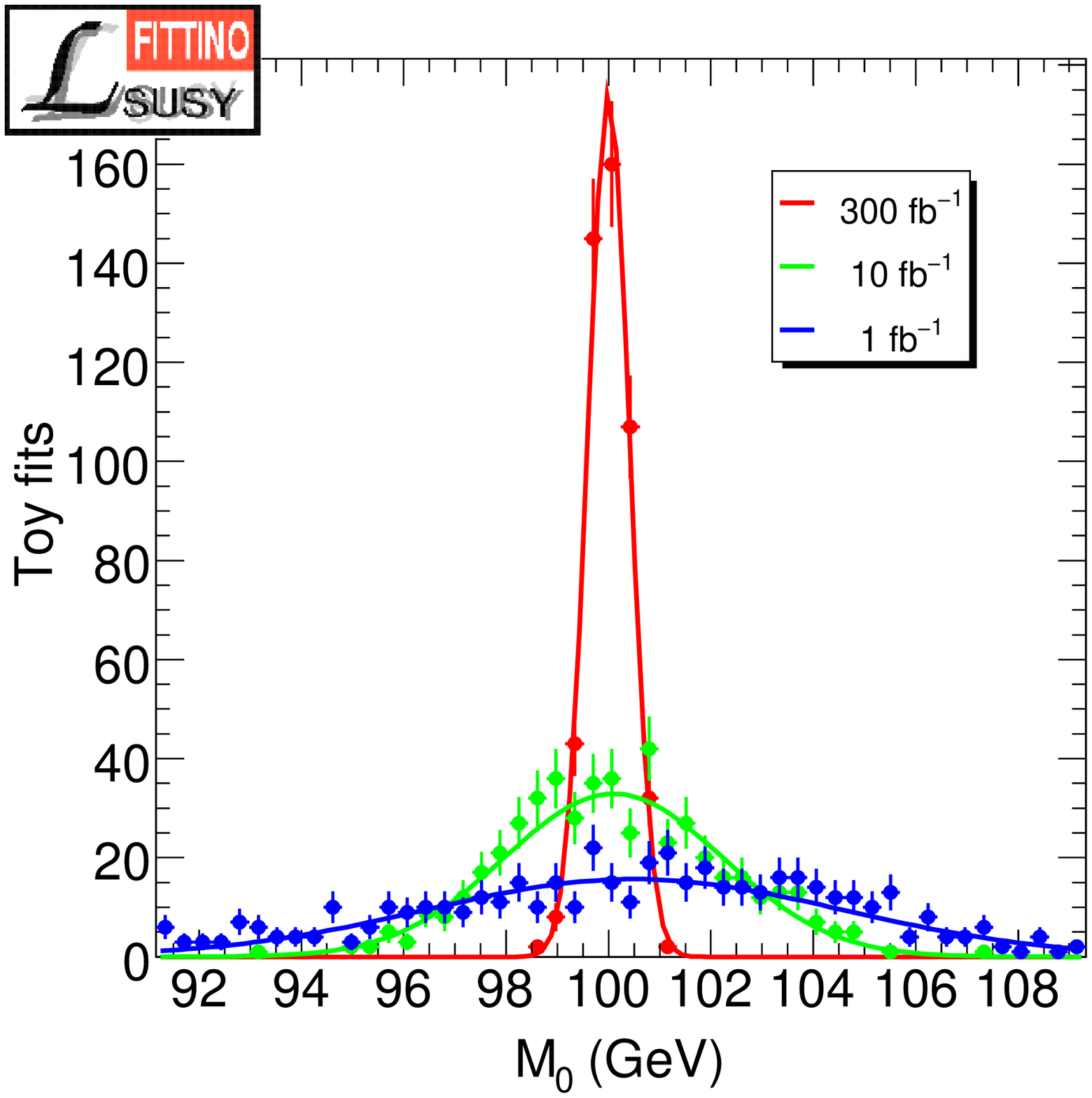}
        \hfill
        \includegraphics[width=0.49\textwidth]{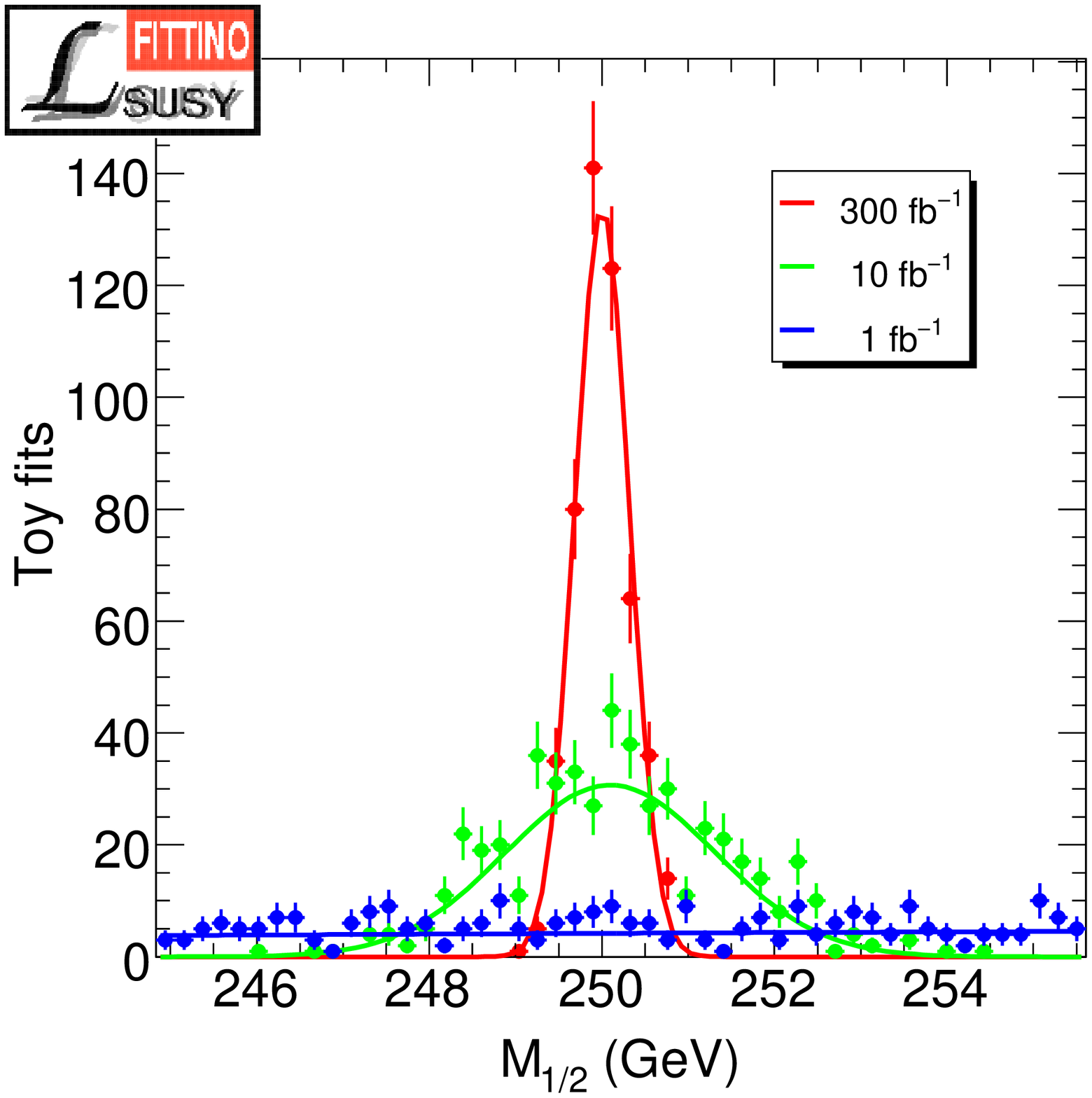}
        \\[1mm]
        \includegraphics[width=0.49\textwidth]{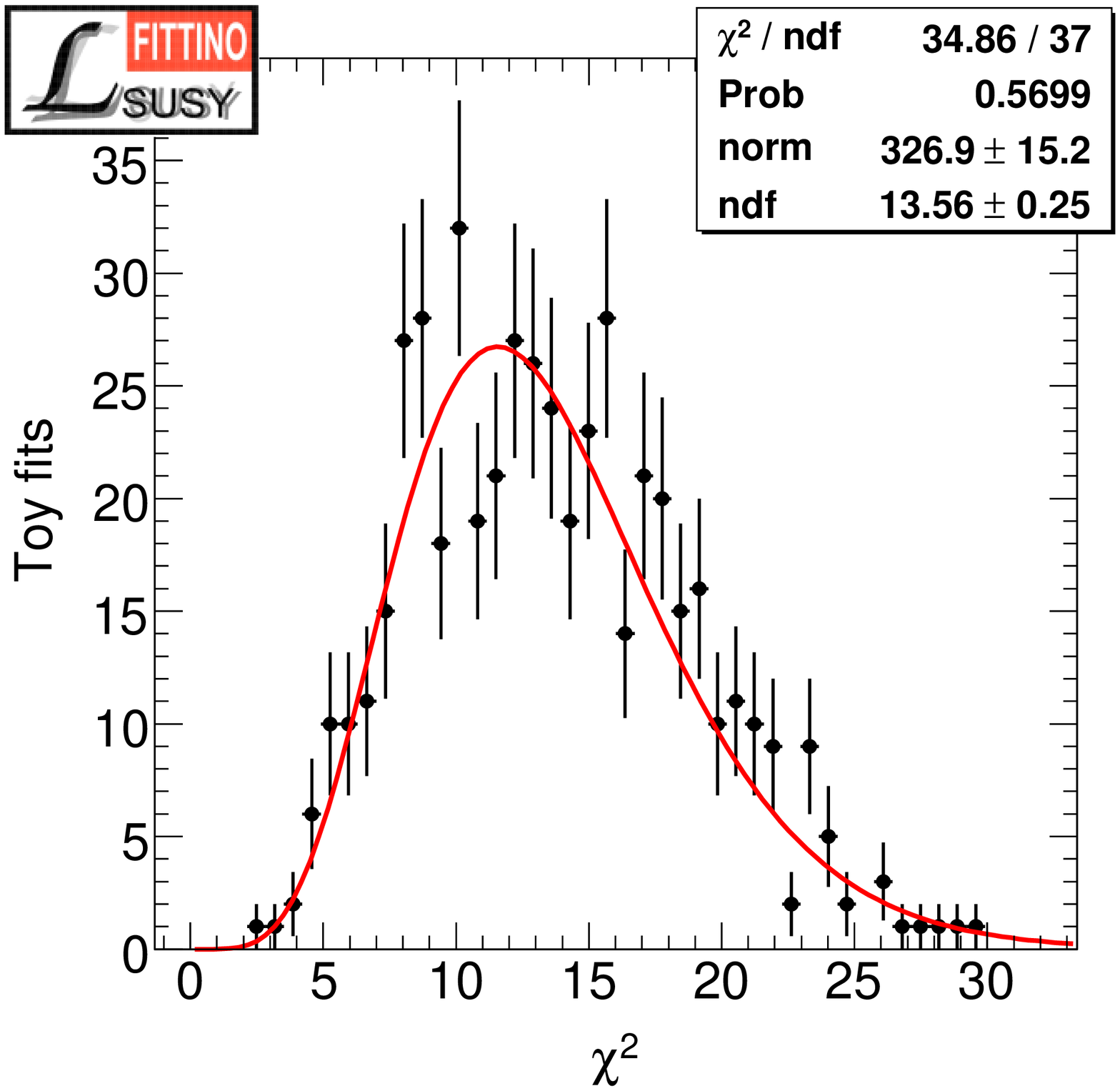}
        % \hfill
        % \includegraphics[width=0.32\textwidth]{figures/Chi2_mSUGRA_LHC10}
        \hfill
        \includegraphics[width=0.49\textwidth]{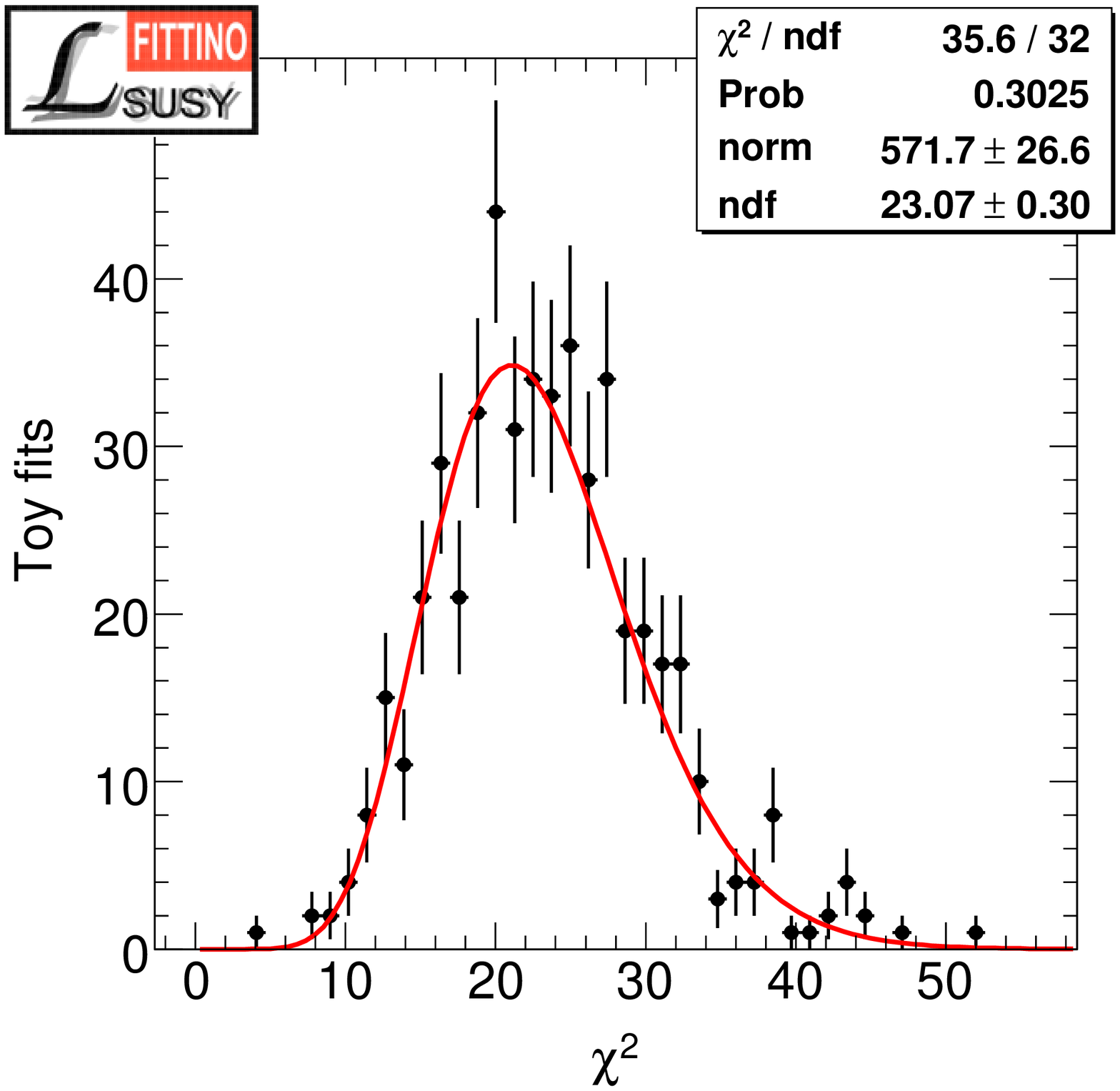}
      \end{minipage}
    \end{center}
    \caption{Results of the Toy Fits to the expected LHC
      ``measurements'' from Table~\ref{tab:inputs} for different
      integrated luminosities. The upper plots show the expected
      distribution of best fit points, exhibiting the strong increase
      in precision for ${\cal
        L}^{\mathrm{int}}\geq10\,\mathrm{fb}^{-1}$.  The lower row of
      plots shows the $\chi^2$ distributions for the Toy Fits with
      integrated luminosities of 1~fb$^{-1}$ and 300~fb$^{-1}$,
      showing very good agreement with the expected $\chi^2$
      distributions for the respective degrees of freedom.}
    \label{fig:results:mSUGRALHC}
  \end{figure*}
\end{center}

The good agreement in collider phenomenology between SPS1a and the
mSUGRA best fit point offers the possibility to use the wealth of
Monte Carlo studies performed for this benchmark point to attempt a
projection of the SUSY model discrimination power and parameter
constraints to the LHC era. This is done by performing fits to Toy
data which have been obtained by smearing the observable values
according to a Gaussian around the nominal SPS1a values as explained
in Section~\ref{sec:techniques}. In the top four plots in
Figure~\ref{fig:results:mSUGRALHC} the distributions for the fitted
parameters from these Toy Fits are shown assuming $\mu > 0$ for
integrated luminosities of 1~fb$^{-1}$, 10~fb$^{-1}$ and
300~fb$^{-1}$. The fitted central values and their uncertainties are
obtained from a Gaussian fit to the parameter distributions. The
corresponding values are listed in
Table~\ref{tab:results:LHConly:mSUGRA:1fb-1},
\ref{tab:results:LHConly:mSUGRA:10fb-1} and
\ref{tab:results:LHConly:mSUGRA:300fb-1}.  The two bottom plots of
Figure~\ref{fig:results:mSUGRALHC} display, as examples, the $\chi^2$
distributions from the Toy Fits for the smallest and largest
considered luminosity.  The fact that the $\chi^2$ distributions are
in very good agreement with the expectations for the respective
degrees of freedom provides confidence that the $\chi^2$ minimisation
algorithm works reliably.

Already with 1~fb$^{-1}$ of LHC data it is possible to constrain the
scalar mass parameter $M_0$ and the gaugino mass parameter $M_{1/2}$
to the level of a few percent due to the already relatively precise
measurements of the endpoints of the $\ell q$ and the $\ell \ell q$
invariant mass spectra. For the determination of $M_{1/2}$ also the
$m_{T2}$ measurement in events with $\tilde{q}_R \to q
\tilde{\chi}_1^0$ decays is important. $\tan \beta$ and $A_0$ are more
difficult to determinee. For $\tan \beta$, 
approximately 40~\% precision and for
$A_0$ only an order of magnitude estimate is obtained. With
1~fb$^{-1}$ the best constraints on these parameters come from
measurements involving third generation particles, in particular from
$m_{tb}^w$.

\begin{table}
  \caption{Result of the fit of the mSUGRA model to the expected LHC
    observables for 1 fb$^{-1}$. 
}
  \label{tab:results:LHConly:mSUGRA:1fb-1}
  \begin{center}
    \begin{tabular}{lrcl}
      \hline\hline
      Parameter     & Best Fit  &       & Uncertainty \\
      \hline
      sign$(\mu)$     &  $+$1       &       &             \\
      $\tan\,\beta$ &   9.1   & $\pm$ &   3.7   \\ % better rms?
      $A_0$ (GeV)   & $-$131.8   & $\pm$ & 742.1     \\ % better rms?
      $M_0$ (GeV)   & 100.2   & $\pm$ &  4.2    \\ % remove fits?
      $M_{1/2}$ (GeV) & 249.7   & $\pm$ &  6.7     \\
      \hline\hline
    \end{tabular}
  \end{center}
\end{table}

The increased precision on $M_0$ for an integrated luminosity of
10~fb$^{-1}$ mainly comes from more precise measurements of the
endpoints of the $\ell q$ spectra. For $M_{1/2}$ also new sensitive
measurements become available with a larger data sample, in particular
$m_{\tilde{g}} - m_{\tilde{\chi}_1^0}$. The measurement of $m_{tb}^w$
still remains important to constrain $\tan \beta$ and $A_0$. In
addition $m_{\ell \ell}^{\mathrm{max}}$ from $\tilde{\chi}_4^0$ decays
provides valuable additional information on $\tan\beta$ and $A_0$ at
10~fb$^{-1}$. For $\tan\beta$ also the ratio of branching fractions
(\ref{eq:brratiochi2decays}) starts to contribute.
\begin{table}
  \caption{Result of the fit of the mSUGRA model to the expected LHC
    observables for 10 fb$^{-1}$.}
  \label{tab:results:LHConly:mSUGRA:10fb-1}
  \begin{center}
    \begin{tabular}{lrcl}
      \hline\hline
      Parameter     & Best Fit &       & Uncertainty \\
      \hline
      sign$(\mu)$     &  $+$1      &       &             \\
      $\tan\,\beta$ &  10.08   & $\pm$ &  0.84   \\
      $A_0$ (GeV)   & $-$98.0    & $\pm$ & 52.9     \\ % remove fits?
      $M_0$ (GeV)   & 100.1    & $\pm$ &  2.1    \\
      $M_{1/2}$ (GeV) & 250.1    & $\pm$ &  1.2    \\
      \hline\hline
    \end{tabular}
  \end{center}
\end{table}

With 300~fb$^{-1}$ of LHC data it will finally be possible to
constrain $M_0$ and $M_{1/2}$ down to (a few) permille level. The
driving factor is an increased precision on $m_{\ell
  q}^{\mathrm{high}}$, $m_{\ell q}^{\mathrm{low}}$ and -- in case of
$M_{1/2}$ -- also on $m_{\ell\ell q}^{\mathrm{max}}$.  Similarly the
improvement on $A_0$ can be traced back to better measurements of
those observables which already provide the best constraints for
10~fb$^{-1}$, namely $m_{\ell \ell}^{\mathrm{max}}$ from
$\tilde{\chi}_4^0$ decays and $m_{tb}^w$. $\tan \beta$ at
300~fb$^{-1}$ is mainly controlled by measurements of
quantity~(\ref{eq:brratiochi2decays}) and the lightest Higgs mass
$m_h$. For $\tan \beta$ ($A_0$) a relative precision of approximately
4~\% (11~\%) is finally achieved from the given list of observables.
\begin{table}
  \caption{Result of the fit of the mSUGRA model to the expected LHC
    observables for 300~fb$^{-1}$.}
  \label{tab:results:LHConly:mSUGRA:300fb-1}
  \begin{center}
    \begin{tabular}{lrcl}
      \hline\hline
      Parameter     & Best Fit   &       & Uncertainty \\
      \hline
      sign$(\mu)$     &   $+$1       &       &             \\
      $\tan\,\beta$ &    9.98    & $\pm$ &  0.35   \\
      $A_0$ (GeV)   & $-$100.2   & $\pm$ & 11.1     \\
      $M_0$ (GeV)   &  100.0   & $\pm$ &  0.39   \\
      $M_{1/2}$ (GeV) &  250.0   & $\pm$ &  0.30   \\
      \hline\hline
    \end{tabular}
  \end{center}
\end{table}

Concerning the most constraining observables mentioned above it should
be noted that they might be very sensitive to small changes of the
input measurements. Therefore they ought to be taken with some care
and should not be generalised without further cross-checks. \newline

\subsubsection{Determination of sign$(\mu)$}\label{sec:results:LHConly:signmu}

\begin{center}
  \begin{figure*}
    \begin{center}
      \includegraphics[width=0.32\textwidth]{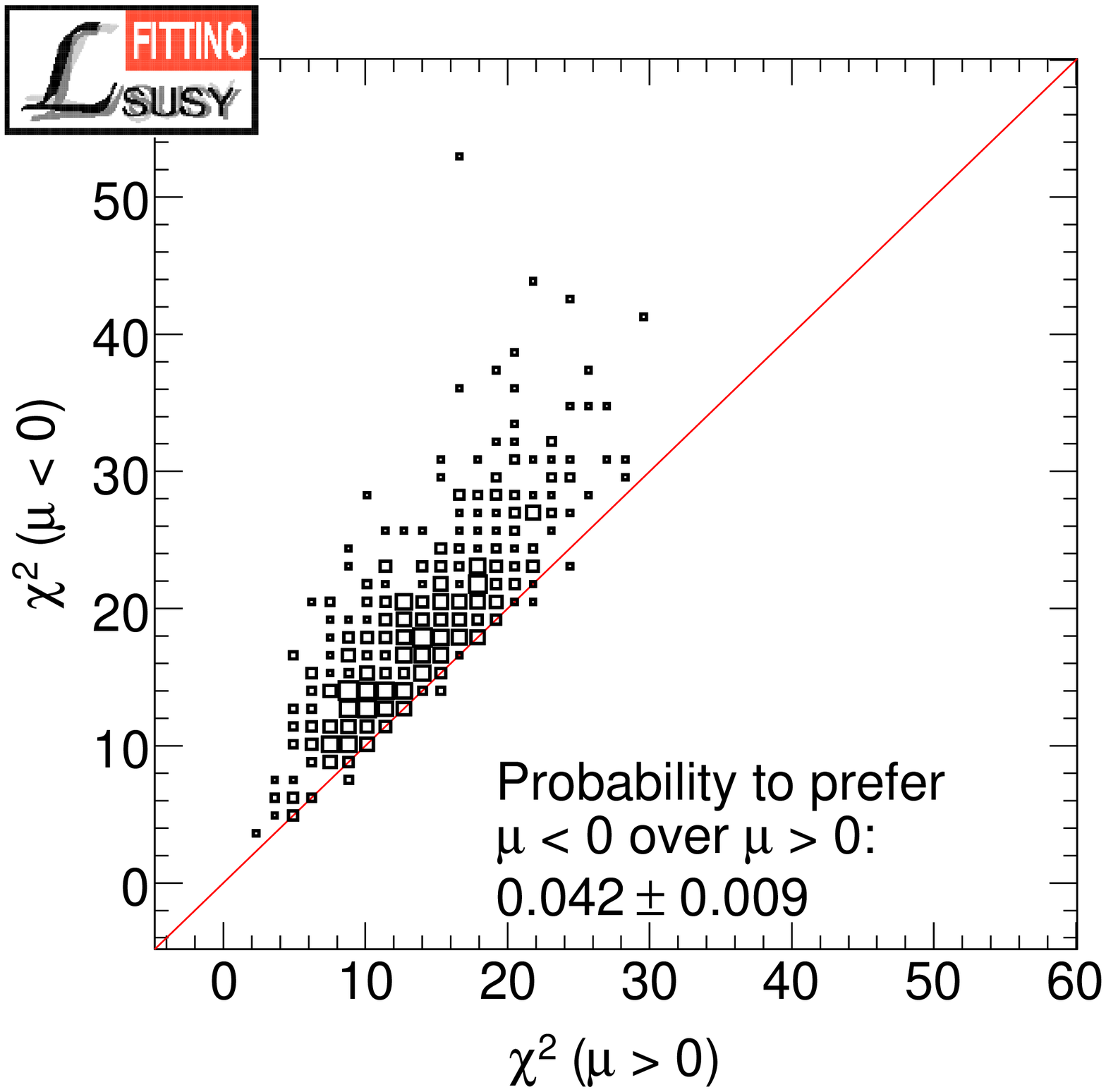}
      \hfill
      \includegraphics[width=0.32\textwidth]{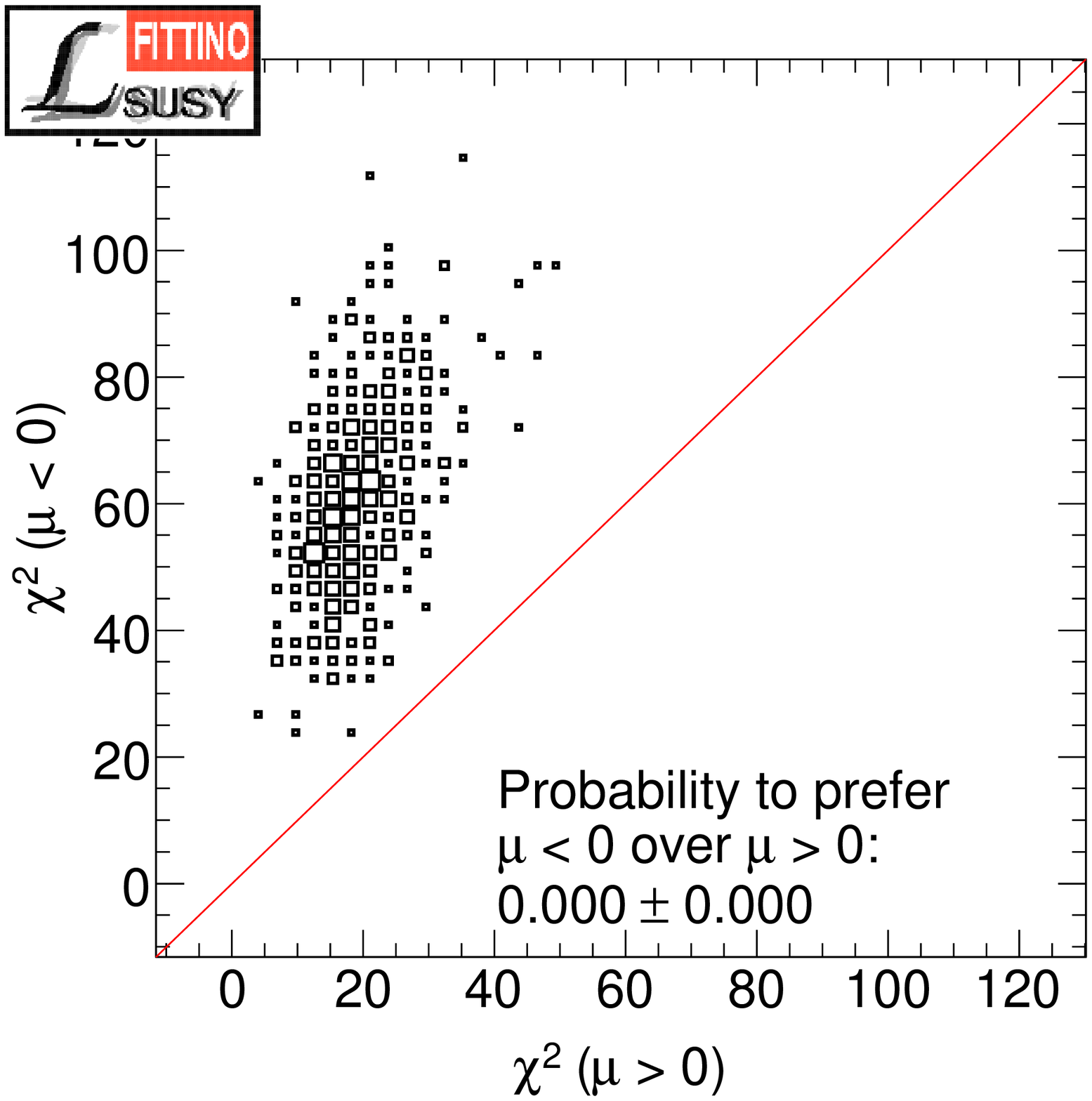}
      \hfill
      \includegraphics[width=0.32\textwidth]{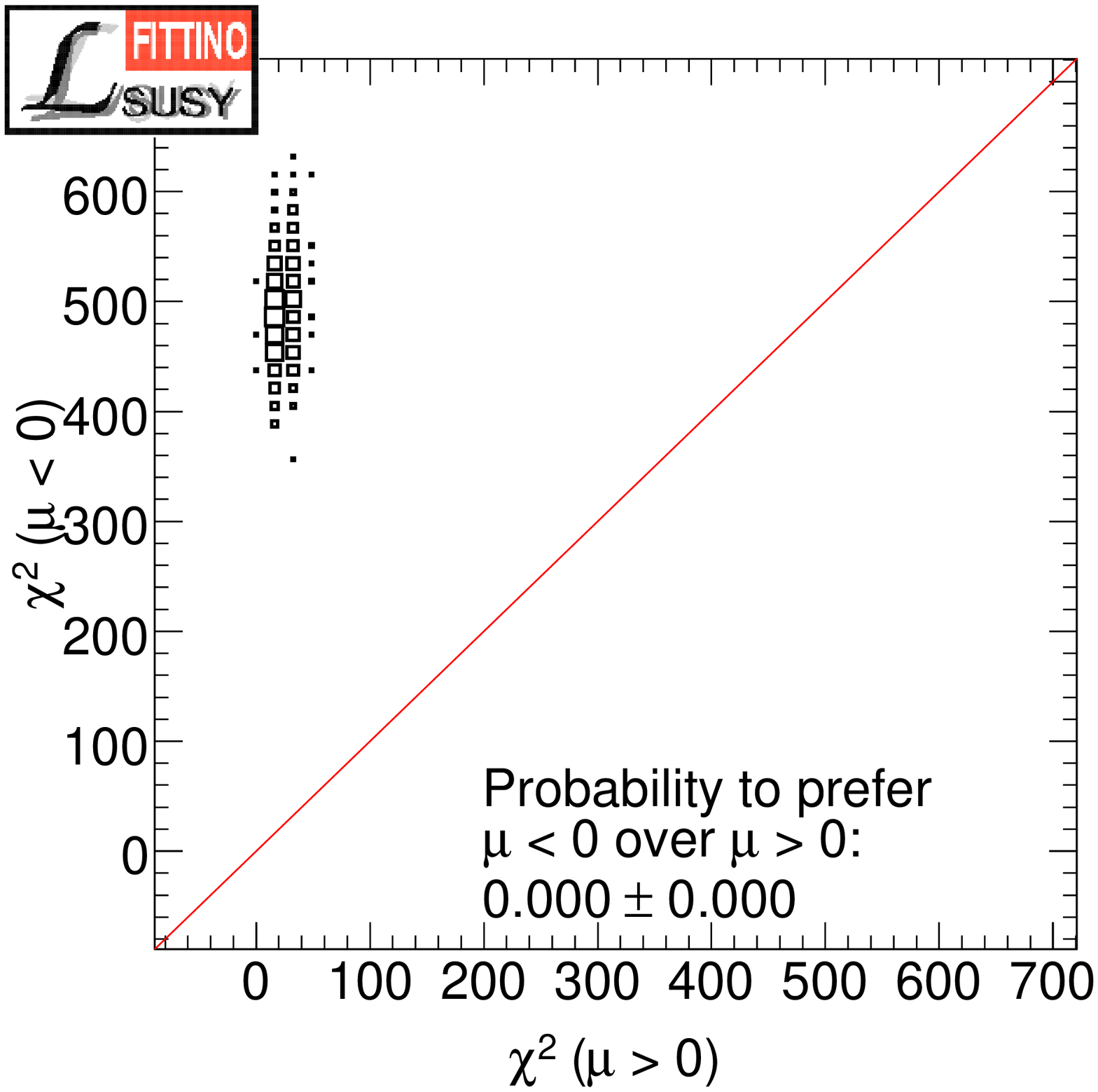}
    \end{center}
    \caption{$\chi^2$ correlations for mSUGRA fits with
      $\mu > 0$ and $\mu < 0$ to the same Toy data set (left/middle/right
      for an integrated LHC luminosity of 1 fb$^{-1}$/10 fb$^{-1}$/300
      fb$^{-1}$).}
    \label{fig:Chi2CorrelationsSignMuLHC}
  \end{figure*}
\end{center}
For the LHC fit results described above, we have not yet discussed how
to find the correct sign of $\mu$. Using the technique described in
Section~\ref{sec:techniques:comparisons}, we also checked how well the
sign of $\mu$ can be determined from LHC data. This is done by
performing Toy Fits for each sign of $\mu$ to the identical set of Toy
data. Figure~\ref{fig:Chi2CorrelationsSignMuLHC} shows the
$\chi^2$ correlations obtained from such fits. If we choose the value for sign
of $\mu$ which yields the best $\chi^2$ for a given set of LHC
measurements we can estimate the probability to make the wrong choice
by counting the number of Toy Fits below the red bisecting line in
Figure~\ref{fig:Chi2CorrelationsSignMuLHC} and normalise it to the
total number of Toy Fits. The corresponding numbers can be read off
the plots. Already with 1~fb$^{-1}$ there is a good chance
to extract sign$(\mu)$ correctly. The probability for a wrong choice
is less than 5~\%. Based on 10~fb$^{-1}$ or more of LHC data,
sign$(\mu)$ can be determined with negligible error
probability. \newline

\subsubsection{First Investigation of Chain Ambiguities}\label{sec:results:LHConly:chainAmbiguities}

\begin{center}
  \begin{figure}
    \begin{center}
      \includegraphics[width=0.32\textwidth]{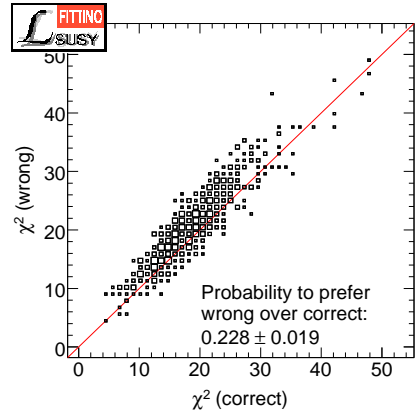}
    \end{center}
    \caption{$\chi^2$ correlations obtained from fits to the same Toy data
      set for two different interpretations of
      $m_{\ell\ell}^{\text{max}}$.}
    \label{fig:results:LHConly:chi2ModelComp}
  \end{figure}
\end{center}

\begin{center}
  \begin{figure*}
    \begin{center}
      \begin{minipage}{0.8\textwidth}
        \includegraphics[width=0.49\textwidth]{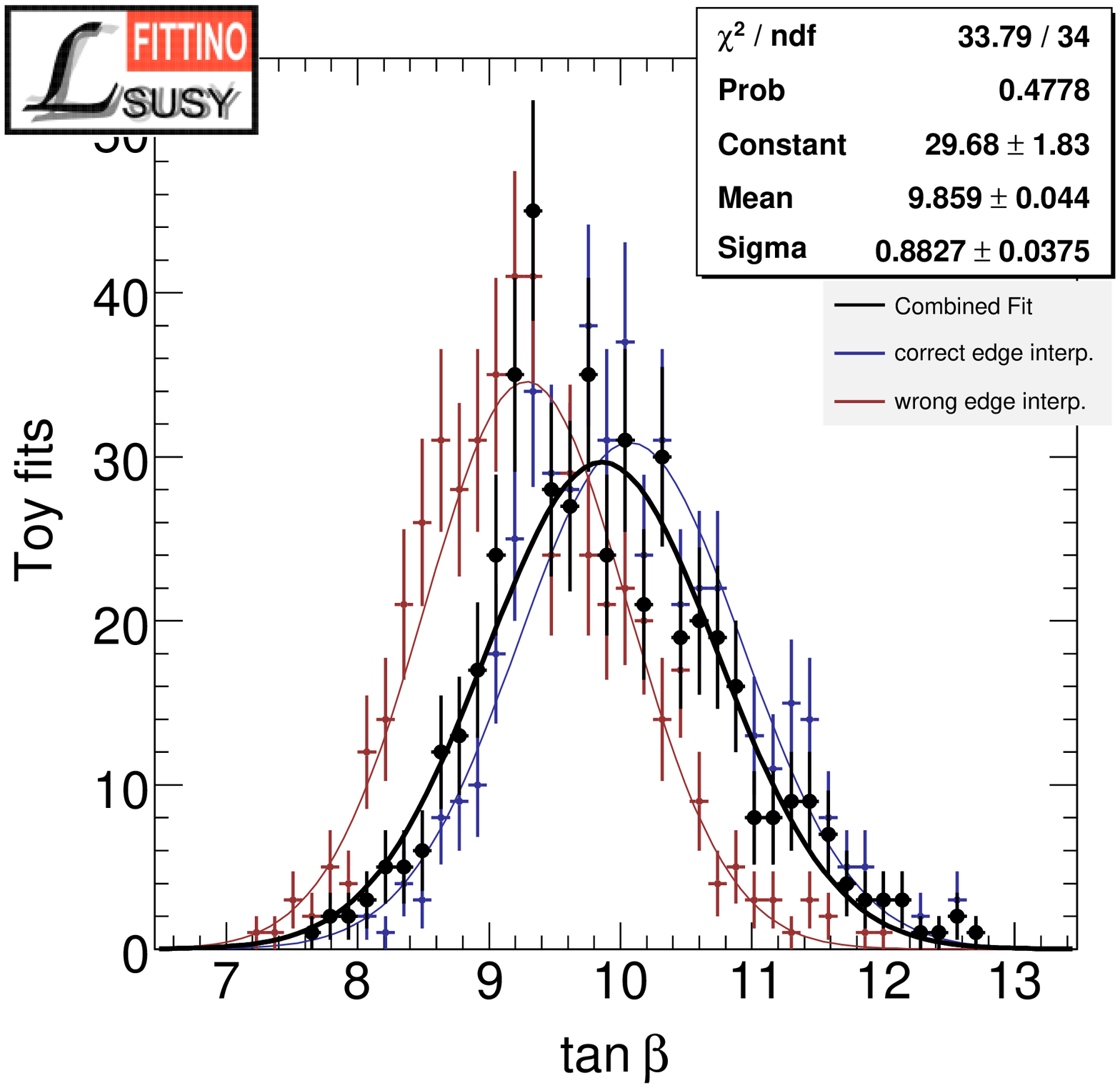}
        \hfill
        \includegraphics[width=0.49\textwidth]{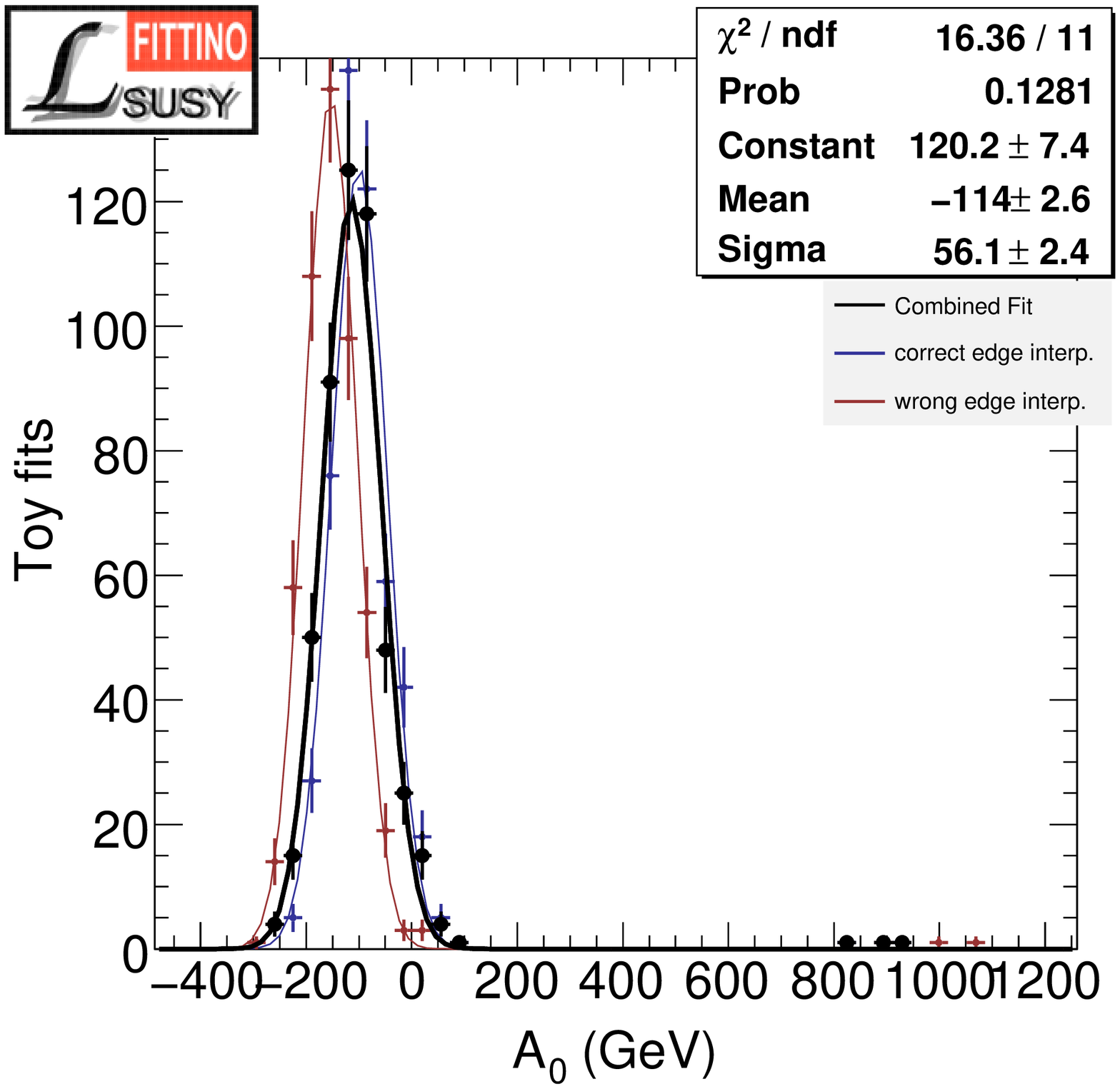}
        \\[1mm]
        \includegraphics[width=0.49\textwidth]{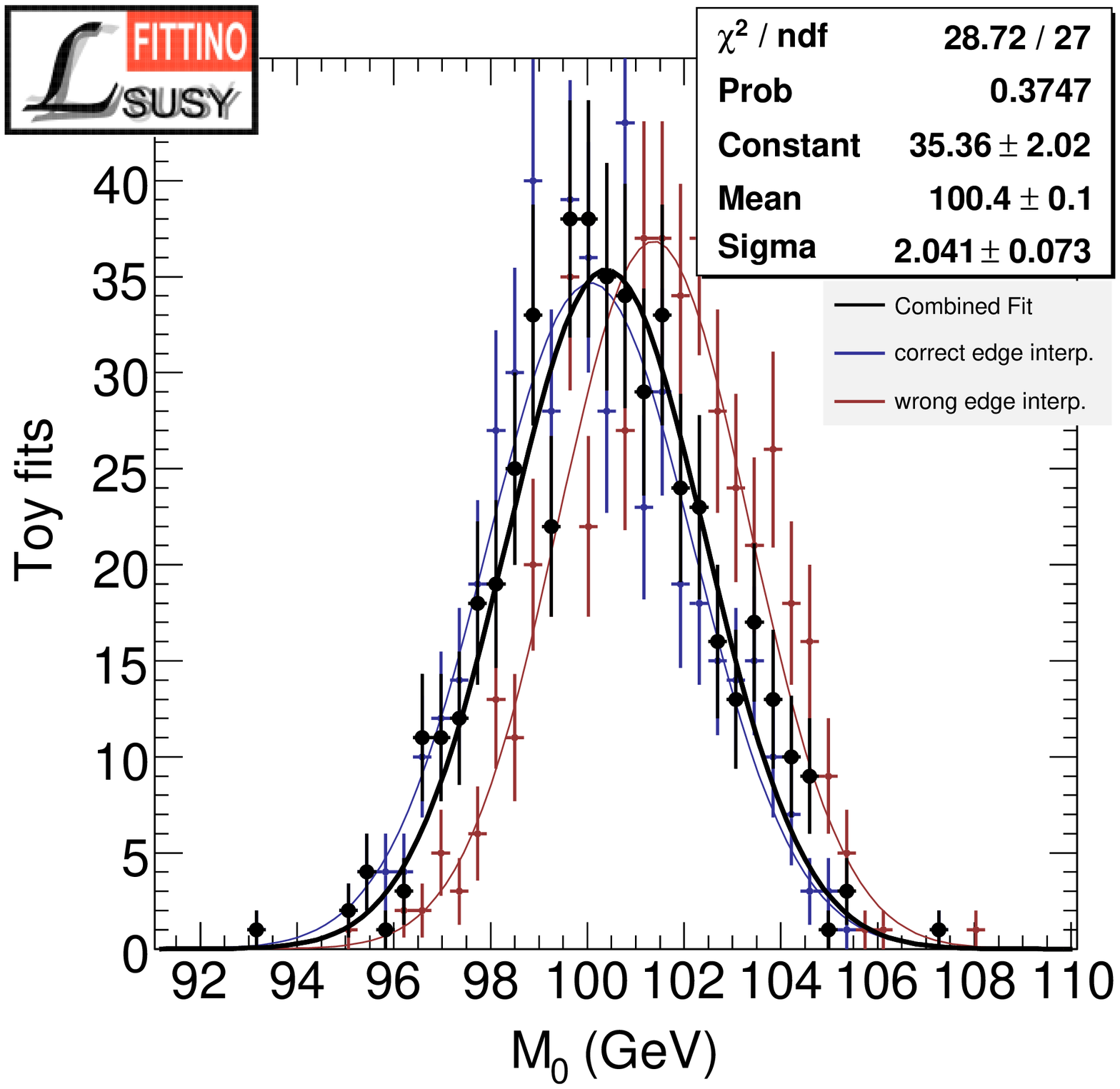}
        \hfill
        \includegraphics[width=0.49\textwidth]{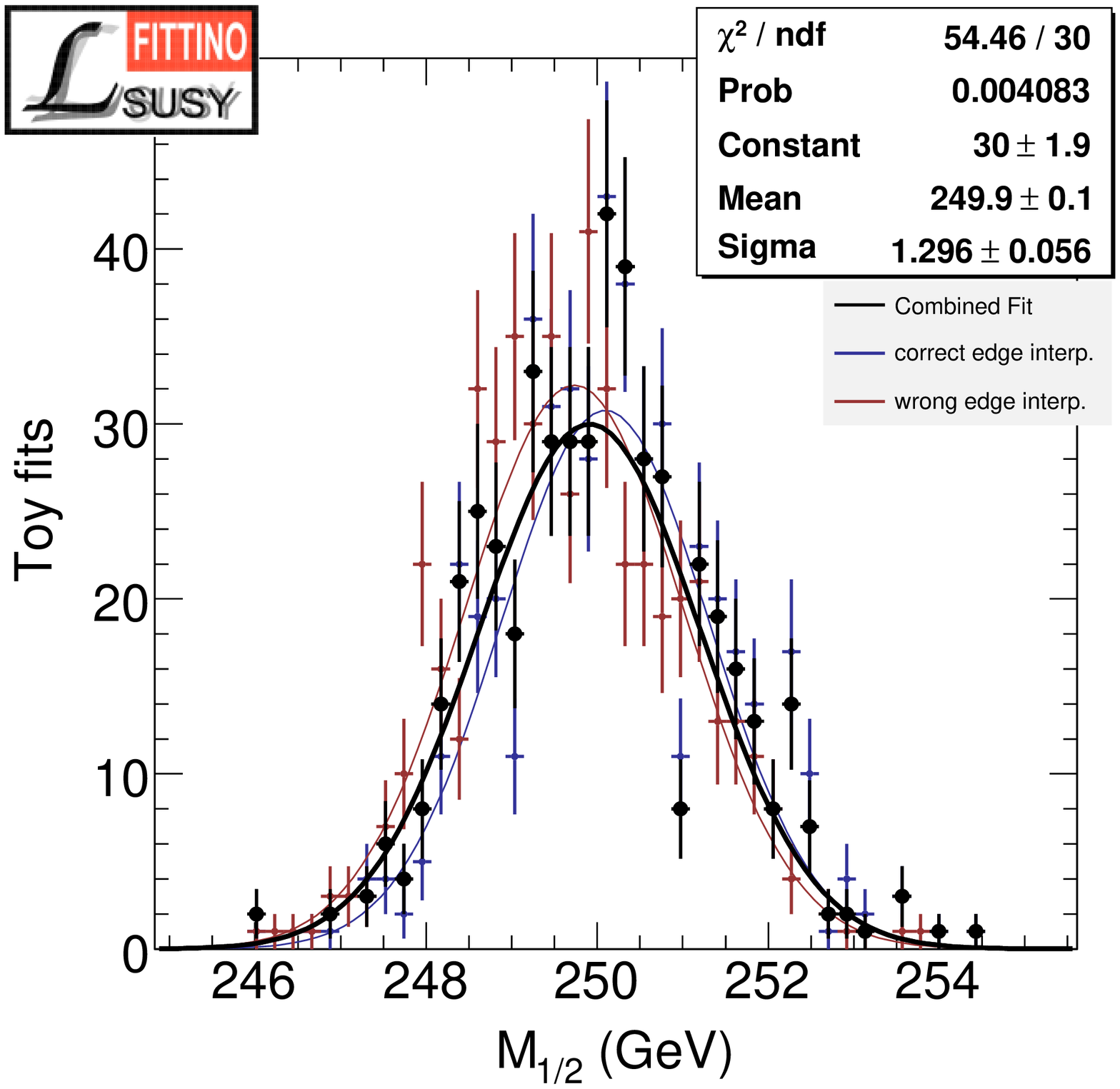}
      \end{minipage}
    \end{center}
    \caption{Parameter distributions (for 10 fb$^{-1}$) of the mSUGRA
      model of two different interpretations of the data, fitted to
      the same Toy data set. The parameter distributions assuming the
      correct endpoint assignment is shown in blue, those using the
      wrong interpretation is indicated in red and the distribution
      which is obtained if always the interpretation with the lowest
      minimal $\chi^2$ is chosen is displayed in black.}
    \label{fig:results:LHConly:parsModelComp}
  \end{figure*}
\end{center}
All the above LHC fit results are based on the idealised assumption
that one knows the contributing decay chain for a given mass spectrum.
In reality this is, of course, not the case. Therefore there are in
general various possible decays which can contribute to a specific
mass spectrum. In the context of this study we do not yet address this
problem systematically.
A fully realistic analysis, taking the full combinatorics for such chain
ambiguities into account, can be performed when data are present. 
Here we constrain ourselves to the case where we check the impact 
of misinterpreting a single observable to test the methodology.
To accomplish this we perform fits to the same Toy data set for
each possible interpretation of a mass spectrum. As an example we
allow for two different interpretations of one particular di-lepton
endpoint as either (correct)
$m_{\ell\ell}^{\text{max}}(m_{\tilde{\chi}_1^0},m_{\tilde{\chi}_4^0},m_{\tilde{\ell}_L})$
or (wrong)
$m_{\ell\ell}^{\text{max}}(m_{\tilde{\chi}_1^0},m_{\tilde{\chi}_4^0},m_{\tilde{\ell}_R})$
for the case of 10~fb$^{-1}$ of integrated luminosity.

Figure~\ref{fig:results:LHConly:chi2ModelComp} shows the $\chi^2$
correlations obtained from fits to the same Toy data set for these two
different interpretations of $m_{\ell\ell}^{\text{max}}$. If one
always chooses the interpretation which yields the smallest minimal
$\chi^2$ the probability to make a wrong decision is approximately
23~\%. Figure~\ref{fig:results:LHConly:parsModelComp} shows the mSUGRA
parameter distributions from Toy Fits assuming the correct endpoint
assignment (blue), the wrong interpretation (red) and the distribution
which is obtained if the one with the lowest minimal $\chi^2$ is
always chosen (black). It is apparent that this chain ambiguity has
some impact on the reconstructed parameters leading to a bias on the
mean and to systematically larger values for the uncertainties on the
parameters, but these effects are rather small compared to the
uncertainty on the parameters. While this observation certainly cannot
be generalised to arbitrary ambiguities in the decay chains, the
principal method can always be applied. Depending on the result of the
$\chi^2$ comparison, the ambiguity can be either translated into
increased parameter errors or certain hypotheses can be discarded if
they yield significantly worse $\chi^2$. \newline

\subsection{Role of Low Energy Observables in the LHC Era}
\label{sec:results:LELHC}

Now we address the question to which extent low energy measurements
still contribute to the determination of SUSY parameters once LHC
results become available. We perform this study for two different SUSY
models, namely mSUGRA and MSSM18. The input observables for these
analyses comprise all the low energy observables listed in
Table~\ref{tab:leobserables} in addition to the LHC observables of
Table~\ref{tab:inputs}. To ensure a consistent set of ``measurements''
for these analyses, nominal SPS1a values are used for the low energy
observables instead of the actually measured values.

\begin{center}
  \begin{figure*}
    \begin{center}
      \begin{minipage}{0.8\textwidth}
        \includegraphics[width=0.49\textwidth]{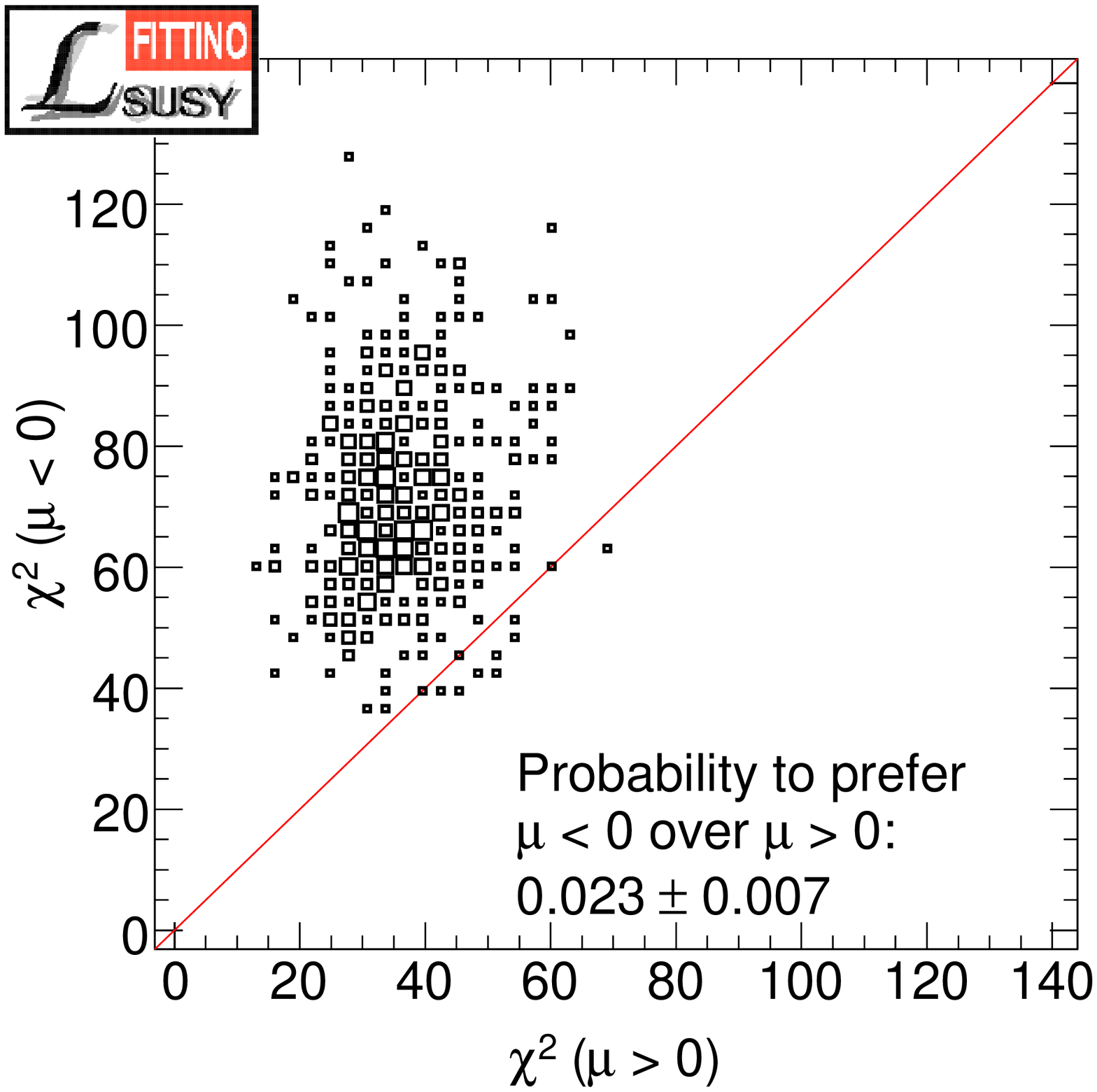}
        \hfill
        \includegraphics[width=0.49\textwidth]{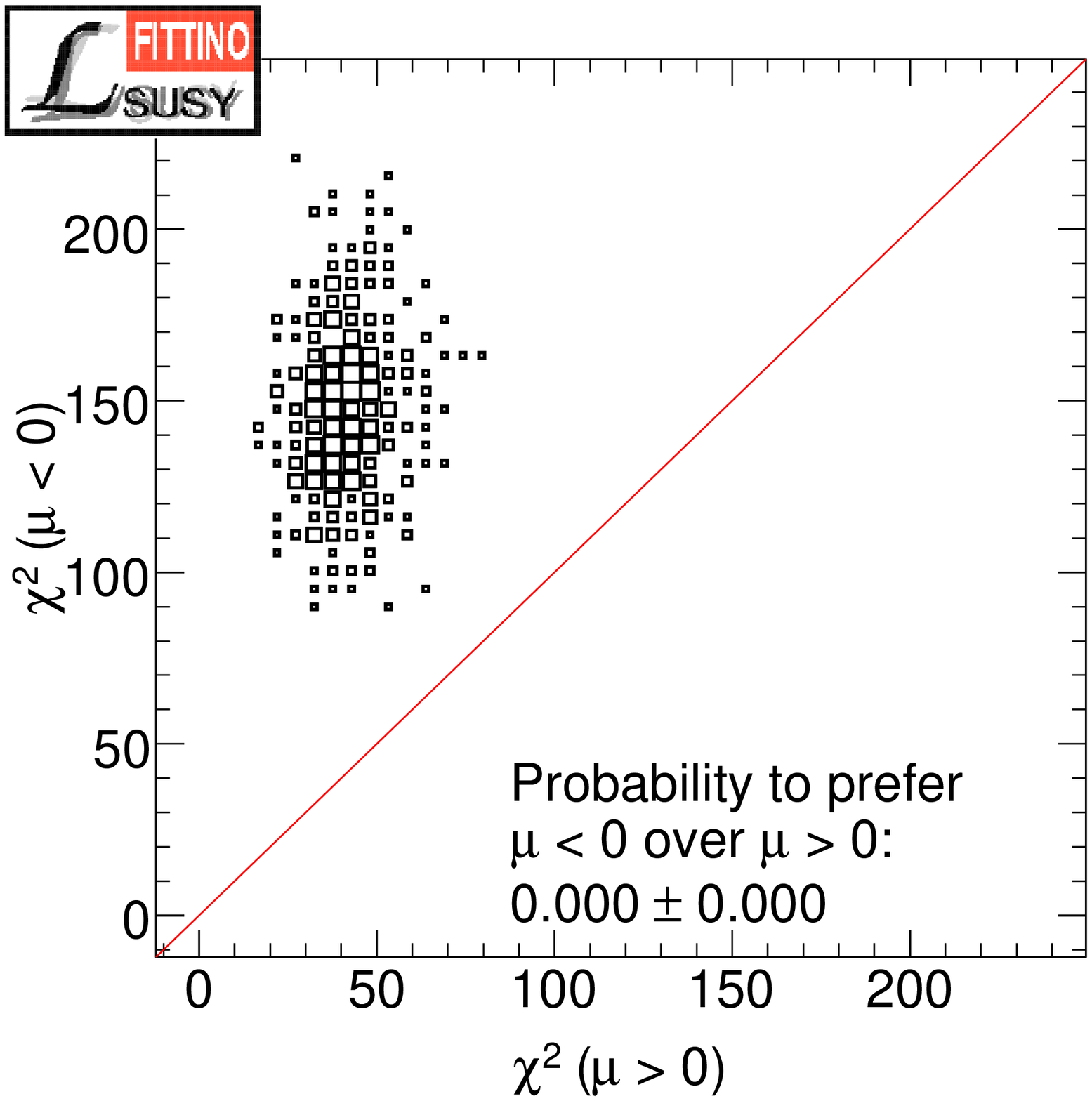}
      \end{minipage}
    \end{center}
    \caption{Discrimination power of the expected low-energy and LHC
      results from Table~\ref{tab:leobserables} and~\ref{tab:inputs} for
      ${\cal L}^{\mathrm{int}}=1\,\mathrm{fb}^{-1}$ (left) and ${\cal
        L}^{\mathrm{int}}=10\,\mathrm{fb}^{-1}$ in a 2-dimensional plot of the
      minimal $\chi^2$ values of simultaneous Toy Fits of two different
      models ($\mathrm{sign}(\mu)=+1$ and $\mathrm{sign}(\mu)=-1$). In
      comparison with Figure~\ref{fig:Chi2CorrelationsSignMuLHC} a clear
      increase in separation power for ${\cal
        L}^{\mathrm{int}}=1\,\mathrm{fb}^{-1}$ is observed.}
    % \label{}
  \end{figure*}
\end{center}

\subsubsection{mSUGRA Fit}\label{sec:results:LELHC:mSUGRA}

\begin{center}
  \begin{figure*}
    \begin{center}
      \begin{minipage}{0.8\textwidth}
        \includegraphics[width=0.49\textwidth]{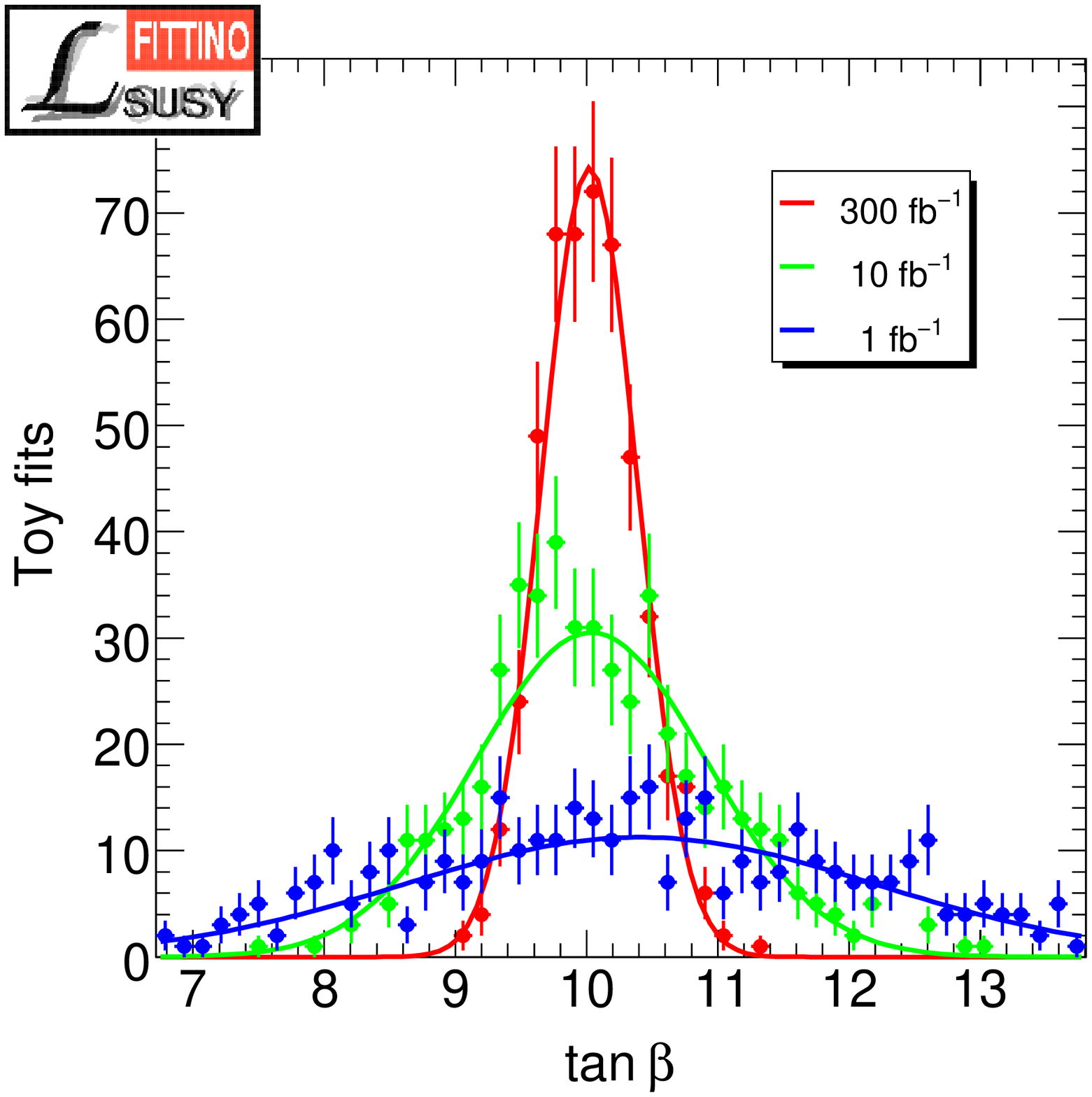}
        \hfill
        \includegraphics[width=0.49\textwidth]{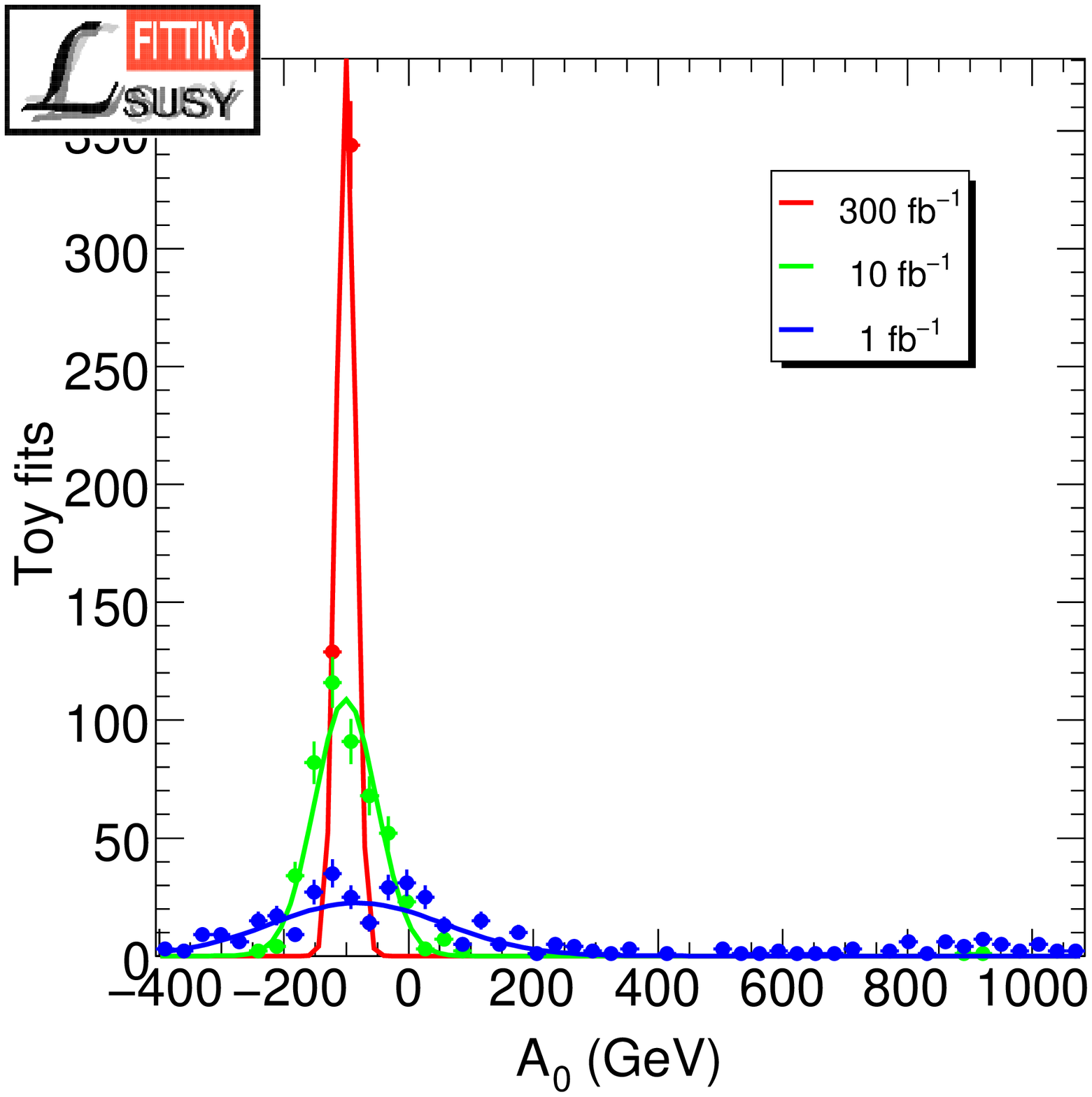}
        \\[1mm]
        \includegraphics[width=0.49\textwidth]{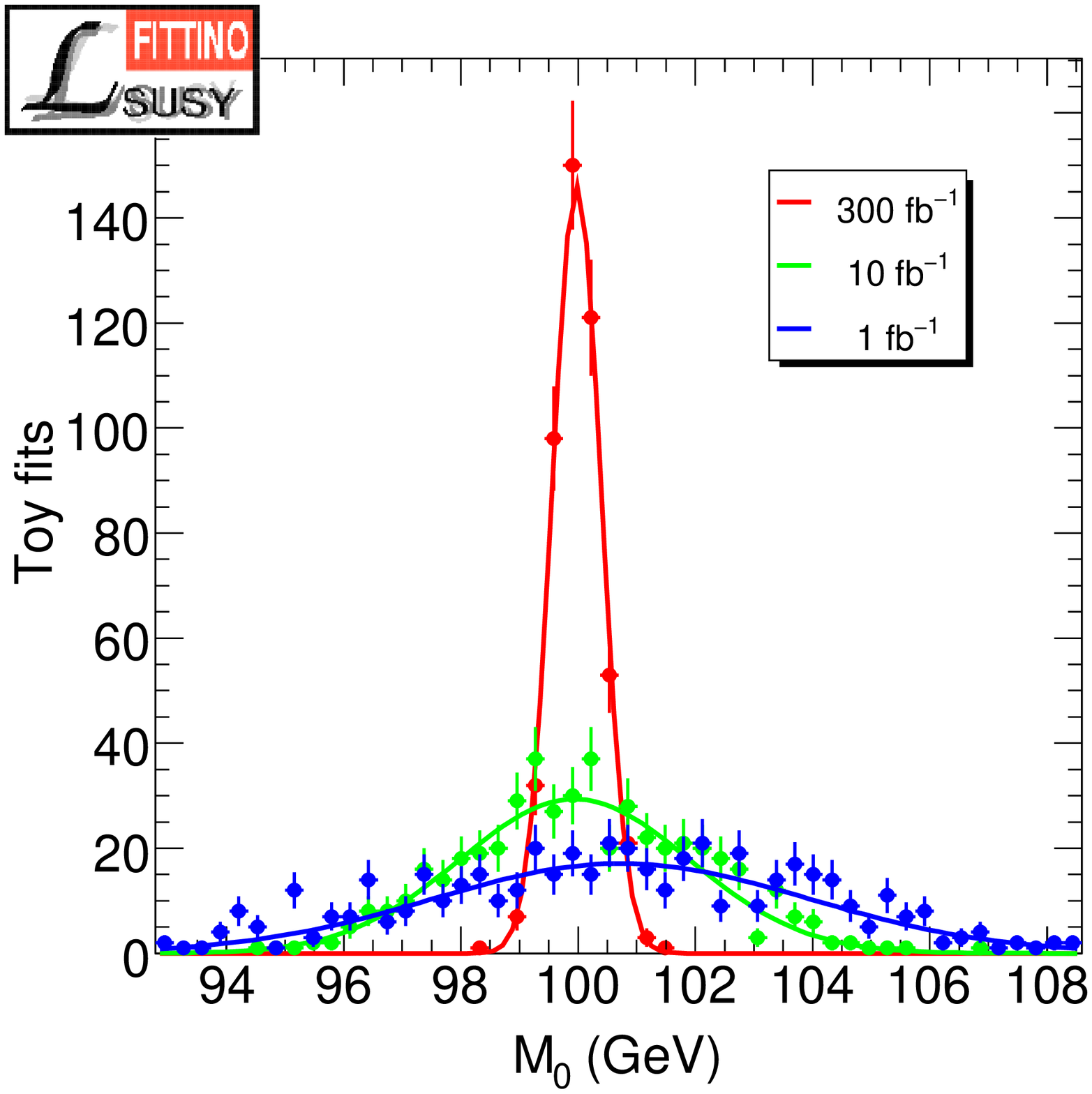}
        \hfill
        \includegraphics[width=0.49\textwidth]{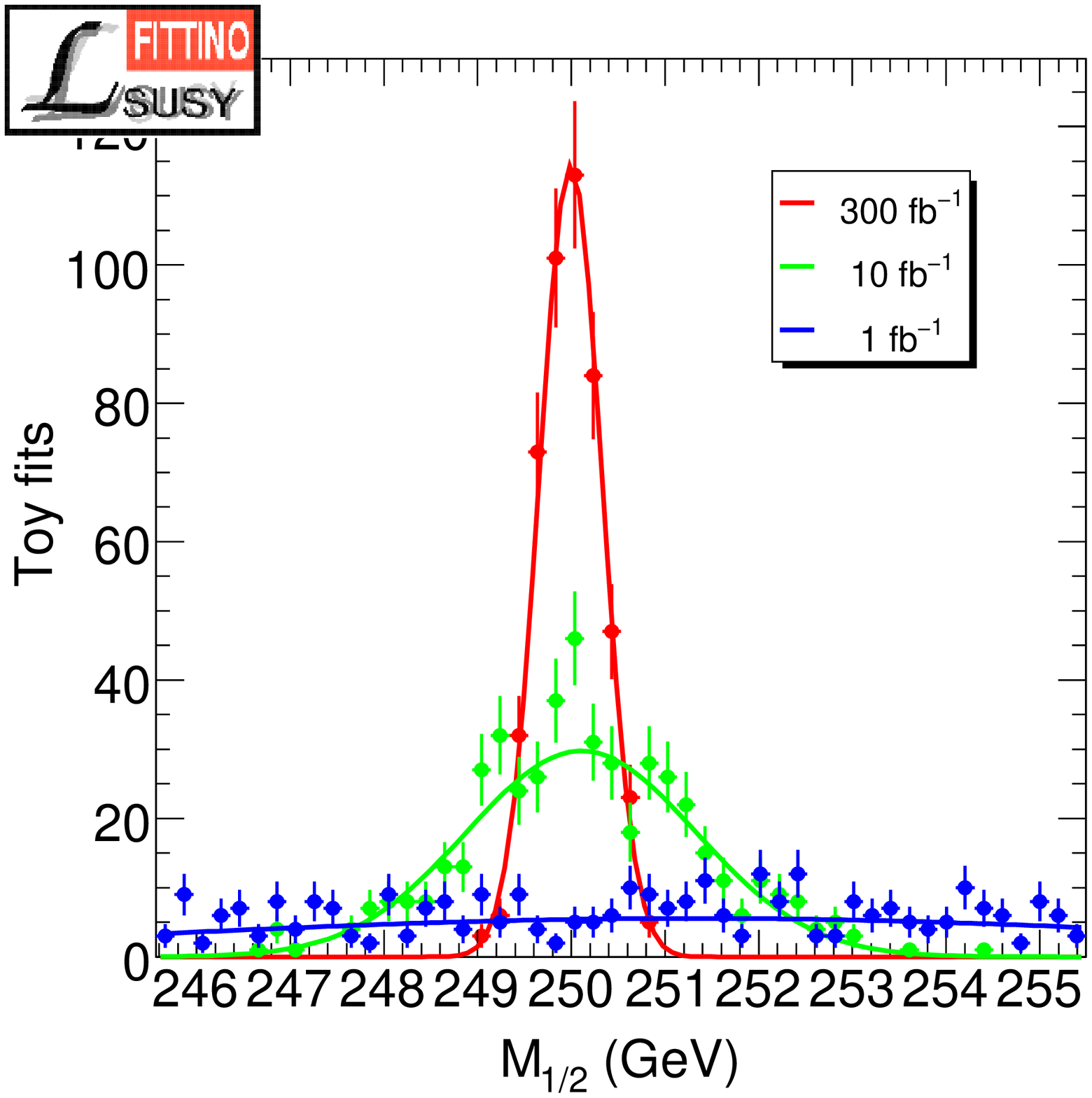}
        \\[1mm]
        \includegraphics[width=0.49\textwidth]{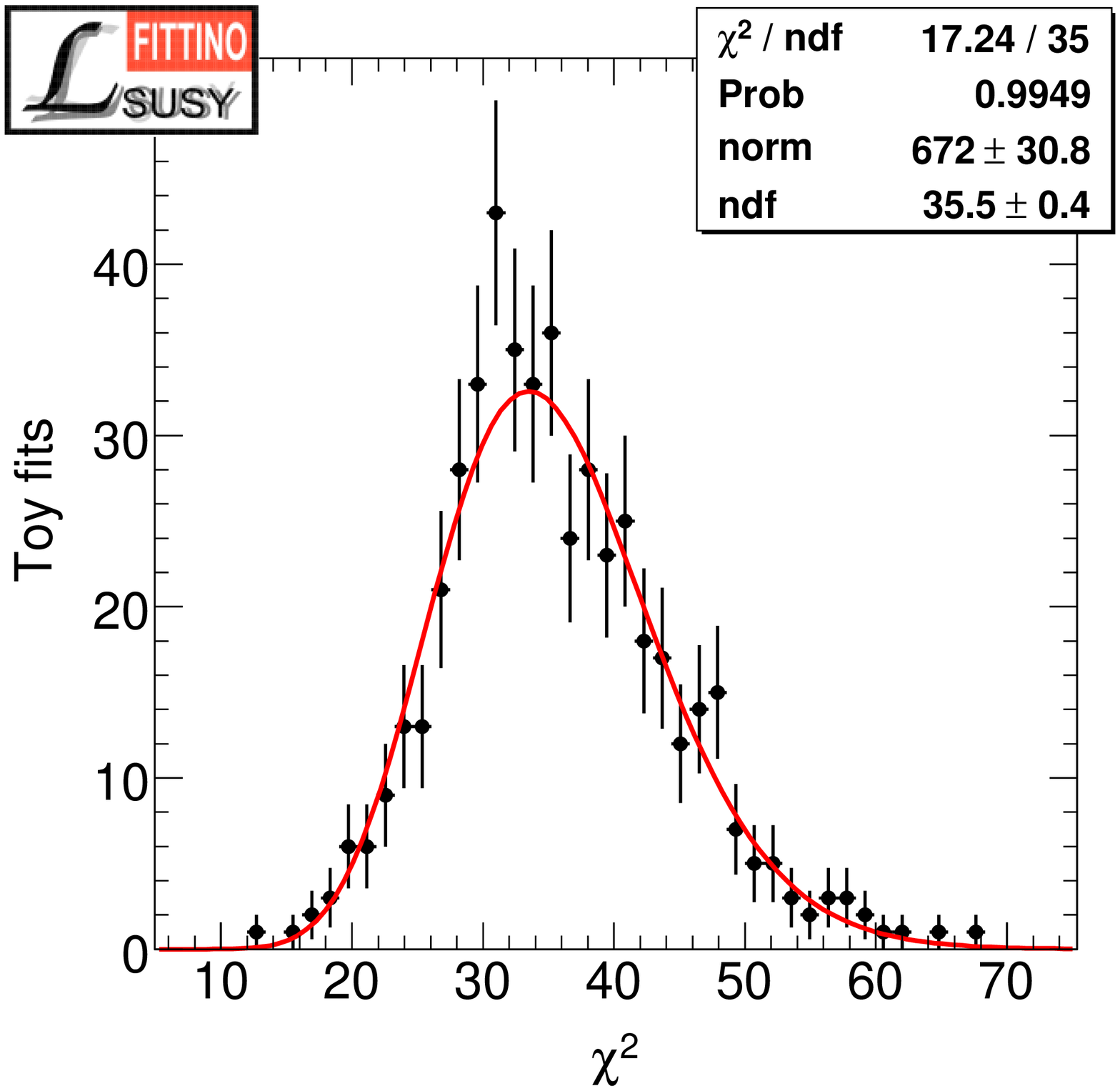}
%        \hfill
%        \includegraphics[width=0.32\textwidth]{figures/Chi2_mSUGRA_LHC_LE10}
        \hfill
        \includegraphics[width=0.49\textwidth]{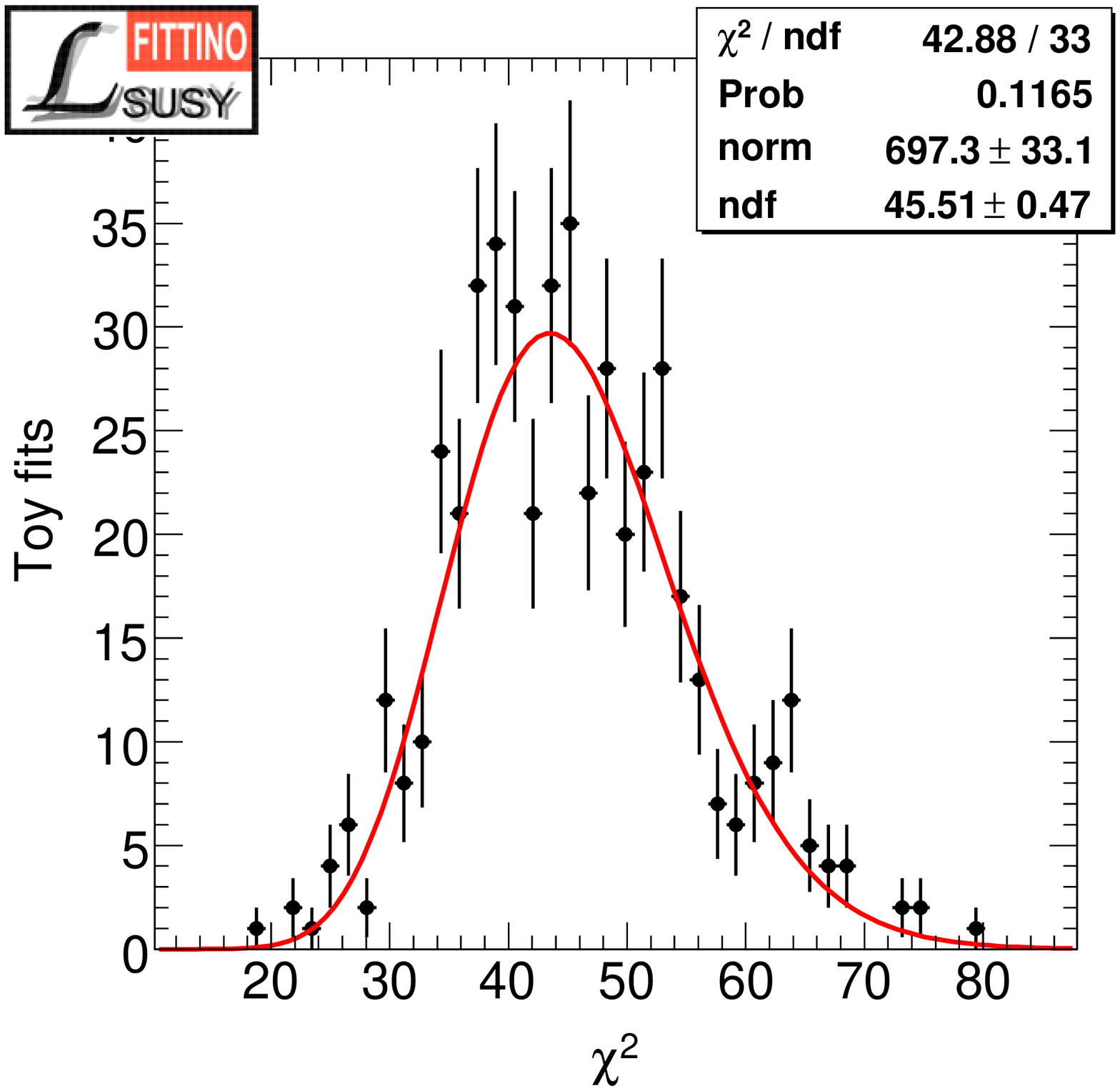}
      \end{minipage}
    \end{center}
    \caption{Results of the Toy Fits to the low-energy measurements
      and the expected LHC results from Table~\ref{tab:leobserables}
      and~\ref{tab:inputs} for different integrated luminosities. The
      upper plots shows the expected distribution of best fit points,
      exhibiting the strong increase in precision for ${\cal
        L}^{\mathrm{int}}\geq10\,\mathrm{fb}^{-1}$.  The lower row of
      plots shows the $\chi^2$ distributions of the Toy Fits with
      integrated luminosities of 1~fb$^{-1}$ and 300~fb$^{-1}$,
      showing very good agreement with the expected degrees of
      freedom. In comparison with Figure~\ref{fig:results:mSUGRALHC},
      a clear increase in precision is observed, especially for ${\cal
        L}^{\mathrm{int}}=1\,\mathrm{fb}^{-1}$ and for the parameters
      $\tan\beta$ and $M_{0}$.}
    \label{results:LELHC:mSUGRA:parDists}
  \end{figure*}
\end{center}
The fit results of the mSUGRA Toy Fits using low energy and LHC
observables for the three different luminosities are shown in
Figure~\ref{results:LELHC:mSUGRA:parDists}. The corresponding fitted
mSUGRA parameters and the corresponding correlation coefficients are
summarised in Tables~\ref{tab:results:LE+LHC:mSUGRA:1fb-1},
\ref{tab:results:LE+LHC:mSUGRA:1fb-1:corr},
\ref{tab:results:LE+LHC:mSUGRA:10fb-1},
\ref{tab:results:LE+LHC:mSUGRA:10fb-1:corr},
\ref{tab:results:LE+LHC:mSUGRA:300fb-1} and
\ref{tab:results:LE+LHC:mSUGRA:300fb-1:corr}. As described in
Section~\ref{sec:results:LHConly} $\tan \beta$ and $A_0$ are rather
weakly constrained by LHC measurements alone at low luminosity. For
these parameters the addition of low energy measurements clearly
improves the situation. At 1~fb$^{-1}$, $(g-2)_{\mu}$ is an important
additional measurement constraining $\tan \beta$ and $A_0$.  For
$M_0$, the cold dark matter relic density $\Omega_{\mathrm{CDM}} h^2$
becomes the most sensitive observable followed by $m_{\ell
  q}^{\mathrm{high}}$, the most important LHC quantity. The precision
on $M_{1/2}$ is still dominated by the LHC ``measurements'' listed in
Section~\ref{sec:results:LHConly}. Nevertheless some improvements are
also achieved for this parameter, mainly due to $(g-2)_{\mu}$.

At 10~fb$^{-1}$ and above the role of low energy measurements is
largely repressed by LHC observables such that the precision on the
mSUGRA parameters for increasing luminosity asymptotically approaches
the precision obtained from LHC observables alone.

\begin{figure}
  \includegraphics[width=0.49\textwidth]{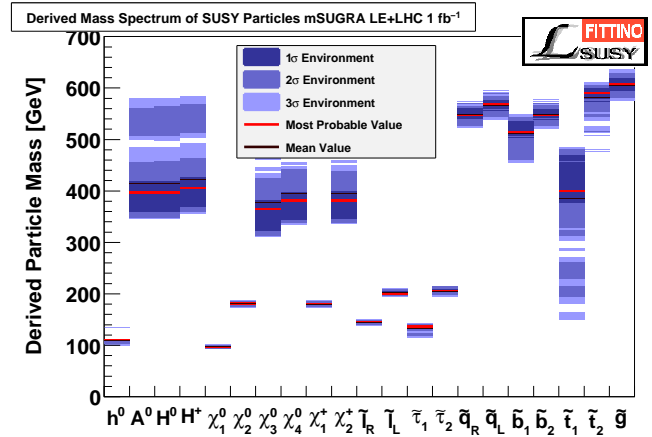}
  \caption{SUSY mass spectrum consistent with the existing low-energy
    measurements from Table~\ref{tab:leobserables} and the expected
    LHC measurements from Table~\ref{tab:inputs} at ${\cal
      L}^{\mathrm{int}}=1\,\mathrm{fb}^{-1}$ for the mSUGRA model. The
    uncertainty ranges represent model dependent uncertainties of
    the sparticle masses and not direct mass measurements. This is
    especially visible for the heavy Higgs states $A,H$ and
    $H^{\pm}$, for which no direct measurement is expected in the
    SPS1a scenario.}
  \label{results:LELHC:mSUGRA:massDist1fb}
\end{figure}

\begin{figure}
\includegraphics[width=0.49\textwidth]{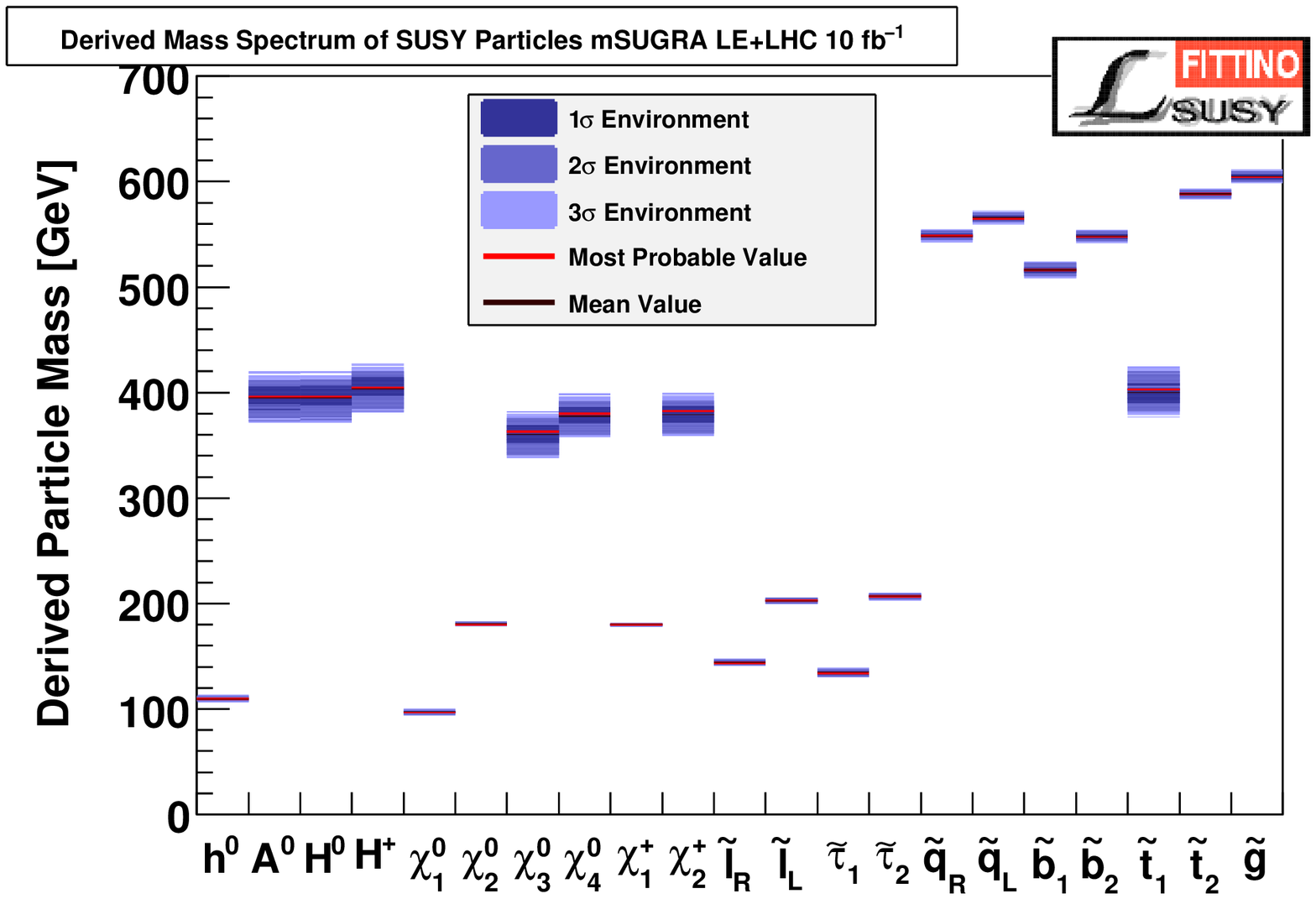}
\caption{SUSY mass spectrum consistent with the existing low-energy
  measurements from Table~\ref{tab:leobserables} and the expected LHC
  measurements from Table~\ref{tab:inputs} at ${\cal
    L}^{\mathrm{int}}=10\,\mathrm{fb}^{-1}$ for the mSUGRA model. The
  uncertainty ranges represent model dependent uncertainties of
  the sparticle masses and not direct mass measurements. With
  respect to Figure~\ref{results:LELHC:mSUGRA:massDist1fb}, a clear
  increase in precision is observed.}
\label{results:LELHC:mSUGRA:massDist10fb}
\end{figure}

\begin{figure}
\includegraphics[width=0.49\textwidth]{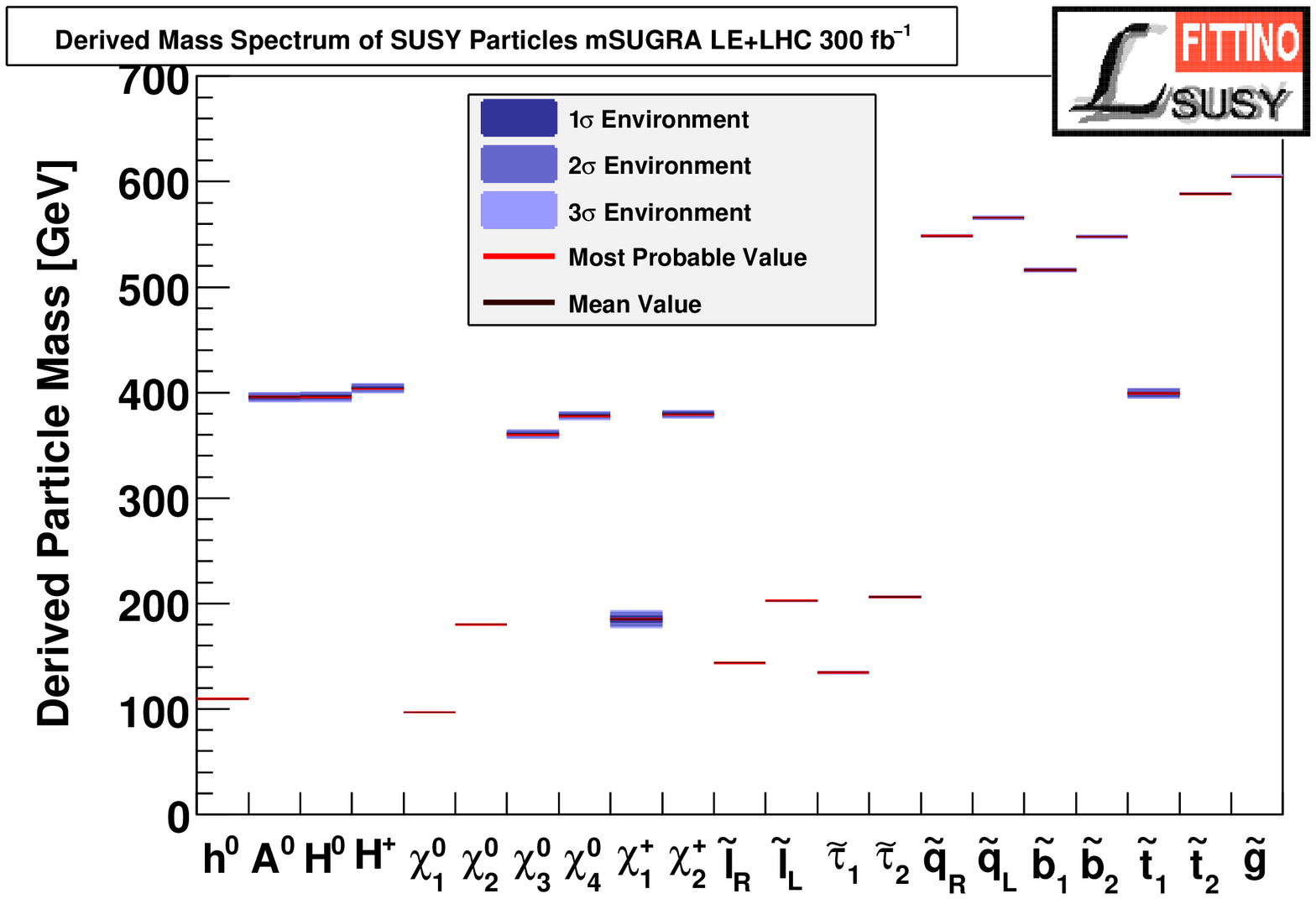}
\caption{SUSY mass spectrum consistent with the existing low-energy
  measurements from Table~\ref{tab:leobserables} and the expected LHC
  measurements from Table~\ref{tab:inputs} at ${\cal
    L}^{\mathrm{int}}=300\,\mathrm{fb}^{-1}$ for the mSUGRA model.
  The uncertainty ranges represent model dependent uncertainties
  of the sparticle masses and not direct mass measurements. With
  respect to Figure~\ref{results:LELHC:mSUGRA:massDist10fb}, again a
  clear increase in precision is observed.}
\label{results:LELHC:mSUGRA:massDist300fb}
\end{figure}

These results can be used to derive the complete SUSY particle mass
spectrum assuming the mSUGRA model and the best fitting
parameters. Figure~\ref{results:LELHC:mSUGRA:massDist1fb} shows the
mass spectrum for an integrated LHC luminosity of 1~fb$^{-1}$ as
obtained from low energy and LHC observables. The respective mean and
most probable values are indicated in black and red. $1\,\sigma$,
$2\,\sigma$ and $3\,\sigma$ uncertainties are indicated by the blue bands.
It should be noted that the masses derived in this way are model
dependent statements and not direct mass measurements. While the
masses of the light Higgs boson, the light gauginos and the sleptons
can already be constrained quite well, the masses of the heavy Higgs
bosons, the heavy gauginos and the squarks are still quite imprecise.
The situation improves significantly if one goes to a luminosity of
10~fb$^{-1}$ (see
Figure~\ref{results:LELHC:mSUGRA:massDist10fb}). Increasing the
luminosity to 300~fb$^{-1}$
(Figure~\ref{results:LELHC:mSUGRA:massDist300fb}) again means a clear
increase in precision with respect to the 10~fb$^{-1}$ result.  This
means that stringent spectroscopic tests of the mSUGRA model will be
possible using the experimentally accessible sparticles and precise
mass predictions are feasible for those SUSY particles which cannot be
directly probed at the LHC.

\begin{center}
  \begin{figure*}[p]
    \begin{center}
      \includegraphics[width=0.32\textwidth,clip]{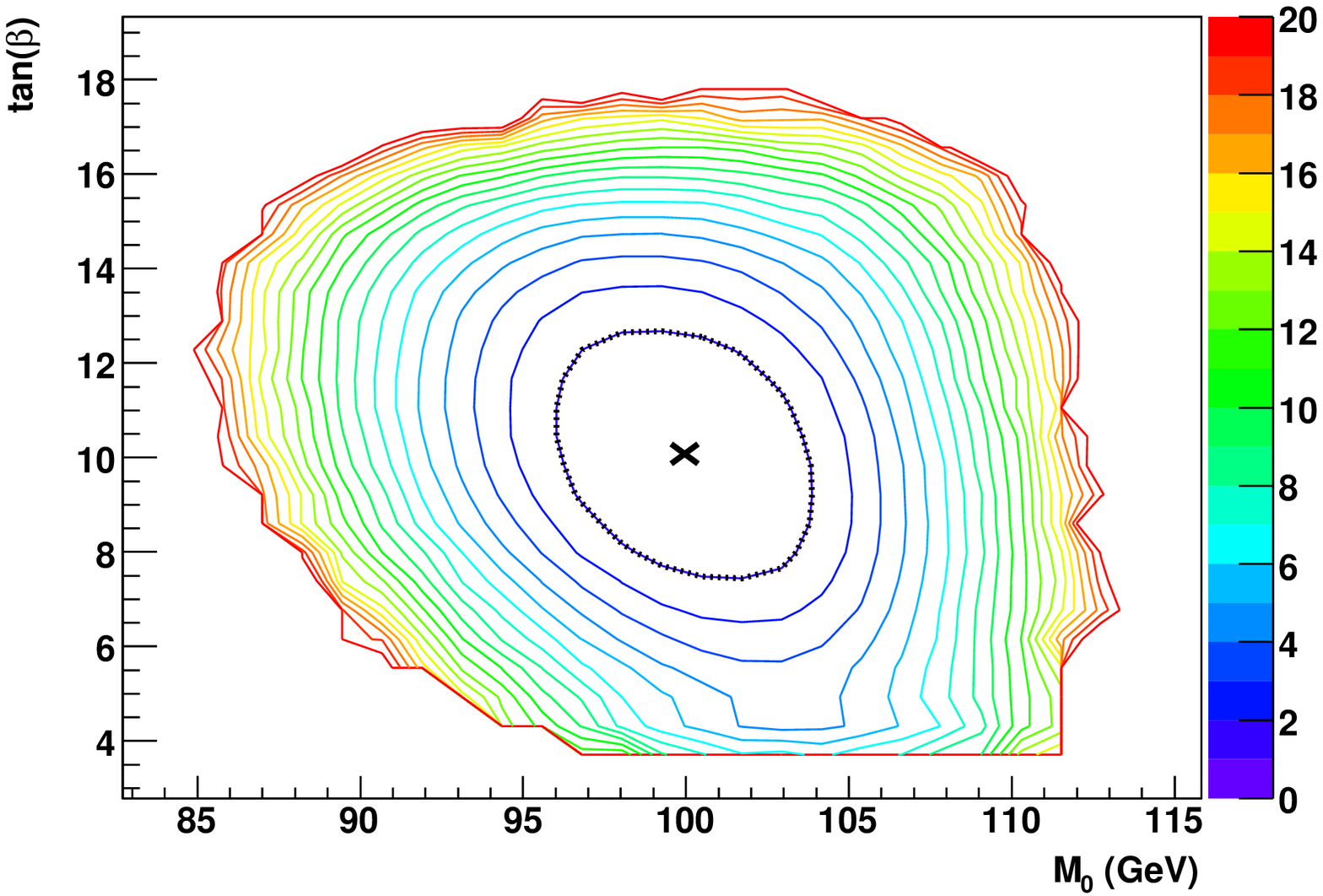}
      \hspace{-0.023\textwidth}
      \includegraphics[width=0.32\textwidth,clip]{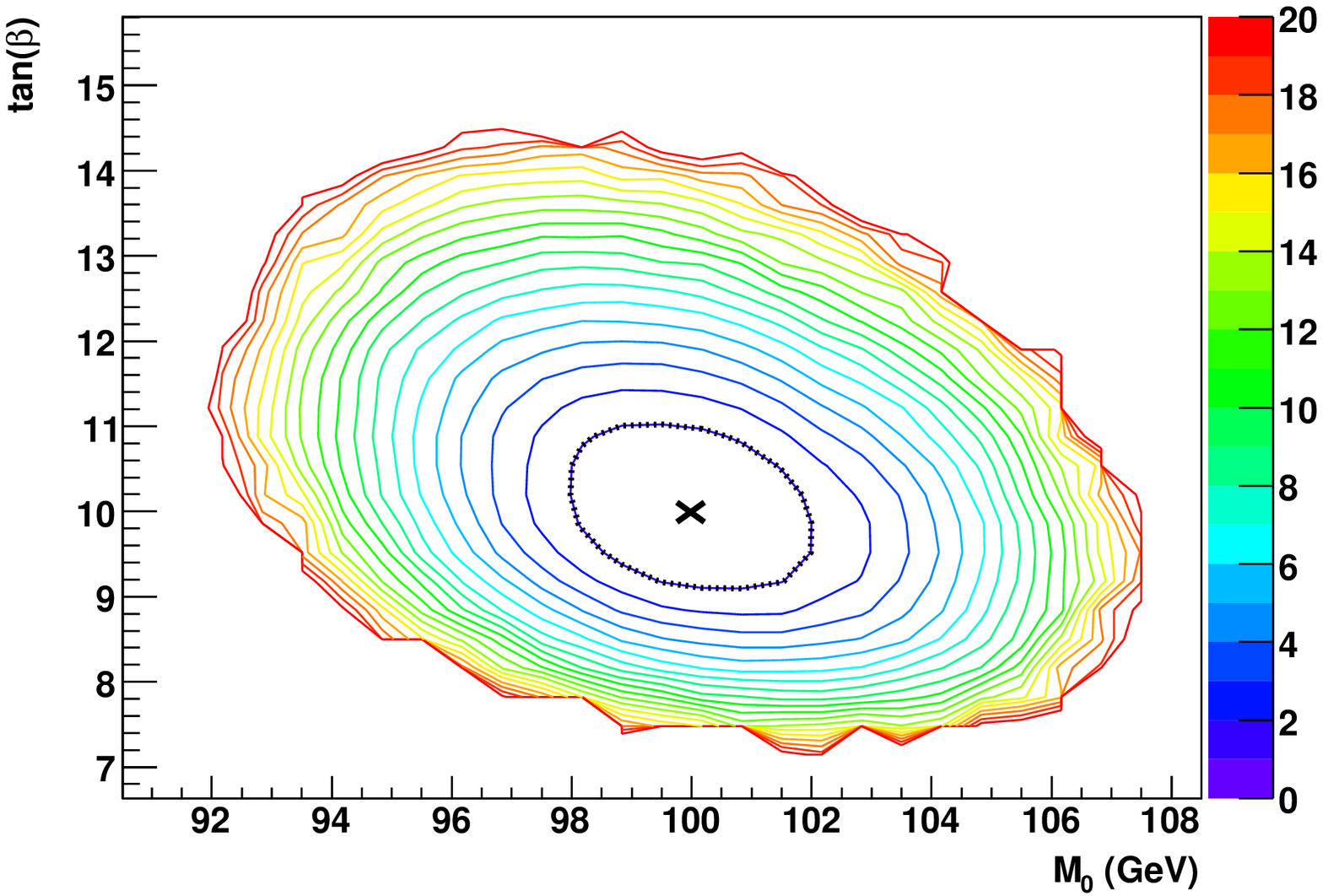}
      \hspace{-0.023\textwidth}
      \includegraphics[width=0.32\textwidth,clip]{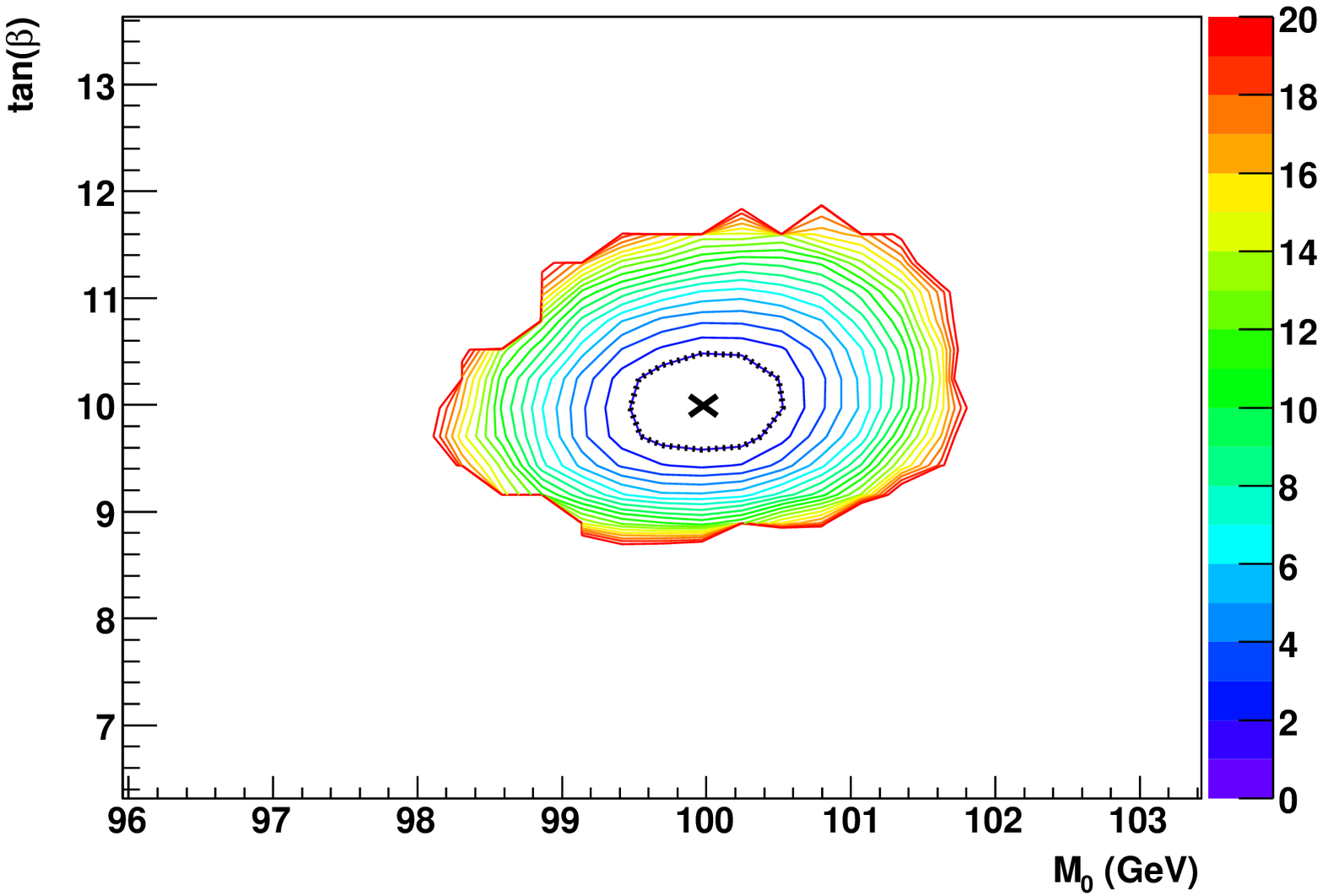}
      \\[-0.55mm]
      \includegraphics[width=0.32\textwidth,clip]{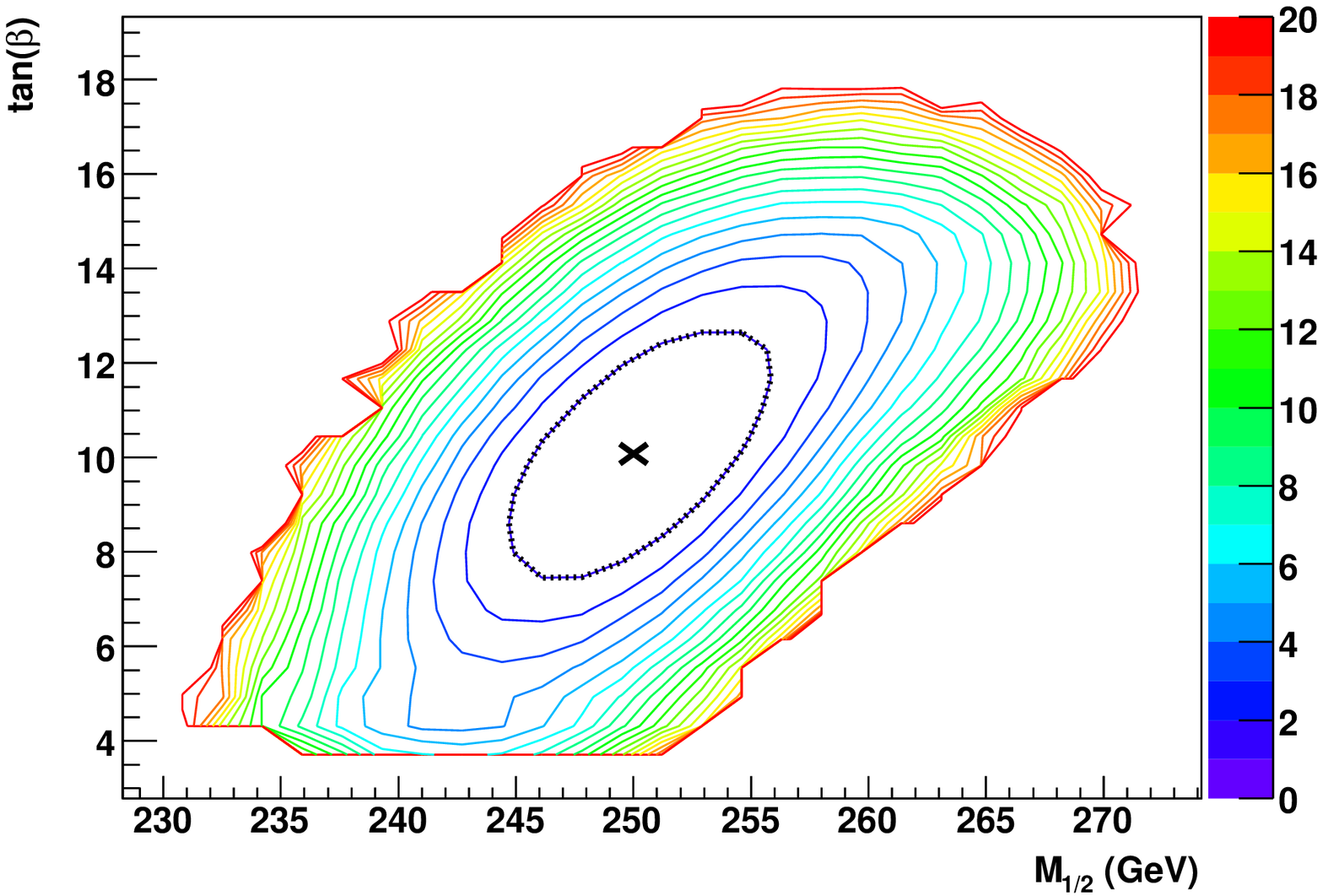}
      \hspace{-0.023\textwidth}
      \includegraphics[width=0.32\textwidth,clip]{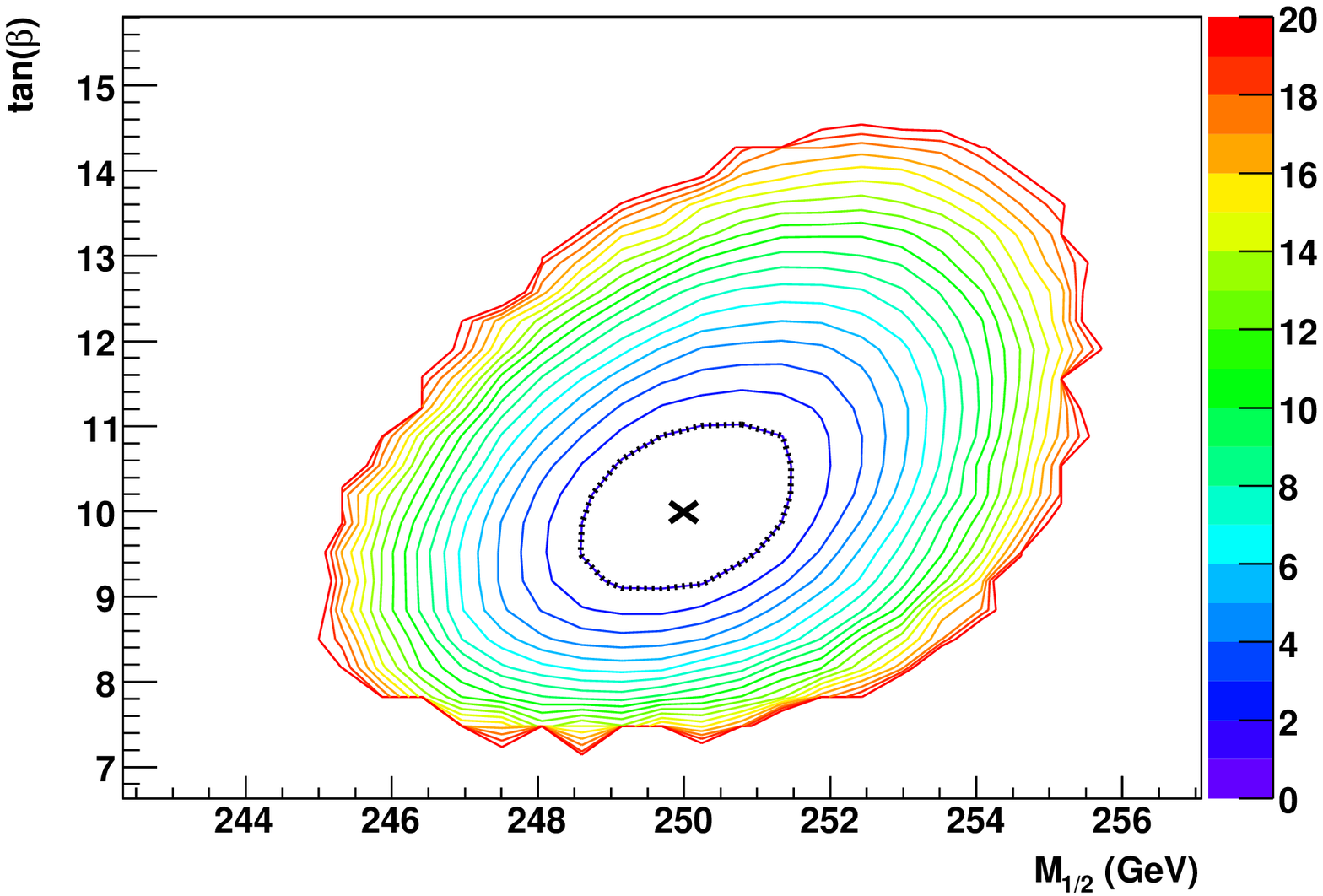}
      \hspace{-0.023\textwidth}
      \includegraphics[width=0.32\textwidth,clip]{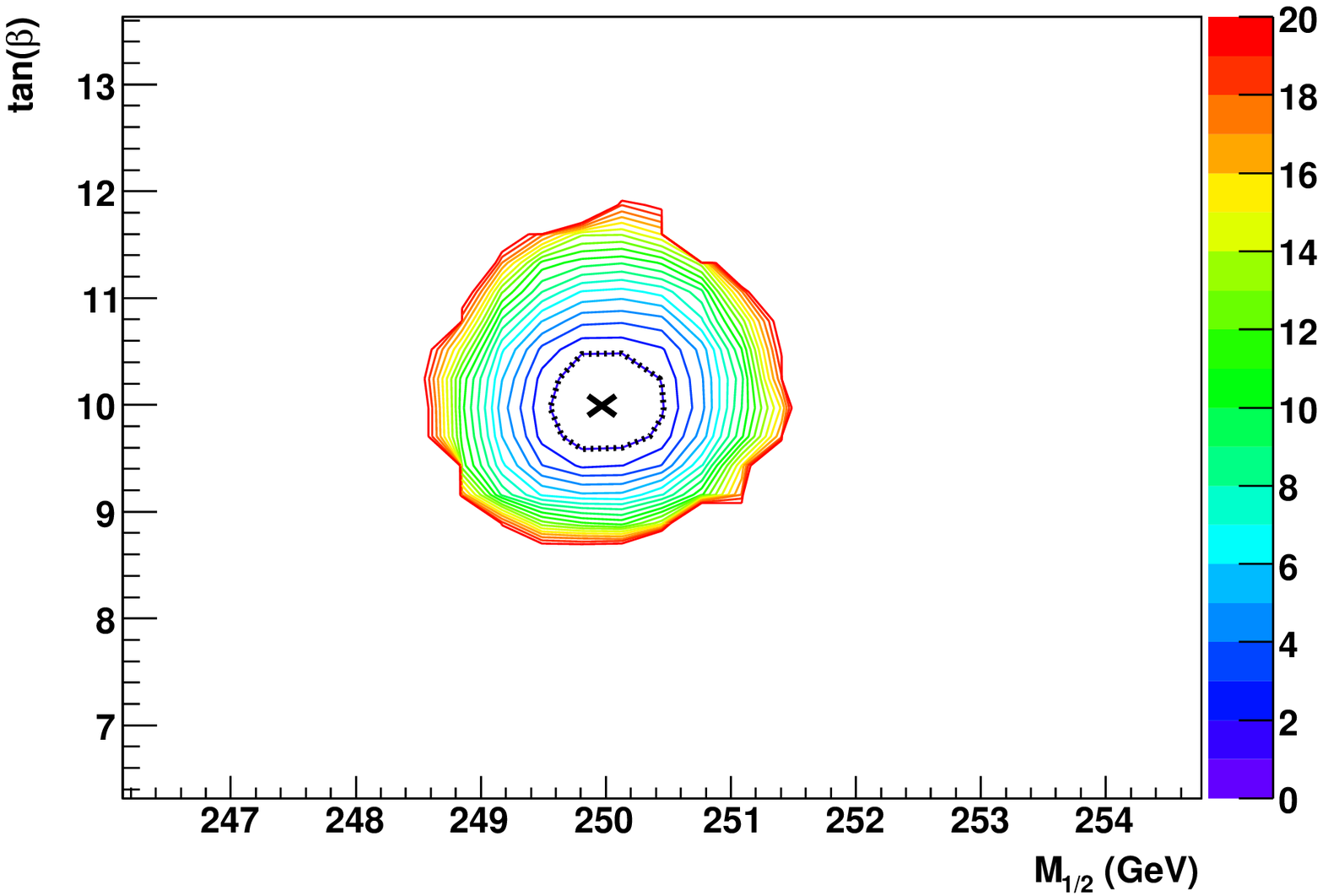}
      \\[-0.35mm]
      \includegraphics[width=0.32\textwidth,clip]{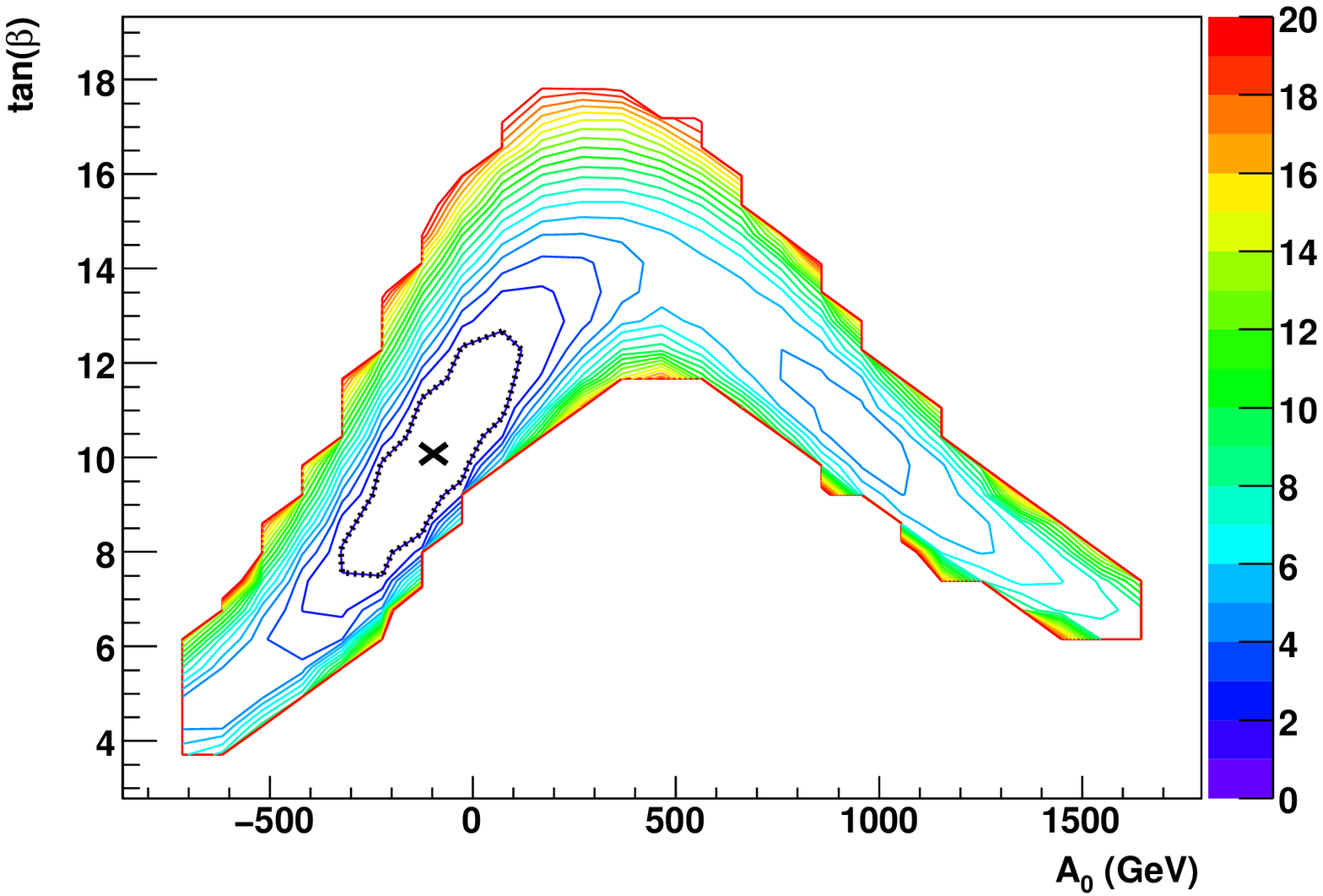}
      \hspace{-0.023\textwidth}
      \includegraphics[width=0.32\textwidth,clip]{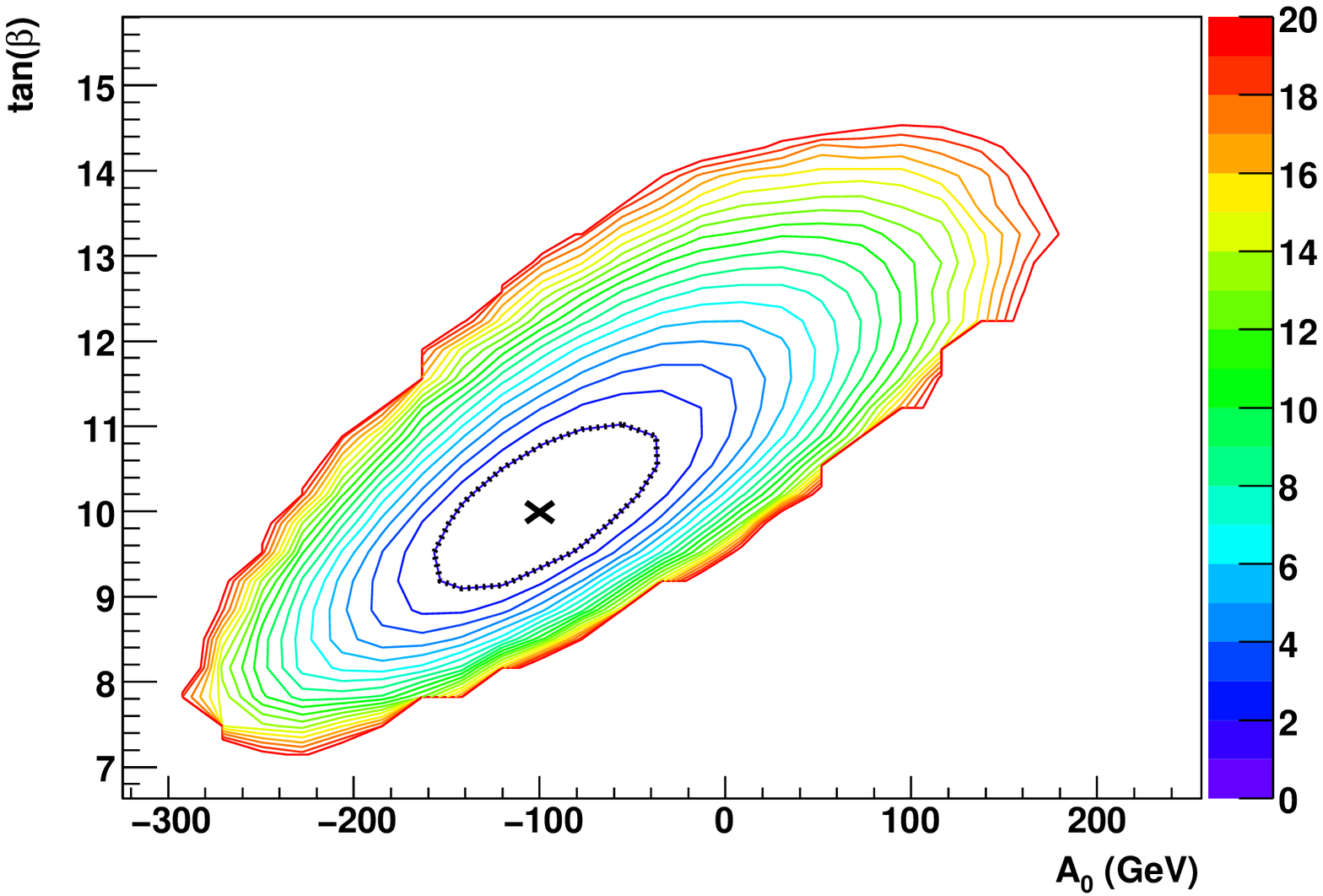}
      \hspace{-0.023\textwidth}
      \includegraphics[width=0.32\textwidth,clip]{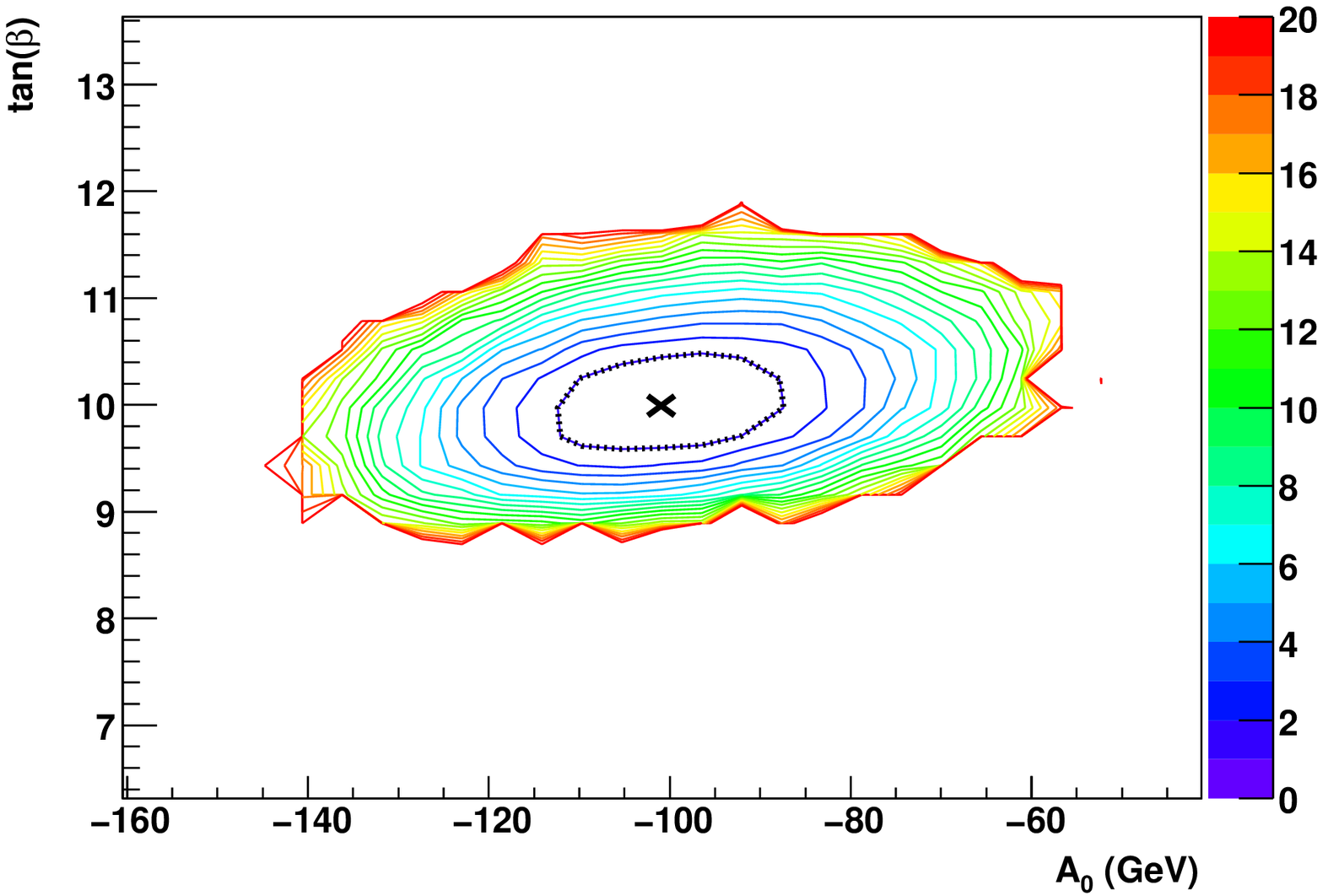}
      \\[-0.55mm]
      \includegraphics[width=0.32\textwidth,clip]{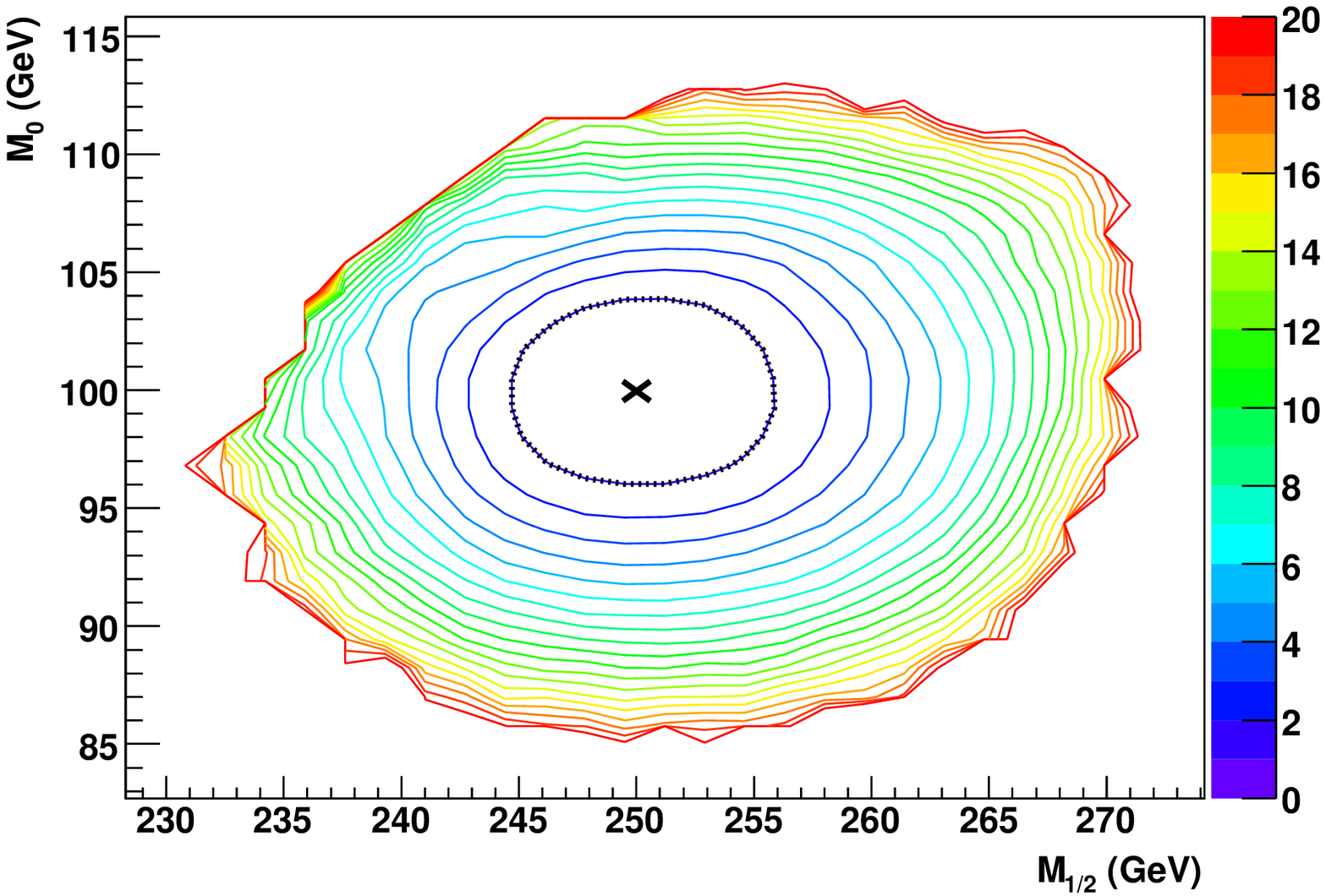}
      \hspace{-0.023\textwidth}
      \includegraphics[width=0.32\textwidth,clip]{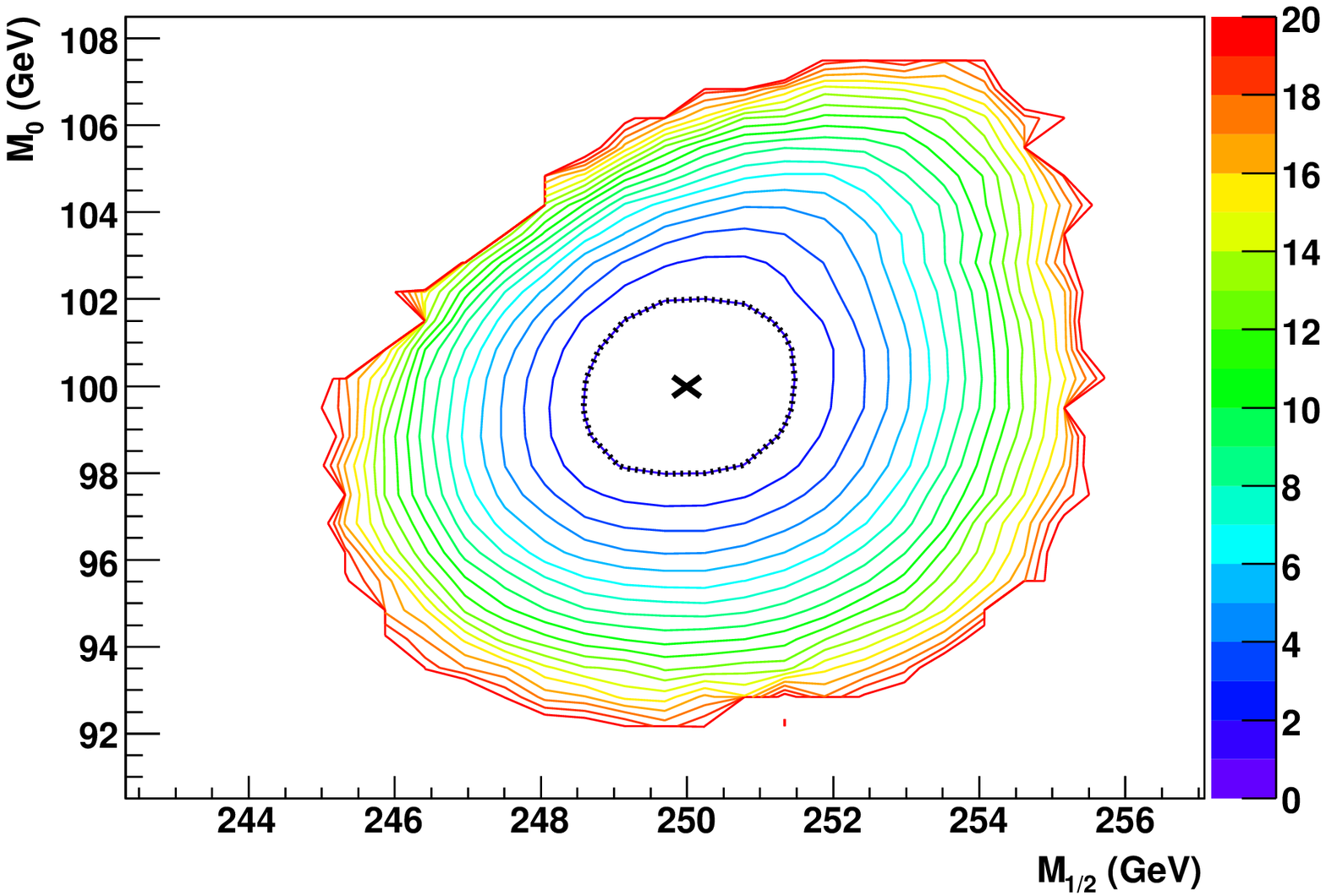}
      \hspace{-0.023\textwidth}
      \includegraphics[width=0.32\textwidth,clip]{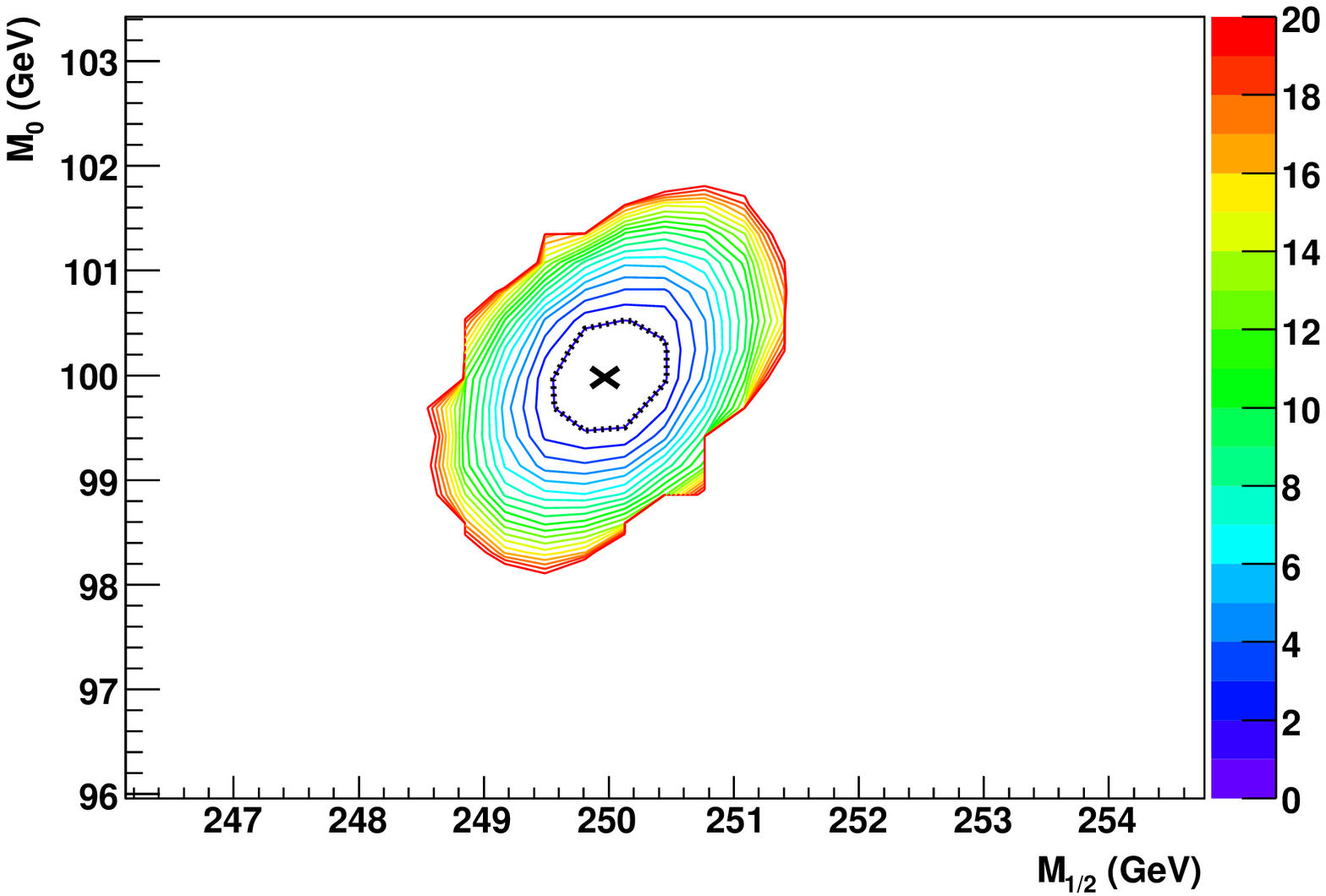}
      \\[-0.55mm]
      \includegraphics[width=0.32\textwidth,clip]{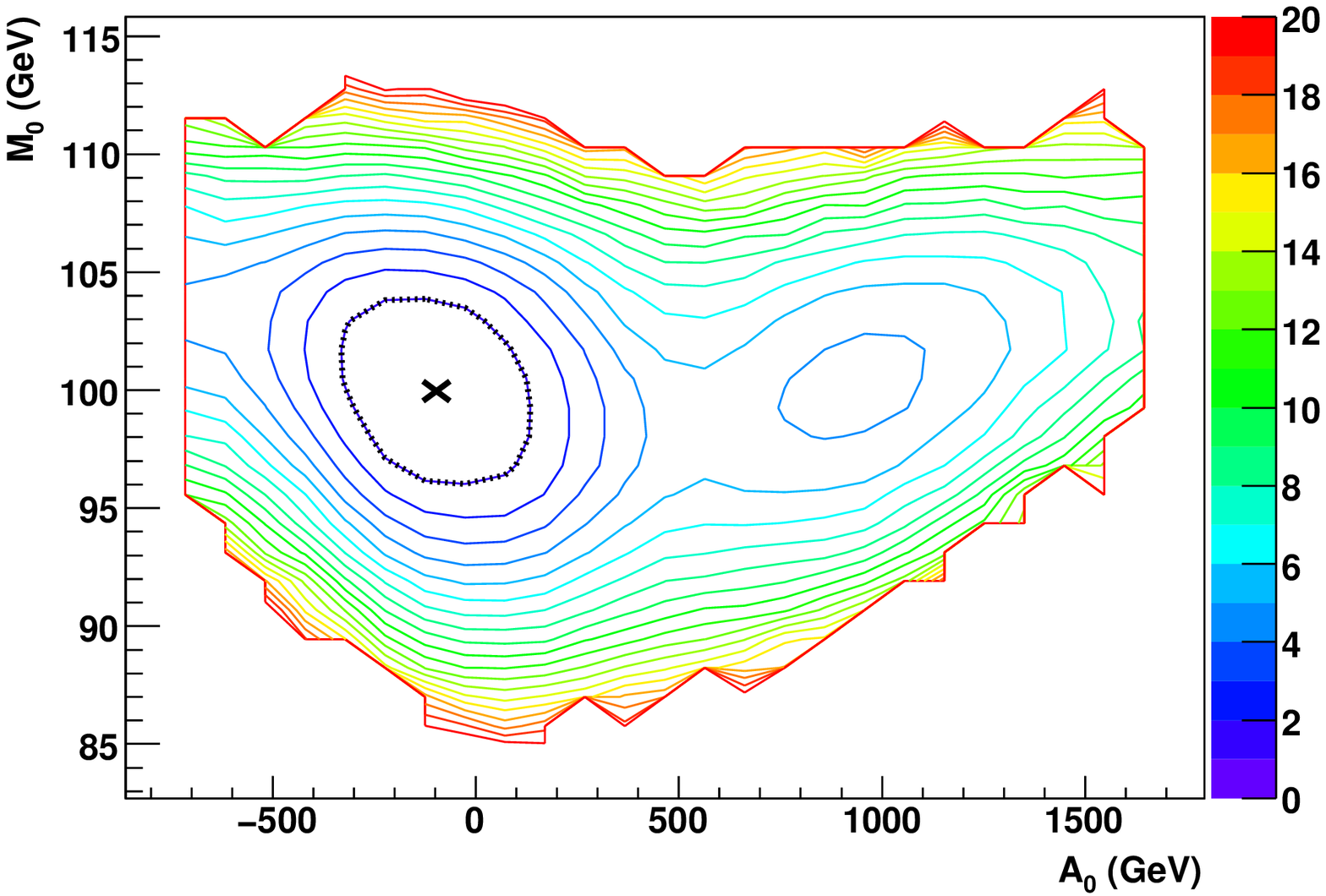}
      \hspace{-0.023\textwidth}
      \includegraphics[width=0.32\textwidth,clip]{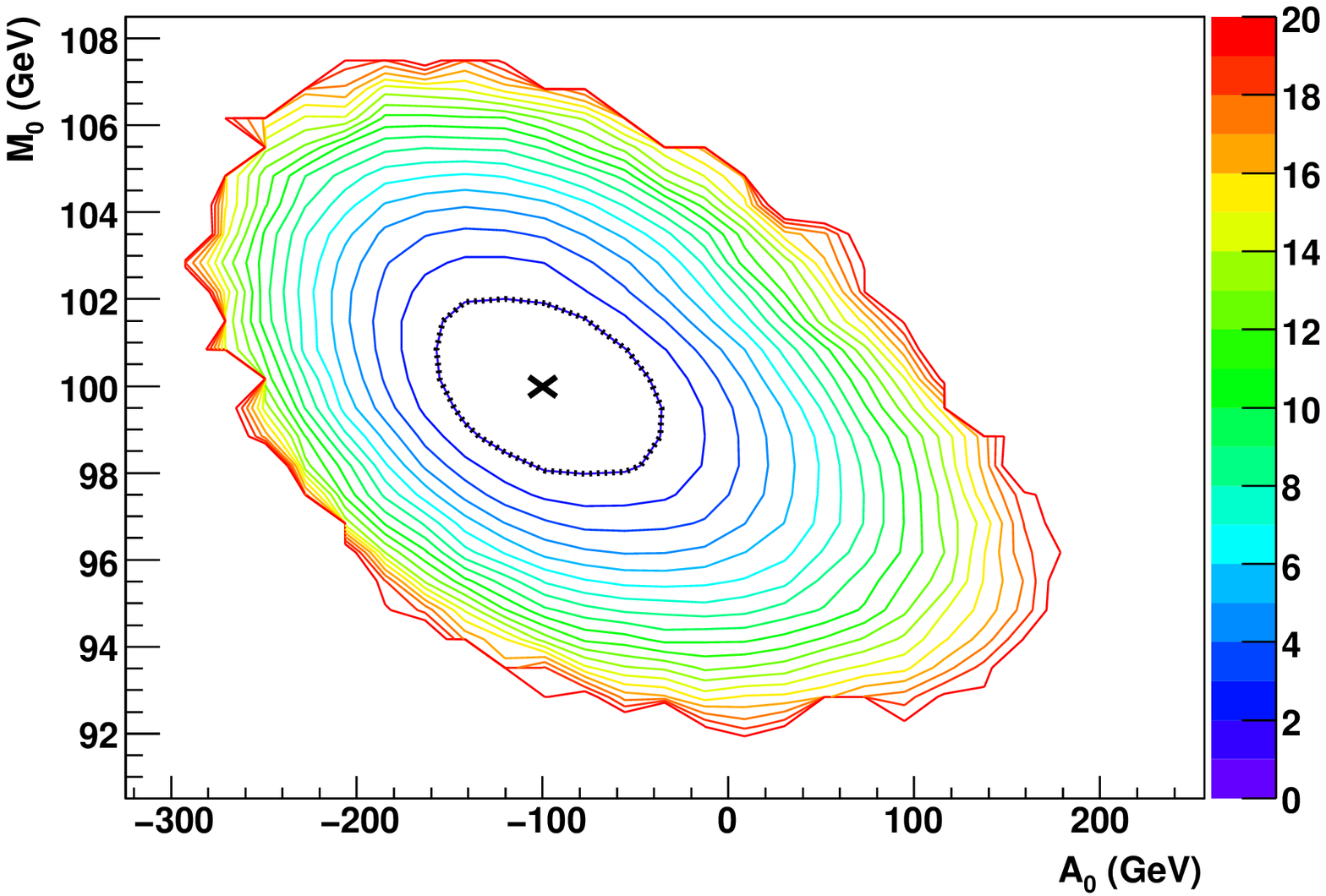}
      \hspace{-0.023\textwidth}
      \includegraphics[width=0.32\textwidth,clip]{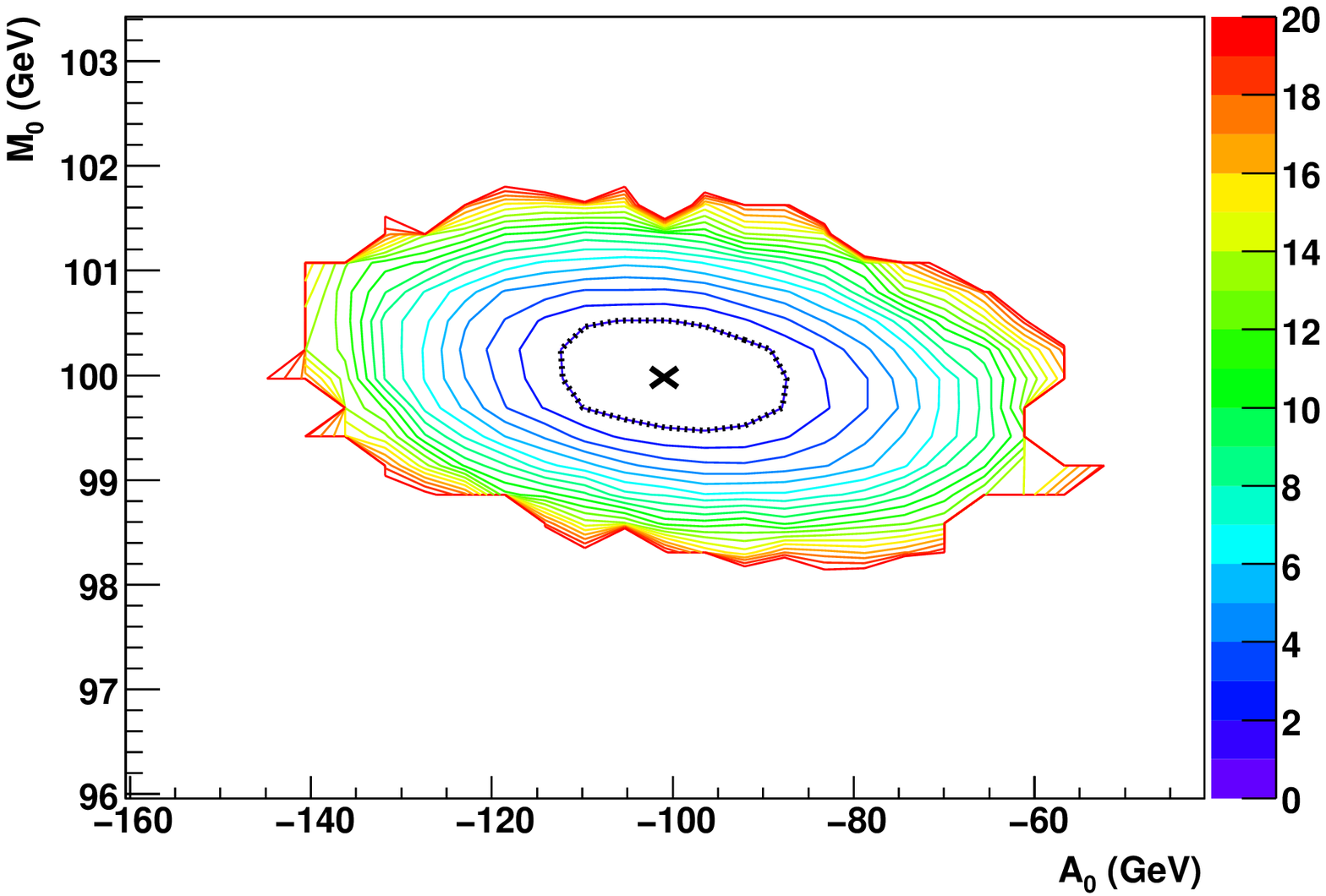}
      \\[-0.55mm]
      \includegraphics[width=0.32\textwidth,clip]{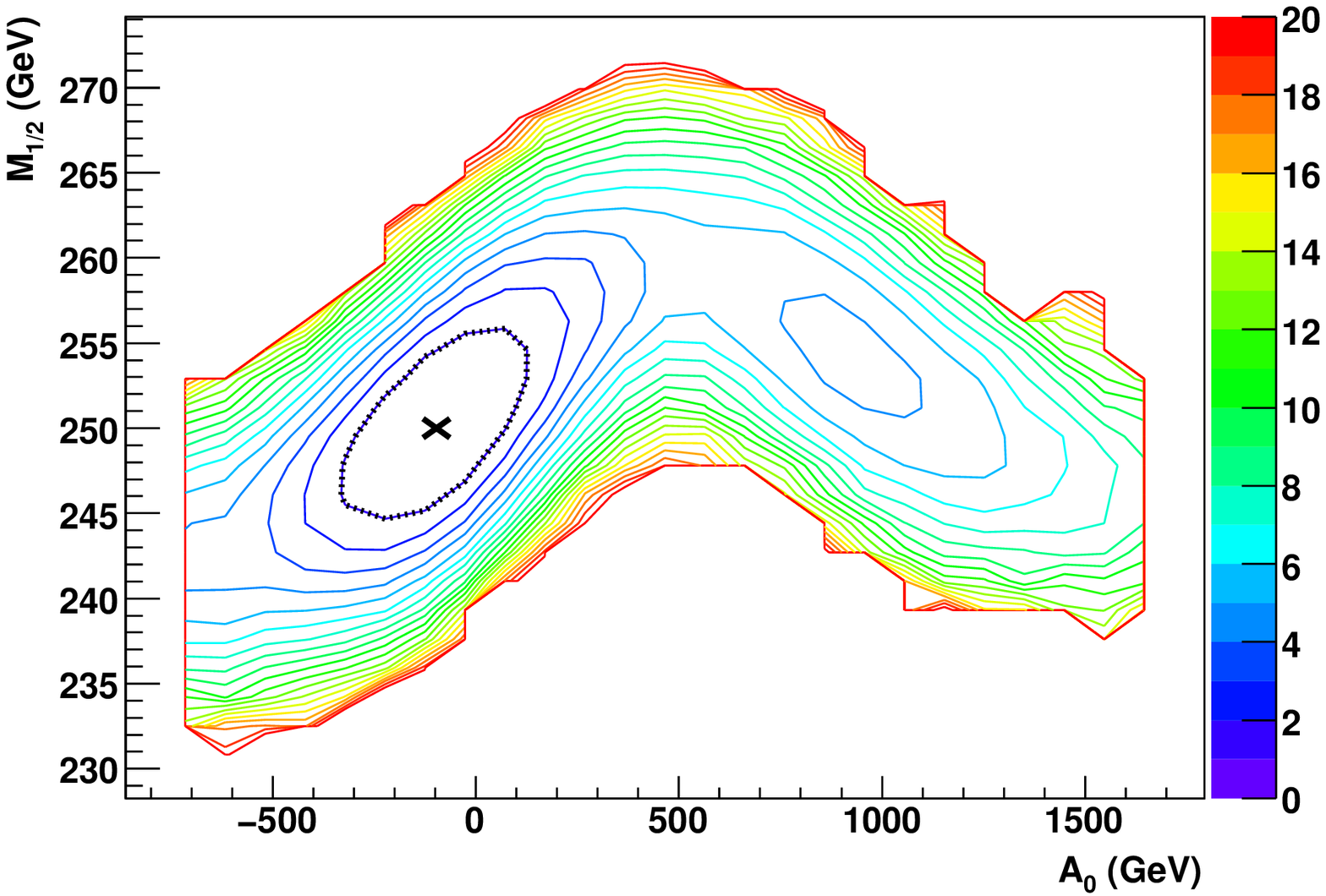}
      \hspace{-0.023\textwidth}
      \includegraphics[width=0.32\textwidth,clip]{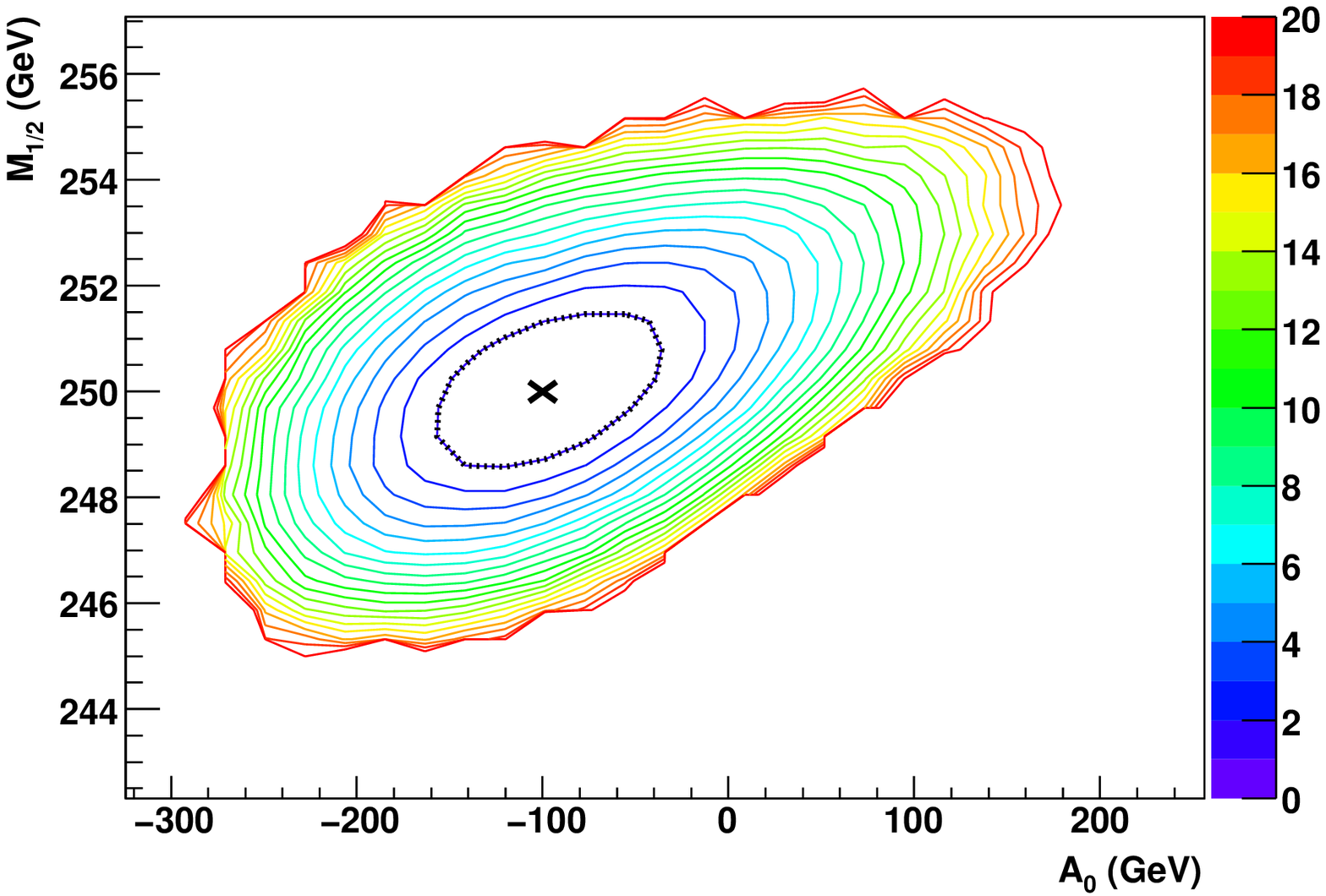}
      \hspace{-0.023\textwidth}
      \includegraphics[width=0.32\textwidth,clip]{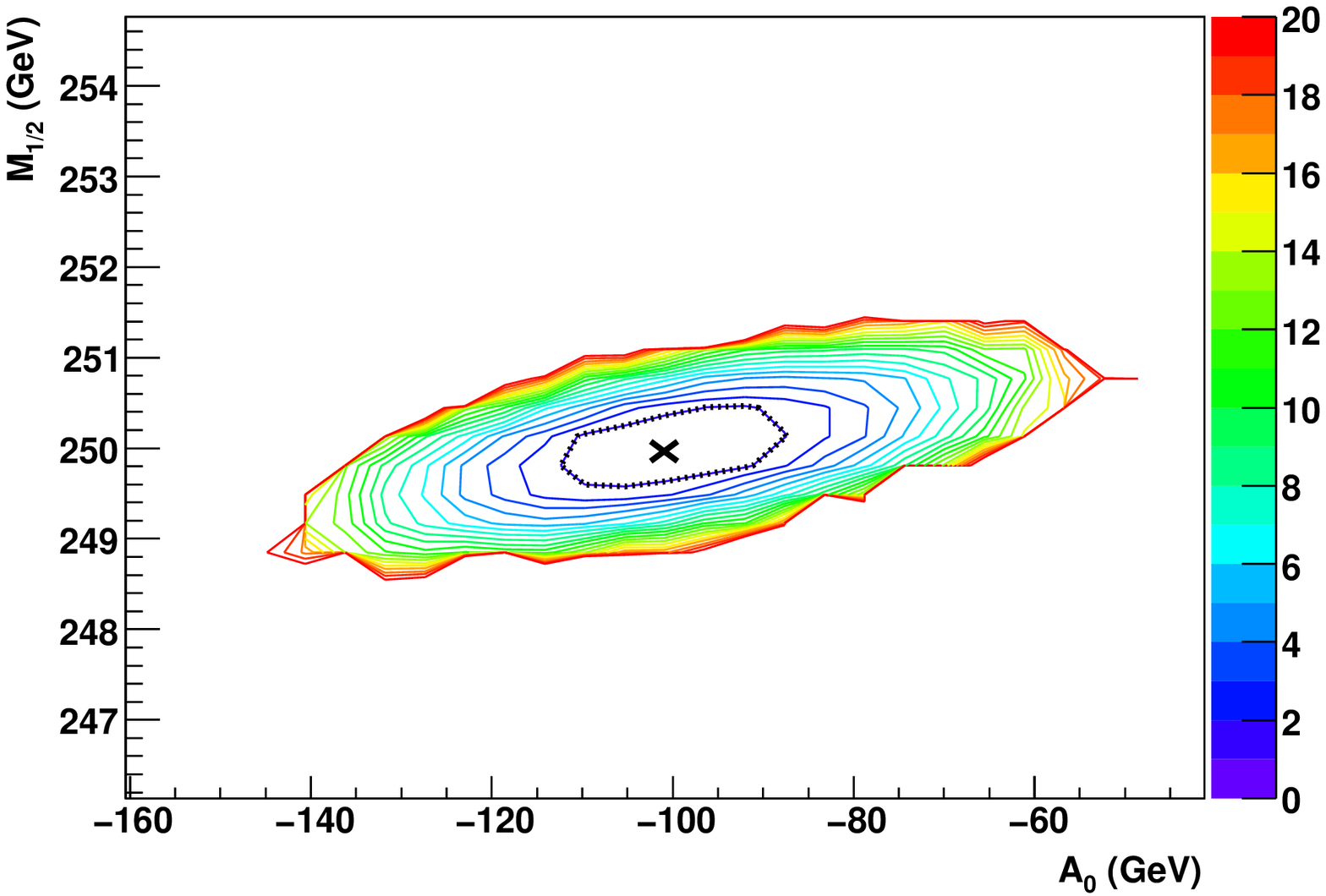}
    \end{center}
    \caption{$\Delta\chi^2 = -2 \ln(\mathcal{L}) + 2
      \ln(\mathcal{L}_{\mathrm{max}})$ contours from Markov Chain for
      the mSUGRA model using observables from
      Tables~\ref{tab:leobserables} and
      \ref{tab:inputs}. $\mathcal{L}$ is the two-dimensional profile
      likelihood and $\mathcal{L}_{\mathrm{max}}$ the global maximum
      of the likelihood. The black dotted contours represent $\Delta
      \chi^2 = 1$ contours. The plots are for integrated LHC
      luminosities of 1~fb$^{-1}$/10~fb$^{-1}$/300~fb$^{-1}$
      (left/middle/right).}
    \label{results:LELHC:mSUGRA:MarkovsFreq}
  \end{figure*}
\end{center}

In addition to Toy Fits we also perform a Markov Chain
analysis. Figure~\ref{results:LELHC:mSUGRA:MarkovsFreq} shows the
quantity $\Delta \chi^2 = -2 \ln(\mathcal{L}) + 2
\ln(\mathcal{L}_{\mathrm{max}})$ for all possible mSUGRA parameter
pairs for the three considered LHC luminosities of
1~fb$^{-1}$/10~fb$^{-1}$/300~fb$^{-1}$ (left/middle/right).
$\mathcal{L}$ is the two-dimensional profile likelihood and
$\mathcal{L}_{\mathrm{max}}$ the global maximum of the likelihood.
The black dotted contours represent $\Delta \chi^2 = 1$ contours. 
The results are in good agreement with those
obtained from the Toy Fits and nicely show the partly strong
correlation between the parameters which is also reflected in
Tables~\ref{tab:results:LE+LHC:mSUGRA:1fb-1:corr},
\ref{tab:results:LE+LHC:mSUGRA:10fb-1:corr} and
\ref{tab:results:LE+LHC:mSUGRA:300fb-1:corr}.

\begin{center}
  \begin{figure*}
    \begin{center}
      \includegraphics[width=0.34\textwidth]{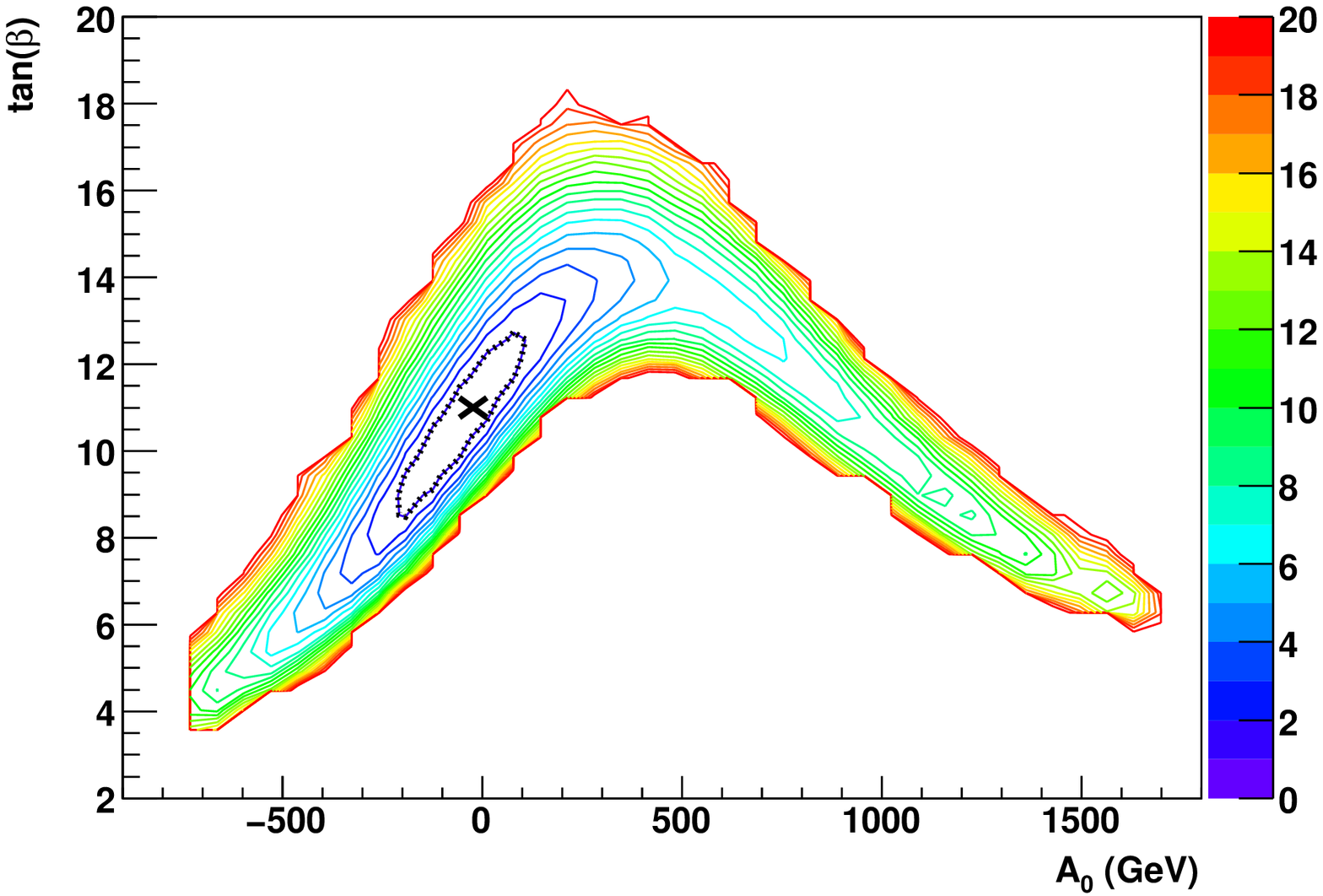}
      \hspace{-0.023\textwidth}
      \includegraphics[width=0.34\textwidth]{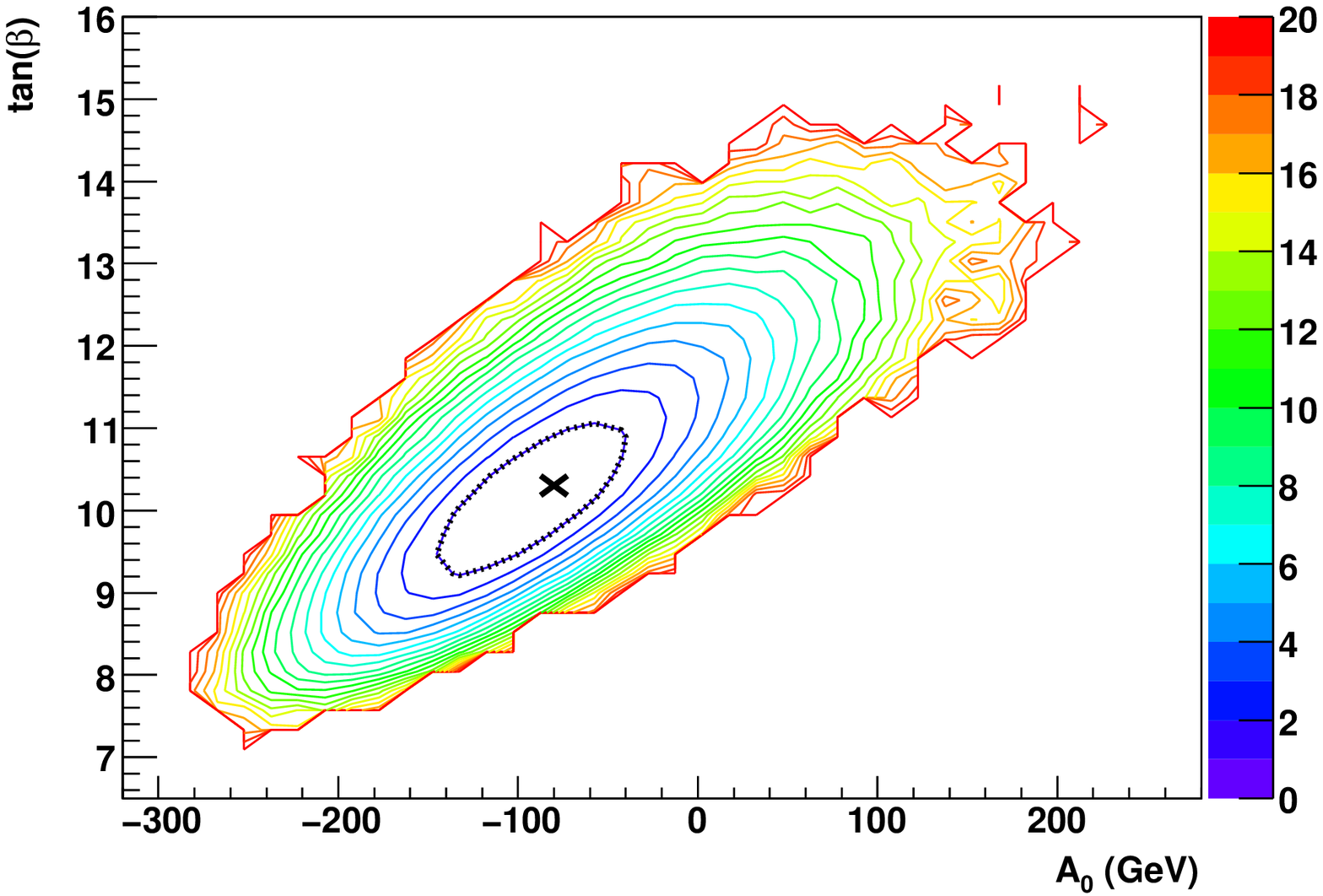}
      \hspace{-0.023\textwidth}
      \includegraphics[width=0.34\textwidth]{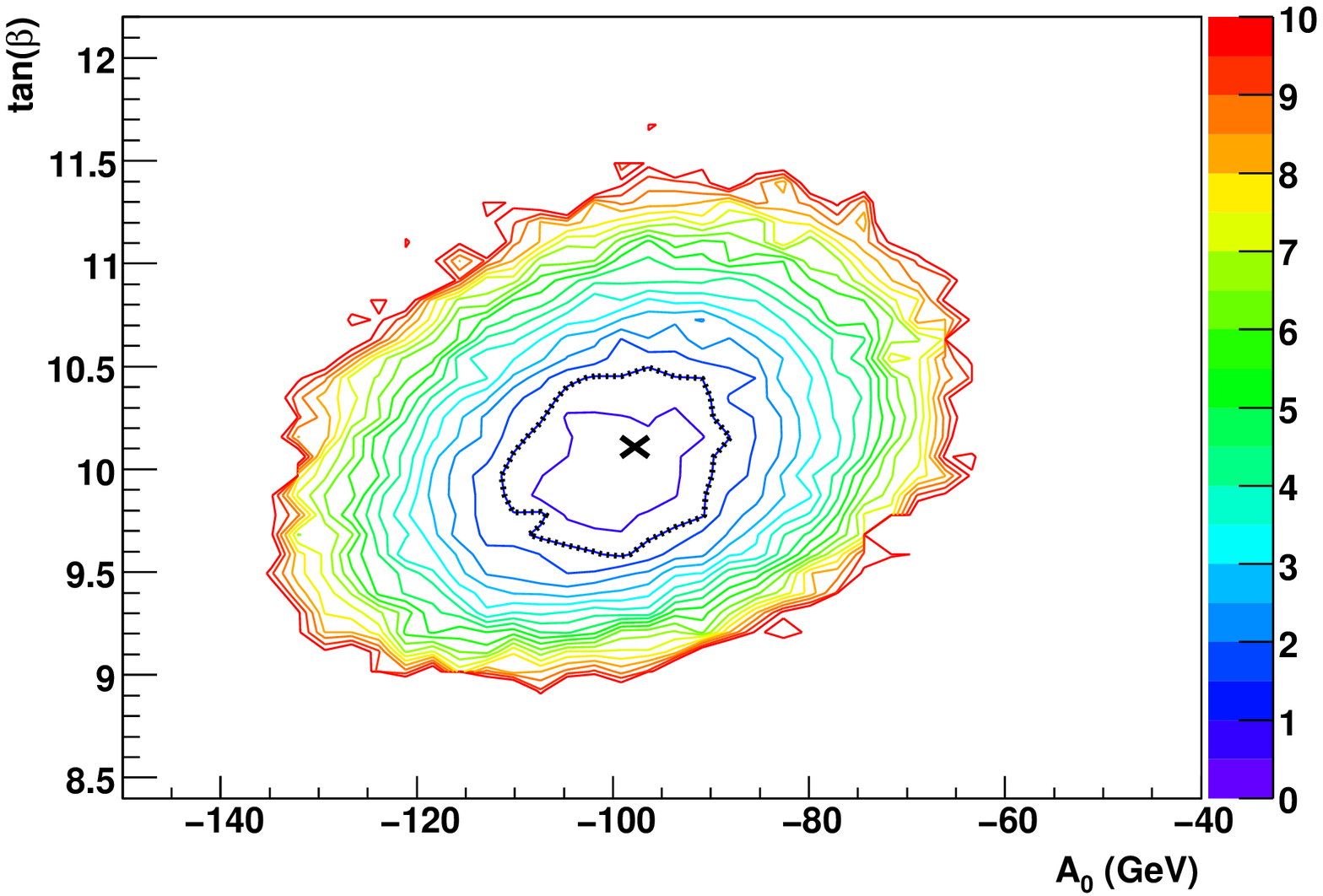}
    \end{center}
    \caption{For comparison with
      Figure~\ref{results:LELHC:mSUGRA:MarkovsFreq}, the outcome of the same
      Markov chain for the parameter pair $A_0$-$\tan \beta$ is shown
      using Bayesian statistics. The lines indicate $\Delta \chi^2 = -2
      \ln(\mathcal{L}) + 2 \ln(\mathcal{L}_{\mathrm{max}})$ contours with
      $\mathcal{L}$ being the marginalised posterior probability (using a
      flat prior probability). The black dotted contours represent $\Delta
      \chi^2 = 1$ contours.}
    \label{results:LELHC:mSUGRA:MarkovsBayes}
  \end{figure*}
\end{center}

For illustrative purposes
Figure~\ref{results:LELHC:mSUGRA:MarkovsBayes} shows the outcome of
the same Markov Chain for the parameter pair $A_0$-$\tan \beta$ using
Bayesian statistics. The lines again indicate $\Delta \chi^2 = -2
\ln(\mathcal{L}) + 2 \ln(\mathcal{L}_{\mathrm{max}})$ contours but
this time $\mathcal{L}$ denotes the marginalised posterior probability
(using a flat prior probability). Compared to the results derived from
the profile likelihood the contour lines are more jagged for the same
Markov Chain length. Apart from these fluctuations good agreement
between the results derived from the marginalised posterior
probability and those from the profile likelihood (shown in
Figure~\ref{results:LELHC:mSUGRA:MarkovsFreq}) is found.

\begin{table}
  \caption{Result of the fit of the mSUGRA model to low energy and LHC
    observables for 1~fb$^{-1}$.}
  \label{tab:results:LE+LHC:mSUGRA:1fb-1}
  \begin{center}
    \begin{tabular}{lrcl}
      \hline\hline
      Parameter     & Best Fit &       & Uncertainty \\
      \hline
      sign$(\mu)$     &  $+$1      &       &             \\
      $\tan\,\beta$ &  10.2    & $\pm$ & 2.3   \\ 
      $A_0$ (GeV)   & $-$76.3    & $\pm$ & 184       \\ 
      $M_0$ (GeV)   & 100.6    & $\pm$ & 3.4   \\
      $M_{1/2}$ (GeV) & 250.2    & $\pm$ & 5.3    \\
      \hline\hline
    \end{tabular}
  \end{center}
\end{table}

\begin{table}
  \caption{Correlation coefficients for the fitted parameters of the
    mSUGRA model to the expected low energy and LHC observables for
    1~fb$^{-1}$.}
  \label{tab:results:LE+LHC:mSUGRA:1fb-1:corr}
  \begin{center}
    \begin{tabular}{lcccc}
      \hline\hline
                     & $\tan\,\beta$  & $A_0$ & $M_0$ & $M_{1/2}$ \\
      \hline
      $\tan\,\beta$  & 1.000    & 0.534  & $-$0.405 & 0.793   \\ 
      $A_0$          & 0.534  & 1.000    & $-$0.184 & 0.493   \\
      $M_0$          & $-$0.405 & $-$0.184 & 1.000    & $-$0.077  \\
      $M_{1/2}$       & 0.793  & 0.493  & $-$0.077 & 1.000     \\
      \hline\hline
     \end{tabular}
  \end{center}
\end{table}

\begin{table}
  \caption{Result of the fit of the mSUGRA model to low energy and LHC
    observables for 10~fb$^{-1}$.}
  \label{tab:results:LE+LHC:mSUGRA:10fb-1}
  \begin{center}
    \begin{tabular}{lrcl}
      \hline\hline
      Parameter     & Best Fit &       & Uncertainty \\
      \hline
      sign$(\mu)$     &  $+$1      &       &             \\
      $\tan\,\beta$ &  10.0    & $\pm$ &  0.79       \\
      $A_0$ (GeV)   & $-$99.1    & $\pm$ & 48.3        \\ % remove events?
      $M_0$ (GeV)   & 100.0    & $\pm$ &  1.9        \\
      $M_{1/2}$ (GeV) & 250.1    & $\pm$ &  1.1        \\
      \hline\hline
    \end{tabular}
  \end{center}
\end{table}

\begin{table}
  \caption{Correlation coefficients for the fitted parameters of the
    mSUGRA model to the expected low energy and LHC observables for
    10~fb$^{-1}$.}
  \label{tab:results:LE+LHC:mSUGRA:10fb-1:corr}
  \begin{center}
    \begin{tabular}{lcccc}
      \hline\hline
                     & $\tan\,\beta$  & $A_0$ & $M_0$ & $M_{1/2}$ \\
      \hline
      $\tan\,\beta$  & 1.000    & 0.805  & $-$0.328 & 0.415  \\ 
      $A_0$          & 0.805  & 1.000    & $-$0.483 & 0.548  \\
      $M_0$          & $-$0.328 & $-$0.483 & 1.000    & 0.241  \\
      $M_{1/2}$       & 0.415  & 0.548  & 0.241  & 1.000    \\
      \hline\hline
     \end{tabular}
  \end{center}
\end{table}

\begin{table}
  \caption{Result of the fit of the mSUGRA model to low energy and LHC
    observables for 300~fb$^{-1}$.}
  \label{tab:results:LE+LHC:mSUGRA:300fb-1}
  \begin{center}
    \begin{tabular}{lrcl}
      \hline\hline
      Parameter     & Best Fit &       & Uncertainty \\
      \hline
      sign$(\mu)$     &  $+$1      &       &             \\
      $\tan\,\beta$ &  10.00    & $\pm$ &  0.36       \\
      $A_0$ (GeV)   & $-$99.1    & $\pm$ & 12.0        \\
      $M_0$ (GeV)   & 100.00   & $\pm$ &  0.39       \\
      $M_{1/2}$ (GeV) & 250.01   & $\pm$ &  0.33       \\
      \hline\hline
    \end{tabular}
  \end{center}
\end{table}

\begin{table}
  \caption{Correlation coefficients for the fitted parameters of the
    mSUGRA model to the expected low energy and LHC observables for
    300~fb$^{-1}$.}
  \label{tab:results:LE+LHC:mSUGRA:300fb-1:corr}
  \begin{center}
    \begin{tabular}{lcccc}
      \hline\hline
                     & $\tan\,\beta$  & $A_0$ & $M_0$ & $M_{1/2}$ \\
      \hline
      $\tan\,\beta$  & 1.000    & 0.356  & 0.178  & 0.134  \\ 
      $A_0$          & 0.356  & 1.000    & $-$0.266 & 0.673  \\
      $M_0$          & 0.178  & $-$0.266 & 1.000    & 0.391  \\
      $M_{1/2}$       & 0.134  & 0.673  & 0.391  & 1.000    \\
      \hline\hline
     \end{tabular}
  \end{center}
\end{table}

\subsubsection{MSSM18}\label{sec:results:LELHC:MSSM18}

So far, we only considered SUSY models with specific assumptions on
the SUSY breaking mechanism, namely mSUGRA and GMSB (for the fits to
low energy measurements).  As shown in
Section~\ref{sec:results:LELHC:mSUGRA} the LHC measurements together
with LE measurements allow to derive tight constraints on the mSUGRA
parameters if sufficient luminosity is accumulated at the LHC. In this
section we investigate if it is possible to relax the strong
constraints imposed on sparticle masses and couplings by the
requirement of a specific breaking scenario. If it will be possible to
measure the parameters of a more general model, like e.~g.~the MSSM18,
at the electro-weak scale, properties of SUSY breaking models could be
investigated in a bottom-up approach~\cite{Blair:2002pg}.

The results of a Markov Chain analysis of the MSSM18 model using LE
and LHC observables with an integrated luminosity of 300~fb$^{-1}$ are
shown in Table~\ref{tab:results:LELHCILC:MSSM18}. Most parameters can
be determined to the level of a few percent, except for third
generation sfermion mass parameters, the trilinear coupling parameters
$X_{\tau} = A_{\tau} - \mu \tan \beta$, $X_{b} = A_{b} - \mu \tan
\beta$ and $X_{t} = A_{t} - \mu \cot \beta$ and the Higgs parameters
$\tan \beta$ and $m_A$. The precision on the Higgs parameters $\tan
\beta$ and $m_A$ suffers from the fact that for the analysed benchmark
point SPS1a the heavy Higgs bosons are not directly accessible at the
LHC.

Whereas for the mSUGRA fit to LE and LHC observables the
impact of LE observables is almost negligible for
300~fb$^{-1}$ of LHC data, the situation for the MSSM18 is different. 
For some parameters the most stringent constraints still come from LE
measurements. The most prominent examples are $\mathcal{B}(B \to
s\gamma)$ and $(g-2)_{\mu}$. $\mathcal{B}(B \to s\gamma)$ is sensitive
to the charged Higgs boson mass $m_{H^{\pm}}$ which in turn is tightly
connected to the fitted SUSY parameter $m_A$. $(g-2)_{\mu}$ provides
the most sensitive constraints on $M_{\tilde{\tau}_L}$ and $X_{\tau}$.

\begin{figure}
  \includegraphics[width=0.49\textwidth]{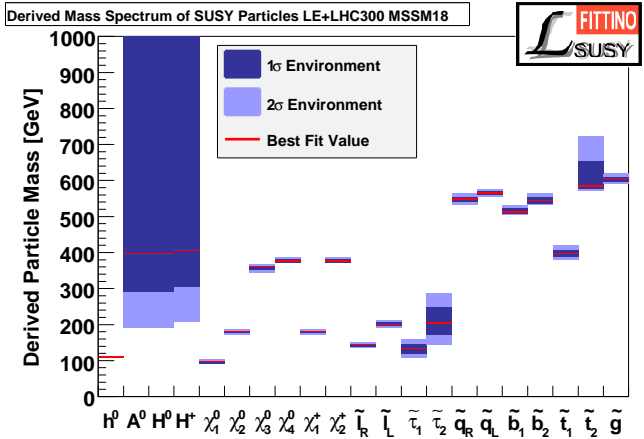}
  \caption{SUSY mass spectrum consistent with the existing low-energy
    measurements from Table~\ref{tab:leobserables} and the expected
    LHC measurements from Table~\ref{tab:inputs} at ${\cal
      L}^{\mathrm{int}}=300\,\mathrm{fb}^{-1}$ for the MSSM18 model.
    The uncertainty ranges represent model dependent uncertainties
    of the sparticle masses and not direct mass measurements.}
  \label{fig:results:LELHC:MSSM18:massDist300fb}
\end{figure}

The obtained MSSM18 fit result can again be translated into a
corresponding sparticle mass spectrum. This spectrum is presented in
Figure~\ref{fig:results:LELHC:MSSM18:massDist300fb}. Again the masses
in this Figure are model dependent predictions and do not represent
direct mass measurements. Compared to the corresponding result for the
more constrained mSUGRA model
(Figure~\ref{results:LELHC:mSUGRA:massDist300fb}), some of the
sparticle masses have significantly larger uncertainties in the
MSSM18.  This is particularly pronounced for the heavy Higgs boson
masses, which are -- as stated above -- not directly accessible at the
LHC for the considered SUSY benchmark point.

Although not studied explicitly for the MSSM18, one may expect that chain
ambiguities may have a larger impact for this model than in the mSUGRA
case. Since MSSM18 has more independently adjustable parameters,
different decay chain interpretations can be more easily
matched with the model due to the increased flexibility.

\subsection{Low-Energy Observables, LHC and Expectations for ILC}\label{sec:results:LELHCILC}

The results of the previous sections show that the expected data of
the LHC allow to obtain rather precise constraints on the mSUGRA
parameters once sufficient luminosity is accumulated. However, for the
MSSM18 scenario, the constraints severely diminish due to the
increased theoretical freedom. The parameter uncertainties typically
increase by a factor of 10 or more. Therefore an extrapolation of the
SUSY parameters from the electro-weak scale to the GUT scale is
afflicted with large uncertainties, if only low-energy, flavour
physics, electro-weak precision, cosmological and LHC observables are
used.

The expected measurements at the International Linear Collider,
however, could dramatically increase the experimental precision of the
measurements of sparticle masses and couplings. In addition to just
increasing the experimental precision, the ILC is also expected to
deliver a wealth of measurements of absolute branching fractions and
cross-sections, many cross-section times branching fraction
measurements, and many model-independent measurements of quantum
numbers and CP-properties. This expected wealth of data, especially
in a SUSY scenario with a rich phenomenology below a mass scale of
500~GeV, as predicted by the present measurements in
Section~\ref{sec:results:LEonly}, will strongly enhance the knowledge
from the LHC due to the expected complementarity of ILC and LHC
results~\cite{Weiglein:2004hn}. 

In this section, first the expected precision on the parameters of the
mSUGRA model is studied, followed by a detailed comparison of the
results of the MSSM18 fit using only LE and LHC data with those
obtained using LE, LHC and ILC data. Finally, the increase in
precision is used to predict the cosmic cold dark matter relic density
$\Omega_{\mathrm{CDM}}h^2$ from collider data, from fits excluding
$\Omega_{\mathrm{CDM}}h^2$ itself from the list of observables.

\subsubsection{mSUGRA}\label{sec:results:LELHCILC:mSUGRA}

\begin{table}
  \caption{Result of the fit of the mSUGRA model to the existing
    measurements and to the expected results from LHC with ${\cal
      L}^{\mathrm{int}}=300\,\mathrm{fb}^{-1}$ and ILC.}
  \label{tab:results:LELHCILC:mSUGRA}
   \begin{center}
      \begin{tabular}{lrrcl}
        \hline\hline
        Parameter & Nominal value & Fit & & Uncertainty \\
        \hline
        $\tan\beta$       &  10   &  9.999   & $\pm$ &  0.050   \\
        $M_{1/2}$ (GeV)          & 250   &  249.999 & $\pm$ &  0.076  \\
        $M_0$ (GeV)            & 100   &  100.003 & $\pm$ &  0.064   \\
        $A_0$ (GeV)            & $-$100  & $-$100.0   & $\pm$ &  2.4   \\
        \hline\hline
      \end{tabular}
    \end{center}
\end{table}

Using the same available and expected measurements as in the fit using
${\cal L}^{\mathrm{int}}=300\,\mathrm{fb}^{-1}$ of LHC luminosity in
Section~\ref{sec:results:LELHC:mSUGRA}, plus the expected ILC
measurements discussed in Section~\ref{sec:ilcobservables}, the fit of
the mSUGRA model to the data of the SPS1a scenario is shown in
Table~\ref{tab:results:LELHCILC:mSUGRA}. The comparison with the
results without ILC in Table~\ref{tab:results:LE+LHC:mSUGRA:300fb-1}
shows the increase in precision by a factor of 5 to 10. However, the
pure increase in precision for the fit of a high scale scenario is not
the only improvement using ILC. First, possible deviations of the SUSY
breaking implemented in Nature from a given GUT-scale SUSY breaking
scenario, involving assumptions on unification, are much more visible
using also ILC data. Second, the high accuracy and especially the
larger variety (covering couplings, mixings, masses, widths and
quantum numbers) and stronger model independence of the measurements
allow to fit more general models of New Physics. This makes it
possible to study the SUSY breaking mechanism using a bottom-up
instead of a top-down approach.

\subsubsection{MSSM18}\label{sec:results:LELHCILC:MSSM18}

As discussed in Section~\ref{sec:results:LELHC:MSSM18}, the fit of the
MSSM parameters at the SUSY breaking scale allows a bottom-up test of
SUSY breaking and is independent of any assumptions about physics at
the GUT scale. Section~\ref{sec:results:LELHC:MSSM18} showed that for
the MSSM18 model, the parameter uncertainties from fits to existing
data and expected LHC data are larger by at least one order of
magnitude with respect to the fits of the mSUGRA scenario.

\begin{table*}
  \caption{Results of the Markov Chain MC analysis of the MSSM18 model
    using low energy observables, expected LHC results for ${\cal
      L}^{\mathrm{int}}=300\,\mathrm{fb}^{-1}$ and ILC.}
  \label{tab:results:LELHCILC:MSSM18}
  \begin{center}
    \begin{tabular}{lrrcll}
      \hline\hline
      Parameter & Nominal value & ILC Fit & & $\sigma_{\mathrm{LE+LHC\,300}}$ & $\sigma_{\mathrm{LE+LHC300+ILC}}$ \\
      \hline
      $M_{\tilde{\ell}_L}$ (GeV)    &     194.31 &  194.315 & $\pm$ & 6.4                & 0.068   \\
      $M_{\tilde{\ell}_R}$ (GeV)    &     135.76 &  135.758 & $\pm$ & 10.5               & 0.071   \\
      $M_{\tilde{\tau}_L} $ (GeV) &   193.52 &  193.46  & $\pm$ & 43.0               & 0.33    \\
      $M_{\tilde{\tau}_R} $ (GeV) &   133.43 &  133.45  & $\pm$ & 38.2               & 0.35    \\
      $M_{\tilde{q}_L}    $ (GeV) &   527.57 &  527.61  & $\pm$ & 3.4                & 0.64    \\
      $M_{\tilde{q}_R}    $ (GeV) &     509.14 &  509.3   & $\pm$ & 9.0                & 9.0     \\
      $M_{\tilde{b}_R}    $ (GeV) &     504.01 &  504.2   & $\pm$ & 33.3               & 2.4     \\
      $M_{\tilde{t}_L}    $ (GeV) &     481.69 &  481.6   & $\pm$ & 15.5               & 1.5     \\
      $M_{\tilde{t}_R}    $ (GeV) &     409.12 &  409.2   & $\pm$ & 103.8              & 1.6     \\
      $\tan\beta          $ &    10      &  10.01 & $\pm$ & 3.3                & 0.29      \\
      $\mu                $ (GeV) &     355.05 &  355.02  & $\pm$ & 6.2                & 0.88    \\
      $X_{\tau}           $ (GeV) & $-$3799.88   & $-$3795.1  & $\pm$ & 3053.5             & 46.6     \\
      $X_{t}              $ (GeV) &  $-$526.62   & $-$526.8   & $\pm$ & 299.2              & 4.7     \\
      $X_{b}              $ (GeV) & $-$4314.33   & $-$4252.1  & $\pm$ & 5393.6             & 728.7     \\
      $M_1                $ (GeV) &     103.15 &  103.154 & $\pm$ & 3.5                & 0.046   \\
      $M_2                $ (GeV) &     192.95 &  192.95  & $\pm$ & 5.5                & 0.11    \\
      $M_3                $ (GeV) &     568.87 &  568.66  & $\pm$ & 6.9                & 1.65     \\
      $m_{A}            $ (GeV) &   359.63   &  360.07  & $\pm$ & $_{-99.3}^{+1181}$ & 1.83     \\
      \hline\hline
    \end{tabular}
  \end{center}
\end{table*}

Table~\ref{tab:results:LELHCILC:MSSM18} shows a comparison of the
parameter uncertainties of the fits of the MSSM18 model using LE data
in combination with ${\cal L}^{\mathrm{int}}=300\,\mathrm{fb}^{-1}$ of
data at the LHC (LE+LHC300) and the latter plus the expected ILC
results (LE+LHC300+ILC). The results of Markov Chain Monte Carlo scans
and Toy Fits are in good agreement, therefore just the Markov Chain
result is shown. For most parameters, the uncertainties decrease by
approximately one order of magnitude. Interestingly, the increase in
precision is not only limited to those parameters which are linked
directly to observables at tree level. For example it is expected that
the uncertainties of the gaugino mass parameters $M_1$ and $M_2$ are
significantly decreased at ILC due to the increased precision on the
$\tilde{\chi}^0_{1/2}$ and $\tilde{\chi}^{\pm}_1$ masses and the
additional information from precise measurements of cross-sections
times branching fractions for different polarisations. Also, the
precision of the heavy Higgs sector parameter $m_{A}$ is expected to
increase dramatically, since the heavy Higgs bosons $A$, $H$ and
$H^{\pm}$ are not expected to be discovered at the LHC in this
scenario, but to be precisely measured at the
ILC~\cite{Weiglein:2004hn}. In contrast to those measurements, no
additional experimental information is obtained on the gluino mass or
the heavier squark masses at ILC. In any case, with the exception of
$M_{\tilde{q}_R}$, all parameter uncertainties improve
dramatically. The reason for this behaviour is the strong decrease of
correlations. For example, the $\tilde{b}_{1/2}$ masses are determined
by $M_{\tilde{q}_L}$ and $M_{\tilde{b}_R}$, but also by the
off-diagonal elements $m_bX_b$ with $X_b=A_b-\mu \tan\beta$. Due to
the strong increase in the determination of $\mu$ and $\tan\beta$ from
the measurements in the Higgs sector (where also $A_b$ plays a role in
loop effects) and the gaugino sector, also the precision of the
parameter $M_{\tilde{b}_R}$ is strongly improved, although no direct
measurement in the sbottom sector is made at the ILC in this
scenario. This example highlights the importance of precision
measurements for the detailed unravelling of the SUSY spectrum, and it
is an example of the complementarity of LHC and ILC.

\begin{figure}
  \includegraphics[width=0.49\textwidth]{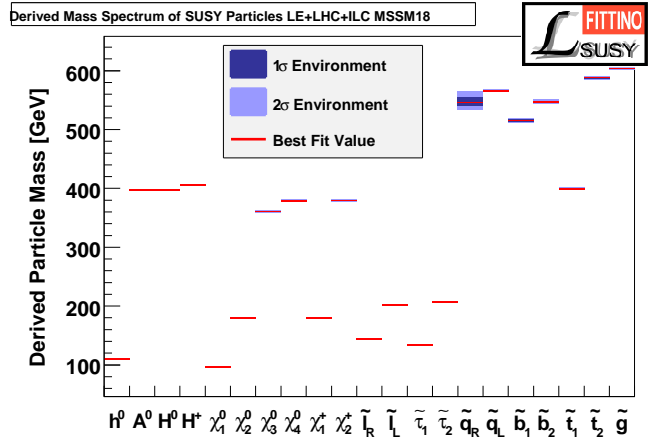}
  \caption{Derived mass distributions of the SUSY particles using existing measurements, expected 
    results from LHC with ${\cal L}^{\mathrm{int}}=300\,\mathrm{fb}^{-1}$ and expected results from ILC.}
  \label{fig:results:LELHCILC:MSSM18:massDist}
\end{figure}

The resulting derived spectrum of sparticle masses is shown in
Figure~\ref{fig:results:LELHCILC:MSSM18:massDist}. It represents a very
strong improvement over the results without ILC in
Figure~\ref{fig:results:LELHC:MSSM18:massDist300fb}. The Higgs sector
exhibits the strongest improvement due to the direct observation of
heavy Higgs states. Apart from the squark mass $m_{\tilde{q}_L}$,
solely governed by the parameter $M_{\tilde{q}_R}$, the uncertainties
of all other derived masses increase dramatically. 

\begin{figure}
  \includegraphics[width=0.49\textwidth]{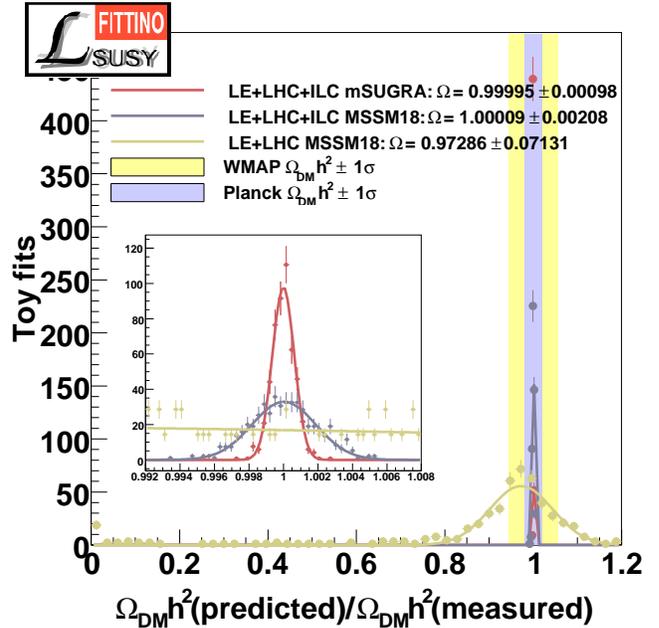}
  \caption{Ratio of the predicted value of $\Omega_{\mathrm{pred}}
    h^2$ to the nominal value of $\Omega_{\mathrm{SPS1a}} h^2$ in the
    SPS1a scenario for a variety of Toy Fits without using $\Omega_{\mathrm{CDM}}
    h^2$ as an observable.}
  \label{fig:results:LELHCILC:MSSM18:OmegaDists}
\end{figure}

As a final test of the agreement between cosmology and collider data,
and as a showcase for the predictive power of precision collider
measurements, additional fits without $\Omega_{\mathrm{CDM}}h^2$ are
performed with and without the use of ILC using the Toy Fit
technique. The resulting predicted values of
$\Omega_{\mathrm{CDM}}h^2$ are shown in
Figure~\ref{fig:results:LELHCILC:MSSM18:OmegaDists} and compared with
the present and expected experimental precision of
$\Omega_{\mathrm{CDM}}h^2$ from the WMAP~\cite{Dunkley:2008ie} and
Planck~\cite{Feng:2005uu} data. The prediction of
$\Omega_{\mathrm{CDM}}h^2$ from collider data without ILC in the
MSSM18 model shows a long non-Gaussian tail down to
$\Omega_{\mathrm{CDM}}h^2=0$. The Gaussian core of the distribution is
one order of magnitude wider than the expected precision from the
Planck satellite. Therefore witout ILC, the relic density constraints
inferable from particle physics within the MSSM18 model do not match
the precision of cosmological measurements.

In contrast to that, the result including ILC for the MSSM18 scenario
achieves a relative precision on
$(\Omega_{\mathrm{CDM}}^{\mathrm{pred}}h^2)/(\Omega_{\mathrm{CDM}}^{\mathrm{meas}}h^2)$
of $0.2\,\%$, which is an order of magnitude more precise than the
expected Planck accuracy. An agreement between the collider result and
the cosmological measurement would provide strong hints that SUSY LSPs
make up the vast majority of dark matter and would allow to make
predictions for direct dark matter search experiments. For
comparison, the achievable accuracy on the relic density is also shown
assuming mSUGRA. The uncertainty is improved again by a factor of two.

In summary, for a SUSY scenario in agreement with the present
cosmological, low-energy and collider data, the ILC would tremendously
improve the theoretical understanding of a SUSY model by improving the
precision of bottom-up determinations of SUSY parameters without
assumptions on unification and breaking mechanisms at the GUT
scale. The precision would ensure that cosmological implications of
New Physics could be predicted with a precision significantly better
than the current and expected cosmological measurements.

%%%%%%%%%%%%%%%%%%%%%%%%%%%%%%%%%%%%%%%%%%%%%%%%%%%%%%%%%%%%%%%%%%%%%%%%%%%%%%%
%%%                               Conclusions                               %%%
%%%%%%%%%%%%%%%%%%%%%%%%%%%%%%%%%%%%%%%%%%%%%%%%%%%%%%%%%%%%%%%%%%%%%%%%%%%%%%%

\section{Conclusions}\label{sec:conclusions}

We have performed a comprehensive study of current and future
uncertainties and correlations of the parameters of supersymmetric
models, i.~e.~the mSUGRA and GMSB model as well as the MSSM18.

For the case of LE data presently available we confirm the results
of~\cite{Buchmueller:2008qe} leading to the conclusion that within the
mSUGRA model, sparticles are predicted to be light enough for an
early discovery at the LHC. In particular, the squark and gluino
masses, which determine the major production cross-sections at the LHC
are below 1~TeV at 68\,\%~CL and below 1.6~TeV at 95\,\%~CL. The most
sensitive measurements are the muon anomalous magnetic moment
$(g-2)_\mu$ and the cold dark matter density
$\Omega_{\mathrm{CDM}}h^2$. For $(g-2)_\mu$ the results rely on the
calculation of the hadronic vacuum corrections based on $e^+e^-$
cross-section data.  Sparticle masses are less constrained for
scenarios where the SM prediction of $(g-2)_\mu$ is closer to its
measured value. This is currently the case for the prediction based on
$\tau$-data for the hadronic vacuum corrections~\cite{Davier:2009ag},
where the heaviest sparticles are expected below 1.4~TeV at 68\,\%~CL
and 3~TeV at 95\,\%~CL.  With no deviation of $(g-2)_\mu$, sparticles
are still constrained to lie below approximately 2 (3.5)~TeV at 68
(95)~\%~CL. A good fit of the data (excluding
$\Omega_{\mathrm{CDM}}h^2$) can also be achieved within GMSB yielding
sparticle masses approximately below 1.2 (2.0)~TeV at 68 (95)~\%~CL.
This result shows that the feature of light sparticles is not
exclusively true within mSUGRA, although it may not be true within the
general MSSM.  Furthermore the LE data and the value of
$\Omega_{\mathrm{CDM}}h^2$ in particular point towards a small mass
difference of the LSP and the NLSP which is $\tilde{\tau}_1$. The mass
of the lightest Higgs boson is predicted to be just above the
exclusion of the LEP experiments.

For the SPS1a parameter point, which provides a phenomenology rather
similar to the region preferred by the LE fit we determined the
prospects for parameter measurements at the LHC for a complete set of
experimentally accessible and well-studied observables. Within mSUGRA,
a coarse determination of the parameters can already be achieved with
an integrated luminosity of 1 fb$^{-1}$. The precision can be
significantly improved when LE data are combined with the early LHC
measurements. For 300 fb$^{-1}$, a precision of better than 1~\% can
be achieved on $M_0$ and $M_{1/2}$. The parameter $\tan\beta$ ($A_0$)
can be determined to 3.5 (11)~\% precision. At high luminosity the
impact of LE data becomes small.

For the MSSM18, parameter determination is significantly more
difficult and requires larger integrated luminosity. Nevertheless, for
300 fb$^{-1}$ a decent determination of all sparticle masses can
be achieved to a few percent precision with the exception of
$\tilde{t}_2$, $\tilde{\tau}_2$ and the heavy Higgs bosons. Here the
inclusion of LE data still has significant impact, in particular in
constraining third generation sparticles. With a linear collider
like the ILC operating at up to 1~TeV, the MSSM18 can be reconstructed
with a precision increased by approximately one order of magnitude.

It should be noted that the bulk region of the MSSM as exemplified in
the SPS1a parameter point is certainly favourable for the prospects of
parameter measurements at both the LHC and the ILC.  For the LHC,
various experimental studies of different parameter points exist,
however no full analysis of more difficult regions exists to date.  In
particular in regions where long decay chains with charged leptons are
suppressed, the reconstruction of SUSY parameters will be
substantially more difficult and imprecise. However, given the
constraints from LE data, such scenarios appear less likely. Given the
smaller mass difference between sleptons and the lightest neutralino
observed in the fit to the LE data when compared with SPS1a, a more
detailed comprehensive experimental study of co-annihilation points at
LHC would be beneficial.

In addition to these quantitative results we proposed some new
methodological approaches to take care of ambiguities in the
assignment of experimental observables to physical final states. It
was shown, that in some cases these ambiguities may be translated into
the uncertainty on the parameters when the ambiguities cannot be
resolved statistically.  We have also shown that the Bayesian and
Frequentist interpretation of Markov Chain Monte Carlo lead to very
similar results for fits including LHC data when flat priors are used
in the Bayesian approach. For fits of LE data only, however, the two
interpretations do not necessarily agree. It is observed that the
Bayesian approach, which includes marginalisation of the hidden
parameters requires a prohibitive amount of computing power. In such
cases, only the Frequentist interpretation is exploited.

In the future, the technologies presented in this paper will be
applied to a larger variety of models and finally to real data from
the LHC. In addition, the proposed treatment of the assignment
ambiguities will be extended to further possible self-consistent
interpretations of the data and the resulting effect on parameter
uncertainties and possible exclusions of assignments will be
evaluated. It also is expected to be important to evaluate the effect
of theoretical uncertainties stemming e.~g.~from missing higher order
effects and differences between different implementations of RGE
running in more detail. In addition to the uncertainties itself, the
evaluation of correlations among theoretical uncertainties could be
relevant.  Finally, if SUSY is realised in Nature, sparticles could be
discovered before the discovery of a SUSY Higgs boson. Therefore, the
implementation of present and future limits on Higgs boson production
in arbitrary models of New Physics, using~\cite{Bechtle:2008jh} could
be important.

% The computer calculations necessary to obtain the results of this
% paper led to an estimated CO$_2$ emission of 17~tons. This clearly
% shows the need for more efficient algorithms, more power-effective
% CPUs, and less CO$_2$-intensive generation of electrical power.

%%%%%%%%%%%%%%%%%%%%%%%%%%%%%%%%%%%%%%%%%%%%%%%%%%%%%%%%%%%%%%%%%%%%%%%%%%%%%%%
%%%                            Acknowledgements                             %%%
%%%%%%%%%%%%%%%%%%%%%%%%%%%%%%%%%%%%%%%%%%%%%%%%%%%%%%%%%%%%%%%%%%%%%%%%%%%%%%%

\begin{acknowledgement}
{\large\bfseries\sffamily Acknowledgements\\[0.5ex]}
The authors are grateful to Werner Porod for his sedulous support with
SPheno and to Oliver Buchm\"uller, Fr\'ed\'eric Ronga, Georg Weiglein
and Sven Heinemeyer for providing a working copy of their program
``Mastercode'' and for helpful discussions. In addition we acknowledge
helpful discussions with Olaf Behnke, Claus Kleinwort and Stefan
Schmitt from the ``Physics at the Terascale'' Helmholtz Alliance
statistics group and Glen Cowan on statistical issues. Concerning
predictions of the cold dark matter relic density from the code
Micromegas, we are thankful for discussions with Genevi\`eve
B\'elanger, Gilbert Moultaka and Sean Bailly. We thank Janet Dietrich
for the preparation of the overlay with the ATLAS discovery potential
plot.

In addition we would like to thank the Helmholtz Alliance ``Physics at
the Terascale'' and the BMBF for providing reliable computing
resources at the National Analysis Facility at DESY.

This work has been supported by the Helmholtz Association under the
fund VH-NG-303 and by the Collaborative Research Grant SFB 676 of the DFG.

\end{acknowledgement}

%%%%%%%%%%%%%%%%%%%%%%%%%%%%%%%%%%%%%%%%%%%%%%%%%%%%%%%%%%%%%%%%%%%%%%%%%%%%%%%
%%%                              Bibliography                               %%%
%%%%%%%%%%%%%%%%%%%%%%%%%%%%%%%%%%%%%%%%%%%%%%%%%%%%%%%%%%%%%%%%%%%%%%%%%%%%%%%

\end{document}